\documentclass[12pt,letterpaper,oneside,openany]{book}
\pdfoutput=1

\usepackage[top=1in, bottom=1in, left=1in, right=1in]{geometry}

\usepackage{sectsty}
\chapterfont{\centering} 
\usepackage{setspace}
\usepackage{appendix}

\usepackage{xcolor}

\usepackage[titles]{tocloft}
\newlength\oldchwidth
\newlength\newchwidth
\newcommand{\labelchapters}{
    \renewcommand{\cftchappresnum}{Chapter~}
    \renewcommand{\cftchapaftersnum}{:~}
    \setlength{\cftchapnumwidth}{\oldchwidth}
    \addtolength\cftchapnumwidth\newchwidth
}

\newcommand{\listappendicesname}{List of Appendices}
\newlistof{appendices}{loa}{\listappendicesname}
\newlength\newappwidth
\usepackage{etoolbox} 

\usepackage{graphicx}
\usepackage{amsmath,amssymb,mathtools}
\usepackage{amsfonts}
\usepackage{physics}
\usepackage{amsthm}
\usepackage{multirow}
\usepackage{cite}

\newcommand{\bea}{\begin{eqnarray}}
\newcommand{\eea}{\end{eqnarray}}
\newcommand{\nn}{\nonumber}
\newcommand{\la}{\label}
\newcommand{\be}{\begin{equation}}
\newcommand{\ee}{\end{equation}}
\newcommand{\ep}{\epsilon}
\newcommand{\cN}{\mathcal{N}}
\newcommand{\cJ}{\mathcal{J}}

\makeatletter
\newcommand*{\rom}[1]{\expandafter\@slowromancap\romannumeral #1@}
\makeatother

\begin{document}

\frontmatter

\pagenumbering{gobble} 

\newgeometry{top=2.5in}
\thispagestyle{empty}

{\begin{center}
\begin{doublespace}
{\LARGE \bf Quantum Anatomy of Supersymmetric Black Holes in AdS Spacetimes}\\[.5cm] 
by\\
\hphantom{} \\
Siyul Lee\\[1cm] 
\end{doublespace}
\begin{singlespace}
A dissertation submitted in partial fulfillment\\
of the requirements for the degree of\\
Doctor of Philosophy\\
(Physics)\\ 
in The University of Michigan\\
2024 
\end{singlespace}
\end{center}

\vspace*{5cm}
\begin{singlespace}
\noindent Doctoral Committee:\\[4mm] 
\hspace*{1cm} Professor Finn Larsen, Chair\\
\hspace*{1cm} Professor Lydia Bieri\\
\hspace*{1cm} Professor James T. Liu\\
\hspace*{1cm} Professor Leopoldo A. Pando Zayas\\
\hspace*{1cm} Professor Jianming Qian
\end{singlespace}}
\restoregeometry

\clearpage


\newgeometry{top=4.2in}

{\begin{center} Siyul Lee\\\vspace{1em}siyullee@umich.edu\\\vspace{1em}ORCID iD: 0000-0001-9764-6809\\\vspace{3.5em}\textcopyright~Siyul Lee 2024
\end{center}}

\restoregeometry

\clearpage

\pagenumbering{roman} 
\setcounter{page}{2}  


\phantomsection
\addcontentsline{toc}{chapter}{Acknowledgements}
\newgeometry{top=2in, left=1in, right=1in}

\begin{center}
\textbf{\Huge Acknowledgements}\\[1.8cm]
\end{center}

I would like to express my heartfelt gratitude toward my advisor, Finn Larsen.
He has taught me to become a physicist from a physics major.
I can truly understand why the advisor is often metaphorized as the academic father.

I am grateful to Jim Liu and Leopoldo Pando Zayas,
for serving both in my preliminary and dissertation committees
to see me grow up and for teaching me in between.
I also thank Lydia Bieri and Jianming Qian for serving in my dissertation committee.

I would like to thank everyone in the high energy theory group
at the University of Michigan.
In particular, I thank Henriette Elvang and Ratindranath Akhoury
for their enthusiastic teaching and advising.
I also thank Karen O'Donovan, Aaron Pierce and James Wells,
former and current postdocs Prudhvi Bhattiprolu, Jose de la Cruz Moreno,
Nick Geiser, Jun Nian and Christoph Uhlemann,
and former and current fellow students Leia Barrowes, Justin Berman, Alan Chen, Marina David, Evan Deddo, Shaghayegh Emami, Nizar Ezroura, Aidan Herderschee, Junho Hong, Sabare Jayaprakash, Callum Jones, Loki Lin, Brian McPeak, Shruti Paranjape, Evan Petrosky, Robert Saskowski and Ben Sheff.

Across oceans, I thank my fantastic collaborators
Jaehyeok Choi, Sangmin Choi, Sunjin Choi,
Eunwoo Lee, Jehyun Lee, Sungjay Lee and Jaemo Park.
I am especially grateful to Professor Seok Kim,
for being a trustworthy mentor for a long time.

I thank good friends in Ann Arbor who have made my life more fruitful,
Yire Ahn, Jungpyo Hong, Kihyuk Hong, Seungug Jae, Seungjun Ki
and Chulwon Lee, among others.

I acknowledge financial support from Leinweber Center for Theoretical Physics,
Rackham Conference Travel Grants and the Rackham Predoctoral Fellowship.
I also acknowledge that a part of this dissertation was supported
through computational resources and services provided by
Advanced Research Computing,
a division of Information and Technology Services
at the University of Michigan, Ann Arbor.

I would like to take this opportunity to thank my family once more.
Without their support, I probably would have diverted to a less fancy job.

Last but most, I thank Seohyeon Byeon. For everything.

\restoregeometry
\clearpage


\phantomsection
\labelchapters 
\tableofcontents
\clearpage

\phantomsection
\addcontentsline{toc}{chapter}{List of Tables}
\listoftables
\clearpage

\phantomsection
\addcontentsline{toc}{chapter}{List of Figures}
\listoffigures
\clearpage


\phantomsection
\addcontentsline{toc}{chapter}{Abstract}
\newgeometry{top=2in, left=1in, right=1in}

\begin{center}
	\textbf{\Huge Abstract}\\[2cm]
\end{center}

The entropy of the universe might decrease if black holes did not have entropy.
Hawking's derivation of black hole temperature and the
first law of thermodynamics suggest that black holes indeed have entropy.
However, Einstein's classical gravity does not allow black holes to have
internal degrees of freedom that entropy implies.
Thus it is the central mission of quantum gravity to uncover the many
quantum microstates of black holes.
Via the AdS/CFT correspondence, black holes in Anti-de-Sitter (AdS)
spacetimes are dual to ensembles of quantum states
in conformal field theories (CFTs).
Recently, the number of supersymmetric quantum states in CFTs
has been counted to successfully account for the
supersymmetric AdS black hole entropy.
We take a step forward by studying properties of such
supersymmetric quantum states dual to black holes.
First, a supersymmetric AdS black hole may exist only if
its charges obey a certain relation.
We reproduce the same relation from a certain ensemble of the
supersymmetric states in CFT.
This gives a heuristic derivation of the supersymmetric black hole
charge constraint for AdS black holes in $3$, $4$, $5$, and $7$ dimensions
from the respective microscopic theories.
Second, we find explicit expressions for the black hole cohomologies
in the weakly coupled 4d maximal Super-Yang-Mills theory
with gauge groups $SU(2)$ and $SU(3)$.
These are connected to the actual microstates
of quantum black holes in the 5-dimensional AdS spacetime.

\restoregeometry 
\clearpage


\mainmatter


\chapter{Introduction}

It took around a century from the beginning of quantum physics
to complete the standard model \cite{ATLAS:2012yve,CMS:2012qbp},
that provides a fairly accurate theoretical framework for
electromagnetic, weak and strong interactions.
Many valuable questions remain, including the dark sectors,
neutrinos, strongly coupled QCD, precision of the electroweak sector,
baryogenesis and naturalness to name a few,
but a major goal of fundamental understanding of Nature
has moved towards quantum gravity in the last several decades.

Unfortunately, Einstein's theory of general relativity is not directly
compatible with the framework of quantum field theory.
Gravity is not renormalizable, meaning that nonsensical divergences
that occur at very high energies cannot be separated from
physics of ordinary energy scale that we experience.
Superstring theory has been developed and established itself
as an improved and plausible theory of quantum gravity,
but it also poses many difficulties, both conceptually and technically.

The AdS/CFT correspondence \cite{Maldacena:1997re} opened
a new window towards understanding gravity.
Of many versions, it states that the quantum gravity theory in the
Anti-de-Sitter space of $d+1$ dimensions is `dual' to a
conformal field theory without gravity in $d$ dimensions.
The Anti-de-Sitter space, AdS in short,
is the maximally symmetric space with constant negative curvature.
It can be considered the vacuum spacetime for a negative cosmological constant. 
The AdS/CFT correspondence is not proven, but is widely accepted
throughout the physics community.
In fact, it is hard to imagine how one would prove it without
completely understanding theories on both sides.
We refer to \cite{Aharony:1999ti,Natsuume:2014sfa,Nastase_2015,Ammon:2015wua}
for reviews of the subject.

One important arena of the AdS/CFT correspondence pertains to
the black hole solutions of the gravity theories
and the conformal field theories with finite temperature.
It has been known that a black hole is a thermal system that has
temperature \cite{Hawking:1975vcx},
so naturally it is dual to a finite temperature `solution' of the
conformal field theory.
This statement is made precise by identification of the
partition functions of both sides of the duality:
\bea\label{Z=Z}
Z_\mathrm{AdS} &=& Z_\mathrm{CFT}~.
\eea

In fact, the black holes are believed to be extremely well suited
for a window to new physics, because of the no-hair theorem
\cite{Israel:1967wq,Israel:1967za,Carter:1971zc,Misner:1973prb}.
According to the no-hair theorem, which is again not a proved theorem
but a conjecture that represents conventional wisdom about gravity,
black holes are completely described by a few macroscopic quantities
such as the energy, angular momenta and charges,
and do not possess any internal degrees of freedom.
Provided the no-hair theorem, black holes allow physicists to
study gravity via a model with few parameters, akin to a toy model.
Therefore, it is a very approachable but interesting goal to
understand black holes in AdS spacetimes through CFTs.

Meanwhile, black holes are extremely interesting objects
that intermingle all areas of physics.
This fact is concisely symbolized by the Hawking temperature of a black hole
\cite{Hawking:1975vcx}
\bea
T &=& \frac{\hbar c^3}{8 \pi G_N k_B M}~,
\eea
that puts together
\begin{eqnarray*}
\hbar &:& \text{Planck's constant,
the fundamental constant of quantum mechanics,} \\
c &:& \text{Speed of light,
the fundamental constant of special theory of relativity,} \\
G_N &:& \text{Newton's constant,
the fundamental constant for general theory of relativity, and} \\
k_B &:& \text{Boltzmann's constant,
the fundamental constant of statistical mechanics.}
\end{eqnarray*}

Black holes originated as the first non-trivial solution to the classical Einstein gravity
\cite{Schwarzschild:1916uq}, but many mysteries regarding the event horizon
have been properly raised and addressed several decades later.
For example, if one throws in a cup of hot tea into the black hole,
the black hole grows slightly larger but the entropy of the universe seems to decrease.
It was then realized that if one attributes to a black hole
a temperature proportional to its surface gravity
and an entropy proportional to the area of its event horizon
\cite{Bekenstein:1972tm,Bekenstein:1973ur,Bekenstein:1974ax},
then all laws of thermodynamics including the first and the second fit perfectly
\cite{Christodoulou:1970wf,Christodoulou:1971pcn,Hawking:1971vc,Penrose:1971uk,Bardeen:1973gs}.

Hawking showed that the attribution of temperature is not merely an analogy
to the laws of thermodynamics, but that the black hole in fact
radiates as if it were a blackbody with that temperature
\cite{Hawking:1974rv,Hawking:1975vcx}.

Similarly, if the Bekenstein-Hawking entropy of the black hole is indeed
the entropy that is known from thermodynamics,
then the black hole must be an ensemble that consists of a corresponding
--- in fact, an enormous --- number of microscopic degrees of freedom.
This is contrary to the no-hair theorem of classical gravity,
that the black hole is completely described by several macroscopic
parameters and does not have any internal structure.
This is what a successful theory of quantum gravity, if any, must address
\cite{tHooft:1990fkf,Susskind:1993ws,Susskind:1994sm,Sen:1995in}.

One of the biggest successes of superstring theory is exactly this.
In the seminal work \cite{Strominger:1996sh}, the authors considered
a particular setup of string theory, namely the type-IIB string theory
compactified on $K3 \times S^1$.
It yields supersymmetric black hole solutions in 5 non-compact dimensions
with finite horizon area, and thus finite Bekenstein-Hawking entropy.
These solutions are realized by D-branes that source the charge of the black hole,
and the number of their bound states was matched with the entropy.

With the advent of the AdS/CFT correspondence,
it looked hopeful that microscopic accounting of the entropy of
AdS black holes from the dual CFTs, which are arguably
more thoroughly understood than systems of strings and branes,
may shed brighter light on quantum gravity.
Supergravity solutions for AdS black holes were found in various
dimensions, including \cite{Gutowski:2004ez,Gutowski:2004yv,
Chong:2005da,Chong:2005hr,Kunduri:2006ek,Wu:2011gq}
for 5 dimensions.
However, despite various attempts including
\cite{Aharony:2003sx,Kinney:2005ej,Romelsberger:2005eg,Berkooz:2006wc,
Janik:2007pm,Grant:2008sk,Berkooz:2008gc,Chang:2013fba},
the success had to wait until recently.

The difficulties encountered in the early attempts and overcome recently,
are strongly tied to a property of the AdS/CFT correspondence
that weakly coupled theory on one side is dual to strongly coupled theory
on the other.
The black holes as supergravity solutions are valid in the quantum gravity
side of the duality, namely the superstring theory,
when the string coupling constant $g_s$ is small and the string length scale
$\ell_s$ is smaller than the length scale of the curvature.
These limits translate in the CFT side of the correspondence to
large gauge group $N\to\infty$ and large 't Hooft coupling $\lambda\to\infty$.
Strongly coupled gauge theories are much harder to work on analytically
than the weakly coupled.
In order to circumvent this difficulty, an index
\cite{Kinney:2005ej,Romelsberger:2005eg},
which is a coupling independent function of a CFT, has been devised.
One can compute the index in the more approachable
weakly coupled field theory but still argue that it counts the
same number of states as in the strongly coupled theory.
However, the index only contains information about the number of bosonic
states minus the number of fermionic states,
while the total number of microstates that accounts for the black hole entropy
should be a sum over both.
It was realized only during the recent advances that it is possible to
faithfully count the total number of microstates using the index by
complexifying the chemical potentials, the variables that the index depends on.
Based on various approaches, entropies of various supersymmetric black holes
in different dimensions of AdS space with different limits and precisions
have been matched with enumerations of microscopic states in the
dual conformal field theories, see
\cite{Benini:2015noa,Benini:2015eyy,Benini:2016rke,Azzurli:2017kxo,
Hosseini:2017fjo,Hosseini:2017mds,Cabo-Bizet:2018ehj,Choi:2018hmj,
Benini:2018ywd,Hosseini:2018dob,Choi:2018fdc,Choi:2019miv,
GonzalezLezcano:2019nca,Choi:2019zpz,Nian:2019pxj,Hosseini:2019iad,
Benini:2019dyp,Choi:2019dfu,GonzalezLezcano:2020yeb,
Murthy:2020rbd,Agarwal:2020zwm,Cabo-Bizet:2020nkr,Copetti:2020dil,
Larsen:2021wnu,Choi:2021rxi,David:2021qaa,Cassani:2022lrk,
Aharony:2024ntg}
among the vast sea of literature.

Extending the remarkable match between the black hole entropy
and the number of microstates in the dual conformal field theory,
we now aim to anatomize the microstates beyond counting them.

First, we study how the quantum numbers of the field theory
microstates match the conserved quantities of the black hole.
All supersymmetric AdS black holes that we study have a property
that their conserved quantities, also known as charges,
obey a certain relation between themselves.
This relation is non-linear and quite non-trivial except in AdS$_3$,
and its interpretation is not yet clear.
It is sometimes linked to the absence of closed timelike curves,
but one clear and simple way to put it is that no regular, supersymmetric
black hole solutions are known away from the constraint.

If the black hole is dual to an ensemble of black hole microstates
in the field theory via \eqref{Z=Z},
then the charges, or quantum numbers, of the microstates
must reproduce the black hole charge constraint.
Apparently, the supersymmetric states in the dual field theories
that are believed to be the black hole microstates according to the counting,
exist all over the charge configuration space and do not obey any particular constraint.
We interpret the supersymmetric black hole charge constraint as a property
of the ensemble, rather than of individual black hole microstates,
and present a heuristic derivation of the charge constraint from
an ensemble of supersymmetric states in the dual field theory.

Second, we look for explicit expressions of black hole microstates
in the field theory language.
Although supergravity black holes are dual to microstates in the large-$N$,
strongly coupled limit of the field theory,
the fact that the black hole entropy is counted by a
coupling-independent quantity suggests that
there are as many analogous states in the weakly coupled field theory
as there are black hole microstates in the strongly coupled theory.
From a different point of view, one may argue that microstates in the
finite-$N$, weakly coupled regime of the field theory
hint towards black hole microstates in the quantum gravity theory,
as opposed to its supergravity approximation.

In this light, we explore the Hilbert space of supersymmetric states in
the 4-dimensional $\cN=4$ Yang-Mills theory with finite gauge group $SU(N)$,
dual to black holes in the 5-dimensional AdS space,
and identify supersymmetric states that corresponds to the black holes.
In the weakly coupled, or perturbative field theories,
there is an established way of assembling basic elements
of the theory to compose the Hilbert space.
In particular, the supersymmetric states can be represented by
cohomologies with respect to the preserved supercharge.
We shall find some of these cohomologies that are dual to the
supersymmetric black hole in AdS$_5$,
but not to the gas of super-graviton particles that are more trivial.

\section{Overview of the Dissertation}

This dissertation is organized as follows.

In chapter \ref{sec:BHs} that belongs to the introductory part,
we introduce black holes in AdS$_3$ and AdS$_5$ spacetimes
in gravity perspective.
We will focus on the macroscopic quantities including the energy,
charges and the entropy that describe the black holes,
and thermodynamic relations between these quantities.
We also illustrate the supersymmetric limits of the black holes,
which will reveal two properties, namely the entropy and the
charge constraint, that will be the target of the next part.
For AdS$_3$ black holes we also discuss its classical stability property.
The AdS$_3$ part of this chapter is largely based on \cite{Larsen:2021wnu}.

The rest of the dissertation is divided into two main parts.
The first part is about the microscopic accounting of the entropy
and the charge constraint of supersymmetric black holes.
The second part is about the black hole cohomology problem,
an attempt to find explicit expressions for supersymmetric black hole microstates
in the language of perturbative field theory.

The first part consists of two chapters:
chapter \ref{sec:entropy} on the entropy and
chapter \ref{sec:cc} on the charge constraint.

We start in section \ref{sec:early} by reviewing the early attempts,
including the introduction of the index that will be the core concept
throughout the dissertation.
As explained in the introduction, complexifying the chemical potentials
in the index has been the key to the recent success.
In section \ref{sec:ccp} we demonstrate how the complex chemical
potentials overcome the issue of boson/fermion cancellation,
in a simplified setup with the $U(1)$ gauge group.
Then in section \ref{sec:EEP} we present a derivation of the
black hole entropy given the index, for AdS$_3$ and AdS$_5$ black holes.
The AdS$_3$ part of this section is based on \cite{Larsen:2021wnu}.

In section \ref{sec:3d} we present a microscopic argument for the
supersymmetric charge constraint of AdS$_3$ black holes.
This section is based on \cite{Larsen:2021wnu}.
In section \ref{sec:gendim} we develop the argument for the AdS$_3$
black holes into a generic prescription for deriving the supersymmetric
charge constraints of higher dimensional AdS black holes heuristically.
Then in sections \ref{sec:5d} through \ref{sec:7d}, we apply the generic prescription
to black holes in AdS$_5$, AdS$_4$ and AdS$_7$.
In section \ref{sec:ccdiscuss}, we discuss some future directions
that may reinforce our derivation to be more complete and rigorous.
The sections on higher dimensions are based on \cite{Larsen:2024xxx}.

The second part consists of three chapters:
chapter \ref{sec:cohoproblem} on formulation of the black hole cohomology problem,
chapter \ref{sec:ngindex} on computing the non-graviton index, and
chapter \ref{sec:constructcoho} on constructing the black hole cohomologies.

We start in section \ref{sec:cohoformulation} by arguing how
the writing of black hole microstates can be turned into
a problem of finding cohomologies with respect to the preserved supercharge.
In section \ref{sec:bmn}, we introduce the BMN sector where
the computations can be done more easily while allowing
nearly as powerful answers as in the full sector.
We define the index as a tool for counting the cohomologies in section \ref{sec:cohoindex},
define the graviton cohomologies to be ruled out from the search of
black hole cohomologies in section \ref{sec:gravitons},
and lay out the strategy for solving this problem in section \ref{sec:cohostrategy}.

Then we explain some ideas for computing the index over
graviton coholomogies in section \ref{sec:gindex}.
Using these ideas, we compute the graviton index,
and therefore the non-graviton index, for the BMN sector of the $SU(2)$ theory,
the full $SU(2)$ theory, the BMN sector of the $SU(3)$ theory and
the BMN sector of the $SU(4)$ theory in sections \ref{sec:ngiSU2bmn} through \ref{sec:ngiSU4}.

In section \ref{sec:cohosu2bmn}, we construct the expressions for
all core black hole cohomologies detected by the index
in the BMN sector of the $SU(2)$ theory.
In section \ref{sec:cohosu2}, we show that there must be a new
black hole cohomology in the $SU(2)$ theory that is not in the BMN sector.
We also discuss the partial no-hair behavior of the black hole cohomologies.
Finally in section \ref{sec:cohosu3}, we construct the expression
for the threshold black hole cohomology in the $SU(3)$ theory.

The second part is based on two papers \cite{Choi:2023znd,Choi:2023vdm}.

We conclude in chapter \ref{sec:conc} with a brief summary
and future directions.

\chapter{Black Holes in AdS Spacetime}\label{sec:BHs}

In this chapter, we review black holes in asymptotically
AdS$_3$ and AdS$_5$ spacetimes as gravitational objects.
The focus will be on their thermodynamic properties,
including energy, temperature, charges, chemical potentials and entropy.
The supersymmetric limits of the black holes will be also important,
as most parts of this thesis will take advantage of the supersymmetry
to study quantum aspects of the black holes.
With the AdS$_3$ black hole as an example, we also comment
on the stability of black holes.

\section{AdS$_3$ Black Holes}\label{sec:AdS3BH}

In this section we consider asymptotically AdS$_3$ black hole solutions
and review its semi-classical properties.
This section is largely based on \cite{Larsen:2021wnu}
in collaboration with Finn Larsen.

The AdS$_3$ black hole derives from rotating black hole solutions \cite{Cvetic:1996xz}
to the 5-dimensional $\mathcal{N}=4$ or $\mathcal{N}=8$ supergravity,
by interpreting it as a system of rotating black strings in 6 dimensions
and taking the decoupling limit \cite{Cvetic:1998xh}.
The local geometry of the black hole is a direct product between
a BTZ black hole solution \cite{Banados:1992wn,Banados:1992gq}
to the $2+1$-dimensional gravity
and a three-sphere $S^3$ with equal but opposite constant curvatures.
The global structure enables rotation of the $S^3$ with respect to the
time in AdS$_3$.

Regardless of its string theoretical origin or the structure of the transverse space,
it is important for our purposes that the solution can be understood as a black hole
in the AdS$_3$ spacetime,
and that it is described by four conserved quantities.
There are energy, or mass, $E$ and angular momentum $J$
from the isometry $SO(2,2)$ of AdS$_3$,
and there are two charges $Q_L$ and $Q_R$
from the isometry $SO(4) \sim SU(2)_L \times SU(2)_R$ of $S^3$.
We consider the isometry of the transverse space as the internal symmetry,
and therefore name the corresponding conserved quantities as charges.
Often throughout this thesis, we will use the term `charges' to collectively
refer to the charges and the angular momenta.

This black hole is known to be dual to the 2-dimensional conformal
field theory with $(4,4)$ supersymmetry \cite{Maldacena:1997re}.
The symmetry algebra of this theory consists of two copies of the
small $\mathcal{N}=4$ Virasoro algebra $\mathfrak{su}(2|1,1)$
\cite{Sevrin:1988ew,Lee:2019uen}.
Each copy includes two Cartans: one related to $\mathfrak{su}(1,1)$,
a half of the 2-dimensional conformal algebra $\mathfrak{so}(2,2)$,
and one related to the $\mathfrak{su}(2)$ R-symmetry.
The four Cartans from both copies are identified with the four Cartans
of the isometry of AdS$_3 \times S^3$,
or with the four conserved quantities of the black hole.
In particular, $E_L \equiv E-J$ and $Q_L$ are identified with one copy
of the algebra, while $E_R \equiv E+J$ and $Q_R$ are with the other.
More precisely, the eigenvalues of Virasoro generators are introduced through
\bea
L_0 - \frac{k_R}{4} = \frac{\epsilon + j}{2}~, \qquad
\tilde{L}_0 - \frac{k_L}{4} = \frac{\epsilon - j}{2}~.
\eea
The constants $k_{L,R}$ are levels of the $SU(2)$ R-currents.
They are related to the central charges as $c_{L,R} = \frac16 k_{L,R}$
by $\cN=4$ supersymmetry.
In the absence of gravitational anomaly, that is when $k = k_L = k_R$,
they are related to the 3-dimensional Newton's constant by \cite{Brown:1986nw}
\bea\label{BrownHenneaux}
6k ~=~ c &=& \frac{3R}{2G_3}~.
\eea
The quantum numbers $\epsilon$, $j$, $q_R$, $q_L$ characterize individual states.
The corresponding macroscopic charges, evaluated as averages
over many states, are denoted $E$, $J$, $Q_R$, $Q_L$.
The unique $SL(2) \times SL(2)$ invariant ground state annihilated by
$L_0$ and $\tilde{L}_0$ has strictly negative energy
$E_{\rm vac}=-\frac{1}{4}(k_R+k_L)$
and corresponds to the AdS$_3$ vacuum.
It is separated by a gap from the black holes which have non-negative 
energy in the CFT$_2$ terminology.

\subsection{Thermodynamics}

The entropy of the black hole as a function of the charges,
or the microcanonical density of states,
contains all essential information about the black hole as a thermal system:
\begin{eqnarray}\label{AdS3S}
S &=&  2\pi \sqrt{ \frac{1}{2} k_R(E+J) - \frac{1}{4} Q^2_R}
+ 2\pi \sqrt{ \frac{1}{2} k_L (E-J) - \frac{1}{4} Q^2_L}~.
\end{eqnarray}
From the entropy (\ref{AdS3S}), the chemical potentials conjugate to the charges
and the temperature conjugate to the energy can be derived
using the first law of thermodynamics,
\bea\label{AdS31stlaw}
T dS = dE - \mu dJ  - \omega_L dQ_L - \omega_R dQ_R~.
\eea
The potentials written as functions of charges $(E,J,Q_L,Q_R)$ are
\bea\label{potascharge}
\beta &=& \pdv{S}{E}  ~=~
\frac12 \sqrt{\frac{2\pi^2k_L}{E-J-\frac{Q_L^2}{2k_L}}}
+ \frac12 \sqrt{\frac{2\pi^2k_R}{E+J-\frac{Q_R^2}{2k_R}}}~, \nn\\
\mu &=& -\frac{1}{\beta} \pdv{S}{J}  ~=~
\frac{-\sqrt{\frac{E-J}{k_L} - \frac{Q_L^2}{2k_L^2}} + \sqrt{\frac{E+J}{k_R} - \frac{Q_R^2}{2k_R^2}}}
{\sqrt{\frac{E-J}{k_L} - \frac{Q_L^2}{2k_L^2}} + \sqrt{\frac{E+J}{k_R} - \frac{Q_R^2}{2k_R^2}}}~, \nn\\
\omega_R &=& -\frac{1}{\beta} \pdv{S}{Q_R}  ~=~
\frac{\sqrt{\frac{E-J}{k_L} - \frac{Q_L^2}{2k_L^2}}}
{\sqrt{\frac{E-J}{k_L} - \frac{Q_L^2}{2k_L^2}} + \sqrt{\frac{E+J}{k_R} - \frac{Q_R^2}{2k_R^2}}}
\cdot \frac{Q_R}{k_R}~, \nn\\
\omega_L &=&  -\frac{1}{\beta} \pdv{S}{Q_L}  ~=~
\frac{\sqrt{\frac{E+J}{k_R} - \frac{Q_R^2}{2k_R^2}}}
{\sqrt{\frac{E-J}{k_L} - \frac{Q_L^2}{2k_L^2}} + \sqrt{\frac{E+J}{k_R} - \frac{Q_R^2}{2k_R^2}}}
\cdot \frac{Q_L}{k_L}~.
\eea
One can invert these relations to write the charges in terms of the potentials:
\bea\label{chargeaspot}
E_R ~\equiv~ E+J &=& \frac{2 k_R}{\beta^2(1-\mu)^2} \left( \pi^2 + \beta^2 \omega^2_R\right)~, \nn\\
E_L ~\equiv~ E-J  &=& \frac{2k_L}{\beta^2(1+\mu)^2} \left( \pi^2 + \beta^2 \omega^2_L\right)~, \nn\\
Q_R&=& \frac{2k_R}{1-\mu}  \omega_R~, \nn\\
Q_L &=& \frac{2k_L}{1+\mu}  \omega_L~.
\eea

The grand canonical partition function in thermodynamics is defined from
the microcanonical ensemble by a weighted sum over microstates:
\bea\label{AdS3defZ}
Z (\beta,\, \mu, \, \omega_R, \, \omega_L) &=& \mathrm{Tr}
\left[e^{- \beta (E-\mu J-\omega_R Q_R - \omega_L Q_L)} \right]~.
\eea
It is a function of the inverse temperature $\beta$ and
chemical potentials $\mu$, $\omega_R$ and $\omega_L$,
which determine the weight with which each microstate with certain charges
$E$, $J$, $Q_R$ and $Q_L$ contributes to the partition function.
It is possible to write the grand canonical partition function from the
information about the thermodynamic system presented thus far:
\bea
\label{eqn:genlnZ}
\log Z &=& S - \beta\left( E - \mu J - \omega_R Q_R - \omega_L Q_L\right) \nn\\
&=& \frac{k_R}{\beta(1-\mu)} \left( \pi^2 + \beta^2 \omega^2_R\right) + \frac{k_L}{\beta(1+\mu)} \left( \pi^2 + \beta^2 \omega^2_L\right)~.
\eea
Although we have not presented in that order,
one may alternatively take \eqref{eqn:genlnZ} as the starting point for describing the
thermodynamic system and derive all quantities from it.
For example, the macroscopic energy and charges as ensemble averages are
\bea\la{eqn:E'gen}
E -\mu J - \omega_R Q_R - \omega_L Q_L & = &
- \frac{\partial\log Z}{\partial\beta} \nn \\
& = & 
\frac{k_R}{\beta^2(1-\mu)} \left( \pi^2 - \beta^2 \omega^2_R\right) +
\frac{k_L}{\beta^2(1+\mu)} \left( \pi^2 - \beta^2 \omega^2_L\right)~, \nn\\
J &=& \frac{1}{\beta}\frac{\partial\log Z}{\partial\mu} \nn\\
&=& \frac{k_R}{\beta^2(1-\mu)^2} \left( \pi^2 + \beta^2 \omega^2_R\right)
- \frac{k_L}{\beta^2(1+\mu)^2} \left( \pi^2 + \beta^2 \omega^2_L\right)~, \nn\\
Q_{L,R} & = & \frac{1}{\beta}\frac{\partial\log Z}{\partial\omega_{L,R}}
~=~ \frac{2k_{L,R}}{1\pm\mu}  \omega_{L,R}~.
\eea
These are equivalent to \eqref{chargeaspot}.

Note that (\ref{AdS3S}) does not make sense unless \cite{Cvetic:1998xh}
\bea\label{AdS3regbound}
E - J - \frac{1}{2k_L}Q^2_L &\geq& 0 ~, \nn\\
E + J - \frac{1}{2k_R}Q^2_R &\geq& 0 ~.
\eea
It implies that there is simply no black hole solution for energy and charges
that violate either of (\ref{AdS3regbound}),
and all of the thermodynamic formulae above have assumed these inequalities.
Saturation of both inequalities corresponds to $\beta \to \infty$
with generic $-1 < \mu < 1$, but leads to zero entropy.
Saturation of only one of the two corresponds to $\beta \to \infty$ with
either $\beta(1\pm \mu)$ kept finite,
as we will elaborate in the next subsection.

\subsection{Supersymmetry}\label{sec:AdS3BHBPS}

Up to this point we did not impose any conditions on the black hole parameters.
We now impose supersymmetry and show that the resulting BPS black holes satisfy {\it two} conditions. 

In the 2d superconformal theory with $(4,4)$ supersymmetry,
there are four $\frac14$-BPS sectors.
Each sector preserves two real supersymmetries that are either
holomorphic ($R$) or anti-holomorphic ($L$),
and that either raise or lower the corresponding R-charge.
We focus without loss of generality on the $\frac14$-BPS sector
which preserves supersymmetries that are anti-holomorphic ($L$)
and raise the R-charge.
Then the unitarity bound from the anticommutator of the supercharges
on individual CFT states in the NS sector is:
\be\la{eqn:unitboundmicro}
\epsilon - j + \frac{1}{2}k_L\geq q_L  ~,
\ee
from which a bound for black hole energy and charges follows:
\be\label{eqn:unitboundmacro}
 E - J + \frac{1}{2}k_L\geq Q_L  ~.
\ee
Microscopic states whose quantum numbers saturate the inequality \eqref{eqn:unitboundmicro} are called chiral primaries.
Unitarity further requires that chiral primaries have $0\leq q_L \leq 2k_L$ \cite{Eguchi:1987sm,Eguchi:1987wf}.

Saturation of the inequality \eqref{eqn:unitboundmacro} is a
necessary condition for a supersymmetric black hole
but it is not sufficient.
Recall that the black hole charges must obey (\ref{AdS3regbound}).
A hypothetical black hole solution that violates this inequality
would have event horizon with imaginary area.
Such geometries are not regular so black holes with these
quantum numbers simply do not exist.
This regularity condition is variously referred to as the
cosmic censorship bound or the condition for absence of closed timelike curves. 

The BPS condition demands that the inequality \eqref{eqn:unitboundmacro}
be saturated but then compatibility with regularity 
\eqref{AdS3regbound} gives
\bea\label{eqn:JL=kL}
Q_L=k_L~.
\eea
This is the charge constraint on BPS black holes in AdS$_3$.
Thus BPS black holes have the same quantum numbers as the particular
chiral primaries situated in the middle of the interval
$0\leq q_L \leq 2k_L$ allowed by unitarity. 

We established that BPS black holes in AdS$_3$ are co-dimension 2
in parameter space:
saturation of {\it two} inequalities (\ref{AdS3regbound}) and
(\ref{eqn:unitboundmacro}) introduces {\it two} relations between
the four parameters $E$, $J$, and $Q_{R,L}$.
Now recall the formulae (\ref{chargeaspot}) that relate the quantum numbers
to potentials, reproduced here for convenience:
\begin{subequations}\la{eqn:chargerep}
\bea
\la{eqn:Egen'}
E &=& \frac{k_R}{\beta^2(1-\mu)^2} \left( \pi^2 + \beta^2 \omega^2_R\right)
+    \frac{k_L}{\beta^2(1+\mu)^2} \left( \pi^2 + \beta^2 \omega^2_L\right)~,
\\
J &=& \frac{k_R}{\beta^2(1-\mu)^2} \left( \pi^2 + \beta^2 \omega^2_R\right) - \frac{k_L}{\beta^2(1+\mu)^2} \left( \pi^2 + \beta^2 \omega^2_L\right)
~,
\\
\la{eqn:JRLgen'}
Q_{L,R} & = & \
 \frac{2k_{L,R}}{1\pm\mu}  \omega_{L,R}~.
\eea
\end{subequations}
In the canonical ensemble the extremal limit amounts to vanishing
temperature $\beta\to\infty$.
However, we must be careful with what remains finite in this limit.

Consider a pair of particular combinations of these charges:
\begin{subequations}\label{eqn:extbound0}
\bea
\label{eqn:extbound1}
E + J - \frac{Q_R^2}{2k_R} &=& \frac{2k_R \pi^2 }{\beta^2 (1-\mu)^2} \geq 0 
~,
\\
\label{eqn:extbound2}
E - J - \frac{Q_L^2}{2k_L} &=& \frac{2k_L \pi^2}{\beta^2 (1+\mu)^2} \geq 0
~.
\eea
\end{subequations}
If one na\"ively takes $\beta\to\infty$ with the chemical potential
$\mu$ finite and generic,
both of these inequalities will be saturated.
However, when the expressions on the left hand sides of
both equations in \eqref{eqn:extbound0} vanish,
the black hole entropy \eqref{AdS3S} will be zero as well.
Therefore, the limit taken this way yields an extremal ``black hole" 
with an event horizon that has vanishing area.
Such a geometry is singular, it is not a black hole solution. 
 
In order to circumvent this obstacle, we need to saturate
only one of the inequalities \eqref{eqn:extbound0}.
We pick the latter without loss of generality.
To avoid also saturating \eqref{eqn:extbound1},
we take $\beta\to\infty$ while rescaling $\mu$ so that
$\tilde\mu \equiv \beta (\mu-1)$ remains finite.
Note that $\tilde{\mu}\leq 0$ because $\mu\leq 1$.
It further follows from \eqref{eqn:JRLgen'} that,
in order to describe black holes with generic values of $Q_R$,
the limit must also take $\omega_R\to 0$ with
$\tilde\omega_R \equiv \beta \omega_R$ kept finite.
In contrast, $\omega_L$ does not require any rescaling, it can be kept finite by itself.

In summary, the extremal limit of a general AdS$_3$ black hole is: 
\be\la{eqn:extlimit}
\text{Extremal limit: } \quad
\begin{cases}
\beta \to \infty ~,& \\
\mu \to 1 & ~ \text{with}~  \tilde\mu \equiv \beta (\mu-1) ~\text{finite,} \\
\omega_R \to 0 & ~ \text{with}~ \tilde\omega_R \equiv \beta \omega_R ~\text{finite,} \\
\omega_L  ~\text{finite.} &
\end{cases}
\ee
This limit was designed so that (\ref{eqn:chargerep}) gives expressions that are finite:
\begin{subequations}\la{eqn:extcharge}
\bea
E &=& 
\frac{k_R}{\tilde{\mu}^2} \left( \pi^2 + \tilde{\omega}^2_R\right) + \frac{k_L}{4} \omega^2_L~,
\\
J &=& 
\frac{k_R}{\tilde{\mu}^2} \left( \pi^2 + \tilde{\omega}^2_R\right) - \frac{k_L}{4} \omega^2_L
~,
\\
Q_{R} & = &  - \frac{2k_{R}}{\tilde{\mu}}  \tilde{\omega}_{R}
~,
\\
Q_{L} & = &  k_{L}\omega_{L}
~.
\eea
\end{subequations}
The explicit sign in the formula for $Q_R$ compensates $\tilde{\mu}<0$
so that the angular momentum $Q_R$ has the same sign as the
rescaled angular velocity $\tilde{\omega}_{R}$, as expected.
These formulae for the conserved charges give the energy as a function of the charges
\begin{equation}
\label{eqn:Eext}
E_{\rm ext} =  J + \frac{1}{2k_L} Q^2_L~.
\end{equation}
This is the ground state energy for these conserved charges.
It saturates \eqref{AdS3regbound} and 
is identified with the extremal black hole mass.
The extremal entropy becomes
\begin{eqnarray}\la{eqn:Sext}
S_{\rm ext} &=& - \frac{2k_R \pi^2}{\tilde{\mu}}  
~=~ 2\pi \sqrt{ \frac{1}{2} k_R(E_{\rm ext}+J) - \frac{1}{4} Q^2_R}  \nonumber \\
& = &  2\pi \sqrt{  k_R J + \frac{k_R}{4k_L} Q^2_L - \frac{1}{4} Q^2_R}~.
\end{eqnarray}
The last equation eliminated the energy using the extremality condition (\ref{eqn:Eext}).

As we have stressed, the extremal black holes are not necessarily supersymmetric. 
As the second and last step of implementing the BPS limit,
we now examine supersymmetry.
Recall from \eqref{eqn:unitboundmacro} that charges of
supersymmetric black holes must saturate the inequality
$$
 E - J - Q_L + \frac{1}{2}k_L\geq 0  ~.
$$
The left hand side can be recast as a sum of two squares  
\be\la{eqn:susybound}
E - J - Q_L + \frac{1}{2}k_L
= \frac{2k_L \pi^2}{\beta^2 (1+\mu)^2} + \frac{k_L}{2} \left(1-\frac{2\omega_L}{1+\mu}\right)^2~,
\ee
using (\ref{eqn:chargerep}).
The first square is precisely \eqref{eqn:extbound2} so it vanishes
in the extremal limit.
In order to saturate the BPS bound \eqref{eqn:unitboundmacro}
the second square must vanish as well so we demand that the potentials satisfy
\be\la{eqn:BPSpotentials}
\varphi \equiv  1+\mu - 2\omega_L = 0 ~,
\ee
in addition to conditions for extremality. We defined the parameter $\varphi$ for future use. 
Since $\mu = 1$ at extremality we must have $\omega_L = 1$ in the BPS limit. 
However, just as the extremal limit is taken with $\tilde\mu \equiv \beta (\mu -1)$ kept finite there is no obstacle to 
taking the BPS limit $\omega_L \to 1$ so $\tilde{\omega}_L \equiv \beta(\omega_L-1)$ remains finite.
The value of $\tilde\omega_L$ is, like $\tilde\mu$ and $\tilde\omega_R$, not constrained.

To summarize, the BPS AdS$_3$ black holes are limits of generic AdS$_3$ black holes as
\begin{equation}
\label{eqn:BPSlimitT}
T = \beta^{-1} \to 0~,
\end{equation}
while the potentials
\begin{equation}
\label{eqn:BPSlimit}
\tilde\mu = \beta (\mu -1)~, \quad \tilde{\omega}_R = \beta \omega_R~, \quad \tilde{\omega}_L = \beta(\omega_L-1)~,
\end{equation}
are kept finite. 
In this limit two inequalities \eqref{eqn:unitboundmacro} and
\eqref{AdS3regbound} are saturated.

The definition of the grand canonical partition function
can be adapted to the BPS limit \eqref{eqn:BPSlimitT}-\eqref{eqn:BPSlimit} as
\bea\label{AdS3defZBPS0}
Z (\beta,\, \mu, \, \omega_R, \, \omega_L) &=&
\mathrm{Tr} \left[e^{- \beta (E-\mu J-\omega_R Q_R - \omega_L Q_L)} \right] \nn\\
&=& e^{\frac12 \beta k_L} \mathrm{Tr}
\left[ e^{- \beta \left(E-J-Q_L+\frac{k_L}{2} \right)
+ \tilde\mu J + \tilde{\omega}_R Q_R + \tilde{\omega}_L Q_L} \right]~.
\eea
The second line manifests that in the BPS limit \eqref{eqn:BPSlimitT}-\eqref{eqn:BPSlimit},
contribution to the partition function from non-BPS states such that
$E - J - Q_L + \frac{1}{2}k_L > 0$ will be suppressed,
and the partition function will have an overall divergent factor $e^{\frac12 \beta k_L}$
which is its sole dependence on $\beta \to \infty$.
This factor can be interpreted as the supersymmetric Casimir energy
\cite{Assel:2015nca}
\bea\label{eqn:SUSYCast}
E_{\rm SUSY} &=& - \frac{1}{2}k_L~,
\eea
that is common to all states. Note that it is not the conventional Casimir energy $E_C = - \frac{1}{4} ( k_L + k_R)$ that enters here and the two notions of Casimir energy agree only when the levels $k_L=k_R$. The Casimir energy appears explicitly because we study
the partition function {\it  defined as a path integral} rather than as a trace over a Hilbert space normalized such that the vacuum contributes unity.

Therefore, the BPS partition function as the limit \eqref{eqn:BPSlimitT}-\eqref{eqn:BPSlimit}
of the grand canonical partition function,
can be written as
\bea\label{AdS3defZBPS}
Z_{\rm BPS}(\beta, \, \tilde\mu, \, \tilde\omega_R, \, \tilde\omega_L) &=&
\lim_{\beta\to\infty} e^{\frac12 \beta k_L} \mathrm{Tr}\left[ e^{- \beta \left(E-J-Q_L+\frac{k_L}{2} \right)
+ \tilde\mu J + \tilde{\omega}_R Q_R + \tilde{\omega}_L Q_L} \right] \nn\\
&=& \left( \lim_{\beta\to\infty} e^{\frac12 \beta k_L} \right)
\mathrm{Tr}_\mathrm{BPS} \left[
e^{\tilde\mu J + \tilde{\omega}_R Q_R + \tilde{\omega}_L Q_L} \right]~.
\eea
In the second line, we have restricted the trace to BPS states only,
as non-BPS states are suppressed by $\beta\to\infty$.

For the black hole system, the BPS partition function is obtained by taking the limit 
\eqref{eqn:BPSlimitT}-\eqref{eqn:BPSlimit} on the grand canonical partition function
(\ref{eqn:genlnZ}):
\bea
\label{AdS3ZBPS}
\log Z_{\rm BPS} &=& \frac12 \beta k_L
- \frac{k_R}{\tilde{\mu}} \left( \pi^2 + \tilde{\omega}^2_R\right) +
k_L  \left( \tilde{\omega}_L - \frac{1}{4} \tilde{\mu} \right) ~.
\eea
The BPS limit of the macroscopic energy and charges can be obtained
by differentiating \eqref{AdS3ZBPS} as in \eqref{eqn:E'gen},
or by taking the BPS limit of the potentials
\eqref{eqn:BPSlimitT}-\eqref{eqn:BPSlimit} from \eqref{eqn:E'gen}:
\begin{subequations}\la{eqn:BPScharge}
\bea
E &=& 
\frac{k_R}{\tilde{\mu}^2} \left( \pi^2 + \tilde{\omega}^2_R\right) + \frac{k_L}{4}~,
\\
J &=& 
\frac{k_R}{\tilde{\mu}^2} \left( \pi^2 + \tilde{\omega}^2_R\right) - \frac{k_L}{4}~,
\\
Q_{R} & = &  - \frac{2k_{R}}{\tilde{\mu}}  \tilde{\omega}_{R} ~,
\eea
\end{subequations}
and notably,
\bea\la{eqn:BPSJL}
Q_{L} & = &  k_{L} ~.
\eea
The extremal black hole entropy \eqref{eqn:Sext} also simplifies further in the BPS limit
\begin{eqnarray}\la{AdS3SBPS}
S_{\rm BPS} & = &  2\pi \sqrt{  k_R \left(J + \frac{1}{4} k_L\right) - \frac{1}{4} Q^2_R}~.
\end{eqnarray}

In the BPS limit, the four macroscopic quantities $E, J, Q_{L,R}$
are parametrized by only two potentials $\tilde\mu$ and $\tilde\omega_R$,
they are independent of the third potential $\tilde\omega_L$.
This confirms the expectation that the parameters of a BPS black hole form
a co-dimension 2 surface in the space of all possible charges.
On the other hand, there really are three independent rescaled potentials 
$\tilde\mu, \tilde\omega_{L,R}$.
This is possible because $\tilde\omega_L$ parametrizes a flat direction
along which the BPS black hole does not change.

\subsection{Stability}

So far we have discussed the AdS$_3$ black hole solutions
that are described by 4 charges $(E,J,Q_L,Q_R)$
or equivalently by 4 chemical potentials $(\beta,\mu,\omega_L,\omega_R)$.
The charges must obey the extremality bound (\ref{AdS3regbound})
as well as the unitarity bound (\ref{eqn:unitboundmacro}),
and we have discussed in detail the saturation of these bounds.
In this subsection we touch on another issue of semi-classical black holes,
namely stability.

A quick way to determine the stability condition is from the
first law of thermodynamics (\ref{AdS31stlaw}):
\bea\label{1stlawrep}
T dS = dE - \mu dJ  - \omega_R dQ_R - \omega_L dQ_L~.
\eea
Consider a particle with generic quantum numbers $(\ep,j,q_R,q_L)$.
We assume that these quantum numbers are infinitesimal compared
to corresponding macroscopic charges of the black hole.
Change of black hole entropy under the emission of such a particle
is proportional to
\bea
T dS = -\ep + \mu j  + \omega_R q_R + \omega_L q_L~.
\eea
If this quantity is positive, the black hole gains entropy by emitting this particle,
and thus it is unstable against decaying into this particle.

There are several candidate stability bounds, depending on
the quantum numbers of particles to which the black hole may emit.

\begin{itemize}
\item First, consider particles with
$(\ep,j,q_R,q_L) \propto (1,0,1,0)$ or $(\ep,j,q_R,q_L) \propto (1,0,0,1)$.
The stability bounds against these particles, as obtained from (\ref{1stlawrep}), are
\bea\label{RLstabilitync}
\omega_R \leq 1~, \qquad \omega_L \leq 1~,
\eea
respectively.
These stability bounds are analogous to those for black holes in
higher dimensions \cite{Kim:2023sig,MinwallaTalk2024}.

\item Next, consider particles with $(\ep,j,q_R,q_L) \propto (1/2,1/2,1,0)$.
In CFT$_2$, these correspond to chiral primaries, because they have
$(L_0, q_R) \propto (1/2,1)$ and $\bar{L}_0 =q_L = 0$.
The black hole stability bound against chiral primaries is
\bea\label{Rstability}
0 \geq \omega_R - \frac{1-\mu}{2} \qquad \Leftrightarrow  \qquad Q_R \leq k_R~.
\eea

\item Similarly, consider particles with $(\ep,j,q_R,q_L) \propto (1/2,-1/2,0,1)$.
In CFT$_2$, these correspond to anti-chiral primaries, because they have
$(\bar{L}_0, q_L) \propto (1/2,1)$ and $L_0 = q_R = 0$.
The black hole stability bound against anti-chiral primaries is
\bea\label{Lstability}
0 \geq \omega_L - \frac{1+\mu}{2} \qquad \Leftrightarrow  \qquad Q_L \leq k_L~.
\eea

\end{itemize}

The question of stability can be rephrased as follows.
Given a black hole with given total charges,
it is unstable if a system with another black hole
and a gas of particles but with same total charges has
bigger entropy than the original system.
Therefore, given a microcanonical system with given total charges,
it is important to find a configuration of a black hole and particles
that maximizes the entropy.
We neglect the entropy of the gas of the particles,
so the problem reduces to maximizing the entropy of the black hole piece.
Phrased in this way, the microcanonical system in question
does not have to be realized by a black hole alone.
For example, one may consider a system with total charges
that violate the extremality bound (\ref{AdS3regbound}).
Then there is no such system that contains only a black hole and no others,
but it is still meaningful to ask what is the most entropic configuration
of a black hole and particles.

In the rest of this subsection, we address this question
while assuming existence of chiral and anti-chiral primaries in the theory.

For our purposes in this subsection,
it is advantageous to use $E_R = E+J$, $E_L = E-J$, $Q_R$ and $Q_L$
as the four charges of a system, emphasizing chirality.
Suppose that a system has total charges $E_{R,\mathrm{tot}}$,
$E_{L,\mathrm{tot}}$, $Q_{R,\mathrm{tot}}$ and $Q_{L,\mathrm{tot}}$.
These need to satisfy the unitarity bound for both holomorphic
and antiholomorphic sectors:
\be\label{eqn:unitboundmacrorep}
 E - J + \frac{1}{2}k_L\geq Q_L  ~, \qquad
 E + J + \frac{1}{2}k_R\geq Q_R  ~,
\ee
but the extremality (\ref{AdS3regbound}) is not imposed.

The system consists of a black hole (bh) and gas of particles (gp),
so $E_{R,\mathrm{tot}} = E_{R,\mathrm{bh}} +E_{R,\mathrm{gp}}$
and similarly for the other charges.
We assume that the entropy of the gas of particles is negligible
to that of the black hole. Therefore, the entropy of the system is
\begin{eqnarray}\label{SfromBH}
S_\mathrm{bh} &=&  2\pi \sqrt{ \frac{1}{2} k_R E_{R,\mathrm{bh}} - \frac{1}{4} Q^2_{R,\mathrm{bh}}}
+ 2\pi \sqrt{ \frac{1}{2} k_L E_{L,\mathrm{bh}} - \frac{1}{4} Q^2_{L,\mathrm{bh}}}~.
\end{eqnarray}
Meanwhile, each chiral and anti-chiral primary carries charges
$(\ep_R,\ep_L,q_R,q_L) \propto (1,0,1,0)$ and $(\ep_R,\ep_L,q_R,q_L) \propto (0,1,0,1)$,
respectively.
We use the proportionality sign because the macroscopic charges of the black hole
scale differently from those of the microscopic particles.
In fact, we expect the macroscopic charges to scale with the large central charge.
It follows that $E_{R,\mathrm{gp}} = Q_{R,\mathrm{gp}}$
and $E_{L,\mathrm{gp}} = Q_{L,\mathrm{gp}}$.
Thus, the contribution to the entropy (\ref{SfromBH}) from the
right (holomorphic) sector is
\begin{eqnarray}\label{SRfromBH}
S_{R,\mathrm{bh}} &=& 
2\pi \sqrt{ \frac{1}{2} k_R E_{R,\mathrm{bh}} - \frac{1}{4} Q^2_{R,\mathrm{bh}}} \nn\\
&=&
2\pi \sqrt{ \frac{1}{2} k_R (E_{R,\mathrm{tot}} - E_{R,\mathrm{gp}})
- \frac{1}{4} (Q_{R,\mathrm{tot}} - E_{R,\mathrm{gp}})^2} \nn\\
&=& \pi \sqrt{2k_L \left(E_{R,\mathrm{tot}}-Q_{R,\mathrm{tot}}+\frac{k_R}{2} \right)
- (Q_{R,\mathrm{tot}} - k_R - E_{R,\mathrm{gp}})^2}~.
\end{eqnarray}
For the entropy to be maximized,
\bea
E_{R,\mathrm{gp}} = \begin{cases}
Q_{R,\mathrm{tot}} - k_R \\ 0
\end{cases} & \Rightarrow &
S_{R,\mathrm{bh}} = \begin{cases}
2 \pi \sqrt{\frac12 k_L \left(E_{R,\mathrm{tot}}-Q_{R,\mathrm{tot}}+\frac{k_R}{2} \right)}
& (Q_{R,\mathrm{tot}} > k_R) \\
2\pi \sqrt{\frac12 k_R E_{R,\mathrm{tot}} - \frac14 Q^2_{R,\mathrm{tot}}}
& (Q_{R,\mathrm{tot}} \leq k_R)
\end{cases} \nn
\eea
The upshot is that if the total charges are such that they violate the stability bound
$Q_{R,\mathrm{tot}} > k_R$, the excess $Q_R$ must be taken up by the particles
so that the black hole sits at the threshold of the stability bound:
$E_{R,\mathrm{bh}} = k_R$.
The same logic applies to the anti-holomorphic (left) sector independently.

We have only considered the chiral and anti-chiral primaries.
However, the descendants do not play a role even if they are included.
This is because emission of a descendant necessarily deprives the black hole
of more energy (either left or right) than a primary with same $Q_R$ or $Q_L$ would,
which necessarily results in the smaller entropy.

\section{AdS$_5$ Black Holes}\label{sec:AdS5BH}

In this section we consider asymptotically AdS$_5$ black hole solutions
and review its semi-classical properties.

Asymptotically AdS$_5$ black holes arise as solutions to type-IIB supergravity
in AdS$_5 \times S^5$ \cite{Gutowski:2004ez,Gutowski:2004yv,Chong:2005da,
Chong:2005hr,Kunduri:2006ek,Wu:2011gq}.
They carry the mass, or energy $E$ and two angular momenta $J_{1,2}$
for the isometry $SO(2,4)$ of AdS$_5$,
and three charges $Q_{1,2,3}$ for the isometry $SO(6)$ of $S^5$.
The black hole solution with all 5 charges independent is known.
Similarly to the AdS$_3$ black holes, we consider the isometry of the
transverse space $S^5$ as an internal symmetry for the AdS$_5$ black hole
and therefore name the corresponding conserved quantities as charges.

The type-IIB theory in AdS$_5 \times S^5$ is known to be dual to the
4-dimensional maximally supersymmetric ($\mathcal{N}=4$) Yang-Mills theory,
in fact it is the original and most well understood case of the
AdS/CFT correspondence \cite{Maldacena:1997re}.
The symmetry group of the $\mathcal{N}=4$ SYM is the 4d $\mathcal{N}=4$
superconformal group $PSU(2,2|4)$, whose maximal bosonic subalgebra is
the 4d conformal group $SU(2,2) \sim SO(4,2)$ times the R-symmetry $SU(4)$.
The three Cartans of the conformal group correspond to the mass
and the two angular momenta of the black hole,
while the three Cartans of the R-symmetry group correspond to the
three charges of the black hole.

Thermodynamic quantities such as energy, temperature, charges, potentials
and entropy and their relations are algebraically complicated.
We shall present only some of those with a simplification by restricting to the
case of equal charges $Q_1=Q_2=Q_3$, following \cite{Larsen:2019oll}.
The goal of this section is to illustrate the followings.
\begin{itemize}
\item Supersymmetric AdS$_5$ black holes are co-dimension 2
in the space of AdS$_5$ black holes.

\item The two conditions for supersymmetry translates to vanishing temperature
and a relation between the charges, that we shall refer to as
the supersymmetric charge constraint.

\item The formulae for the entropy of the black hole,
as well as for the supersymmetric charge constraint,
are known without restriction to $Q_1=Q_2=Q_3$.
\end{itemize}

\subsection{Thermodynamics}

As anticipated, we restrict to the AdS$_5$ black holes with equal charges
$Q \equiv Q_1=Q_2=Q_3$.
This does not qualitatively alter the main arguments of this section.
It is algebraically convenient to express the mass and the three charges
(two angular momenta $J_1$ and $J_2$, and one synchronized charges $Q$)
of the AdS$_5$ black holes using four auxiliary parameters $(r_+, q, a,b)$ as
\begin{align}\label{AdS5charges}
        E & = \frac{\pi}{4G_5}\frac{m(2(1 - a^2)+2(1 - b^2)-(1 - a^2)(1 - b^2))
        +2qab((1 - a^2)+(1 - b^2))}{(1 - a^2)^2(1 - b^2)^2}~,\nn\\
        Q & = \frac{\pi}{4G_5}\frac{q}{(1 - a^2)(1 - b^2)}~,\nn\\
        J_1 & = \frac{\pi}{4G_5}\frac{2ma+qb(1+a^2g^2)}{(1 - a^2)^2(1 - b^2)}~,\nn\\
        J_2 & = \frac{\pi}{4G_5}\frac{2mb+qa(1+b^2g^2)}{(1 - a^2)(1 - b^2)^2}~.
\end{align}
Here $G_5$ is five-dimensional Newton's gravitational constant,
we have set $g=\ell_5^{-1}=1$ where $g$ is the coupling of gauged supergravity
and $\ell_5$ is the AdS$_5$ radius, and
\bea\label{minr+}
2m &=& \frac{(r_+^2+a^2)(r_+^2+b^2)(1+g^2r_+^2)+q^2+2abq}{r_+^2}~.
\eea

The entropy of the black hole can be expressed using the same parameters:
\begin{eqnarray}\label{AdS5Sinab}
S &=& 2 \pi\cdot \frac{\pi}{4 G_5}\frac{(r_+^2+a^2)(r_+^2+b^2)+abq}{(1-a^2)(1-b^2)r_+}~.
\end{eqnarray}
Implicitly via the auxiliary parameters, \eqref{AdS5charges}-\eqref{AdS5Sinab}
yield the entropy as a function of the mass and the charges.
This defines the microcanonical density of states,
which contains all essential information about the black hole as a thermal system.
From the entropy (\ref{AdS5Sinab}), the chemical potentials conjugate to the charges
and the temperature conjugate to the energy can be derived
using the first law of thermodynamics as in (\ref{potascharge}):
\bea\label{AdS51stlaw}
T dS = dE - \Phi dQ  - \Omega_1 dJ_1 - \Omega_2 dJ_2~.
\eea
For example, the temperature can be derived as
\bea\label{AdS5Tascharge}
T &=& \left( \pdv{S}{E} \right)^{-1}  ~=~
\frac{r_+^4[1+(2r_+^2+a^2+b^2)]-(ab+q)^2}{2\pi r_+[(r_+^2+a^2)(r_+^2+b^2)+abq]}~.
\eea

\subsection{Supersymmetry}\label{sec:AdS5BHBPS}

The general AdS$_5$ black holes introduced in the last subsection
have independent thermodynamic quantities $(E, Q, J_1, J_2)$.
Now we discuss the conditions that these black holes become supersymmetric.

Unitarity guarantees that their mass and the charges satisfy 
\bea\label{AdS5unitbound}
E - \left( 3 Q + J_1 + J_2\right)  \geq 0~,
\eea
where the coefficient 3 stands for the three redundant charges $Q \equiv Q_1=Q_2=Q_3$.
The black hole is supersymmetric, i.e. it is $\frac{1}{16}$-BPS
when its charges saturate this inequality:
\bea\label{mqjbps}
E^* - 3 Q^* - J_1^* - J_2^* &=& 0~.
\eea
We use the starred symbols $(E^*, Q^*, J_1^*, J_2^*)$ instead of $(E, Q, J_1, J_2)$
when we stress that the variables refer to the BPS case.

Collecting \eqref{AdS5charges}, the left hand side of \eqref{AdS5unitbound}
can be written in terms of the
auxiliary parameters as
\begin{equation}
\label{eqn:Mmrelation}
E - \left( 3 Q + J_1 + J_2\right)  =
\frac{\pi}{4G_5} \frac{3 + (a+b) - ab}{(1-a)(1+a)^2(1-b)(1+b)^2}
\left[ m - q(1+a+b)\right]~,
\end{equation}
where $m$ is a placeholder for the expression \eqref{minr+} in terms of $r_+$.
The coefficient in front of the square bracket is always positive,
so the BPS condition \eqref{mqjbps} is equivalent to the
following relation between the BPS (starred) quantities:
\bea\label{BPS condition}
q^*&=&\frac{m^*}{1+a+b}~.
\eea

A very interesting result of \cite{Larsen:2019oll} is that the factor in the square bracket
of \eqref{eqn:Mmrelation}, when $m$ is replaced by the corresponding expression
in terms of $r_+$ via \eqref{minr+}, can be reorganized as
\bea\label{AdS5sumofsq}
m - q(1+a+b) \!&\!=\!&\!
\frac{r_+^2 (q - q^*)^2+ \left( ((1 + a+ b)^2 + r_+^2 ) (r_+^2 - r^{*2}) - (1 + a + b) (q - q^*) \right)^2}
{2r_+^2 \left( (1 + a + b)^2 + r_+^2 \right)}~, \nn\\
\eea
where
\bea\label{AdS5BPSqr}
q^* = (a+b)(1+a)(1+b)~, \qquad r^* \equiv r_+ = \sqrt{a+b+ab}~.
\eea
Note that the right hand side of (\ref{AdS5sumofsq}) is a sum of two squares.
It thus amplifies the BPS condition \eqref{mqjbps},
which is a single relation between real auxiliary parameters,
into two relations $q=q^*$ and $r_+ = r^*$.

As a result of the two relations between the four auxiliary
parameters that describe the AdS$_5$ black holes,
the BPS black hole is parametrized by only two remaining auxiliary parameters $(a,b)$.
The energy and the charges \eqref{AdS5charges} of BPS black holes are
\bea\label{AdS5BPScharges}
E^* &=& \frac{\pi}{4G_5}\frac{\left(3(a+b)-(a^3+b^3)-ab(a+b)^2 \right)}{(1-a)^2 (1-b)^2}~, \nn\\
Q^* &=& \frac{\pi}{4G_5} \frac{a+b}{(1-a) (1-b)}~, \nn\\
 J_1^* &=& \frac{\pi}{4G_5}\frac{(a+b) (2a +b+ab)}{(1-a)^2 (1-b)}~, \nn\\
 J_2^* &=& \frac{\pi}{4G_5}\frac{(a+b) (a+2b+ab)}{(1-a) (1-b)^2}~.
 \eea
These expressions satisfy the BPS condition (\ref{mqjbps}) for any $(a,b)$, as they must.

We highlight two consequences for the BPS black holes.

First, the temperature of a BPS black hole vanishes.
The temperature of the black hole \eqref{AdS5Tascharge} can be
rewritten using the BPS values \eqref{AdS5BPSqr} of parameters $q^*$ and $r^*$ as
\bea\label{eqn:Tapx}
T &=& \frac{ \left[1 + 3(a+b) + (a^2 + b^2 + 3ab) \right] (r_+^2 - r^{*2})
-  (1 + a+b) \, (q - q^*)}{\pi r^* q^*}~.
\eea
This expression makes it clear that the temperature vanishes for BPS black holes,
for which $q=q^*$ and $r_+ = r^*$.

Second, the charges $Q^*$, $J_1^*$ and $J_2^*$ of a BPS black hole
obeys a constraint among themselves.
This can be seen from the fact that the three charges \eqref{AdS5BPScharges}
are parametrized by only two variables.
For future reference, we present a more general charge constraint
where $Q_{1,2,3}$ are not identified.
\bea\label{5dccgrav}
&& \left( Q_1Q_2Q_3 + \frac{\pi}{4G_5}J_1J_2 \right) \nn\\
&=& \left( Q_1+Q_2+Q_3 + \frac{\pi}{4G_5}\right)
\left( Q_1Q_2 + Q_2Q_3 + Q_3Q_1 - \frac{\pi}{4G_5}(J_1+J_2) \right)~.
\eea
The relation between the three charges in \eqref{AdS5BPScharges}
is obtained by simply setting $Q=Q_1=Q_2=Q_3$ in \eqref{5dccgrav}.

The BPS limit of AdS$_5$ black holes is in complete analogy with that of
AdS$_3$ black holes introduced in the previous section.
Supersymmetry requires saturation of a linear bound between the energy and the charges
that derives from unitarity, namely \eqref{eqn:unitboundmacro} and \eqref{AdS5unitbound}.
However, parametrization of the black hole solutions is non-linear in such a way that
the saturation of the unitary bound translates to a vanishing sum of two squares,
namely \eqref{eqn:susybound} and \eqref{AdS5sumofsq},
when written in terms of black hole parameters.
Therefore the supersymmetry of the black hole is amplified into two relations:
vanishing of temperature and charge constraints \eqref{eqn:JL=kL} and \eqref{5dccgrav}.

The entropy of the BPS black hole can be also obtained by substituting
the BPS values \eqref{AdS5BPSqr} for the auxiliary parameters in \eqref{AdS5Sinab}:
\bea\label{AdS5SBPSab}
S^* &=& 2 \pi \cdot \frac{\pi}{4 G_5}\frac{a+b}{(1-a) (1-b)}\sqrt{a+b+ab}~.
\eea
Ideally, we would like to have a formula for the entropy as a function of its charges,
which was not feasible in \eqref{AdS5Sinab} where we did so only implicitly via
the auxiliary parameters.
For the BPS black holes, this can be done, i.e. $a$ and $b$ in \eqref{AdS5SBPSab}
can be replaced by the charges $Q^*$, $J_1^*$ and $J_2^*$ via \eqref{AdS5BPScharges}.
Note that there is not a unique way to do so, because the charges parametrize
$a$ and $b$ redundantly, up to the relation \eqref{5dccgrav}.
A particularly nice expression has been found in \cite{Kim:2006he},
which in fact applies to a more general BPS black holes whose $Q_{1,2,3}$ are not identified:
\begin{eqnarray}\label{AdS5S}
S^* &=& 2\pi\sqrt{Q_1^*Q_2^*+Q_2^*Q_3^*+Q_3^*Q_1^*
-\frac{\pi}{4G_5} \left(J_1^* + J_2^* \right)}~.
\end{eqnarray}
%

\clearpage

\part{Entropy and Charges of Supersymmetric Black Holes}
\chapter{Black Hole Entropy from the Index}\label{sec:entropy}

In the first main part of this thesis that consists of this and the next chapter,
we derive two important properties of the supersymmetric black holes in
AdS space, namely the entropy and the charge constraint,
from the dual conformal field theories.

As we reviewed in chapter \ref{sec:BHs} for AdS$_3$ and AdS$_5$ black holes,
the supersymmetric AdS black holes have large entropy,
and their charges cannot take arbitrary values but must obey one constraint.
The asymptotically AdS black holes in gravity theories are known to be dual to
ensembles of quantum states in superconformal field theories in
one fewer dimensions \cite{Maldacena:1997re}.
Therefore, reproducing the properties of the black hole from the
microstates in the field theory
will shed light on understanding the gravity through quantum theories.

In this chapter, we address that the entropy of the black holes can be
accounted for by degeneracy of quantum states in the field theory.
We will first introduce the index, a powerful tool for enumerating
quantum states in supersymmetric field theories.
We will review some early attempts on using the index to count the black hole microstates,
then demonstrate how the difficulties encountered were overcome recently.
This chapter will conclude with a review of the entropy extremization principle,
which derives the black hole entropy by treating the index as a partition function
and performing Legendre transformation.

\section{Early Attempts}\label{sec:early}

Given the central position of the 5-dimensional type-IIB supergravity
and the 4-dimensional maximally supersymmetric
Yang-Mills theory in the AdS/CFT correspondence \cite{Maldacena:1997re},
it is not surprising that there have been many attempts to account for the
entropy of the AdS$_5$ black holes from the gauge theory.
In this section, we review some important developments
\cite{Aharony:2003sx,Kinney:2005ej,Romelsberger:2005eg}
that have paved the way for the later progress.

\subsection{The Superconformal Index}\label{sec:earlyindex}

Perhaps the most important development towards microscopic accounting
of the AdS black hole entropy is the introduction of the index,
also referred to as the superconformal index
\cite{Kinney:2005ej,Romelsberger:2005eg}.
It can be understood as a special case of the grand canonical partition function
for the Hilbert space of the theory, that has the remarkable property of being
invariant under continuous deformations of the theory,
thus allowing one to learn about the strongly coupled theory
by studying the weakly coupled theory.

Consider the $\cN=4$ Super-Yang-Mills theory in 4 dimensions.
Its symmetry group is the 4d $\cN=4$ superconformal group $PSU(2,2|4)$,
that consists of the 4d conformal group $SO(4,2) \sim SU(2,2)$,
the R-symmetry group $SU(4)$, and fermionic generators that transform under
both bosonic groups and complete the graded Lie group.

As we have mentioned in the context of AdS$_5$ black holes in section \ref{sec:AdS5BH},
we define the three Cartans of the conformal group as $E$, $J_1$ and $J_2$.
$E$ corresponds to the timelike part and therefore plays the role of energy,
and $J_{1,2}$ corresponds to each factor of
$SU(2)$ in the $SU(2) \times SU(2) \sim SO(4)$ Lorentz group.
We also define the three Cartans of the R-symmetry group as $Q_{1,2,3}$,
in such a way that they correspond to rotations within orthogonal 2-planes
among 6-dimensional rotations $SO(6) \sim SU(4)$.

In group theoretic contexts, it is often useful to use the Dynkin basis
instead of the orthogonal bases introduced in the previous paragraph.
The Dynkin basis $(E, j_1, j_2, R_1, R_2, R_3)$ is linearly related to the
orthogonal basis $(E, J_1, J_2, Q_1, Q_2, Q_3)$ above by
\bea
&& J_1 = \frac{j+\bar{j}}{2}~, \qquad \qquad ~~~~~~~
J_2 = \frac{j-\bar{j}}{2}~, \nn\\
&& Q_1 = R_2 + \frac{R_1+R_3}{2}~, \qquad
Q_2 = \frac{R_1+R_3}{2}~, \qquad
Q_3 = \frac{R_1-R_3}{2}~.
\eea
The energy $E$ is common to the two bases.

Every state in the Hilbert space, or every local operator of the theory,
must be grouped into representations of the symmetry algebra.
Therefore, it is always possible to find a basis of the states/operators
that diagonalize all 6 Cartans of the symmetry algebra.
Since we will always assume this diagonalization,
we do not distinguish notations for the symmetry operators
and for the corresponding eigenvalues.
So every state/operator has definite values of six quantum numbers
$(E, J_1, J_2, Q_1, Q_2, Q_3)$,
or equivalently $(E, j_1, j_2, R_1, R_2, R_3)$ in the Dynkin basis,
and it is possible to define the grand canonical partition function of the theory
as the following trace over the Hilbert space:
\bea\label{4dN4GCPF}
Z (\beta,\, \Delta_I,\, \omega_i) &\equiv& \mathrm{Tr}
\left[e^{- \beta E} e^{\Delta_I Q_I+\omega_i J_i} \right]~.
\eea
The partition function is a function of 6 variables:
$\beta$ which is usually understood as the inverse temperature,
and five chemical potentials $\Delta_{1,2,3}$ and $\omega_{1,2}$.
The role of the chemical potentials, as well as the inverse temperature,
is to weigh different states within the partition function.
We shall often refer to the factors $e^{-\beta}$, $e^{\Delta_I}$ and $e^{\omega_i}$
as fugacities.

The supersymmetric AdS$_5$ black holes discussed in section \ref{sec:AdS5BHBPS}
preserve $\frac{1}{16}$ of the supersymmetries, so are referred to as $\frac{1}{16}$-BPS.
Similarly, we expect the dual microstates in the gauge theory to preserve the
same amount of supersymmetries.
There are 32 Hermitian supersymmetry generators in $PSU(2,2|4)$:
$16$ Poincar\'e supercharges $Q^i_\alpha$, $\overline{Q}_{i\dot\alpha}$ and 
$16$ conformal supercharges $S_{i\alpha}$, $\overline{S}^i_{\dot\alpha}$,
that are Hermitian conjugates of the Poincar\'e supercharges in a radially quantized theory.
$i=1,2,3,4$ is the fundamental or anti-fundamental index for the
$SU(4)$ R-symmetry and $\alpha$ and $\dot\alpha$ are the doublet indices
for the Lorentz group $SU(2)_L\times SU(2)_R\sim SO(4)$.
We choose 2 of them, $Q\equiv Q^4_-$ and $S = Q^\dag\equiv S_4^-$,
as the preserved supercharges.
An important commutation relation among the $PSU(2,2|4)$ algebra is
\begin{equation}\label{QSalgebra}
2\{Q,Q^\dag\}=E-(Q_1+Q_2+Q_3+J_1+J_2)~,
\end{equation}
for our choice of the preserved supercharges.

For any state $|\psi \rangle$ of the SYM, the norm of $Q |\psi \rangle$
must be non-negative.
It follows that the eigenvalues of the $\frac{1}{16}$-BPS states must obey
\bea\label{SYMBPSbound}
E &\geq& Q_1+Q_2+Q_3+J_1+J_2~.
\eea
Moreover, the $\frac{1}{16}$-BPS states $|\psi_\mathrm{BPS} \rangle$
of the SYM are annihilated by the chosen supercharge:
$Q |\psi_\mathrm{BPS} \rangle = 0$.
It follows that the eigenvalues of the $\frac{1}{16}$-BPS states saturate \eqref{SYMBPSbound}:
\bea\label{SYMBPScond}
E &=& Q_1+Q_2+Q_3+J_1+J_2~.
\eea

We can adapt the grand canonical partition function \eqref{4dN4GCPF}
to the BPS states that satisfy \eqref{SYMBPScond}.
First rewrite \eqref{4dN4GCPF} as
\bea\label{4dN4GCPFBPS1}
Z (\beta,\, \Delta_I,\, \omega_i) &=& \mathrm{Tr}
\left[e^{- \beta (E-Q_1-Q_2-Q_3-J_1-J_2)}
e^{\tilde{\Delta}_I Q_I+ \tilde{\omega}_i J_i}  \right]~.
\eea
where $\tilde{\Delta}_I \equiv \Delta_I-\beta$ and $\tilde{\omega}_i \equiv \omega_i-\beta$.
Then, take the limit $\beta \to \infty$ while keeping the redefined chemical potentials
$\tilde{\Delta}_I$ and $\tilde{\omega}_i$ finite.
As a result, non-BPS states that do not saturate \eqref{SYMBPSbound}
will be suppressed,
and effectively the trace will sum only over the $\frac{1}{16}$-BPS states.
\bea\label{4dN4GCPFBPS2}
Z_\mathrm{BPS} (\tilde{\Delta}_I,\, \tilde{\omega}_i) &=&
\lim_{\beta \to \infty} \mathrm{Tr} \left[e^{- \beta (E-Q_1-Q_2-Q_3-J_1-J_2)}
e^{\tilde{\Delta}_I Q_I+ \tilde{\omega}_i J_i} \right] \nn\\
&=&
\mathrm{Tr}_\mathrm{BPS} \left[
e^{\tilde{\Delta}_I Q_I+ \tilde{\omega}_i J_i} \right]
\eea
This BPS partition function depends only on 5 chemical potentials.
Dependence on one fewer variables reflects that the $\frac{1}{16}$-BPS states
have only 5 independent charges due to \eqref{SYMBPScond}.

Now let us introduce another way to restrict the trace to the BPS states.
Consider a generic state $|\psi \rangle$ that has certain eigenvalues
$E$, $Q_I$ and $J_i$.
Since all states of the theory must organize into representations of the
symmetry algebra, it follows that another state $Q |\psi \rangle$ must also
exist in the Hilbert space.
Our choice of $Q$ is such that the eigenvalues of $Q |\psi \rangle$ is
$E + \frac12$, $Q_I + \frac12$ and $J_i - \frac12$.
Note from \eqref{SYMBPSbound} that if $|\psi \rangle$ does not saturate
the BPS bound, nor does $Q |\psi \rangle$:
the value of $E-Q_1-Q_2-Q_3-J_1-J_2$ is the same for both states.
Due to the nilpotency of the fermionic operator $Q$, $Q$ may be applied to
$|\psi \rangle$ only once.
All non-BPS states in the theory must appear in such a pair:
$|\psi \rangle$ and $Q |\psi \rangle$,
while the $\frac{1}{16}$-BPS states appears alone because
$Q |\psi_\mathrm{BPS} \rangle = 0$.
Therefore, if one tunes the chemical potentials in \eqref{4dN4GCPFBPS1} such that
\bea\label{makeitindex}
e^{\frac{\tilde{\Delta}_1+\tilde{\Delta}_2+\tilde{\Delta}_3-\tilde{\omega}_1-\tilde{\omega}_2}{2}}
&=& -1~,
\eea
then contributions to \eqref{4dN4GCPFBPS1} from $|\psi \rangle$ and
from $Q |\psi \rangle$ exactly cancel each other,
so \eqref{4dN4GCPFBPS1} receives contributions only from $\frac{1}{16}$-BPS states
and the $\beta$-dependence automatically vanishes.
The grand canonical partition function defined as such is the index,
also known as the superconformal index to emphasize that this is
an adaptation of the Witten index \cite{Witten:1982df}
to the superconformal field theory.
\bea\label{4dN4index1}
\mathcal{I} (\tilde{\Delta}_I,\, \tilde{\omega}_i) &=&
\mathrm{Tr} \left[e^{- \beta (E-Q_1-Q_2-Q_3-J_1-J_2)}
e^{\tilde{\Delta}_I Q_I+ \tilde{\omega}_i J_i} \right] \nn\\
&=&
\mathrm{Tr}_\mathrm{BPS} \left[
e^{\tilde{\Delta}_I Q_I+ \tilde{\omega}_i J_i} \right]~,
\hspace{1.4cm} \mathrm{where} ~~~
e^{\frac{\tilde{\Delta}_1+\tilde{\Delta}_2+\tilde{\Delta}_3-\tilde{\omega}_1-\tilde{\omega}_2}{2}}
= -1~.
\eea

It is important to notice that, due to the condition \eqref{makeitindex},
the index is a function of only 4 independent chemical potentials,
one fewer than the partition function \eqref{4dN4GCPFBPS2}
with a similar definition,
so the index contains less information than the partition function.
For example, suppose there are two BPS states in the Hilbert space
whose charges $Q_I$ differ by $\frac{n}{2}$ and $J_i$ by $-\frac{n}{2}$,
where $n$ is an integer.
If $n$ is even, the index will not distinguish contributions from both states
and only tell us that there are two states along a line in the 5-dimensional charge space.
If $n$ is odd, contributions from both states will cancel each other
and the index will not notice these states in any way even though they are BPS states.

Despite this loss of information, the index has a remarkable advantage
over the partition function, in that it is coupling independent.
A superconformal field theory may receive continuous deformations
due to interactions, and as a result the spectrum of the Hilbert space is shifted.
In this process, it is possible that a BPS state is lifted by an anomalous dimension
and become non-BPS.
The grand canonical partition function \eqref{4dN4GCPFBPS2} will change
under this process, as a state suddenly drops out of the range of summation.
However, such a process may only happen in a limited manner.
Since all states in the Hilbert space must organize into representations of
$PSU(2,2|4)$ in any case, such transitions between BPS and non-BPS states
may occur only if there is a set of representations that contain a BPS state
that is continuously isomorphic to a set of representations that do not.
Such limited relations between sets of representations are called
recombination rules, see \cite{Cordova:2016emh} for an extensive list of them.

The index \eqref{4dN4index1}, unlike the partition function \eqref{4dN4GCPFBPS2},
is invariant under the recombinations.
If a BPS state may be lifted, i.e. it is not protected under recombination rules,
then it must be that there is another BPS state whose charges differ from
the former as in the example of the previous paragraph with odd $n$,
and that they must participate in the recombination rule together.
This is to guarantee that a lift from the BPS state $|\psi\rangle$ has its superpartner
$Q |\psi\rangle$ somewhere in the Hilbert space.
The index had not depended on these two BPS states even though they were BPS states,
so it does not change under the recombination process.
From this, it is clear that the loss of information for the index is precisely
what gives it the powerful property of coupling independence.

Historically, the minus sign on the right hand side of \eqref{makeitindex}
has been replaced with an explicit factor $(-1)^F$ where $F$ is a fermion number operator.
That is, the index was originally defined as
\bea\label{4dN4index2}
\mathcal{I} (\tilde{\Delta}_I,\, \tilde{\omega}_i) &=&
\mathrm{Tr} \left[(-1)^F 
e^{\tilde{\Delta}_I Q_I+ \tilde{\omega}_i J_i} \right]~,
\hspace{1.4cm} \mathrm{where} ~~~
e^{\frac{\tilde{\Delta}_1+\tilde{\Delta}_2+\tilde{\Delta}_3-\tilde{\omega}_1-\tilde{\omega}_2}{2}}
= 1~.
\eea
The two definitions are ultimately equivalent because the
$(-1)^F$ factor can be replaced by $e^{2 \pi i J_1}$ among many other
possible choices \cite{Copetti:2020dil,Cassani:2021fyv},
effectively shifting $\tilde\omega_1$ by $2 \pi i$
and flipping the sign on the right hand side of \eqref{makeitindex}.
However, the modern definition \eqref{4dN4index1} is more suggestive
that the chemical potentials may be complex numbers,
which has been the key to the recent advance as will be reviewed later in this chapter.

\subsection{The Index as a Matrix Integral}\label{sec:earlymm}

Taking advantage of the coupling independence of the index,
attempts have been made to count the number of microstates of the
weakly coupled $\mathcal{N}=4$ SYM to account for the AdS$_5$ black hole
entropy, which is dual to the strongly coupled $\mathcal{N}=4$ SYM
with a large-$N$ gauge group $SU(N)$.
We review an early attempt of \cite{Kinney:2005ej} where the index
has been computed using a unitary matrix model.

The free $\mathcal{N}=4$ SYM consists of six real scalars,
eight fermions and a gauge field.
Each of them transforms in the adjoint representation of the gauge group $SU(N)$:
\begin{eqnarray}\label{N4SYMfields}
  \textrm{vector}&:&A_\mu\sim A_{\alpha\dot\beta}\ , \hspace{4.3cm}
  (\mu=1,2,3,4\ , \ \alpha=\pm\ ,\ \dot\beta=\dot{\pm}) \nn\\
  \textrm{scalar}&:&\Phi_{ij}(=-\Phi_{ji})\ ,\ \overline{\Phi}^{ij}\sim
  {\textstyle \frac{1}{2}}\epsilon^{ijkl}\Phi_{kl}
  \ ,\qquad (i,j,k,l=1,2,3,4)\nn\\
  \textrm{fermion}&:&\Psi_{i\alpha}\ ,\ \overline{\Psi}^i_{\dot\alpha}\ .
\end{eqnarray}
As explained above \eqref{QSalgebra},
$\alpha,\dot\alpha$ are the doublet indices for the Lorentz group,
$\mu$ is the vector index,
superscripts $i,j$ are for the fundamental representation of the $SU(4)$ R-symmetry,
while the subscripts are for the anti-fundamental representation.
Of these, 3 scalars $\bar\phi^m=\overline{\Phi}^{4m}$ where $m=1,2,3$,
3 chiralini $\psi_{m+}=-i\Psi_{m+}$,
2 gaugini $\bar\lambda_{\dot\alpha}=\overline{\Psi}^4_{\dot\alpha}$
and the gauge field $f_{++} = ({\sigma^{\mu \nu}}_{++})F_{\mu\nu}$
are $\frac{1}{16}$-BPS at the free, i.e. zero-loop $\mathcal{O}(g_\mathrm{YM}^0)$, level.
The $\frac{1}{16}$-BPS states are those whose charges satisfy \eqref{SYMBPScond}.

Each field also has a spacetime argument, and a field localized at different
points in the spacetime are considered separate degrees of freedom.
Equivalently, a field and its derivatives at the origin are separate degrees of freedom.
Therefore, any numbers of 4 derivatives $\partial_{\alpha \dot\alpha}$
in the 4d spacetime may act on each field.
Of these, only 2 derivatives $\partial_{+ \dot\alpha}$
preserve the $\frac{1}{16}$-BPSness,
i.e. commute with the preserved supercharge $Q$.

Finally, the free fields obey equations of motion.
Of these, only one equation of motion concerning gaugini:
\be\label{gauginoeom}
  \partial_{+\dot\alpha}\bar\lambda^{\dot\alpha}=0\ \Leftrightarrow\ 
  \partial_{+[\dot\alpha}\bar\lambda_{\dot{\beta}]}=0 ~,
\ee
is $\frac{1}{16}$-BPS.

These define the $\frac{1}{16}$-BPS letters:
the 9 free fields and any numbers of 2 derivatives acting on them,
modulo the gaugino equation of motion.

The free fields belong to a single super-representation of $PSU(2,2|4)$,
known as the free vector multiplet.
It is a representation that has 6 real scalars as its superconformal primaries,
and consists of descendants that can be obtained by acting the
$PSU(2,2|4)$ generators on the primary.
The descendants include other free fields, by action of supercharges,
and their derivatives, by action of momentum operators,
modulo the equation of motions which is derived from the symmetry algebra.
This representation has the name $B_1\overline{B}_1 [0;0]^{[0,1,0]}_1$
following notation of \cite{Cordova:2016emh}, to which we refer for
extensive information on representations of superconformal algebras.
$[0;0]$ indicates that the superconformal primary is a singlet under the Lorentz group,
$[0,1,0]$ indicates that it is a representation ${\bf 6}$ of the
R-symmetry group $SU(4) \sim SO(6)$,
and the subscript $1$ indicates its conformal dimension.
The letters and the subscripts $B_1$ and $\overline{B}_1$ indicate the
structure of the representation.
We summarize the $\frac{1}{16}$-BPS contents of the representation
$B_1\overline{B}_1 [0;0]^{[0,1,0]}_1$,
as well as their charges in both bases, in Table \ref{5dfmtable}.
In the Table, the charges of each entry were displayed in two bases:
the Dynkin basis $(j, \bar{j},R_1,R_2,R_3)$ and the orthogonal basis
$(J_1,J_2,Q_1,Q_2,Q_3)$. They are related by
\bea\label{Dynkinortho}
&& J_1 = \frac{j+\bar{j}}{2}~, \qquad \qquad ~~~~~~~
J_2 = \frac{j-\bar{j}}{2}~, \nn\\
&& Q_1 = R_2 + \frac{R_1+R_3}{2}~, \qquad
Q_2 = \frac{R_1+R_3}{2}~, \qquad
Q_3 = \frac{R_1-R_3}{2}~,
\eea
and $E$ is common in both bases.

\begin{table}\catcode`\-=12
\begin{center}
\begin{tabular}{| c | c | c | c c c c c | c c c c c |}
\hline
& Bosonic Rep. & $E$ & $j$ & $\bar{j}$ & $R_1$ & $R_2$ & $R_3$ &
$J_1$ & $J_2$ & $Q_1$ & $Q_2$ & $Q_3$ \\
\hline
\hline
\multirow{9}{*}{Free fields} & \multirow{3}{*}{$[0;0]^{[0,1,0]}_1$} & 1 & 0 & 0 & 0 & 1 & 0 & 0 & 0 & 1 & 0 & 0 \\
& & 1 & 0 & 0 & 1 & $-1$ & 1 & 0 & 0 & 0 & 1 & 0 \\
& & 1 & 0 & 0 & 1 & 0 & $-1$ & 0 & 0 & 0 & 0 & 1 \\
\cline{2-13}
& \multirow{3}{*}{$[1;0]^{[0,0,1]}_\frac32$} & $\frac32$ & 1 & 0 & 0 & 0 & 1 &
$\frac12$ & $\frac12$ & $\frac12$ & $\frac12$ & $-\frac12$ \\
& & $\frac32$ & 1 & 0 & 0 & 1 & $-1$ &
$\frac12$ & $\frac12$ & $\frac12$ & $-\frac12$ & $\frac12$ \\
& & $\frac32$ & 1 & 0 & 1 & $-1$ & 0 &
$\frac12$ & $\frac12$ & $-\frac12$ & $\frac12$ & $\frac12$ \\
\cline{2-13}
& \multirow{2}{*}{$[0;1]^{[1,0,0]}_\frac32$} & $\frac32$ & 0 & 1 & 1 & 0 & 0 &
$\frac12$ & $-\frac12$ & $\frac12$ & $\frac12$ & $\frac12$ \\
& & $\frac32$ & 0 & $-1$ & 1 & 0 & 0 &
$-\frac12$ & $\frac12$ & $\frac12$ & $\frac12$ & $\frac12$ \\
\cline{2-13}
& $[2;0]^{[0,0,0]}_2$ & 2 & 2 & 0 & 0 & 0 & 0 &
1 & 1 &0 & 0 & 0 \\
\hline
\hline
Eq. of motion & $[1;0]^{[1,0,0]}_\frac52$ & $\frac52$ & 1 & 0 & 1 & 0 & 0 &
$\frac12$ & $\frac12$ & $\frac12$ & $\frac12$ & $\frac12$ \\
\hline
\hline
\multirow{2}{*}{Derivatives} & \multirow{2}{*}{$[1;1]^{[0,0,0]}_1$} & 1 & 1 & 1 & 0 & 0 & 0 & 1 & 0 & 0 & 0 & 0 \\
& & 1 & 1 & $-1$ & 0 & 0 & 0 & 0 & 1 & 0 & 0 & 0 \\
\hline
\end{tabular}
\caption{\label{5dfmtable}
Components of the BPS operators in the free vector multiplet
$B_1 \bar{B}_1 [0;0]^{[0,1,0]}_1$.
The first $9$ rows are free fields, followed by the equation of motion
and $2$ derivatives.}
\end{center}
\end{table}

The BPS letters freely generate the Fock space.
That is, a product of arbitrary numbers of each bosonic BPS letter,
times a product of either 0 or 1 of each fermionic BPS letter
is a BPS operator included in the Fock space.
Note that there are infinite numbers of BPS letters:
corresponding to each free field, any of its derivatives is a new BPS letter.
Furthermore, each field transforms as an adjoint representation of the
gauge group $SU(N)$,
so a field should be understood as a set of $N^2-1$ independent degrees of freedom
each with its own gauge charges.
Finally, only the gauge singlets are considered the physical degrees of freedom.
Therefore, the Hilbert space of the $\cN=4$ SYM is a projection
onto gauge singlets of the BPS Fock space.

Now, let us translate the structure of the Hilbert space into the index
\eqref{4dN4index2} defined in the previous subsection.

Suppose there is a bosonic BPS letter.
Let its contribution to the trace be $x_B \equiv e^{\frac{\tilde{\Delta}_1
+\tilde{\Delta}_2+\tilde{\Delta}_3-\tilde{\omega}_1-\tilde{\omega}_2}{2}}$.
The contribution from an arbitrary number of the letter is
\bea\label{countB}
1+x_B+x_B^2 + \cdots &=& \frac{1}{1-x_B} \nn\\
&=& \exp[-\log (1-x_B)] \nn\\
&=& \exp\left[ \sum_{n=1}^\infty \frac{x_B^n}{n} \right] \nn\\
&\equiv& \mathrm{PE}[x_B]~.
\eea
In the last line we have defined the Plethystic exponential.
Instead, suppose there is a fermionic BPS letter,
and let its contribution to the trace be $x_F$.
Here we pull out the $(-1)^F$ factor explicitly, using the definition of the index
as in \eqref{4dN4index2}.
Therefore, the contribution from all allowed number of the letter is
\bea\label{countF}
1-x_F &=& \exp[\log (1-x_F)] \nn\\
&=& \exp\left[ -\sum_{n=1}^\infty \frac{x_F^n}{n} \right] \nn\\
&\equiv& \mathrm{PE}[-x_F]~.
\eea
Note that the plethystic exponential follows the rule of ordinary exponentials:
$\mathrm{PE}[-x_F] = \left( \mathrm{PE}[x_F] \right)^{-1}$ and
$\mathrm{PE}[x_1 + x_2] = \mathrm{PE}[x_1] \times \mathrm{PE}[x_2]$.
Therefore, the index over the BPS Fock space will be a Plethystic exponential
of the sum over all bosonic BPS letters minus the sum over all fermionic BPS letters:
\bea
\mathcal{I}_\mathrm{Fock} &=&
\mathrm{PE} \left[ \sum_\text{bosonic letters} x_B
- \sum_\text{fermionic letters} x_F \right]
\eea

Let us temporarily introduce the gauge fugacities $e^{i\alpha_a}$ ($a=1,\cdots,N$)
conjugate to the gauge charges $\zeta_a$ into the index,
that we will shortly project away.
Furthermore, let us replace the chemical potentials $e^{\tilde\Delta_I}$ and
$e^{\tilde\omega_i}$ in favor of the fugacities
\bea
x^2 = e^{\tilde\Delta_1}~, \quad y^2 = e^{\tilde\Delta_2}, \quad z^2 = e^{\tilde\Delta_3}~,
\quad p^2 = e^{\tilde\omega_1}~, \quad q^2 = e^{\tilde\omega_2}~.
\eea
So we define the index over the Fock space as
\bea\label{4dN4GCPFBPS5}
\mathcal{I}_\mathrm{Fock} (x,y,z,p,q; \zeta_{a}) &=&
\mathrm{Tr}_\mathrm{Fock}
\left[(-1)^F x^{2Q_1} y^{2Q_2} z^{2Q_3} p^{2J_1} q^{2J_2}
\prod_a e^{i\alpha_a \zeta_a} \right]~,
\eea
where $\frac{xyz}{pq} = 1$.
From Table \ref{5dfmtable}, we can read off the appropriate factors
that correspond to $x_B$ or to $x_F$ of the BPS letters.
Also including the gauge factor, we have
\bea\label{Fockindex}
\mathcal{I}_\mathrm{Fock} (x,y,z,p,q; \zeta_{a}) &=&
\mathrm{PE} \left[ f(x,y,z,p,q) \cdot \chi^{adj.}(\alpha_{a}) \right]~,
\eea
where
\bea\label{spindex}
f(x,y,z,p,q) &=& \frac{x^2+y^2+z^2-xyzpq
\left(\frac{1}{x^2} + \frac{1}{y^2} + \frac{1}{z^2} + \frac{1}{p^2} + \frac{1}{q^2} - 1\right)
+ p^2q^2}{(1-p^2)(1-q^2)} \nn\\
&=& 1- \frac{(1-x^2)(1-y^2)(1-z^2)}{(1-p^2)(1-q^2)}
\eea
is the single particle index, and
\bea
\chi^{adj.}(\alpha_{a}) &=& \sum_{a,b=1}^N e^{i(\alpha_a-\alpha_b)}
\eea
is the character of the $U(N)$ adjoint representation.
Note the role of the equation of motion and the derivatives in \eqref{spindex}:
the $-1$ inside the parenthesis and the geometric series of $p^2$ and $q^2$.
For the second line of \eqref{spindex}, we used $xyz=pq$.

Finally, the index is a projection of \eqref{Fockindex} onto gauge singlets.
The projection can be done by integrating over the gauge fugacities with the Haar measure:
\bea\label{Focktoindex}
\mathcal{I} (x,y,z,p,q) &=& \oint d\mu[\alpha_{a}]~
\mathcal{I}_\mathrm{Fock} (x,y,z,p,q; \alpha_{a})~,
\eea
where
\bea\label{Haarmeasure}
\oint d\mu[\alpha_{a}] &=& \frac{1}{N!} \cdot 
\int_0^{2\pi} \prod_{a=1}^N \frac{d\alpha_a}{2\pi} \cdot
\prod_{a,b=1}^N \left( 1-e^{i(\alpha_a-\alpha_b)} \right) \nn\\
&=& \frac{1}{N!} \cdot 
\int_0^{2\pi} \prod_{a=1}^N \frac{d\alpha_a}{2\pi} \cdot
\mathrm{PE} \left[ - \sum_{a,b=1}^N e^{i(\alpha_a-\alpha_b)} \right]
\eea
Collecting \eqref{Fockindex}-\eqref{Haarmeasure},
one obtains the following matrix integral formula for the index of the $\mathcal{N}=4$ SYM:
\bea\label{4dN4index3}
\mathcal{I} (x,y,z,p,q) \!&\!=\!&\! \frac{1}{N!} 
\int_0^{2\pi} \prod_{a=1}^N \frac{d\alpha_a}{2\pi} \cdot
\mathrm{PE} \left[ -(1- f(x,y,z,p,q)) \cdot \sum_{a,b=1}^N e^{i(\alpha_a-\alpha_b)} \right]~, \\
\!&\!=\!&\! \frac{1}{N!}
\int_0^{2\pi} \prod_{a=1}^N \frac{d\alpha_a}{2\pi} \cdot
\exp \left[ -\sum_{n=1}^\infty \sum_{a,b=1}^N \frac{1}{n}
\left(1- f(x^n,y^n,z^n,p^n,q^n) \right) \cdot e^{in(\alpha_a-\alpha_b)} \right]~. \nn
\eea

\eqref{4dN4index3} is an integral over $N$ angles, but we take $N$ to be very large
because the validity of supergravity approximation of the string theory,
i.e. the small string length scale limit, is related to the large-$N$ limit
via the AdS/CFT correspondence.
Since the large number of integration variables $\alpha_a$ appear
symmetrically in the integrand, it is useful to think of a distribution of $N$ variables
within the range $[0,2\pi)$ instead of their individual values.
Let $\rho(\alpha)$ be the distribution function of the angles $\alpha_a$,
normalized such that $\int_0^{2\pi} d\theta \rho(\alpha) = 1$.
\eqref{4dN4index3} becomes a functional integral over the distribution function:
\bea\label{indexaseffaction}
\mathcal{I} (x,y,z,p,q) &=& \frac{1}{N!} \int[d\rho] e^{-S[\rho(\alpha)]}~,
\eea
where $S[\rho(\alpha)]$ can be thought of as an effective action functional:
\bea
S[\rho(\alpha)] &=& N^2 \sum_{n=1}^\infty
\int_0^{2\pi} d\alpha_1 d\alpha_2 \rho(\alpha_1) \rho(\alpha_2) e^{in(\alpha_1-\alpha_2)}
\cdot \frac{1}{n} \left( 1-f(x^n,y^n,z^n,p^n,q^n) \right) \nn\\
&=& N^2 \sum_{n=1}^\infty \frac{1}{n} \left( 1-f(x^n,y^n,z^n,p^n,q^n) \right)
\cdot |\rho_n|^2~,
\eea
where
\bea
\rho_n &\equiv& \int_0^{2\pi} d\alpha \rho(\alpha) e^{in\alpha}~,
\eea
is a Fourier coefficient of the distribution function $\rho(\alpha)$.

Since $\rho(\alpha)$ has been normalized so that it is independent of $N$,
$S[\rho(\alpha)]$ is proportional to $N^2$.
Thus, as $N\to\infty$, the functional integral \eqref{indexaseffaction} will be
strongly dominated by a function $\rho(\alpha)$ that minimizes
the effective action $S[\rho(\alpha)]$.

Meanwhile, note from (\ref{spindex}) that
\bea
1-f(x^n,y^n,z^n,p^n,q^n) &=& \frac{(1-x^{2n})(1-y^{2n})(1-z^{2n})}{(1-p^{2n})(1-q^{2n})}
~>~ 0~,
\eea
as long as $0 < x,y,z,p,q < 1$.
The assumption that each fugacity is smaller than $1$ is natural,
since otherwise the index defined as a trace over infinite dimensional
Hilbert space would be divergent.
For example, suppose $x>1$.
Among the bosonic BPS letters (\ref{spindex}),
there is one that contributes $x^2$,
then the contribution from arbitrary numbers of this letter would be divergent,
$x^2 + x^4 + x^6 + \cdots$.

Therefore, the minimum of the action $S[\rho(\alpha)]$ corresponds to
all Fourier coefficients vanishing: $\rho_1 = \cdots = 0$.
This indicates a constant function $\rho(\alpha)$, or the uniform distribution of the
eigenvalues $\alpha_a$, also known as the confined phase
\cite{Gross:1980he,Wadia:1980cp,Witten:1998zw}.
Importantly, $S[\rho(\alpha)] = 0$ for such a distribution,
and therefore the index scales as
\bea\label{indexN0}
\mathcal{I} (x,y,z,p,q) &=& e^{\mathcal{O}(N^0)}~.
\eea

Recall from \eqref{AdS5BPScharges} that the charges $Q_I$ and $J_i$
of the AdS$_5$ black holes all scale as $\frac{\pi}{4G_5}$,
and therefore the entropy (\ref{AdS5S}) also scales as such $\frac{\pi}{4G_5}$.
This factor is related to the rank of the gauge group $N$
via the AdS/CFT correspondence:
\bea
\frac{\pi}{4G_5} = \frac{N^2}{2}~.
\eea
Therefore, it is expected for a successful microscopic accounting of the black hole entropy,
that the index scales as $e^{\mathcal{O}(N^2)}$.
In this sense \eqref{indexN0} does not account for the black hole entropy.

\cite{Kinney:2005ej} went further to evaluate the right hand side of (\ref{indexN0}),
and showed that the result corresponds to a gas of supergravitons.
The gas of supergravitons will be discussed in some detail in section \ref{sec:gravitons}.
However, both here and there, the important point is that the gas of supergravitons
do not exhibit a large enough degeneracy to account for the microscopics of
the black hole entropy, and our goal is to find contributions
other than the supergravitons.

As noted in \cite{Kinney:2005ej}, it is not a contradiction that the index
\eqref{indexN0} does not capture the black hole entropy.
The index counts the BPS states only up to cancellations between
bosons and fermions whose charges differ in certain direction.
For this reason, there have been various efforts to study the BPS operators themselves
--- not just counting them --- that are counted by the index, for example
\cite{Berkooz:2006wc,Grant:2008sk,Berkooz:2008gc,Chang:2013fba},
but those have not been fully successful either.
The last part of this thesis addresses progress in this direction.

\section{Amplifying the Index by Complex Chemical Potentials}\label{sec:ccp}

The AdS black hole entropy has finally been accounted for by the
number of microstates in the gauge theories only very recently.
Initiated by successes in magnetically charged black holes in AdS$_4 \times S^7$
from the topologically twisted index and supersymmetric localization
\cite{Benini:2015noa,Benini:2015eyy,Benini:2016rke,Azzurli:2017kxo,Hosseini:2017fjo}
and hinted by \cite{Hosseini:2017mds},
a major success for the AdS$_5$ black holes has been made in
\cite{Cabo-Bizet:2018ehj,Choi:2018hmj,Benini:2018ywd},
followed by many contributions for AdS black holes in various dimensions.
See \cite{Hosseini:2018dob,Choi:2018fdc,Choi:2019miv,
GonzalezLezcano:2019nca,Choi:2019zpz,Nian:2019pxj,Hosseini:2019iad,
Benini:2019dyp,Choi:2019dfu,GonzalezLezcano:2020yeb,
Murthy:2020rbd,Agarwal:2020zwm,Cabo-Bizet:2020nkr,Copetti:2020dil,
Larsen:2021wnu,Choi:2021rxi,David:2021qaa,Cassani:2022lrk,
Aharony:2024ntg}
among many others.

Although different methods have been explored,
the key difference that distinguishes the modern approach from the early attempts
introduced in section \ref{sec:early} was to let the chemical potentials
take complex values.
It turns out that, by attributing appropriate complex values to the
chemical potentials in \eqref{4dN4index1} while still respecting the
condition $e^{\frac{\tilde{\Delta}_1+\tilde{\Delta}_2+\tilde{\Delta}_3
-\tilde{\omega}_1-\tilde{\omega}_2}{2}} = -1$
for the index, it is possible to reduce the effect of cancellations between
bosonic and fermionic pairs dramatically,
so much that the index scales as $e^{\mathcal{O}(N^2)}$.
It is not just the scaling of the index, but its actual value in some subleading
orders in various limits and approximations that were matched with the
black hole entropy, for example \cite{GonzalezLezcano:2020yeb,
David:2021qaa,Cassani:2022lrk}.

In this section, we illustrate how a complex fugacity can `amplify'
the index by reducing the effect of cancellations.
We consider the index for an abelian theory with $U(1)$ gauge group,
and unrefine the fugacities as far as possible so the index becomes
a function of a single variable.
Coefficients of the series expansion of the index oscillate,
in a pattern that has essentially been observed in \cite{Agarwal:2020zwm}.
We show that they can be made to add up constructively by giving the fugacity
a resonating phase.

For this section, we take the following simplified definition of the index:
\bea\label{4dN4indexur}
\mathcal{I} (x) &=&
\mathrm{Tr} \left[(-1)^F x^{2Q_1+2Q_2+2Q_3+3J_1+3J_2} \right]
~\equiv~ 
\mathrm{Tr} \left[(-1)^F x^{\cJ} \right]
\eea
where we have defined the `overall' charge
\be\label{defcJ}
\cJ \equiv 2Q_1+2Q_2+2Q_3+3J_1+3J_2~,
\ee
for the last equality.
\eqref{4dN4indexur} is nothing more than the index \eqref{4dN4index2}
where we have taken $x^2 = e^{\Delta_1} = e^{\Delta_2}= e^{\Delta_3}$
and $x^3 = e^{\omega_1} = e^{\omega_2}$.
Note that this is compatible with the condition
$e^{\frac{\tilde{\Delta}_1+\tilde{\Delta}_2+\tilde{\Delta}_3
-\tilde{\omega}_1-\tilde{\omega}_2}{2}} = 1$.

One can compute the index on a computer using the formula \eqref{4dN4index3}, 
except that there is no matrix integral since we take the $U(1)$ gauge group,
so simply
\bea\label{4dN4indexU1}
\mathcal{I} (x) &=&
\exp \left[ -\sum_{n=1}^\infty \frac{1}{n} \left(1- f(x^n) \right) \right]~, \nn
\eea
where
\bea\label{spindexU1}
f(x) &=& \frac{3x^2 - 2x^3 - 3x^4 + 2x^6}{(1-x^3)^2} \nn\\
&=& 1- \frac{(1-x^2)^3}{(1-x^3)^2}~.
\eea
For generic $N$, even for moderate values of $N$,
the matrix integral for projecting out the gauge singlets is the bottleneck for
computing the index as a series expansion \cite{Murthy:2020rbd,Agarwal:2020zwm}.
So for $N=1$, the computation simplifies greatly and one can series expand
$\mathcal{I} (x)$ until very high orders of $x$.
For discussions in this section, we have computed until $x^{10000}$.
Also, as can be anticipated from \eqref{spindexU1}, the series expansion is
regular: there is no fractional powers of $x$.
The fractional powers of $x$ and their analogues in various setups
have confused many researchers.
For example, in \cite{Copetti:2020dil} the authors used our $x^3$ as their $x$,
and as a result their $\log x$ acquired a period $6 \pi i$,
causing confusions with branch cuts.

\begin{figure}
\begin{center}
\includegraphics[width=0.49\textwidth]{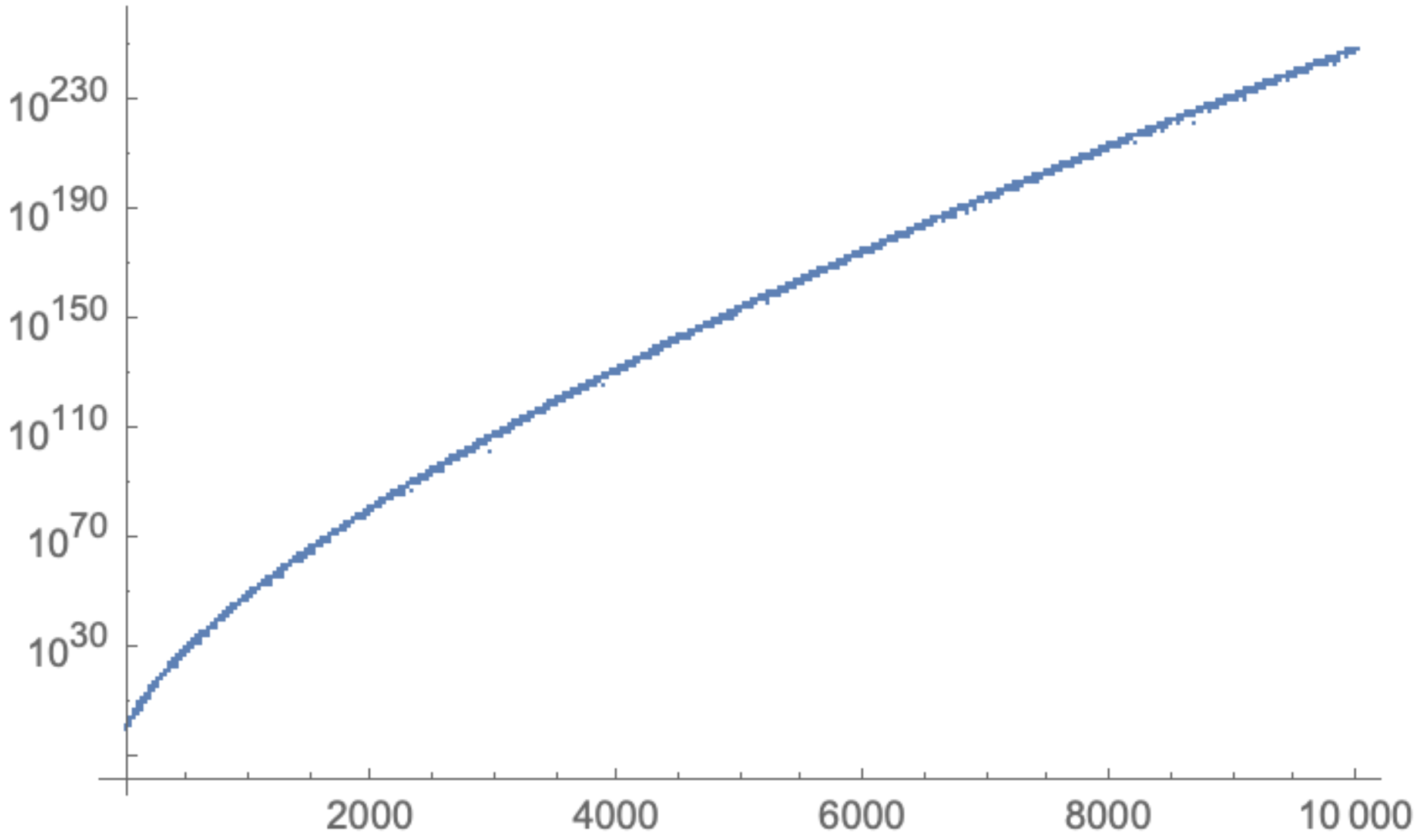}
\includegraphics[width=0.49\textwidth]{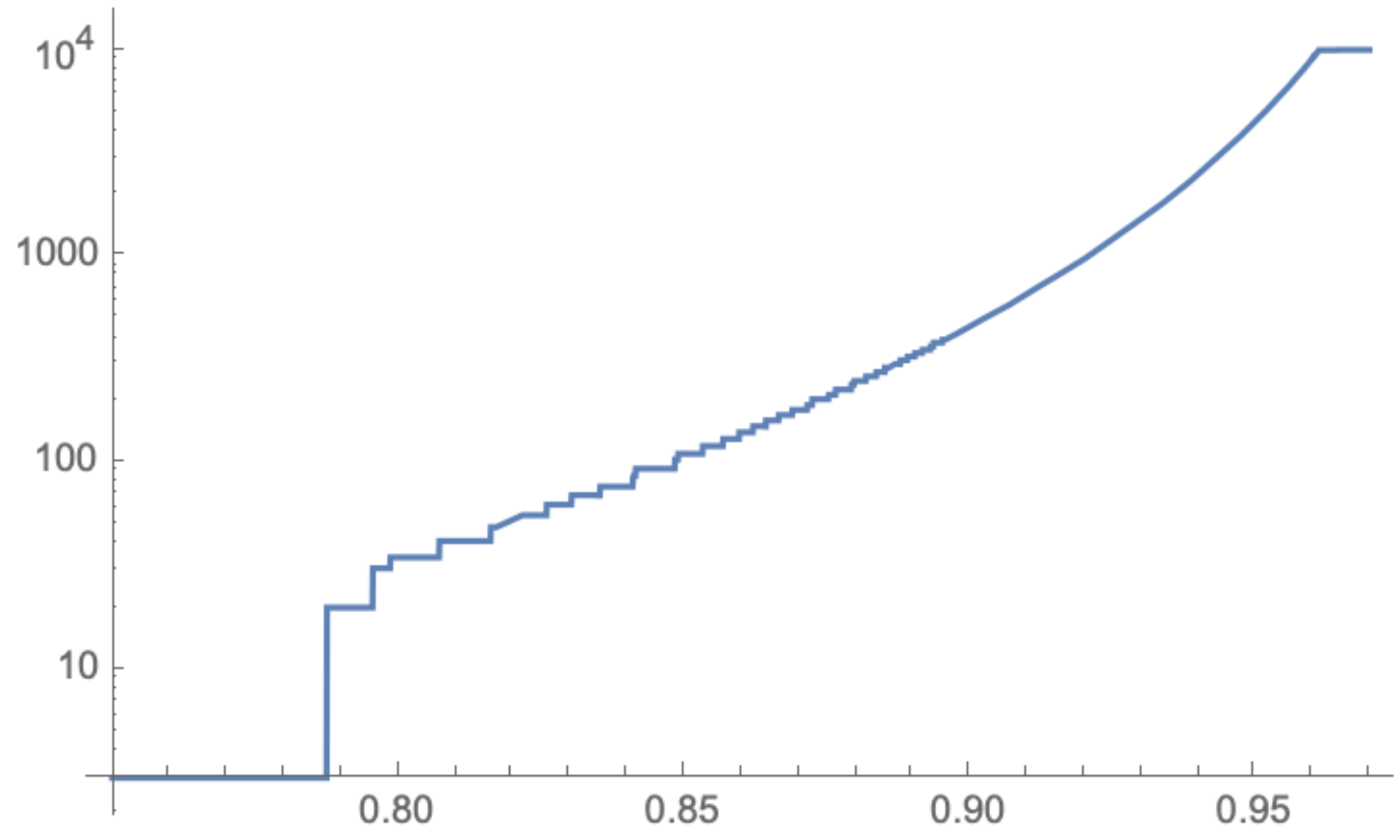}
\caption{\label{Fig:U1}
Index coefficients $|\Omega_\cJ|$ for each $0 \leq \cJ \leq 10000$ (left),
$\cJ_\mathrm{max} (x)$ that maximizes $|\Omega_\cJ \, x^\cJ|$
for given $0.75<x<0.97$ (right). Both drawn in log scale.}
\end{center}
\end{figure}

Let us denote the coefficients of the series expansion by $\Omega_\cJ$:
\bea
\mathcal{I} (x) &=& \sum_{j=0}^\infty \Omega_\cJ x^\cJ~.
\eea
All $\Omega_\cJ$ are integers, not necessarily non-negative because
fermions contribute negatively.
We plot the growth of $|\Omega_\cJ|$ in the left panel of Figure \ref{Fig:U1},
in logarithmic scale.
It is concave down in the logarithmic scale, indicating that the growth of
$|\Omega_\cJ|$ with $\cJ$ is sub-exponential.
As a result, for any given $0<x<1$, $|\Omega_\cJ \, x^\cJ|$ initially grows with $\cJ$
but after a certain point where the slope of $\log \Omega_\cJ$ compensates
the negative $\log x^\cJ = - \cJ \log (1/x)$, it starts to attenuate.
Therefore, for each $0<x<1$ there exists some value $\cJ_\mathrm{max} (x)$
that maximizes $|\Omega_\cJ \, x^\cJ|$.
This value of $\cJ$, as a function of $x$, is plotted in the right panel of Figure \ref{Fig:U1},
again in logarithmic scale.
For $x \lesssim 0.79$, $\Omega_2 \, x^2 = 3x^2$ is the largest contribution to the index.
Such small values of $x$ yield non-generic situations and are uninformative.
For higher values of $x$, $\cJ_\mathrm{max} (x)$ grows super-exponentially,
and for $x \gtrsim 0.96$, $\cJ_\mathrm{max} (x)$ is larger than $10^4$,
so it is not captured in Figure \ref{Fig:U1}.

\begin{figure}
\begin{center}
\includegraphics[width=0.75\textwidth]{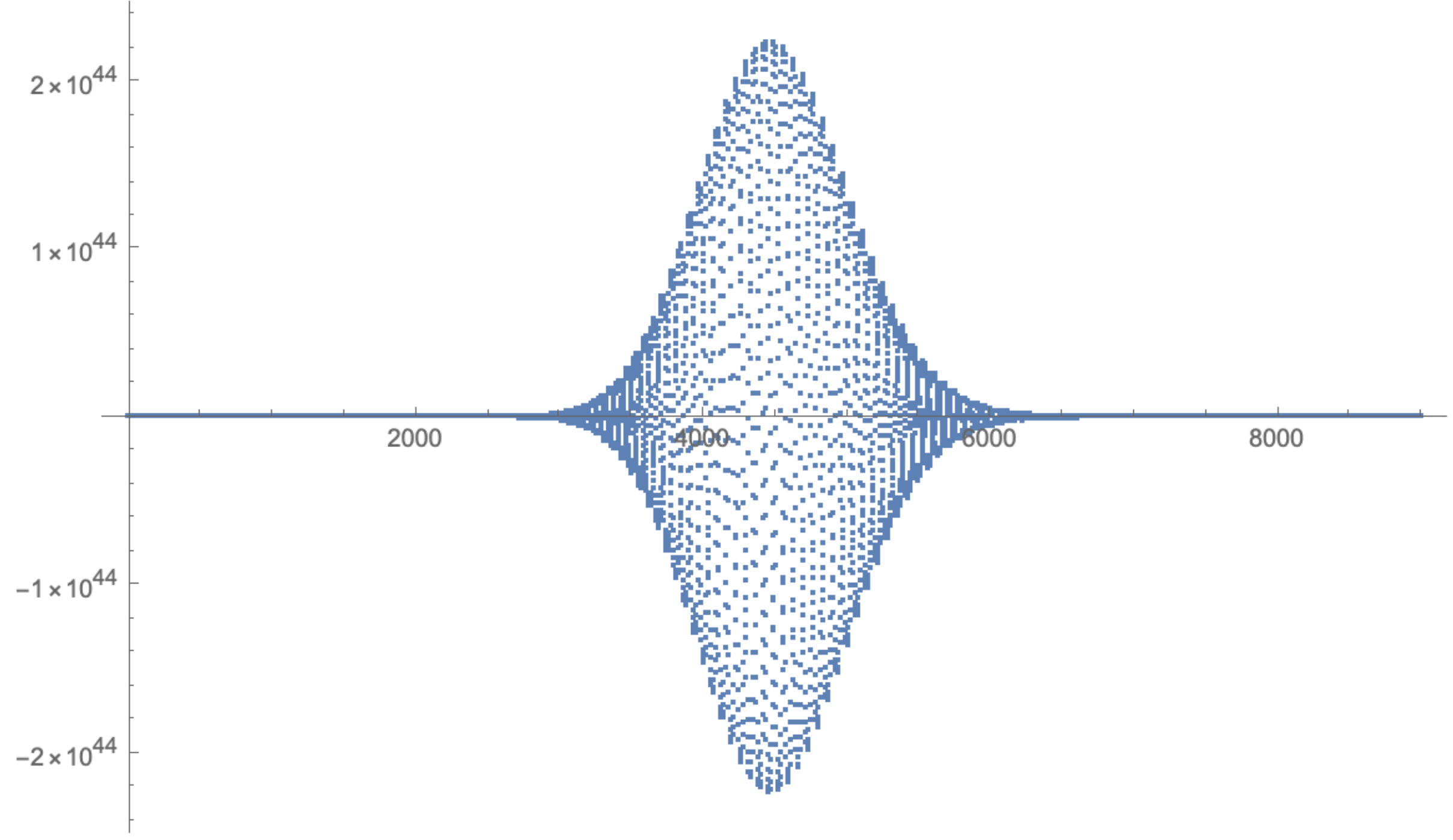}
\caption{\label{Fig:U14500} $\Omega_\cJ \, x^\cJ$ where $x=0.95$,
for each $\cJ$, in linear scale.}
\end{center}
\end{figure}

Now let us take $x=0.95$.
As one can read from Figure \ref{Fig:U1},
$|\Omega_\cJ \, 0.95^\cJ|$ is maximized around $\cJ \approx 4500$.
We plot $\Omega_\cJ \, 0.95^\cJ$, not its magnitude but with sign,
in Figure \ref{Fig:U14500}.
At its maximum which is around $\cJ \approx 4500$,
the magnitude $|\Omega_\cJ \, 0.95^\cJ| \approx 2 \times 10^{44}$.
However, if one looks at the individual coefficients around $\cJ \approx 4500$,
the sign and even the magnitude of each $\Omega_\cJ \, 0.95^\cJ$ oscillates wildly.
If one sums over the contributions from different $\cJ$ simply by
$\sum_\cJ \Omega_\cJ \, 0.95^\cJ$ to compute the index as a single number,
there will be massive cancellations between adjacent terms.

\begin{figure}
\begin{center}
\includegraphics[width=0.49\textwidth]{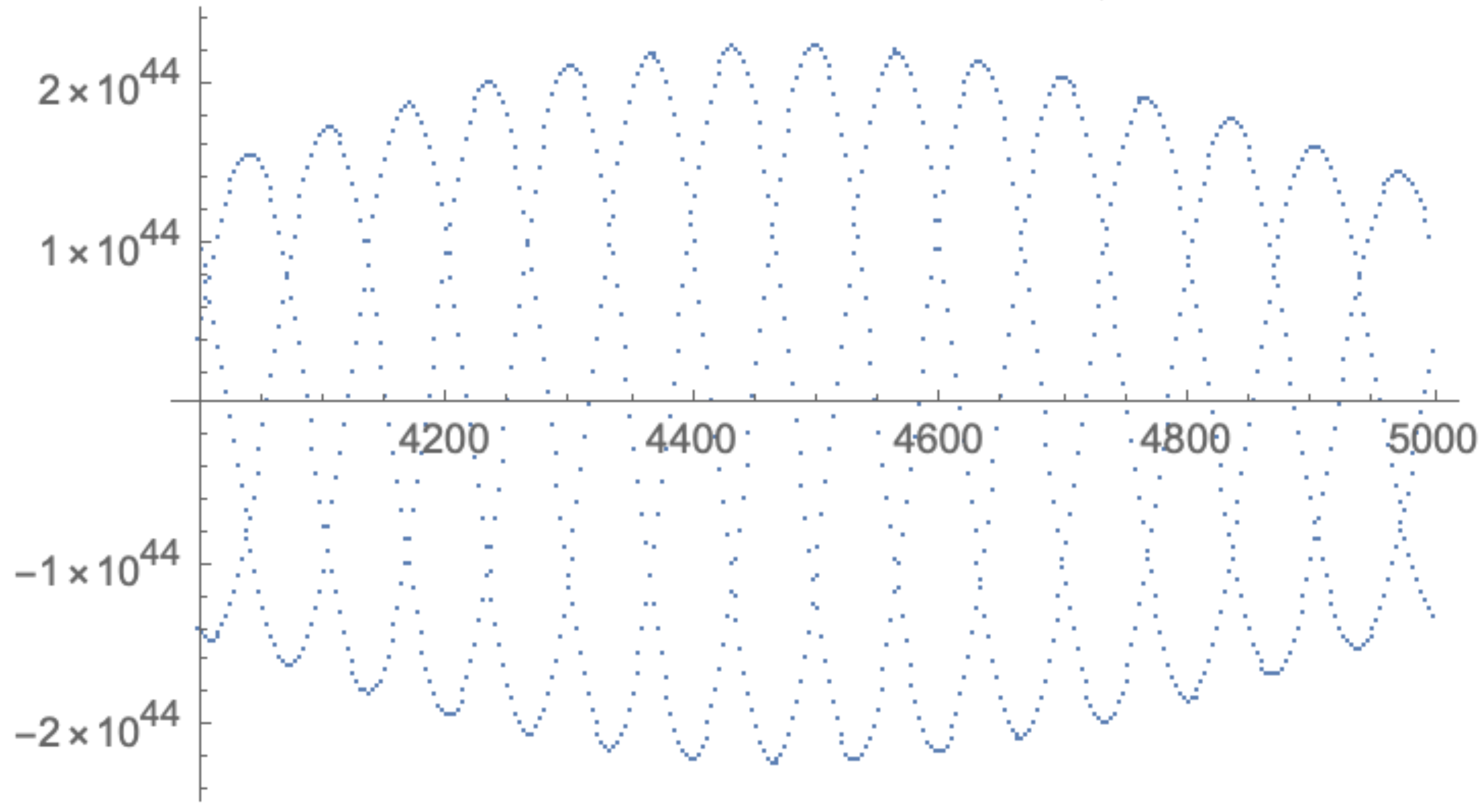}
\includegraphics[width=0.49\textwidth]{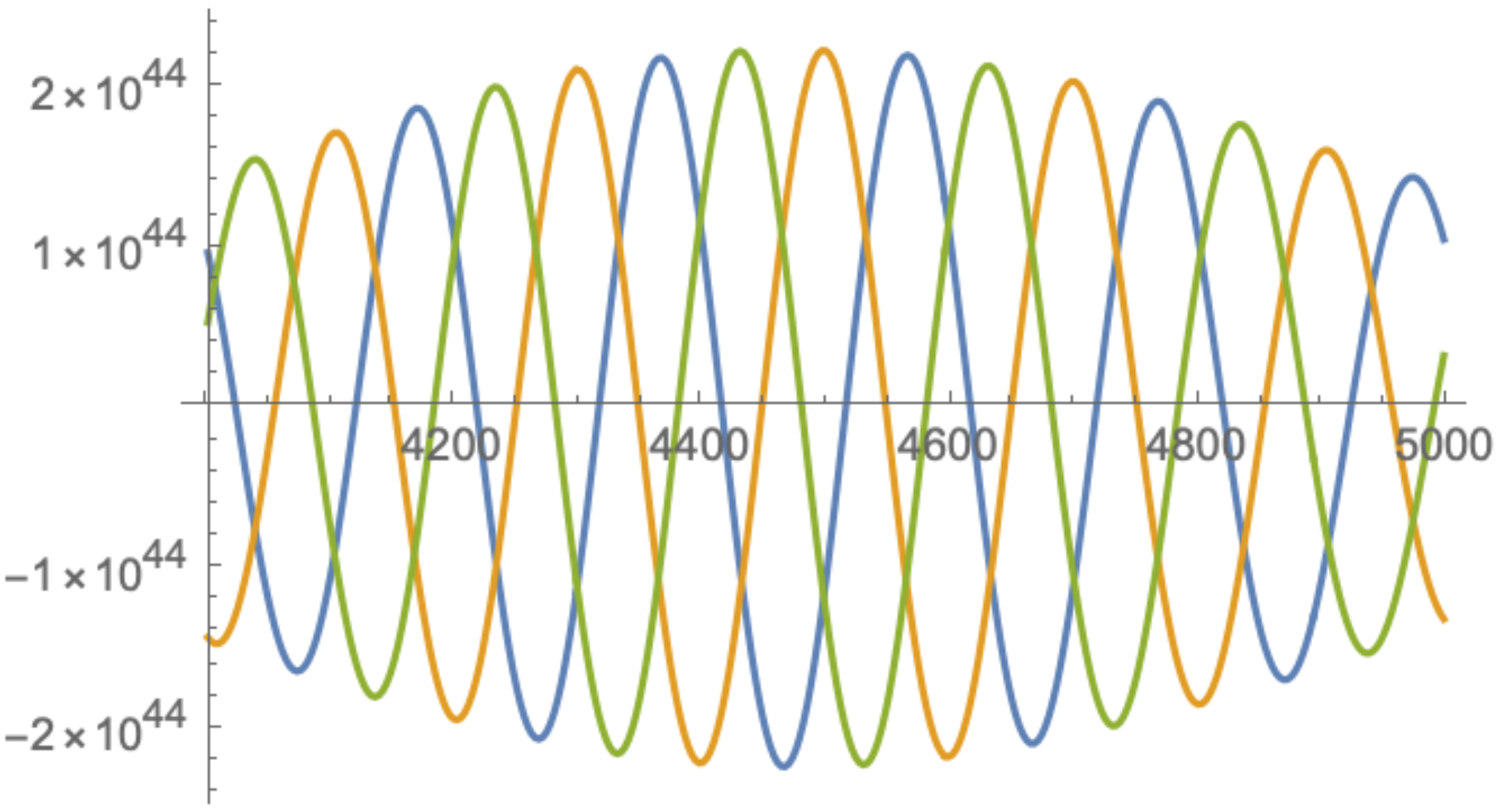}
\caption{\label{Fig:U14500zoom}
Figure \ref{Fig:U14500} zoomed into $4000 < \cJ < 5000$ (left),
the same figure color-coded according to $\cJ$ mod 3 (right).}
\end{center}
\end{figure}

However, if one zooms in closely into around $\cJ \approx 4500$,
a clear pattern of oscillation arises, see the left panel of Figure \ref{Fig:U14500zoom}.
The pattern is such that three sine waves are superimposed,
equally spaced between each other.
The right panel of Figure \ref{Fig:U14500zoom} is designed to illustrate
this pattern. It is the same plot as the left,
but dots corresponding to different values of $\cJ$ mod 3 are color-coded respectively.
Clearly, the values of $\Omega_\cJ \, 0.95^\cJ$ for same values of $\cJ$ mod 3
form a sine wave.

\begin{figure}
\begin{center}
\includegraphics[width=0.75\textwidth]{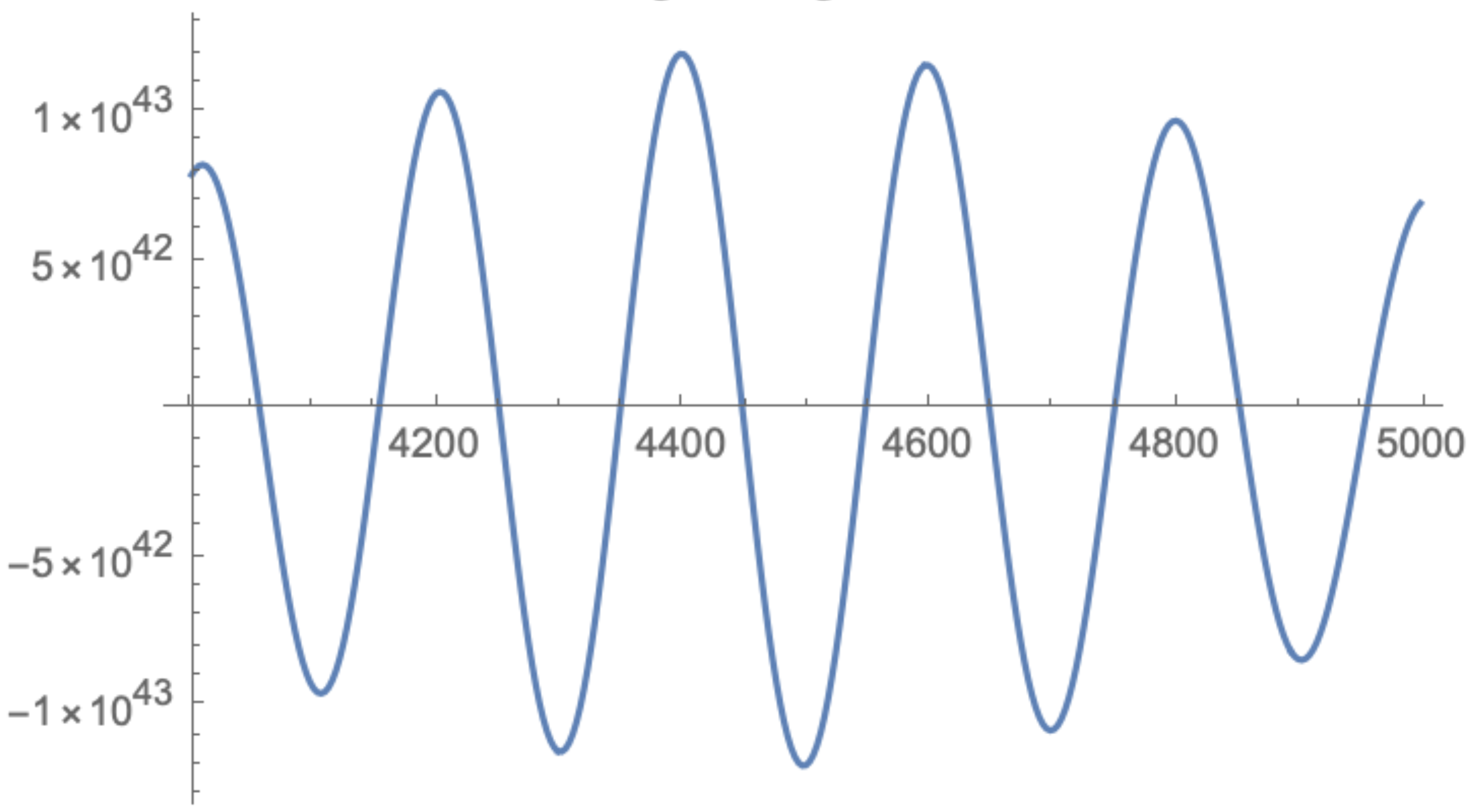}
\caption{\label{Fig:dessum}
$\Omega_\cJ \, x^\cJ + \Omega_{\cJ+1} \, x^{\cJ+1} + \Omega_{\cJ+2} \, x^{\cJ+2}$
where $x=0.95$, for each $\cJ$.}
\end{center}
\end{figure}

This pattern of oscillation has a direct implication.
For any $\cJ$, consider the sum of the three adjacent contributions to the index:
$\Omega_\cJ \, 0.95^\cJ + \Omega_{\cJ+1} \, 0.95^{\cJ+1} + \Omega_{\cJ+2} \, 0.95^{\cJ+2}$.
The three terms are basically the three points of an equilateral triangle
centered at the origin in the complex plane, then projected onto one axis.
Therefore, the sum over the three terms will be much smaller than each individual term.
In fact, Figure \ref{Fig:dessum} illustrates this fact: the sum is indeed smaller by
more than an order of magnitude than individual terms.

\begin{figure}
\begin{center}
\includegraphics[width=0.49\textwidth]{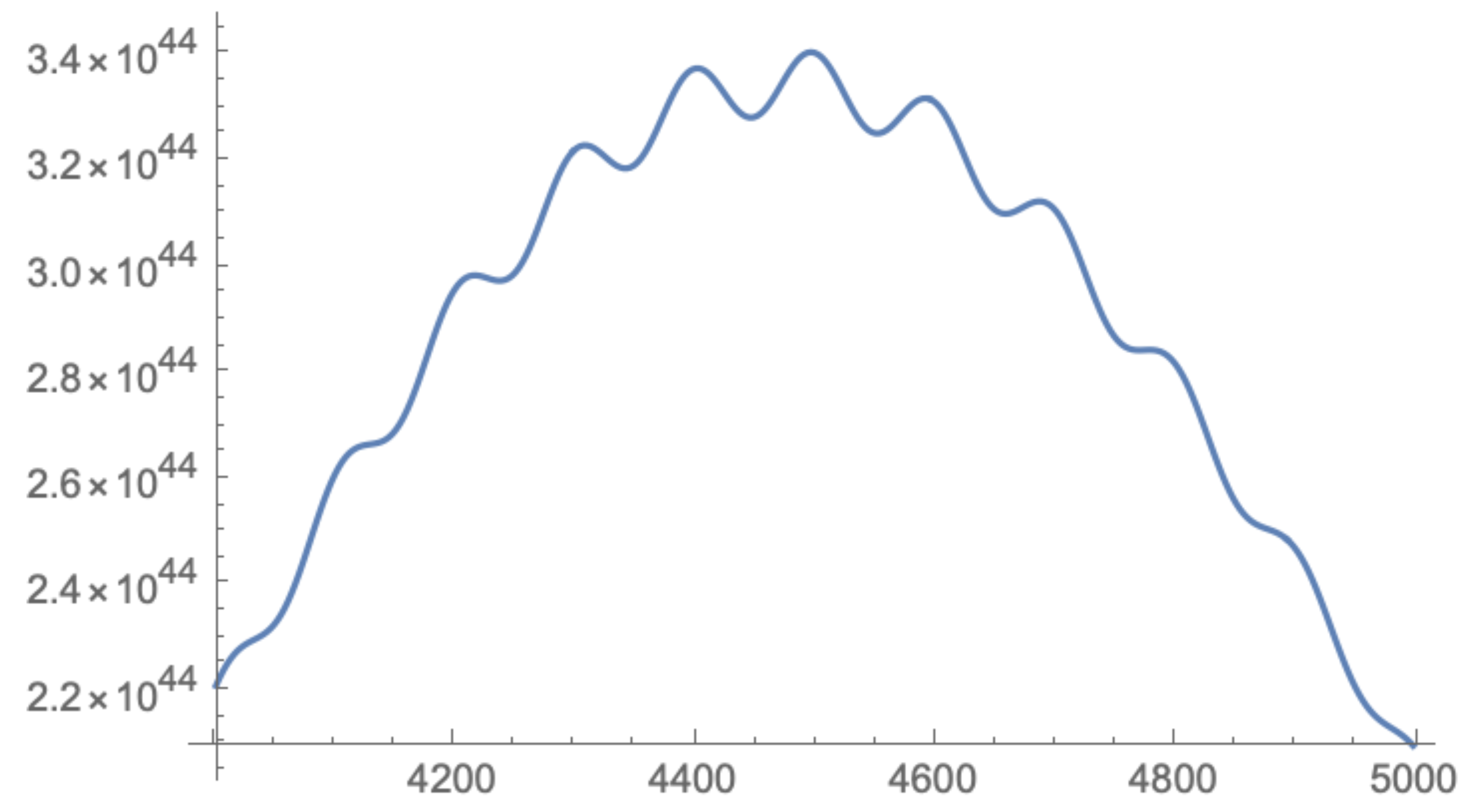}
\includegraphics[width=0.49\textwidth]{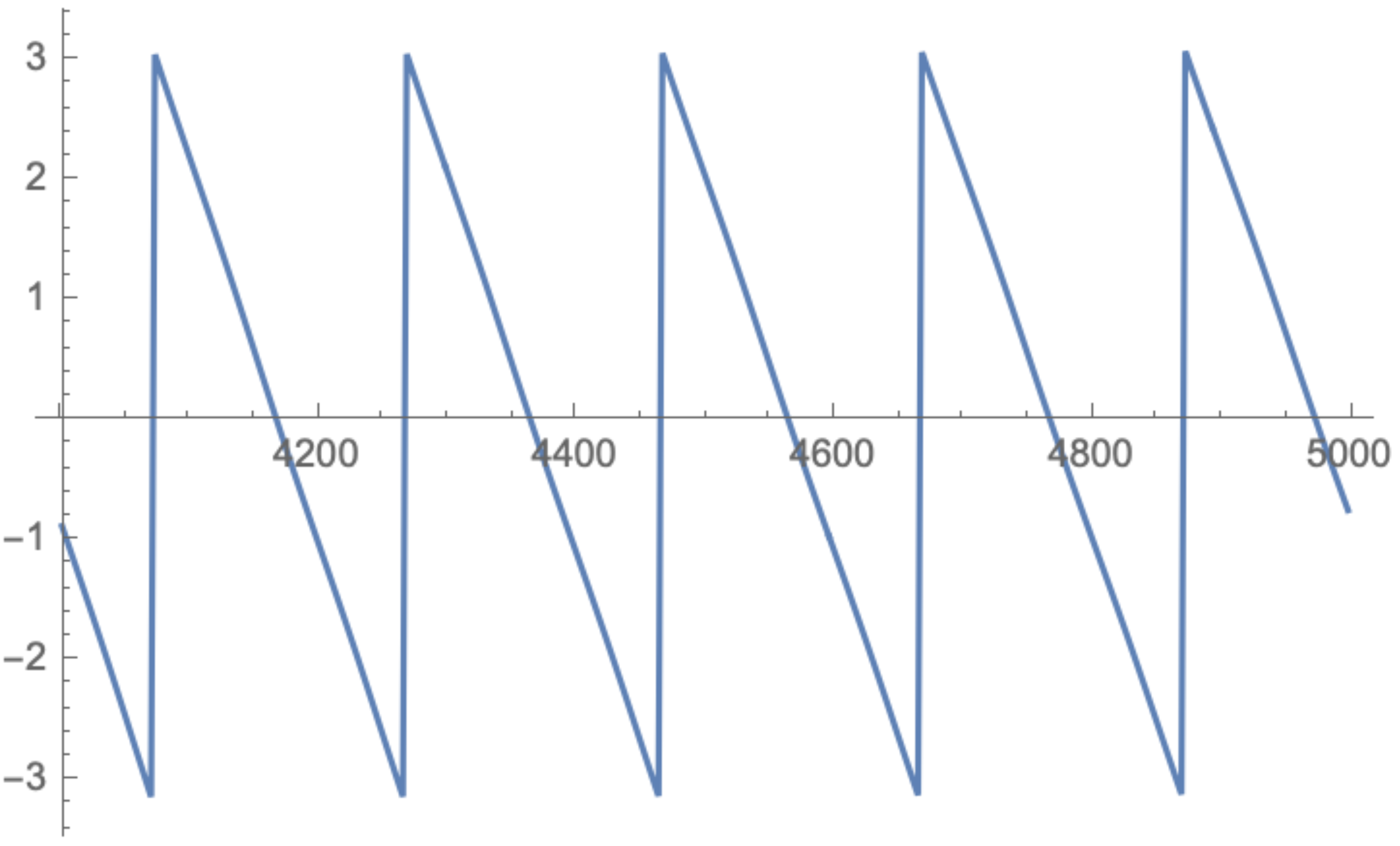}
\caption{\label{Fig:consum}
$\Omega_\cJ \, x^\cJ + \Omega_{\cJ+1} \, x^{\cJ+1} + \Omega_{\cJ+2} \, x^{\cJ+2}$
where $x = 0.95 \cdot e^{i\left(\frac{2\pi}{3}\right)}$, for each $\cJ$.
The magnitude (left) and the phase (right).}
\end{center}
\end{figure}

There is a simple way to turn this destructive sum into a constructive one.
If we give $x$ a complex phase of $\frac{2\pi}{3}$,
the three points of the triangle will gather around one of them.
We plot the magnitude and the phase of the sum
\bea\label{3consum}
\Omega_\cJ \, x^\cJ + \Omega_{\cJ+1} \, x^{\cJ+1} + \Omega_{\cJ+2} \, x^{\cJ+2}~,
\quad \text{where}~~x = 0.95 \cdot \exp \left( \frac{2\pi i}{3} \right)~,
\eea
in Figure \ref{Fig:consum}.
The magnitude of this sum is now consistently larger than individual term,
indicating a constructive sum of the three terms.

This is not the end, however. The right panel of Figure \ref{Fig:consum}
shows that the sum \eqref{3consum} over the three terms,
while being a constructive sum over three adjacent terms,
also slowly oscillate in its phase.
This is because of the sine wave pattern shown in Fig \ref{Fig:U14500zoom}:
the equilateral triangle slowly rotates as a whole.
Therefore, although the phase $\frac{2\pi}{3}$ ensures constructive sum
over 3 adjacent terms, such sums may destruct each other between
a more distant values of $\cJ$.

Fortunately, the rate of rotation of the phase of \eqref{3consum}
is close to a constant.
So it is possible to cancel this rotation effect by adjusting the phase of $x$
slightly from $\frac{2\pi}{3}$.
For $x = 0.95$, the ideal phase is found to be
\bea\label{idealx}
x = 0.95 \cdot \exp i\left( \frac{2\pi}{3} + 0.0313458 \right)~.
\eea

\begin{figure}
\begin{center}
\includegraphics[width=0.49\textwidth]{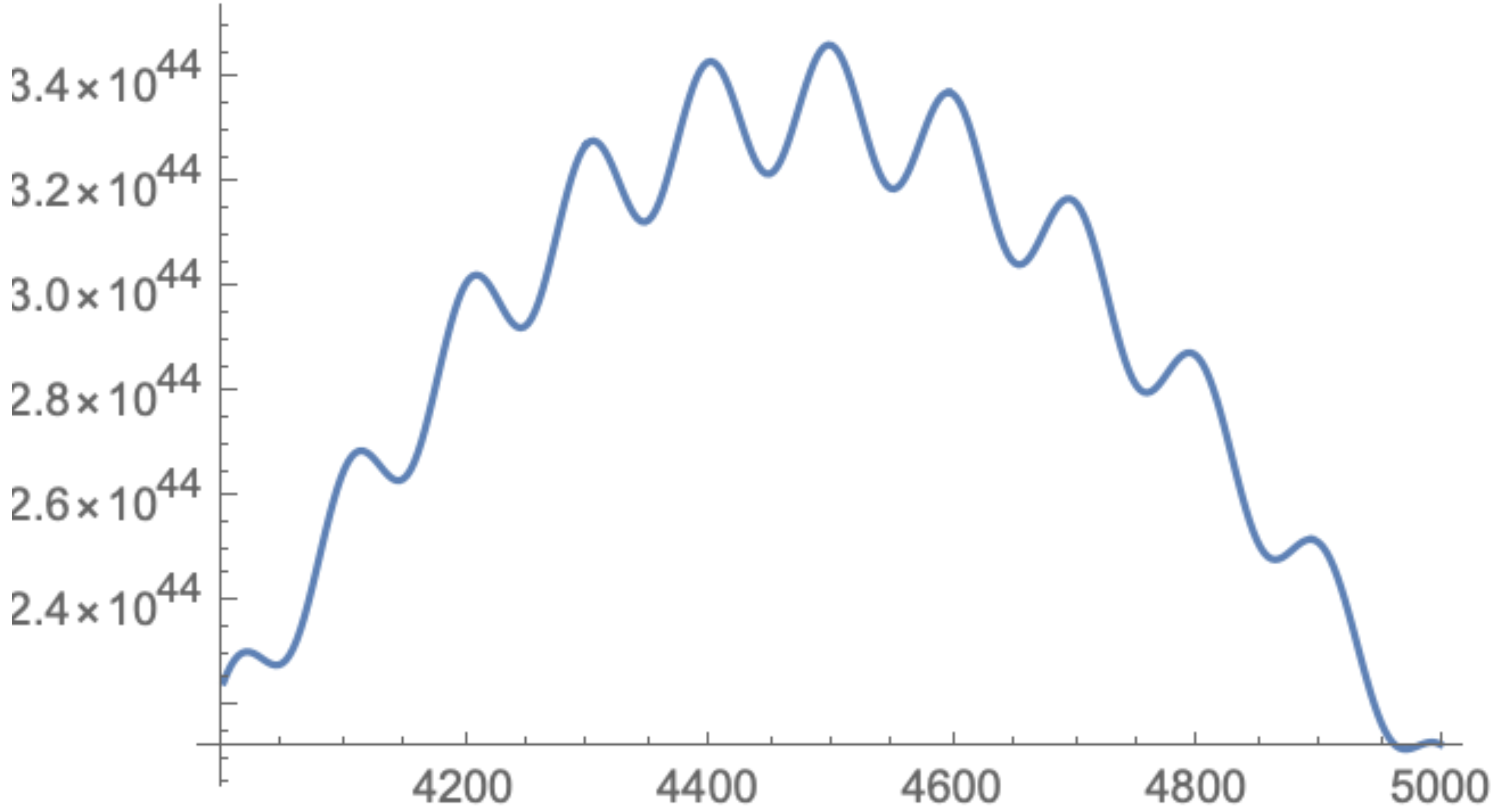}
\includegraphics[width=0.49\textwidth]{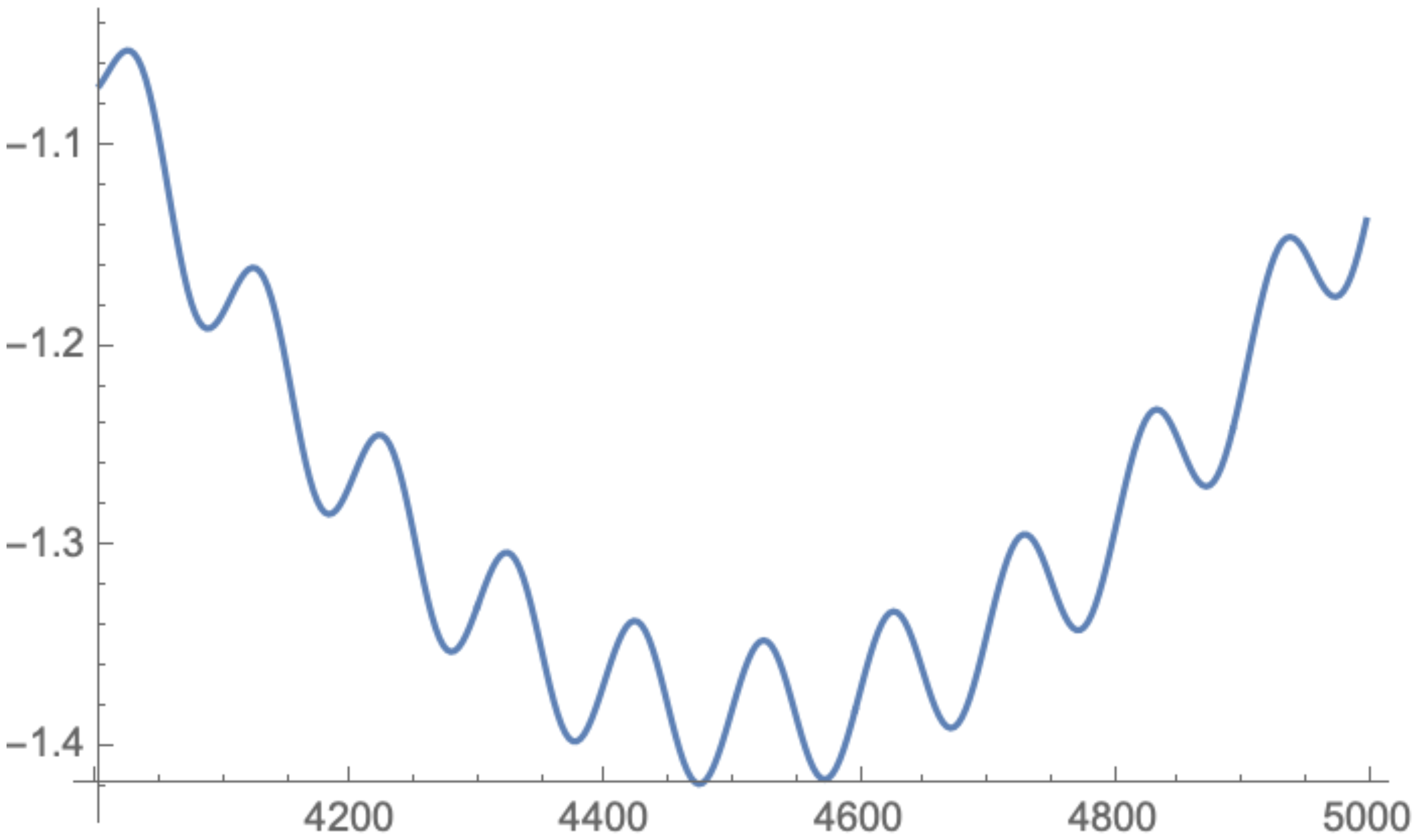}
\caption{\label{Fig:optisum}
$\Omega_\cJ \, x^\cJ + \Omega_{\cJ+1} \, x^{\cJ+1} + \Omega_{\cJ+2} \, x^{\cJ+2}$
where $x = 0.95 \cdot e^{i\left(\frac{2\pi}{3} + 0.0313458\right)}$, for each $\cJ$.
The magnitude (left) and the phase (right).}
\end{center}
\end{figure}

\eqref{3consum} with this ideal phase of \eqref{idealx},
both its magnitude and its phase, is plotted in Figure \ref{Fig:optisum}.
The magnitudes are similar to that in Figure \ref{Fig:consum},
because both are a constructive sum over 3 adjacent terms,
but in Figure \ref{Fig:optisum} the phase of the sum is also close to
being stationary.
Therefore, with the optimal phase of $x$ given by \eqref{idealx},
finally the sum over all contributions $\sum_\cJ \, \Omega_\cJ x^\cJ$
to the index will add up constructively.

As a result of this careful tuning of the phase of $x$ given $|x| = 0.95$,
the numerical value of the index evaluates to
\bea\label{idealxindex}
\mathcal{I} \left( 0.95 \cdot e^{i\left( \frac{2\pi}{3} + 0.0313458 \right)} \right)
&=& 1.33 \times 10^{47}~.
\eea
This is a very sensible result considering that each $\cJ$ around 4500
contributes $\sim 10^{44}$ in magnitude,
and there are $\sim 10^3$ orders that contribute to the index significantly,
recall Figure \ref{Fig:U14500}.
This rough comparison that the sum is close to a magnitude of each term
times the number of significant terms,
supports that the sum in the index has been completely constructive.
For comparison, we note the numerical value of the index for real $x=0.95$:
\bea
\mathcal{I} \left( 0.95 \right)
&=& 2.1 \times 10^{31}~.
\eea
Its smallness compared to \eqref{idealxindex} shows that for real $x$,
the destructiveness of the sum is extremely thorough.

The phase of $x$ needed to optimize the sum,
which was $\frac{2\pi}{3} + 0.0313458$ for $|x| = 0.95$,
is a function of $|x|$.
In fact, for higher $\cJ$, the phase approaches $\frac{2\pi}{3}$.
We expect that the phase indeed converge to $\frac{2\pi}{3}$ as $\cJ \to \infty$.
This expectation is aligned with the new Cardy limit discussed in \cite{Kim:2019yrz}.
In the rest of this section, let us make a quantitative comparison.

It was found in \cite{Kim:2019yrz} that as $x \to 1 \cdot e^{i\left( \frac{2\pi}{3} \right)}$,
what is called the new Cardy limit,
the asymptotic behavior of the index is given by
\bea\label{newCardyI}
\log \mathcal{I} = \frac{8(5a-3c)}{27\omega^2} (-\pi i + \omega)^3
+ \frac{8\pi^2 (a-c)}{3\omega^2} (-\pi i + \omega)~,
\eea
where $a$ and $c$ are the central charges of the 4d superconformal field theory,
and $\omega$ translates to our $x$ via $x^3 = e^{-\omega}$,
so it is a small parameter.
It was also found that the index \eqref{newCardyI} is extremized,
for a given $|x|$, when the phase of $x$ is such that
\bea\label{newCardyphase}
\mathrm{Re} \, \omega = \sqrt{3} \cdot \mathrm{Im} \, \omega
\quad \leftrightarrow \quad
- \log |x| = \sqrt{3} \cdot \left( \mathrm{arg}(x) - \frac{2\pi}{3} \right)
\eea
Under this condition, the magnitude of the extremized index is given by
\bea\label{newCardyIext}
\mathrm{Re} \, (\log \mathcal{I}) &=&
\frac{2\pi^2 (3c-2a)}{9\sqrt{3} (\mathrm{Im} \, \omega)^2}
\eea

\begin{figure}
\begin{center}
\includegraphics[width=0.49\textwidth]{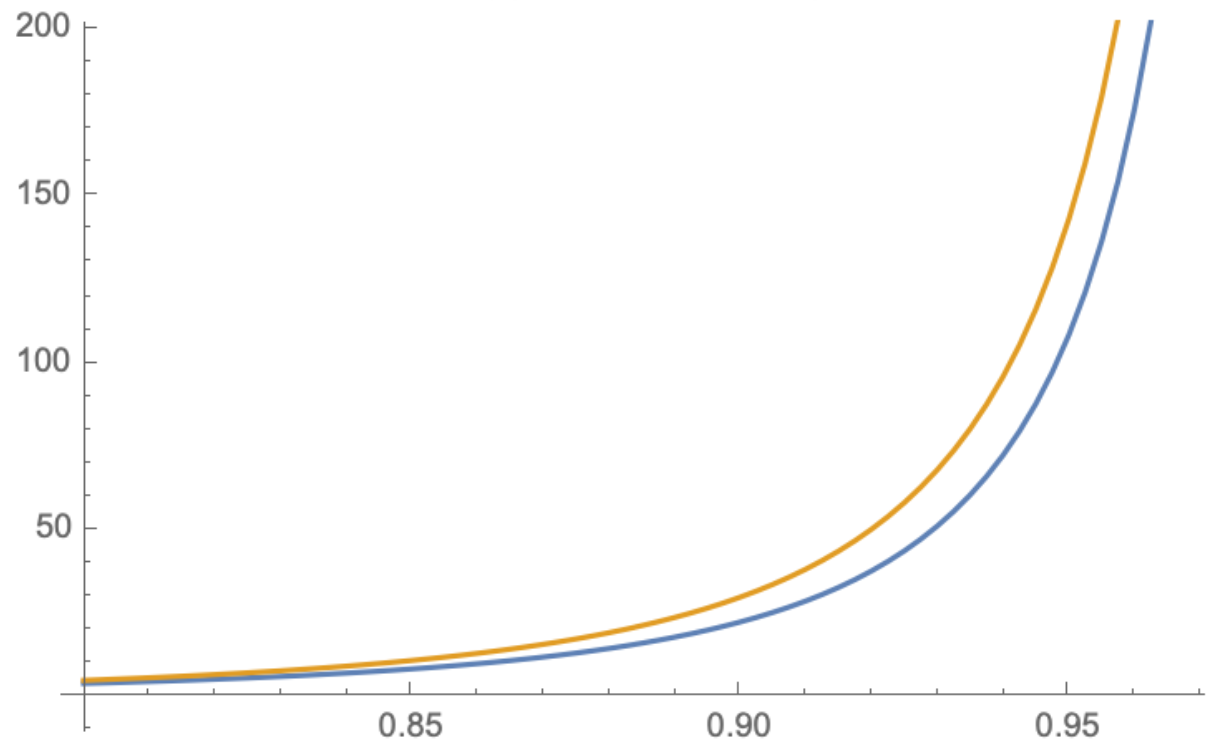}
\includegraphics[width=0.49\textwidth]{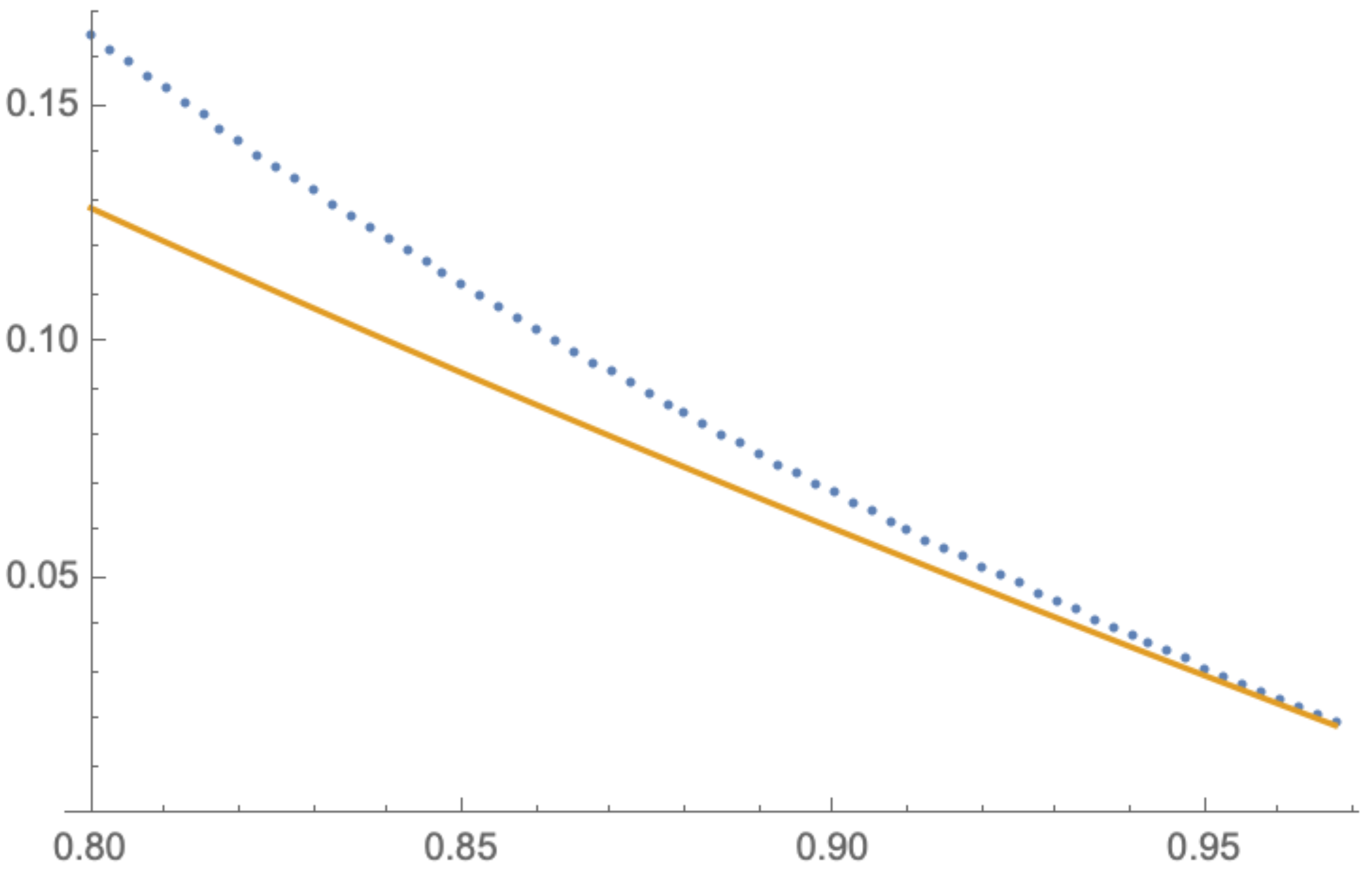}
\caption{\label{Fig:newCardycompare} Left:
$\log \mathcal{I}_{U(1)}$ for the optimal phase of $x$ given $|x|$.
Right: the optimal phase of $x$ in excess of $\frac{2\pi}{3}$ given $|x|$.
Blue lines represent the numerical $U(1)$ index and
yellow lines represent corresponding expectations \eqref{newCardyIext} and \eqref{newCardyphase}.}
\end{center}
\end{figure}

In Figure \ref{Fig:newCardycompare}, we plot the expected results
\eqref{newCardyIext} for the value of the index (left) and
\eqref{newCardyphase} for the extra phase from $\frac{2\pi}{3}$ (right)
for $3c-2a=1$.
In blue solid and dotted lines, we plot the corresponding results
from the numerical analysis of the $U(1)$ index.
The comparison on the left panel is not perfect,
but we believe that it demonstrates the $\frac{1}{\omega^2}$ behavior
that was expected in the new Cardy limit \cite{Kim:2019yrz}
as opposed to the $\frac{1}{\omega}$ behavior that had been
expected from the `old' Cardy limit \cite{DiPietro:2014bca}.

\section{The Entropy Extremization Principle}\label{sec:EEP}

It was demonstrated in the previous section that it is possible to overcome
the cancellations between bosons and fermions to obtain the microcanonical
degeneracy faithfully from the index.
One way or another, the index is obtained as a function of
complex chemical potentials.
In this section, we review the derivation of the black hole entropy
from the index, often referred to as the entropy extremization principle.
This derivation has universality across dimensions, including
AdS$_3$ \cite{Larsen:2021wnu}, AdS$_4$ \cite{Benini:2015eyy,Benini:2016rke},
AdS$_5$ \cite{Hosseini:2017mds,Choi:2018hmj} and
AdS$_7$ \cite{Hosseini:2018dob,Choi:2018hmj},
but we only illustrate it in AdS$_3$ and AdS$_5$,
in respective subsections.

\subsection{AdS$_3$}\label{sec:EEP3}

We first present the entropy extremization principle for AdS$_3$ black holes.
AdS$_3$ is special from higher dimensions,
in line with CFT$_2$ being special from CFTs in higher dimensions.
For the AdS$_3$ black holes, the grand canonical partition function,
as opposed to the index, is known, as reviewed in section \ref{sec:AdS3BH}.
Recall that grand canonical partition function is more general than the index
in that it depends on one more chemical potentials.
Furthermore, the partition function is known beyond its BPS limit,
whereas the index only contains information about the BPS states.
However, in this subsection we shall focus on the index and reproduce
the entropy of BPS black holes \eqref{AdS3SBPS} in AdS$_3$ spacetime from the index.
This subsection is based on \cite{Larsen:2021wnu}.

The grand canonical partition function was defined in \eqref{AdS3defZ},
as a trace over all states: 
\begin{eqnarray}
\label{AdS3defZrep}
Z (\beta,\, \mu, \, \omega_R, \, \omega_L) &=& \mathrm{Tr}
\left[e^{- \beta (E-\mu J-\omega_R Q_R - \omega_L Q_L)} \right] \nn\\
&=& e^{\frac12 \beta k_L} \mathrm{Tr}
\left[ e^{- \beta \left(E-J-Q_L+\frac{k_L}{2} \right)
+ \tilde\mu J + \tilde{\omega}_R Q_R + \tilde{\omega}_L Q_L} \right]~.
\end{eqnarray}
In the second line we have simply replaced the chemical potentials
with their rescaled versions with tildes using \eqref{eqn:BPSlimit}.
In subsection \ref{sec:AdS3BHBPS} we isolated the BPS states by taking
$\beta \to \infty$ with the rescaled potentials kept finite.
This gave the BPS partition function (\ref{AdS3defZBPS}): 
\bea\label{AdS3defZBPSrep}
Z_{\rm BPS}(\beta, \, \tilde\mu, \, \tilde\omega_R, \, \tilde\omega_L)
&=& \left( \lim_{\beta\to\infty} e^{\frac12 \beta k_L} \right)
\mathrm{Tr}_\mathrm{BPS} \left[
e^{\tilde\mu J + \tilde{\omega}_R Q_R + \tilde{\omega}_L Q_L} \right]~.
\eea
%

In this subsection, we study the supersymmetric index,
also known as the elliptic genus in CFT$_2$, rather than the partition function. 
Recall from section \ref{sec:earlyindex}, in particular around \eqref{makeitindex},
that the index is a special case of the partition function where the chemical potentials
are such that a pair $|\psi \rangle$ and $Q |\psi \rangle$ contributes
equal but opposite weights.
We choose the supercharge $Q$ in the anti-holomorphic (left) sector of CFT$_2$,
in line with saturation of \eqref{eqn:unitboundmacro}.
This condition for the index is equivalent to the complex constraint
\be\la{eqn:indcond}
\tilde\mu - 2\tilde\omega_L = 2 \pi i ~,
\ee
on the potentials.
Thus, the index is the grand canonical partition function \eqref{AdS3defZrep}
under the condition \eqref{eqn:indcond}.
We also pull out the Casimir energy factor $e^{\frac12 \beta k_L}$ from the index.
It is conventional to omit this overall factor from the definition of the supersymmetric index,
or of the elliptic genus.

To summarize, the index is
\bea\label{eqn:defI}
\mathcal{I} &\equiv& \left. e^{\beta E_{\rm SUSY}} Z
\right|_{\tilde\omega_L = \frac{\tilde\mu}{2} - i \pi} \nonumber \\
&=& \left. \mathrm{Tr}_\mathrm{BPS} \left[
e^{\tilde\mu J + \tilde{\omega}_R Q_R + \tilde{\omega}_L Q_L} \right]
\right|_{\tilde\omega_L = \frac{\tilde\mu}{2} - i \pi}~,
\eea
where $E_{\rm SUSY}= -\frac12 k_L$ was given in \eqref{eqn:SUSYCast}.
Going from the first to the second line,
we used the aforementioned cancellation for non-BPS states
to get rid of the $\beta$-dependent term in the trace of the second line of
\eqref{AdS3defZrep} and restrict the trace to BPS states only,
then cancelled the Casimir energy factor.
Note that $\beta \to \infty$ is not needed in the definition of the index,
but $\beta$-dependence was eliminated by the cancellation that is
enabled by \eqref{eqn:indcond}.

The BPS partition function $Z_{\rm BPS}$ \eqref{AdS3defZBPSrep}
depends on \emph{three} independent potentials:
$\tilde\mu$ and $\tilde\omega_{L,R}$, apart from the formal
$\left. e^{-\beta E_{\rm SUSY}} \right|_{\beta \to \infty}$ factor.
On the other hand the index $\mathcal{I}$ depends on only \emph{two}
independent parameters due to \eqref{eqn:indcond}
which we take as $\tilde\mu$ and $\tilde\omega_R$.

We can compute the index for supersymmetric black holes in AdS$_3$
explicitly by starting from the general partition function (\ref{eqn:genlnZ}), 
introducing tilde potentials through \eqref{eqn:BPSlimit},
and then imposing the index constraint (\ref{eqn:indcond}):
\bea
\label{eqn:ind}
\log \mathcal{I} &=&
- \frac{k_L}{2}\beta + \frac{k_R}{\beta(1-\mu)} \left( \pi^2 + \beta^2 \omega^2_R\right)
+ \frac{k_L}{\beta(1+\mu)} \left( \pi^2 + \beta^2 \omega^2_L\right) \nn\\
&=& - \frac{k_L}{2}\beta -\frac{k_R}{\tilde\mu} \left( \pi^2 + \tilde\omega^2_R\right)
+ \frac{k_L}{\tilde\mu + 2\beta} \left(\pi^2 + (\tilde\omega_L +\beta)^2\right) \nn\\
&=& -\frac{k_R}{\tilde\mu} \left( \pi^2 + \tilde\omega^2_R\right)
+ \frac{k_L}{4} (\tilde\mu -4 \pi i ) \nn\\
&=&-\frac{k_R}{\tilde\mu} \left( \pi^2 + \tilde\omega^2_R\right)
+ \frac{k_L}{\tilde\mu} (\pi^2+\tilde\omega_L ^2 )~.
\eea
We present the manipulations in detail to highlight that they are exact,
the dependence on $\beta$ disappears without any limit taken, as anticipated.
The final expression with the constraint (\ref{eqn:indcond}) implied
agrees with the BPS partition function (\ref{AdS3ZBPS}), again as anticipated.
A simpler but less illuminating route to the formula for the index
given in the last line of \eqref{eqn:ind} is to evaluate 
the partition function and take the {\it high} temperature limit $\beta\to 0$
with the tilde variables kept fixed.
In other words, the last line of \eqref{eqn:ind} follows from the second line
by taking $\beta=0$.
This uses the $\beta$-independence of the index rather than showing it.

The computation illustrates how the index \eqref{eqn:defI} and the
BPS partition function (\ref{AdS3defZBPSrep}) are closely related, 
yet they are different in significant ways such that they complement one another:
\begin{itemize}
\item
The BPS partition function restricts the trace to the chiral primary states by
an explicit limit $\beta \to \infty$.
In contrast, the index is independent of $\beta$,
the limit $\beta \to \infty$ is possible but not mandatory.
This is one aspect of the index being protected under
continuous deformations of the theory, while the BPS partition function is not.
\item
The supersymmetric index is defined with chemical potentials constrained by
(\ref{eqn:indcond}) or else it is not protected under continuous deformations.
In contrast, the BPS partition function keeps all three potentials
$\tilde\mu$ and $\tilde\omega_{R,L}$ independent.
It is possible to focus on variables that satisfy the constraint,
but the general case incorporates more information about the theory. 
\item
The supersymmetric index is defined with the supersymmetric Casimir energy
stripped off, while the partition function retains it. 
\end{itemize}

In the non-chiral case $k_L=k_R=k$ in the absence of gravitational anomaly,
we can recast our result for the index \eqref{eqn:ind} as
\be\la{eqn:lnZindex2}
 \log \mathcal{I} = k \frac{\tilde{\omega}_1\tilde{\omega}_2}{\tilde{\mu}}~, 
\ee
by choosing the basis $\tilde{\omega}_{L,R} =
\frac12 ( \tilde{\omega}_1 \pm  \tilde{\omega}_2)$ for the potentials. 
This result is aligned with the form of the index that plays a central role
in discussions of black hole entropy in higher dimensional AdS spaces,
as we will see in \eqref{4dN4logI} for AdS$_5$.

Whereas we have derived the supersymmetric index (\ref{eqn:ind})
for AdS$_3$ black holes by imposing a complex condition (\ref{eqn:indcond})
on the more general BPS partition function (\ref{eqn:genlnZ}),
in higher dimensional AdS spaces it is only the index that can be reliably computed.
In that context a procedure to extract the entropy and the charge constraint
of supersymmetric black holes directly from the index has been developed
\cite{Hosseini:2017mds}.
We now apply this procedure to the AdS$_3$ case and show that it reproduces
the results derived from the BPS partition function in section \ref{sec:AdS3BH}.

The claim that is now standard in higher dimensional AdS spaces is that
we can process the index as if it was an ordinary partition function.
According to this prescription\cite{Hosseini:2017mds},
the black hole entropy is given by the Legendre transform of the index (\ref{eqn:ind}), 
subject to the complex constraint (\ref{eqn:indcond}).
That is, the entropy function is defined by
\bea\label{eqn:Sfunction0}
S ( \tilde\mu, \tilde\omega_{R} , \tilde\omega_{L} ) &\equiv&
\log \mathcal{I}  - \tilde{\omega}_L Q_L - \tilde{\omega}_R Q_R - \tilde{\mu} J~,
\eea
and we extremize this function subject to (\ref{eqn:indcond}).
This can be done efficiently by introducing the Lagrange multiplier $\Lambda$
that enforces the condition \eqref{eqn:indcond}, thus extremizing
\bea\la{eqn:Sfunction}
S &=& S ( \tilde\mu, \tilde\omega_{R} , \tilde\omega_{L} )
- \Lambda (\tilde{\mu} - 2\tilde{\omega}_L - 2\pi i) \nn\\[5pt]
&=&
\frac{k_L \left( \tilde{\omega}_L^2 + \pi^2\right) - k_R \left( \tilde{\omega}_R^2 + \pi^2\right)}
{\tilde{\mu}}
 -  \tilde{\omega}_L Q_L  -  \tilde{\omega}_R Q_R - \tilde{\mu} J
 - \Lambda (\tilde{\mu} - 2\tilde{\omega}_L - 2\pi i)~, \quad
\eea
with respect to the potentials $\tilde\mu$, $\tilde\omega_{R,L}$
and the Lagrange multiplier $\Lambda$.

$S$ is homogeneous of degree one in the potentials
$\tilde\mu$, $\tilde\omega_{R,L}$, except 
for $2\pi i \Lambda$ which is constant,
and for the terms proportional to $\pi^2$ which are homogeneous of degree minus one. 
Keeping track of the inhomogeneous terms, the extremization conditions give
\be
0 = \left( \tilde{\omega}_L\partial_{\tilde{\omega}_L} 
+ \tilde{\omega}_R\partial_{\tilde{\omega}_R}
+ \tilde{\mu}\partial_{\tilde{\mu}}\right) S
= S - 2\pi i \Lambda + \frac{2\pi^2 (k_R-k_L)}{\tilde{\mu}}~,
\ee
so that
\be\la{eqn:Sv1}
S = 2\pi i \Lambda  - \frac{2\pi^2 (k_R-k_L)}{\tilde{\mu}}~.
\ee
The second term vanishes only when $k_R=k_L$.
It represents a novel refinement when compared to analogous computations
in higher dimensional AdS spaces.  

The individual entropy extremization conditions are
\begin{subequations}\la{eqn:ext3}\bea
\la{eqn:extwL}
\partial_{\tilde{\omega}_L} S &=& k_L \frac{2\tilde{\omega}_L}{\tilde{\mu}} + (2\Lambda - Q_L) = 0 ~,\\
\la{eqn:extwR}
\partial_{\tilde{\omega}_R} S &=& - k_R \frac{2\tilde{\omega}_R}{\tilde{\mu}} -  Q_R = 0 ~,\\
\la{eqn:extmu}
\partial_{\tilde{\mu}} S&=& - \frac{k_L \left( \tilde{\omega}_L^2 + \pi^2\right) - k_R \left( \tilde{\omega}_R^2 + \pi^2\right)}{\tilde{\mu}^2}  
- (\Lambda + J) = 0 ~.
\eea\end{subequations}
Using the constraint \eqref{eqn:indcond}, the first equation gives
\be \la{eqn:extmu2}
k_L \frac{\tilde{\mu} - 2\pi i }{\tilde{\mu} } = Q_L -2 \Lambda
~~\Rightarrow~~
\frac{\pi i k_L}{\tilde{\mu}} = \Lambda - \frac{1}{2}(Q_L-k_L) ~.
\ee
The entropy function therefore becomes 
\be\la{eqn:Sv2}
S = 2\pi i \left[ \Lambda  + \frac{i\pi}{\tilde{\mu}}(k_R - k_L) \right]
= 2\pi i\left[   \frac{k_R}{k_L} \Lambda - \frac{1}{2k_L} (k_R-k_L)(J_L-k_L) \right] \equiv 2 \pi i \Lambda_{\rm eff}~,
\ee
where we defined
\be\la{eqn:defLeff}
\Lambda_{\rm eff} = \frac{k_R}{k_L}  \Lambda - \frac{1}{2k_L} (k_R-k_L)(Q_L-k_L) ~.
\ee
Rewriting the last extremization condition \eqref{eqn:extmu} using the others (\ref{eqn:extwL}-\ref{eqn:extwR})
and the expression for $\tilde{\mu}$ \eqref{eqn:extmu2} we find
\be\la{eqn:L}
 - \frac{1}{k_L} (\Lambda - \frac{1}{2}Q_L)^2 + \frac{1}{4k_R} Q_R^2 - (\Lambda + J)
 - \frac{1}{k^2_L} (k_R - k_L) \left(\Lambda - \frac{1}{2}(Q_L - k_L) \right)^2= 0 ~,
\ee
which we reorganize into a quadratic equation for $ \Lambda_{\rm eff}$:
\be\la{eqn:Leff}
 \Lambda^2_{\rm eff} - (Q_L-k_L) \Lambda_{\rm eff}    + \frac{1}{4}(Q_L - k_L)^2
+ k_R \left(J+\frac{Q_L}{2} - \frac{k_L}{4} \right)   - \frac{1}{4} Q_R^2   = 0 ~.
\ee
Selecting the root with negative imaginary part we find the extremized entropy function in terms of charges: 
\be\la{eqn:Sfinal}
S = 2\pi i \Lambda_{\rm eff} =
2 \pi \sqrt{k_R \left(J+\frac{Q_L}{2} - \frac{k_L}{4} \right) - \frac{Q_R^2}{4}} + \pi i (Q_L-k_L)~.
\ee

For BPS black holes in higher dimensional AdS the standard prescription posits that
charges must be constrained such that the extremized entropy function is real
\cite{Hosseini:2017mds,Choi:2018hmj}.
Applying this rule in AdS$_3$ as well, we find 
\be
Q_L =  k_L ~,
\ee
in agreement with the charge constraint \eqref{eqn:BPSJL} that
we inferred from gravitational considerations.
Only after fixing the charges this way, the entropy function \eqref{eqn:Sfinal}
is real with the value 
\be\la{eqn:Sv3}
S_{\rm BPS} = 2\pi \sqrt{ k_R \left( J + \frac14 k_L \right) - \frac14 Q_R^2}~, 
\ee
in agreement with the entropy \eqref{AdS3SBPS} of a BPS black hole in AdS$_3$ . 

In summary, in this subsection we defined the supersymmetric index for
the AdS$_3$ black holes, and applied the entropy extremization procedure
to recover thermodynamic properties from the index \eqref{eqn:ind}.
The computation is novel in that the index \eqref{eqn:ind} used here is more refined than
the version \eqref{eqn:lnZindex2} that is directly analogous to higher dimensional cases.

\subsection{AdS$_5$}\label{sec:EEP5}

In this subsection, we demonstrate the entropy extremization principle for
AdS$_5$ black holes \cite{Hosseini:2017mds,Choi:2018hmj}.
In fact, in this context was the principle first established \cite{Hosseini:2017mds}
based on a similar extremization principle used for AdS$_4$ black hole entropy
\cite{Benini:2015eyy,Benini:2016rke}.
Shortly after, the index required for this principle to reproduce the AdS$_5$
black hole entropy was computed
\cite{Cabo-Bizet:2018ehj,Choi:2018hmj,Benini:2018ywd}.

As we reviewed in section \ref{sec:earlyindex},
the AdS$_5$ black holes and the dual 4d $\mathcal{N}=4$ SYM are described by
six quantum numbers, namely the energy or the scaling dimension $E$,
two angular momenta $J_{1,2}$, and three charges $Q_{1,2,3}$.
The index of the 4d $\mathcal{N}=4$ SYM is defined by \eqref{4dN4index1}:
\bea\label{4dN4index1rep}
\mathcal{I} (\tilde{\Delta}_I,\, \tilde{\omega}_i) &=&
\mathrm{Tr} \left[e^{- \beta (E-Q_1-Q_2-Q_3-J_1-J_2)}
e^{\tilde{\Delta}_I Q_I+ \tilde{\omega}_i J_i} \right] \nn\\
&=&
\mathrm{Tr}_\mathrm{BPS} \left[
e^{\tilde{\Delta}_I Q_I+ \tilde{\omega}_i J_i} \right]~,
\eea
where $\tilde{\Delta}_I \equiv \Delta_I-\beta$ and $\tilde{\omega}_i \equiv \omega_i-\beta$
satisfy $e^{\frac{\tilde{\Delta}_1+\tilde{\Delta}_2+\tilde{\Delta}_3-\tilde{\omega}_1
-\tilde{\omega}_2}{2}} = -1$,
which is the condition that ensures independence on $\beta$
as well as the coupling independence of the index.
This condition is realized by
\bea\label{4dN4indcond}
\tilde\Delta_1 + \tilde\Delta_2 + \tilde\Delta_3 - \tilde\omega_1 - \tilde\omega_2 = 2\pi i~.
\eea

The index \eqref{4dN4index1rep} with complex chemical potentials subject to
\eqref{4dN4indcond} has been computed in various limits and approximations.
For example, \cite{Choi:2018hmj} took the Cardy-like limit $|\omega_i| \ll 1$ while
\cite{Benini:2018ywd} took two equal angular momenta for the Bethe Ansatz approach.
The regime of applicability has broaden in \cite{Choi:2021rxi} but not completely.
In one way or another, the leading term in large $N$ of the index is evaluated to be
\be\la{4dN4logI}
\log \mathcal{I} = -\frac{N^2}{2} \frac{\tilde\Delta_1 \tilde\Delta_2 \tilde\Delta_3}
{\tilde\omega_1 \tilde\omega_2}~.
\ee
Let us now derive the AdS$_5$ black hole entropy from this index,
following the entropy extremization principle \cite{Hosseini:2017mds},
in which the index is treated as an ordinary grand canonical partition function
that yields the entropy via Legendre transformation.

First, the entropy function is defined by
\bea\label{4dN4Sfn0}
S ( \tilde\Delta_I, \, \tilde\omega_i ) &\equiv&
\log \mathcal{I}  - \sum_I \tilde\Delta_I Q_I - \sum_i \tilde\omega_i J_i~,
\eea
and we extremize this function subject to (\ref{4dN4indcond}).
This can be done efficiently by introducing the Lagrange multiplier $\Lambda$
that enforces the condition \eqref{4dN4indcond}, thus extremizing
\bea\la{4dN4Sfn}
S &=& S ( \tilde\Delta_I, \, \tilde\omega_i ) 
- \Lambda \left( \sum_I \tilde\Delta_I - \sum_i \tilde\omega_i - 2\pi i \right) \nn\\
&=&
-\frac{N^2}{2} \frac{\tilde\Delta_1 \tilde\Delta_2 \tilde\Delta_3}
{\tilde\omega_1 \tilde\omega_2}
- \sum_I \tilde\Delta_I Q_I - \sum_i \tilde\omega_i J_i
- \Lambda \left( \sum_I \tilde\Delta_I - \sum_i \tilde\omega_i - 2\pi i \right)~,
\eea
with respect to the potentials $\tilde\Delta_I$, $\tilde\omega_i$
and the Lagrange multiplier $\Lambda$.

The entropy function is homogeneous of degree one in the potentials
except for the $2\pi i \Lambda$ which is constant.
Therefore,
\bea
0 = \left( \sum_I \tilde\Delta_I \partial_{\tilde\Delta_I}
+  \sum_i \tilde\omega_i \partial_{\tilde\omega_i} \right) S
= S - 2\pi i \Lambda~,
\eea
so that
\be\label{4dN4SinL}
S = 2 \pi i \Lambda~.
\ee
This is analogous to \eqref{eqn:Sv1}, where the $k_R - k_L$ term
does not have a higher dimensional analogue.

The individual entropy extremization conditions are
\bea
- \tilde\Delta_I \partial_{\tilde\Delta_I} S &=&
\frac{N^2}{2} \frac{\tilde\Delta_1 \tilde\Delta_2 \tilde\Delta_3}
{\tilde\omega_1 \tilde\omega_2}
+ \tilde\Delta_I (Q_I + \Lambda ) ~=~ 0~, \nn\\
\tilde\omega_i \partial_{\tilde\omega_i} S &=&
\frac{N^2}{2} \frac{\tilde\Delta_1 \tilde\Delta_2 \tilde\Delta_3}
{\tilde\omega_1 \tilde\omega_2}
+ \tilde\omega_i (- J_i + \Lambda ) ~=~ 0~,
\eea
It follows that
\bea
\tilde\Delta_I &=& - \frac{1}{Q_I + \Lambda} \cdot
\frac{N^2}{2} \frac{\tilde\Delta_1 \tilde\Delta_2 \tilde\Delta_3}
{\tilde\omega_1 \tilde\omega_2}~, \nn\\
\tilde\omega_i &=& - \frac{1}{-J_i + \Lambda} \cdot
\frac{N^2}{2} \frac{\tilde\Delta_1 \tilde\Delta_2 \tilde\Delta_3}
{\tilde\omega_1 \tilde\omega_2}~,
\eea
so
\bea\label{4dN4Lcubic}
\frac{N^2}{2} \frac{\tilde\Delta_1 \tilde\Delta_2 \tilde\Delta_3}
{\tilde\omega_1 \tilde\omega_2}
&=&
\frac{N^2}{2} \cdot \left( -\frac{N^2}{2} \frac{\tilde\Delta_1 \tilde\Delta_2 \tilde\Delta_3}
{\tilde\omega_1 \tilde\omega_2} \right)^{3-2} \cdot
\frac{(-J_1+\Lambda)(-J_2+\Lambda)}{(Q_1+\Lambda)(Q_2+\Lambda)(Q_3+\Lambda)}~, \nn\\
\Rightarrow \quad 0 &=&
(Q_1+\Lambda)(Q_2+\Lambda)(Q_3+\Lambda)
+ \frac{N^2}{2} (-J_1+\Lambda)(-J_2+\Lambda)~.
\eea
This is a cubic equation on $\Lambda$.

Now, the entropy extremization principle posits that charges must be constrained
such that the extremized entropy function is real
\cite{Hosseini:2017mds,Choi:2018hmj}.
This requires, via \eqref{4dN4SinL}, that $\Lambda$ must be purely imaginary.
Since the coefficients of the cubic equation \eqref{4dN4Lcubic} are all real,
it must be that terms in orders $\Lambda^3$ and $\Lambda^1$,
and terms in orders $\Lambda^2$ and $\Lambda^0$,
must separately add up to zero.
Therefore,
\bea\label{4dN4Ltwo}
\Lambda^3 +
\left(Q_1Q_2 + Q_2Q_3 + Q_3Q_1 - \frac{N^2}{2} (J_1+J_2) \right) \Lambda &=& 0~, \\
\left(Q_1+Q_2+Q_3 + \frac{N^2}{2} \right) \Lambda^2 +
\left( Q_1Q_2Q_3 + \frac{N^2}{2} J_1J_2 \right) &=& 0~.
\eea
One consequence of both equations of \eqref{4dN4Ltwo} is
\bea\label{4dN4cceep}
&& Q_1Q_2Q_3 + \frac{N^2}{2} J_1J_2 \nn\\
&=& \left(Q_1Q_2 + Q_2Q_3 + Q_3Q_1 - \frac{N^2}{2} (J_1+J_2) \right)
\left(Q_1+Q_2+Q_3 + \frac{N^2}{2} \right)~,
\eea
and taking one of the solutions of the first equation to make $S= 2\pi i \Lambda$ positive,
we have for the extremized entropy function,
\bea\label{4dN4Seep}
S &=& 2\pi \sqrt{Q_1Q_2 + Q_2Q_3 + Q_3Q_1 - \frac{N^2}{2} (J_1+J_2)}~.
\eea

The rank (plus one) of the gauge group $N$ in the 4d $\mathcal{N}=4$ SYM
is related to the 5-dimensional Newton's gravitational constant by
\cite{Maldacena:1997re}
\bea\label{4dN4NvsG}
N^2 = \frac{\pi}{2G_5}~.
\eea
Through this dictionary, \eqref{4dN4cceep} is the constraint \eqref{5dccgrav}
between the charges $Q_I$ and $J_i$ that a supersymmetric black hole must satisfy,
and \eqref{4dN4Seep} is the entropy \eqref{AdS5S} of the supersymmetric
AdS$_5$ black hole.

To summarize, the entropy extremization principle directs us to extremize the
entropy function \eqref{4dN4Sfn0} determined from the index,
and demand that the complex result be real.
As a result, one obtains the value of the extremized entropy function
as well as a constraint between the charges.
The former gives the entropy of the supersymmetric black hole,
while the latter gives its charge constraint.

\chapter{The Supersymmetric Charge Constraints}\label{sec:cc}

In this chapter we turn to another important property of
supersymmetric AdS black holes, the charge constraint.
We present a heuristic derivation of the constraint from the dual field theories
as a relation between macroscopic charges of an ensemble
that is classically generated by free fields, up to a uniform rescaling of all charges.
After a brief introduction, we shall first present the derivation for AdS$_3$ black holes.
Motivated by this example, we develop a generic prescription
to derive the charge constraint from the field theories
and apply to the AdS$_5$, AdS$_4$ and AdS$_7$ black holes
in the subsequent sections.
We obtain the correct functional form of the fully refined charge constraint
in each of the dimensions.
We conclude this chapter by discussing various shortcomings
and implications of our arguments.
Section \ref{sec:3d} is based on \cite{Larsen:2021wnu} and
the rest of this chapter is based on \cite{Larsen:2024xxx},
both in collaboration with Finn Larsen.

As we have addressed in the previous section,
the entropies of supersymmetric AdS black holes in different dimensions
have been matched with the number of supersymmetric states
in their dual conformal field theories.
The supersymmetric states are counted using the index,
which can be understood as a special case of the grand canonical
partition function for the ensemble of supersymmetric microstates.

Importantly, the supersymmetric index inevitably depends on one less chemical potentials
than there are independent charges, recall the discussion after \eqref{4dN4index1}.
In other words, the index as a grand canonical partition function does not
distinguish microstates along the direction in the space of conserved charges
generated by the preserved supercharge.
This fundamentally prevents the index from addressing the charge constraint,
which contains information about the location in the space of charges.
Indeed, the charge constraint is surprising from the CFT side of the duality,
because numerous local supersymmetric operators exist also for charges that violate the constraint.

Curiously, the charge constraint emerges in the microscopic accounting of the
supersymmetric black hole entropy from the condition that the extremum of the
complex entropy function is real \cite{Hosseini:2017mds,Choi:2018hmj,Choi:2018fdc}.
This is suggestive, but given the intrinsic shortcoming of the index,
it does not provide a satisfying microscopic explanation.

In section \ref{sec:3d}, we make a proposal for the microscopic origin of the charge constraints
for supersymmetric AdS$_3$ black holes.
In any unitary supermultiplet of the small $\mathcal{N}=4$ super-Virasoro algebra,
whether short or long, all weights appear in pairs.
The two weights in each pair are separated in the charge configuration space
along the direction of the preserved supercharge $Q$,
and the $R$-charges of the two weights average to $k$, the level of the $SU(2)_R$ algebra.
This is precisely the condition that an extremal BTZ black hole is supersymmetric.
This proposal goes beyond the scope of the index,
in that the index does not see both states in a pair individually.

Generalization of this argument to higher dimensions is not straightforward.
Superconformal algebras in higher dimensions are not as large and constraining
as the super-Virasoro algebra in CFT$_2$,
so they are consistent with more diverse multiplet structures.
Moreover, in AdS$_{d+1}$ with $d>2$, the constraints on conserved charges
that we want to illuminate are non-linear and highly non-trivial.

In sections \ref{sec:gendim} through \ref{sec:7d},
we offer a heuristic derivation of the charge constraints for higher dimensions.
In each dimension, we start from the free multiplet of the corresponding superconformal algebra.
We then construct a grand canonical partition function that depends on
as many chemical potentials as there are charges,
thereby overcoming the fundamental limitation of the index.
We define a supersymmetric ensemble that gives equal weight to all states
along the direction generated by the supercharge,
and compute the macroscopic charges of the ensemble.
This procedure gives the correct functional form of the fully refined charge constraint
in AdS$_5$, AdS$_4$, and AdS$_7$. The major heuristic element of our computation
is the number of free multiplets in the theory, which we simply put in by hand.
For example, for the $SU(N)$ SYM in $d=4$, we need $\frac{1}{2}N^2$ free multiplets,
compared with $N^2$ in a genuinely free theory.
This number sets the scale of all conserved charges.

\section{AdS$_3$}\label{sec:3d}

In this section we discuss BTZ black holes in AdS$_3 \times S^3$ that are
dual to 2d CFT with $(4, 4)$ superconformal symmetry.
This section is based on section 5 of \cite{Larsen:2021wnu}.

\subsection{The BTZ Black Hole and its Charge Constraint}\label{sec:3dBH}

Thermodynamic properties of the AdS$_3$ black holes as well as
their BPS limit have been reviewed in section \ref{sec:AdS3BH}.
Let us briefly recall relevant information.
The BTZ black holes in AdS$_3 \times S^3$ carry an energy $E$
and an angular momentum $J$ that both arise from the isometry of AdS$_3$,
as well as two charges $Q_L$ and $Q_R$ associated with the isometry
$SU(2)_L \times SU(2)_R$ of $S^3$.

One choice of $\frac14$-BPS sector in this theory corresponds to energy that saturates the unitarity bound:
\bea\label{3dunitbound}
E &\geq& J + Q_L - \frac{k_L}{2}~.
\eea
In this formula $k_L$ is the level of the $SU(2)_L$ current which, because of ${\cal N}=4$ supersymmetry, is related to the central charge as $c_L = 6k_L$. On the other hand, all black hole solutions in AdS$_3 \times S^3$ satisfy the extremality bound,
\bea\label{3dextbound}
E &\geq& J - \frac{Q_L^2}{2k_L}~,
\eea
which is saturated at vanishing temperature.
This formula is entirely gravitational, but we have simply expressed Newton's constant $G_3$
in terms of the level $k_L$ using the Brown-Henneaux formula for  
the central charge \cite{Brown:1986nw}.

A BTZ black hole can only be $\frac14$-BPS if it saturates both of (\ref{3dunitbound}) and (\ref{3dextbound}). That is only possible if
\bea\label{3dcc}
Q_L &=& k_L~.
\eea
This is the charge constraint on supersymmetric AdS$_3$ black holes. In a charge sector that violates \eqref{3dcc} there are no supersymmetric black holes. 

\subsection{Multiplets of CFT$_2$ with $(4, 4)$ Supersymmetry}

The dual 2d CFT has $(4, 4)$ supersymmetry.
Its superconformal algebra factorizes into two independent copies of super-Virasoro algebra,
and includes a bosonic subgroup $SO(2,2) \times SU(2)_L \times SU(2)_R$
that matches the isometry of AdS$_3 \times S^3$.
To make progress, we first review representations of one chiral copy of the
small $\mathcal{N}=4$ super-Virasoro algebra with $SU(2)$ R-symmetry
\cite{Eguchi:1987sm,Eguchi:1987wf,Sevrin:1988ew,Eguchi:1988af}.

The maximal bosonic subalgebra of the small $\mathcal{N}=4$ super-Virasoro algebra
has two Cartans: $L_0$ of the Virasoro algebra and $Q_L$ of the $SU(2)$ R-symmetry.
Every weight in a representation of the super-Virasoro algebra can be chosen
to diagonalize the Cartans of the bosonic subalgebra,
so it is described by its $L_0$ and $Q_L$ eigenvalues $(h,q_L)$.
Each unitary representation of the super-Virasoro algebra is labeled by
the $L_0$ and $Q_L$ eigenvalues $(h,q_L)$ of its superconformal primary. 
Given a superconformal primary, the entire contents of the multiplet is determined,
as the descendants are obtained by applying various operators of the algebra to the primary.
We can focus on states that satisfy NS boundary conditions,
because representations in the Ramond sector are isomorphic through spectral flow by a half-integral unit.

There are two qualitatively different types of representations:
long multiplets whose primary has $h > q_L$ and
short multiplets whose primary has $h = q_L$.
The allowed values for $q_L$ of the primary are $0,1, \cdots, k_L-1$ for long multiplets,
and $0,1, \cdots, k_L$ for short multiplets.
The content of either type of representation can be summarized by its character
defined by ${\rm Tr} ~q^{L_0} y^{Q_L}$, a function of two fugacities $q$ and $y$.
The characters of the two types of representations are \cite{Eguchi:1987wf}:
\bea\label{3dN4char}
\text{Long : } {\rm ch}_{h,q_L}(q,y) \!&\!=\!&\! q^h F^{NS} 
\sum_{m=-\infty}^{\infty}  \left( y^{2(k_L+1)m +q_L+1}-y^{-2(k_L+1)m -q_L-1} \right) 
\frac{q^{(k_L+1)m^2+(q_L+1)m}}{y-y^{-1}} ~, \nn\\
\text{Short : } ~~\chi_{q_L}(q,y) \!&\!=\!&\! q^{\frac{q_L}{2}} F^{NS} \!
\sum_{m=-\infty}^{\infty}  \! \left(\! \frac{y^{2(k_L+1)m +q_L+1}}
{(1+yq^{m+\frac12})^2} \!-\! \frac{y^{-2(k_L+1)m -q_L-1}}{(1+y^{-1}q^{m+\frac12})^2} \!\right)\!
\frac{q^{(k_L+1)m^2+(q_L+1)m}}{y-y^{-1}} ~, \nn\\
\eea
where
\bea\label{3dFNS}
F^{NS} = \prod_{n\geq1}
\frac{\left( 1+yq^{n-\frac12} \right)^2 \left( 1+y^{-1}q^{n-\frac12} \right)^2}
{(1-y^2 q^n)(1-q^n)^2(1-y^{-2} q^n)} ~,
\eea
accounts for the action of creation operators,
i.e. the negative frequency modes $\{ G_{r<0}\}$ and $\{ L_{n<0}, J^i_{n<0}\}$
of the four fermionic and four bosonic generators. 

\subsection{The Supersymmetric Ensemble and the Charge Constraint}

The long and short multiplets discussed in the previous subsection are the only
unitary representations of the small $\cN=4$ super-Virasoro algebra.
Since the supersymmetry algebra of the 2d $(4, 4)$ theory is a direct sum of
two copies of the small $\cN=4$ super-Virasoro algebra,
any representation thereof is a direct product between two representations
of the small $\cN=4$ super-Virasoro algebra.
Therefore, the microscopic duals of AdS$_3$ black holes
must also organize themselves into such representations.

The long and short multiplet characters \eqref{3dN4char} both
exhibit the following property:
\bea\la{3dZ2even}
\chi_{q_L}(q,y) &=& \chi_{q_L}(q,q^{-1}y^{-1}) \cdot  q^{k_L} y^{2k_L} ~, \nn \\
{\rm ch}_{h,q_L}(q,y) &=& {\rm ch}_{h,q_L}(q,q^{-1}y^{-1}) \cdot  q^{k_L} y^{2k_L} ~.
\eea
This shows that, within any representation, a weight with $(h,q_L)$ is always
paired with another one with $(h+k_L-q_L, 2k_L-q_L)$.
To see this, suppose that there is a weight with $(h,q_L)$
either in the short or the long multiplet.
This contributes to the right hand side of \eqref{3dZ2even} by
\be\label{3swap}
q^h (q^{-1}y^{-1})^{q_L} \cdot q^k y^{2k_L} = q^{h+k_L-q_L} y^{2k_L-q_L} ~.
\ee
Then it follows from the equation that the character must contain
a term $q^{h+k_L-q_L} y^{2k_L-q_L}$, which can only be true if
a weight $(h+k_L-q_L, 2k_L-q_L)$ belonged to the multiplet.

The pair is characterized by the R-charges being mirrored about $k_L$
and the conformal weights \emph{in excess of} the unitarity bound
$L_0 - \frac12 Q_L$ being the same:
\bea\label{3hinexcess}
h - \frac{1}{2}q_L = (h+k_L-q_L) - \frac12 (2k_L-q_L)~.
\eea
Therefore, if both weights contribute equally to the grand canonical partition function,
then the macroscopic charge $Q_L$ obtained as a statistical average over the ensemble
will necessarily be $k_L$.

The condition that the weights $(h,q_L)$ and $(h+k_L-q_L, 2k_L-q_L)$ in the pair
contribute equally to the grand canonical partition function is
\bea\label{3susyens}
yq^{1/2} &=& 1~.
\eea
Therefore, when this condition is satisfied, the average $\langle Q_L \rangle =k_L$.
Indeed, explicit computation shows that
\begin{equation}
\langle Q_L \rangle = y\frac{\partial \log Z}{\partial y}\Big|_{y=q^{-\frac{1}{2}}} = k_L~, 
\end{equation}
for any partition function that is a product of characters satisfying \eqref{3dZ2even}.
We refer to grand canonical partition function with $yq^{1/2} = 1$ as the supersymmetric ensemble.  

Geometrically, the condition $yq^{1/2} = 1$ defining the supersymmetric ensemble
means all quantum states along a straight line in the $(h,q_L)$ plane are counted equally.
This is precisely the direction generated by the preserved supercharge,
corresponding to one of the factors in the numerator of (\ref{3dFNS}).

The definition of the supersymmetric ensemble is reminiscent of imposing
$yq^{1/2} = -1$, the substitution that turns the grand canonical partition function
into the index, or the elliptic genus.
With the condition $yq^{1/2} = -1$, two microstates related by the supercharge
$Q$ contribute equal magnitude, but with opposite signs.
Therefore, the only non-vanishing contributions are from short multiplets
where the primary is annihilated by the supercharge.
Moreover, combinations of short multiplets along the direction of the supercharge
combine to a long multiplet
\begin{equation}
\chi_{h,q_L-1}(q,y) + 2\chi_{h,q_L}(q,y) + \chi_{h,q_L+1}(q,y)  = {\rm ch}_{h,q_L}(q,y)\Big|_{h=\frac{1}{2}q_L}~,
\end{equation}
and also cancel automatically in the index.
With these cancellations, the index is unable to assign relative probabilities
to the charges in this direction, and so it cannot account for the constraint.
In contrast, the supersymmetric ensemble avoids massive cancellations and controls
the direction generated by the supercharge by taking the average over all configurations.
This prescription reproduces the charge constraint $\langle Q_L\rangle = k_L$ (\ref{3dcc})
that is satisfied for all supersymmetric black holes in AdS$_3 \times S^3$.

\section{Prescription for Higher Dimensions and Summary}\label{sec:gendim}

We now turn to the charge constraints in higher dimensions.
Before we present specific examples,
let us briefly outline the generic prescription for a heuristic derivation of
the charge constraints from dual CFTs across dimensions.
We shall follow this prescription in sections \ref{sec:5d} through \ref{sec:7d}
to obtain charge constraints of AdS$_5$, AdS$_4$ and AdS$_7$ black holes
from respective CFTs.

AdS black holes are dual to ensembles of quantum states in a
superconformal field theory in one fewer dimensions that all preserve
the same amount of supersymmetry as the black holes.
The local operators in the dual theory organize themselves into
representations of the applicable superconformal algebra.
Our starting point is the field content of the free representation,
which provides the basic building blocks of the CFTs.
It consists of free fields, both bosons and fermions, as well as derivatives that generate
conformal descendants, and equations of motion that impose physical conditions.
In a free CFT the particle number operator is well-defined,
and so the free fields correspond to single particle states.
There are infinitely many, because an arbitrary number of derivatives may act on the fields.

Every single particle state can be chosen as eigenstates of the Cartan generators
of the bosonic subalgebra.
The corresponding eigenvalues are the conformal dimension $E$,
angular momenta $J_i$, and the R-symmetry charges $Q_I$,
where the ranges of $i$ and $I$ depend on the dimension and on the amount of supersymmetry.
The totals of the microscopic quantum numbers for the entire ensemble give
the $E$, $J_i$, and $Q_I$ that we identify with the black hole charges.
We only pick microscopic states that individually preserve the same supersymmetries as the black hole.
These single particle BPS states are referred to as BPS letters.
Since they are annihilated by the chosen supercharges, $Q |\psi \rangle = 0$,
the BPS states must saturate the unitarity bound $\{ Q, Q^\dag \} \geq 0$.
The superalgebra expresses the left hand side as a sum over the bosonic Cartan operators,
so, schematically,
\bea\label{QQscheme}
\mathrm{BPS:} \quad \{ Q, Q^\dag \} = E - \sum_{i} J_i - \sum_{I} Q_I = 0~.
\eea
The quantum numbers of the BPS letters must satisfy the equality on the right,
giving a linear BPS relation between the energy and the other conserved charges.\footnote{
Depending on normalization of the charges, one or more terms in the sum
\eqref{QQscheme} may contain numerical coefficients that differ from one. 
An example is \eqref{7dunitbound} for the 6d (2, 0) superconformal algebra.
Throughout this paper, we use the notation of \cite{Cordova:2016emh},
to which we refer for details on the algebra and representations.}

The grand canonical partition function is the trace over all quantum states,
with weights assigned to each state by chemical potentials that couple
to the conserved charges.
We define it with an explicit restriction to BPS states:
\bea\label{defZ}
Z &\equiv& \mathrm{Tr}_\mathrm{BPS} \left[ e^{-\beta \{ Q, Q^\dag \}}
e^{\sum_i \omega_i J_i + \sum_I \Delta_I Q_I} \right] \nn\\
&=& \mathrm{Tr}_\mathrm{BPS} \left[ e^{\sum_i \omega_i J_i + \sum_I \Delta_I Q_I} \right]~.
\eea
The second line is because the superalgebra \eqref{QQscheme} gives
$\{ Q, Q^\dag \} = 0$.
This removes the dependence on conformal dimension,
but the partition function retains dependence on all chemical potentials
$\omega_i$ and $\Delta_I$, there are as many of them as there are charges.
Therefore, it is sensitive to the distribution of microstates along all directions in the charge space.

It is useful to define the grand canonical partition function over the BPS letters only.
This gives the single particle BPS partition function $Z_\mathrm{sp}$.
However, in a quantum field theory, general states belong to a multiparticle Fock space
that is generated by the single particle states in the usual way,
with occupation numbers restricted by fermion or boson statistics.
In the free theory any quantum number of a multiparticle state, including its energy,
is the sum over the corresponding single particle quantum numbers.
Therefore, the BPS partition function $Z$ over the entire BPS Hilbert space
can be derived from the single particle BPS partition function $Z_\mathrm{sp}$,
by taking combinatorics into account.

For example, for a single particle bosonic or fermionic BPS state
that yields the single particle partition function $x_B$ or $x_F$,
the partition function for the full Fock space is
\bea\label{countB}
1+x_B+x_B^2 + \cdots &=& \frac{1}{1-x_B}~, 
\eea
and $1 + x_F$, respectively.
If there are $N_B$ bosonic and $N_F$ single particle BPS states,
each of which yields the single particle partition function $x_{B,i}$ and $x_{F,j}$,
the full partition function becomes
\be
\label{countBF}
Z = \frac{\prod_{j=1}^{N_F} (1+x_{F,j})}{\prod_{i=1}^{N_B}(1-x_{B,i})}~. 
\ee
These formulae are simply the standard Bose-Einstein and Fermi-Dirac distributions
from elementary statistical physics,
but expressed in a notation commonly used when discussing supersymmetric indices.
In our prescription, we compute the multiparticle partition function as a
simple exponential of the single particle partition function: 
\bea\label{assumeexp}
Z &=& e^{Z_\mathrm{sp}} = \exp\left(\sum_{i=1}^{N_B}x_{B,i} + \sum_{j=1}^{N_F}x_{F,j}\right)~. 
\eea
This is the limit of classical statistical physics.
It is justified when the occupation number for any single particle state is so small
that it is likely to be either $0$ or $1$.
This assumption may be realized by the large number of gauge degrees of freedom
for each single particle state.
An improved treatment of such gauge degrees of freedom would project
onto gauge singlets at the end, and that we do not do.

Given the grand canonical partition function $Z$ for the full Hilbert space,
we can derive the macroscopic charges as ensemble averages in a standard manner.
\eqref{defZ} gives
\bea\label{genericcharge}
Q_J &=& \frac{\mathrm{Tr}_\mathrm{BPS}
\left[ Q_J \cdot e^{\sum_i \omega_i J_i + \sum_I \Delta_I Q_I} \right]}
{\mathrm{Tr}_\mathrm{BPS}
\left[ e^{\sum_i \omega_i J_i + \sum_I \Delta_I Q_I} \right]}
= \pdv{\Delta_J} \log Z~, \nn\\
J_j &=& \frac{\mathrm{Tr}_\mathrm{BPS}
\left[ J_j \cdot e^{\sum_i \omega_i J_i + \sum_I \Delta_I Q_I} \right]}
{\mathrm{Tr}_\mathrm{BPS}
\left[ e^{\sum_i \omega_i J_i + \sum_I \Delta_I Q_I} \right]}
= \pdv{\omega_j} \log Z~.
\eea
These formulae express all the charges in terms of an equal number of chemical potentials.

\begin{table}
\begin{center}
\begin{tabular}{| c | c | c | c | c | c | c | c |}
\hline
$D$ & Charges & Constraint & SCA & Free multiplet & $N$ & $G_D$ 
\\
\hline
\hline
4 & $J,~Q_{1,2,3,4}$ & (\ref{4dcc}) & 3d $\mathcal{N}=8$ & $B_1 [0]^{[0,0,1,0]}_{1/2}$
& $\frac{\sqrt{2}}{3} N^{\frac32} = 1$ & $G_4 = \frac12$ 
\\
\hline
5 & $J_{1,2},~Q_{1,2,3}$ & (\ref{5dcc}) & 4d $\mathcal{N}=4$ & $B_1 \bar{B}_1 [0;0]^{[0,1,0]}_1$
& $\frac{1}{2}N^2=1$ & $G_5 = \frac{\pi}{4}$ 
\\
\hline
7 & $J_{1,2,3},~Q_{1,2}$ & (\ref{7dcc}) & 6d (2, 0) & $D_1 [0,0,0]^{[1,0]}_2$
& $\frac{2}{3}N^3 =1$ & $G_7 = \frac{\pi^2}{8}$  
\\
\hline
\end{tabular}
\caption{\label{summary} Summary of sections \ref{sec:5d}--\ref{sec:7d}.}
\end{center}
\end{table}

Denoting by $Q$ the supercharge that is preserved by the black hole
and by the dual BPS states,
we now impose a linear relation between the chemical potentials such that
quantum states that differ by the charges of $Q$ are given the same weight.
The statistical computations of macroscopic charges \eqref{genericcharge} were done
prior to this stage, and included the gradient of the partition function in the
direction along the constraint between the potentials.
Therefore, the computation reflects the dependence of the partition function on all chemical potentials.

After the constraint on the chemical potentials is imposed,
the statistical formulae \eqref{genericcharge} express all macroscopic charges
in terms of one variable less than there are charges.
Equivalently, the charges that are realized form a co-dimension one surface
in the space of all charges, i.e. they satisfy a constraint.
We find that the constraint on charges arrived at this way,
from the field content of the microscopic theory, has the same highly non-trivial form
as the non-linear charge constraint of the supersymmetric black holes. 

The black hole charges that satisfy the non-linear constraint in the gravitational theory,
are in units of Newton's gravitational coupling constant. 
In contrast, in its simplest form, the microscopic computation considers a single free field.
Our results are incomplete, because we do not determine the relative scale of the charges
in the two computations.
Comparison between the computations gives a value for Newton's constant or,
equivalently, for the effective number of free fields.
The summary in Table \ref{summary} records these values.

\section{AdS$_5$}\label{sec:5d}

In this section we discuss how the charge constraint of supersymmetric,
rotating and charged black holes in AdS$_5$,
emerges from its dual $\mathcal{N}=4$ Super-Yang-Mills theory in 4d.
We follow the prescription outlined in section \ref{sec:gendim}.

\subsection{The Black Hole and the Charge Constraint}\label{sec:5dBH}

Asymptotically AdS$_5$ black holes arise as solutions to type-IIB supergravity
in AdS$_5 \times S^5$ \cite{Gutowski:2004ez,Gutowski:2004yv,Chong:2005da,
Chong:2005hr,Kunduri:2006ek,Wu:2011gq}.
They carry the mass $E$ and two angular momenta $J_{1,2}$
for the isometry $SO(2,4)$ of AdS$_5$,
and three charges $Q_{1,2,3}$ for the isometry $SO(6)$ of $S^5$.
The black hole solution with all $6$ conserved quantities independent is known.

The black hole is supersymmetric when the unitarity bound between the mass and the charges
\bea\label{5dunitbound}
E \geq J_1 + J_2 + Q_1 + Q_2 + Q_3~,
\eea
is saturated. We have set the AdS$_5$ radius $\ell_5 = 1$.
Importantly, saturation is possible only when the charges obey an additional relation
\cite{Kim:2006he,Choi:2018hmj,Larsen:2019oll}
\bea\label{5dcc}
&& \left( Q_1Q_2Q_3 + \frac{N^2}{2}J_1J_2 \right) \nn\\
&=& \left( Q_1+Q_2+Q_3 + \frac{N^2}{2}\right)
\left( Q_1Q_2 + Q_2Q_3 + Q_3Q_1 - \frac{N^2}{2}(J_1+J_2) \right)~.
\eea
We have traded the 5d Newton's constant $G_5$
into the field theory variable $N$ via 
$$
\frac{1}{2}N^2 = \frac{\pi\ell^3_5}{4G_5}~,
$$
for future convenience, but we stress that the origin of the charge constraint (\ref{5dcc})
is purely gravitational. (\ref{5dcc}) is the charge constraint for supersymmetric AdS$_5$ black holes.

We also present an unrefined ($J_1=J_2=J$ and $Q_1=Q_2=Q_3=Q$,
where $Q$ should not be confused with the preserved supercharge) version
of (\ref{5dcc}) that is more approachable, but still quite non-trivial:
\bea\label{5dccur}
\left( Q^3 + \frac{N^2}{2}J^2 \right)
- \left( 3Q + \frac{N^2}{2}\right)
\left( 3Q^2 - N^2J \right) &=& 0~. \nn\\
\eea

\subsection{The 4d $\mathcal{N}=4$ Free Vector Multiplet}\label{sec:5dmul}

The charged, rotating AdS$_5$ black holes introduced in the previous subsection
are dual to quantum states in the $\mathcal{N}=4$ Super-Yang-Mills theory in 4d.
In this subsection we introduce the free vector multiplet of the 4d $\cN=4$
superconformal algebra $\mathfrak{psu}(2,2|4)$ that generates the single particle states.

Local operators can be organized into super-representations of the
4d $\cN=4$ superconformal algebra. 
A super-representation consists of a superconformal primary and its descendants.
Following the notation of \cite{Cordova:2016emh}, we identify representations
by the Dynkin labels of the superconformal primary under the maximal bosonic subalgebra: 
$$
[j; \bar{j}]^{[R_1, R_2, R_3]}_E~.
$$
Here $E$ is the conformal weight, $j$, $\bar{j}$ are the integer-quantized Dynkin labels
for the $SU(2) \times SU(2)$ Lorentz group, and $R_{1,2,3}$ are the Dynkin labels
for the $SU(4)$ R-symmetry group.

The black hole charges used in section \ref{sec:5dBH} refer to the $SO(2,4) \times SO(6)$
isometry group of the AdS$_5 \times S^5$ geometry.
They are charges of $SO(2)$ rotations in orthogonal $2$-planes.
The orthogonal basis are related to the Dynkin basis as: 
\bea\label{Dynkinorthorep}
&& J_1 = \frac{j+\bar{j}}{2}~, \qquad \qquad ~~~~~~~
J_2 = \frac{j-\bar{j}}{2}~, \nn\\
&& Q_1 = R_2 + \frac{R_1+R_3}{2}~, \qquad
Q_2 = \frac{R_1+R_3}{2}~, \qquad
Q_3 = \frac{R_1-R_3}{2}~.
\eea
The energy $E$ is common to the two bases.
We further note that $[R_1, R_2, R_3]$ are $SU(4)$ Dynkin labels,
not to be confused with $SO(6)$ Dynkin labels that are related via $R_1 \leftrightarrow R_2$.
In our conventions $[1,0,0]$ is {\bf 4} (fundamental of $SU(4)$ but spinor of $SO(6)$)
and $[0,1,0]$ is {\bf 6} (fundamental of $SO(6)$ but antisymmetric tensor of $SU(4)$).

The supersymmetric black holes discussed in section \ref{sec:5dBH}
preserve $\frac{1}{16}$ of the supersymmetry, so they correspond to
BPS states that are annihilated by $2$ out of $32$ Hermitian supercharges.
We choose $Q$ and $Q^\dag$ that obey the algebra
\bea\label{5QSalgebra}
2\{ Q, Q^\dag\} &=&
E - \left( j + \frac32 R_1 + R_2 + \frac12 R_3 \right) \nn\\
&=& E-(Q_1+Q_2+Q_3+J_1+J_2) \geq 0~,
\eea
which plays the role of \eqref{QQscheme} in the generic prescription.
As explained in section \ref{sec:gendim},
any field component can be identified with a weight in a representation,
and so it is an eigenstate with respect to the operators $E$, $Q_I$ and $J_i$.
It is BPS if the corresponding eigenvalues saturate \eqref{5QSalgebra}.

In 4d superconformal theories, a field $[j; \bar{j}]^{[R_1, R_2, R_3]}_E$ is a free field
if at least one of $j$ and $\bar{j}$ is zero and, in addition, $E = 1+\frac{j+\bar{j}}{2}$.
There is one multiplet of the 4d $\mathcal{N}=4$ superconformal algebra that contains
a free field: the free vector multiplet, $B_1 \bar{B}_1 [0;0]^{[0,1,0]}_1$.
All that we need is Table \ref{5dfmtable}, where we summarize the BPS content
of the free vector multiplet, i.e. all weights in the multiplet that saturate the
unitarity bound (\ref{5QSalgebra}). 

There are $9$ field components that satisfy the BPS condition.
The BPS bosons are $3$ of the $6$ scalars in the theory, 
and $1$ of the $2$ gauge field components.
The fermions are, in the language of $\cN=1$ supersymmetry, $3$ chiralini and $2$ gaugini.
The entry below the first double line is an equation of motion that relates the two gaugini.
It should be counted as a ``negative" field that serves to cancel some gaugini operators
with derivatives acting on them.
There are equations of motion for other free fields as well, but this component
of the gaugino equation of motion is the only one that is consistent with the BPS condition.
The last two entries in Table \ref{5dfmtable} are derivatives that may act on any of the fields,
and on the equation of motion, to produce BPS descendants.
The gradient operator has $4$ components in $4$ dimensions,
but only $2$ preserve the BPS-ness of the field.
The $9-1=8$ free fields and their derivatives generate the entire list of
supersymmetric operators in the free vector multiplet.
From a bulk point of view, these are the single particle BPS states.

\subsection{The Supersymmetric Ensemble}

Given the exhaustive list of single particle BPS states generated by
the supersymmetric operators in Table \ref{5dfmtable},
we can now define a grand canonical partition function $Z_\mathrm{sp}$
over the single particle states.
Rather than the chemical potentials as in \eqref{defZ},
we use fugacities $(p,q,x,y,z)$ that are related by
\bea
e^{\omega_1} = p^2~, \qquad
e^{\omega_2} = q^2~, \qquad
e^{\Delta_1} = x^2~, \qquad
e^{\Delta_2} = y^2~, \qquad
e^{\Delta_3} = z^2~, 
\eea
and so define the single particle BPS partition function by
\bea\label{5ddeff}
Z_\mathrm{sp} &\equiv&
\mathrm{Tr}_\mathrm{BPS} \left[ p^{2J_1} q^{2J_2} x^{2Q_1} y^{2Q_2} z^{2Q_3} \right]~.
\eea
The maneuver doubling the exponents avoids fractional powers,
although the subtle feature of non-analyticity and ``second sheet''
\cite{Cassani:2021fyv} is not relevant to our purpose.

We read off the single particle partition function $Z_\mathrm{sp}$ from Table \ref{5dfmtable}.
The sum over the weights of the $8=9-1$ free fields gives
\bea
x^2+y^2+z^2+xyzpq
\left(\frac{1}{x^2} + \frac{1}{y^2} + \frac{1}{z^2} + \frac{1}{p^2} + \frac{1}{q^2} -1 \right)
+ p^2q^2~.
\eea
Any number of the two derivatives that preserve the BPS condition can act on
each of the free fields, and on the equation of motion.
Each derivative contributes a factor of $p^2$ or $q^2$,
so we need a geometric sum over these.
We then find the single particle BPS partition function
\bea\label{5dsppf}
Z_\mathrm{sp} &=& \frac{x^2+y^2+z^2+xyzpq
\left(\frac{1}{x^2} + \frac{1}{y^2} + \frac{1}{z^2} + \frac{1}{p^2} + \frac{1}{q^2} - 1\right)
+ p^2q^2}{(1-p^2)(1-q^2)}~.
\eea
According to our prescription discussed in section \ref{sec:gendim}, the full partition function
is equal to the exponential of the single particle partition function \eqref{5dsppf}:
\bea\label{5dlogZ}
Z &\equiv& \exp[ Z_\mathrm{sp}]~.
\eea

From this grand canonical partition function,
we obtain the macroscopic charges as ensemble averages in the standard manner.
Changing variables $(\Delta_I, \omega_i) \, \to \, (p,q,x,y,z)$, \eqref{genericcharge} becomes
\bea
2Q_1 = x \pdv{x} \log Z~, \qquad 2J_1 = p \pdv{p} \log Z~,
\eea
and analogously for the charges with different indices.
The charges obtained from (\ref{5dlogZ}) in this way are
\bea\label{5dchgpotrel}
Q_1 &=& \frac{2x^2+xyzpq
\left(-\frac{1}{x^2} + \frac{1}{y^2} + \frac{1}{z^2} + \frac{1}{p^2} + \frac{1}{q^2} - 1\right)
}{2(1-p^2)(1-q^2)}~, \\
J_1 &=& \frac{2p^2(q^2+x^2+y^2+z^2) + xyzpq(1+p^2)\left(
\frac{1}{x^2} + \frac{1}{y^2} + \frac{1}{z^2} - \frac{1}{p^2} + \frac{1}{q^2} - 1\right)
+4xyzpq }{2(1-p^2)^2(1-q^2)}~, \nn
\eea
and similarly for the permutations.
\eqref{5dchgpotrel} express the $5$ average charges of the ensemble in terms of $5$ potentials. 
BPS states are populated throughout the five-dimensional charge space,
not just on some specific hypersurface thereof.
Thus, the $5$ average charges may take generic values without any particular constraint as well,
as the $5$ potentials are varied.

We now define the supersymmetric ensemble as a grand canonical ensemble
where the operators that are separated in the charge space along
the direction of the preserved supercharge are weighed equally.
The preserved supercharge $Q$ carries quantum numbers
$(E, J_1, J_2, Q_1, Q_2, Q_3) =
\left( \frac12, -\frac12, -\frac12, \frac12, \frac12, \frac12 \right)$,
so the supersymmetric ensemble corresponds to the relation
\bea\label{5dsusyens}
\frac{xyz}{pq} &=& 1~,
\eea
between the fugacities.
For the supersymmetric ensemble satisfying (\ref{5dsusyens}),
there is one relation between the five charges (\ref{5dchgpotrel}):
\bea\label{5dccmicro}
\left( Q_1Q_2Q_3 + J_1J_2 \right)
- \left( Q_1+Q_2+Q_3 + 1\right)
\left( Q_1Q_2 + Q_2Q_3 + Q_3Q_1 - J_1-J_2 \right) &=& 0~. ~~~~~~~
\eea
This is precisely the supersymmetric AdS$_5$ black hole charge constraint (\ref{5dcc}) with
\bea\label{5dNvalue}
\frac{1}{2}N^2 = 1 \quad \leftrightarrow \quad  \frac{\pi\ell^3_5}{4G_5} = 1~.
\eea
Equivalently, the statistical constraint \eqref{5dccmicro} agrees with the
macroscopic constraint (\ref{5dcc}) if macroscopic charges are in units of $\frac12 N^2$. 
A truly free $SU(N)$ theory would have $N^2$ identical copies of the free fields.
We interpret the remaining relative factor $\frac12$ as a reduction that is due to interactions,
but we claim no quantitative understanding of this factor.
This feature non-withstanding, our computation establishes the functional dependence
on charges of the constraint (\ref{5dcc}) from the combinatorics of free fields. 

The unrefined charges are defined by taking $x=y=z$ and $p=q$ in (\ref{5dchgpotrel}):
\bea\label{5dchgpotrelur}
Q &=& \frac{2x^2 + 2 x^3 + xp^2 - x^3p^2}{2(1-p^2)^2}~, \nn\\
J &=& \frac{3x(1+x)^2p^2 + (2+3x-x^3)p^4}{2(1-p^2)^3}~. 
\eea
It follows automatically that, upon picking the supersymmetric ensemble $x^3=p^2$, 
these charges satisfy the unrefined charge constraint (\ref{5dccur}) with $\frac12 N^2 = 1$.

\section{AdS$_4$}\label{sec:4d}

In this section we derive the charge constraint for the supersymmetric AdS$_4$ black holes.
The AdS$_4$ theory and its dual CFT$_3$ have features that are absent in
AdS$_5$/CFT$_4$, such as magnetic charges and the Chern-Simons term.
Such complications are not directly relevant to our computation.
We find the charge constraint of the supersymmetric, rotating and electrically charged
black holes in AdS$_4$ from the free hypermultiplet of
the 3d $\mathcal{N}=8$ superconformal algebra.

\subsection{The Black Hole and the Charge Constraint}\label{sec:4dBH}

Asymptotically AdS$_4$ black holes arise as solutions to the
4d gauged supergravity theories \cite{Chong:2004na,Hristov:2019mqp}.
They carry the mass $E$ and an angular momentum $J$ for the
isometry $SO(2,3)$ of AdS$_4$,
and four electric charges $Q_{1,2,3,4}$ for the isometry $SO(8)$ of $S^7$.
The solution with the four electric charges pairwise equal ($Q_1=Q_3$ and $Q_2=Q_4$)
was found in \cite{Chong:2004na}, and the most general solution
with all four electric charges independent was found in \cite{Hristov:2019mqp}.

The black hole is supersymmetric when the unitarity bound between
the mass and the charges
\bea\label{4dunitbound}
E \geq J + \frac12 \left( Q_1 + Q_2 + Q_3 + Q_4 \right)~,
\eea
is saturated.
However, the saturation is possible only when the charges obey the additional relation
\cite{Choi:2018fdc,Hristov:2019mqp}\footnote{Although the solution with
all four electric charges independent was found in \cite{Hristov:2019mqp},
its charge constraint had been correctly conjectured earlier \cite{Choi:2018fdc},
based on the solution with pairwise equal charges \cite{Chong:2004na}
and the structure of the entropy function.}
\bea\label{4dcc}
(\mathbb{Q}_3)^2 - (\mathbb{Q}_1) (\mathbb{Q}_2) (\mathbb{Q}_3)
+ (\mathbb{Q}_1)^2 (\mathbb{Q}_4) &=& 0~,
\eea
where we have used the shorthand notation
\bea
(\mathbb{Q}_1) &\equiv& Q_1+Q_2+Q_3+Q_4~, \nn\\
(\mathbb{Q}_2) &\equiv& Q_1Q_2+Q_1Q_3+Q_1Q_4 + Q_2Q_3 + Q_2Q_4 + Q_3Q_4 + \frac{2N^3}{9}~, \nn\\
(\mathbb{Q}_3) &\equiv& Q_1Q_2Q_3 + Q_1Q_2Q_4 + Q_1Q_3Q_4 + Q_2Q_3Q_4 - \frac{4N^3}{9}J~, \nn\\
(\mathbb{Q}_4) &\equiv& Q_1Q_2Q_3Q_4 + \frac{2N^3}{9} J^2~.
\eea
In the formulae above, we set the AdS$_4$ radius $\ell_4 = 1$.
We traded the 4d Newton's constant for the field theory variable $N$ via
$$N^{\frac32} = \frac{3}{2\sqrt{2}G_4}~,$$
for future convenience,
but we stress that the origin of the charge constraint (\ref{4dcc}) is purely gravitational.
(\ref{4dcc}) is the supersymmetric AdS$_4$ black hole charge constraint.

To make the formulae more approachable and to make the connection to the literature,
we also present the charge constraint (\ref{4dcc}) with pairwise equal electric charges
(see e.g. \cite{Choi:2018fdc})
\bea\label{4dccurp}
Q_1Q_2 (Q_1+Q_2)^2 - (Q_1+Q_2) \cdot \frac{2N^3}{9} J - \frac{2N^3}{9} J^2 &=& 0~.
\eea
as well as the version with all four electric charges equal (see e.g. \cite{Larsen:2020lhg}):
\bea\label{4dccur}
4 Q^4 - 2Q \cdot \frac{2N^3}{9} J - \frac{2N^3}{9} J^2 &=& 0~.
\eea
The formulae simplify greatly, but they remain quite nontrivial. 
The unrefined charge $Q$ in this formula should not be confused with the preserved supercharge.

\subsection{The 3d $\mathcal{N}=8$ Free Hypermultiplet}\label{sec:4dmul}

In this subsection we present the free hypermultiplet of the 3d $\cN=8$
superconformal algebra, from which the AdS$_4$ charge constraint
(\ref{4dcc}) will be derived in the next subsection.

The 3d $\mathcal{N}=8$ superconformal algebra has maximal bosonic subalgebra
$\mathfrak{so}(2,3) \oplus \mathfrak{so}(8)$, matching the isometry of AdS$_4 \times S^7$.
Local operators in the theory are organized into representations of this subalgebra.
A super-representation of the 3d $\mathcal{N}=8$ superconformal algebra is
uniquely specified by the Dynkin labels of its superconformal primary.
Following the notation of \cite{Cordova:2016emh},
we write representations of the bosonic subalgebra as
$$[j]^{[R_1, R_2, R_3, R_4]}_E~,$$
where $E$ is the conformal weight, $j$ is the integer-quantized $SO(3)$ Dynkin label,
and $[R_1,R_2, R_3, R_4]$ are the $SO(8)$ Dynkin labels so that $[1,0,0,0]$ is the vector {\bf 8}.

The black hole charges used in section \ref{sec:4dBH} refer to the
orthogonal basis that is related to the Dynkin basis as
\bea
&& J= \frac{j}{2} ~, \\
&& Q_1 = R_3 + R_2 + \frac{R_1+R_4}{2}~, ~~
Q_2 = R_2 + \frac{R_1+R_4}{2}~, ~~
Q_3 = \frac{R_1+R_4}{2}~, ~~
Q_4 = \frac{R_1-R_4}{2}~. \nn
\eea
This relation between the orthogonal and the Dynkin bases of $SO(8)$
differs from the more conventional one by $R_1 \leftrightarrow R_3$.
We have exploited the $S_3$ outer automorphism of $SO(8)$ to match
the convention (\ref{4dunitbound}) with that of \cite{Cordova:2016emh}.

A (not necessarily the highest) weight $[j]^{[R_1, R_2, R_3, R_4]}_E$
is annihilated by our choice of supercharge $Q$ if it saturates the unitarity bound
\bea\label{4dunitboundmicro}
E &\geq& \frac12 j + R_1 + R_2 + \frac12 R_3 + \frac12 R_4 \nn\\
&=& J + \frac12 \left( Q_1 + Q_2 + Q_3 + Q_4 \right)~,
\eea
that every weight must satisfy. Such weights correspond to local BPS operators.

\begin{table}\catcode`\-=12
\begin{center}
\begin{tabular}{| c | c | c | c c c c c | c c c c c |}
\hline
& Bosonic Rep. & $E$ & $j$ & $R_1$ & $R_2$ & $R_3$ & $R_4$ &
$J$ & $Q_1$ & $Q_2$ & $Q_3$ & $Q_4$ \\
\hline
\hline
\multirow{8}{*}{Free fields} & \multirow{4}{*}{$[0]^{[0,0,1,0]}_{\frac12} $} & $\frac12$ & 0 & 0 & 0 & 1 & 0 & 0 & 1 & 0 & 0 & 0 \\
& & $\frac12$ & 0 & 0 & 1 & $-1$ & 0 & 0 & 0 & 1 & 0 & 0 \\
& & $\frac12$ & 0 & 1 & $-1$ & 0 & 1 & 0 & 0 & 0 & 1 & 0 \\
& & $\frac12$ & 0 & 1 & 0 & 0 & $-1$ & 0 & 0 & 0 & 0 & 1 \\
\cline{2-13}
& \multirow{4}{*}{$[1]^{[0,0,0,1]}_1$} & 1 & 1 & 0 & 0 & 0 & 1 &
$\frac12$ & $\frac12$ & $\frac12$ & $\frac12$ & $-\frac12$ \\
& & 1 & 1 & 0 & 1 & 0 & $-1$ &
$\frac12$ & $\frac12$ & $\frac12$ & $-\frac12$ & $\frac12$ \\
& & 1 & 1 & 1 & $-1$ & 1 & 0 &
$\frac12$ & $\frac12$ & $-\frac12$ & $\frac12$ & $\frac12$ \\
& & 1 & 1 & 1 & 0 & $-1$ & 0 &
$\frac12$ & $-\frac12$ & $\frac12$ & $\frac12$ & $\frac12$ \\
\hline
\hline
Derivative & $[2]^{[0,0,0,0]}_1$ & 1 & 2 & 0 & 0 & 0 & 0 & 1 & 0 & 0 & 0 & 0 \\
\hline
\end{tabular}
\caption{\label{4dfmtable}
Components of the BPS operators in the free hypermultiplet $B_1 [0]^{[0,0,1,0]}_{\frac12}$.
The first $8$ rows are free fields, followed by one derivative that preserves BPS.}
\end{center}
\end{table}

In 3d superconformal theories, a field $[j]^{[R_1, R_2, R_3, R_4]}_E$
is free if $j \leq 1$ and $E = \frac{j+1}{2}$.
There are two multiplets of the 3d $\mathcal{N}=8$ superconformal algebra
that contain a free field \cite{Cordova:2016emh}.
The free hypermultiplets $B_1 [0]^{[0,0,1,0]}_{\frac12}$ and
$B_1 [0]^{[0,0,0,1]}_{\frac12}$ are related by a $Z_2$ subgroup
of the outer automorphism of $SO(8)$,
so we can choose $B_1 [0]^{[0,0,1,0]}_{\frac12}$ without loss of generality.
The rest of this section would be reproduced with minimal relabeling
had we chosen otherwise.

In Table \ref{4dfmtable} we summarize all weights in this free hypermultiplet that
saturate the unitarity bound (\ref{4dunitboundmicro}).
There are $8$ free fields: $4$ scalars and $4$ spinors.
There is no equation of motion that is compatible with the BPS condition.
The last entry is a derivative that can act on any of the fields and so produce its BPS descendants.
Note that out of 3 derivatives in 3 dimensions, only 1 preserves the BPS-ness of the field. 
The 8 free fields and their derivatives are the exhaustive list
of supersymmetric operators in the free hypermultiplet.

\subsection{The Supersymmetric Ensemble}

We now compute the single particle BPS partition function as
a trace over the free hypermultiplet states given in Table \ref{4dfmtable},
with fugacities $(p,x,y,z,w)$ conjugate to each charge:
\bea\label{4dsppf}
Z_{\rm sp} &\equiv& {\rm Tr}_{\rm BPS} \left[ p^{2J} x^{2Q_1} y^{2Q_2} z^{2Q_3} w^{2Q_4} \right]\nn\\
&=& \frac{x^2+y^2+z^2+w^2+ pxyzw \left( \frac{1}{x^2} + \frac{1}{y^2} + \frac{1}{z^2}
+ \frac{1}{w^2} \right) }{1-p^2}~.
\eea
It is the derivative that gives rise to the geometric series in $p^2$.
(\ref{4dsppf}) is the single particle partition function.

The grand canonical partition function over the full Hilbert space
is given by the ordinary exponential of the single particle partition function:
$Z \equiv \exp[Z_{\rm sp}]$.
We then compute the macroscopic charges as statistical averages. They are
\bea\label{4dchgpotrel}
Q_1 \!&\!=\!&\! \frac{pxyzw \left( \frac{1}{x^2} + \frac{1}{y^2} + \frac{1}{z^2}
+ \frac{1}{w^2} \right) + 2x^2 - \frac{2 p xyzw}{x^2} }{2 (1-p^2)} ~, \\
J \!&\!=\!&\! \frac{pxyzw \!\left( \frac{1}{x^2} + \frac{1}{y^2} + \frac{1}{z^2}
+ \frac{1}{w^2} \right)\!}{2 \left(1-p^2\right)} + \frac{p^2 \!\left( x^2+y^2+z^2+w^2
+ p xyzw \!\left(\frac{1}{x^2}+\frac{1}{y^2}+\frac{1}{z^2}+\frac{1}{w^2}\right)
\!\right)\!}{\left(1-p^2\right)^2}~, \nn
\eea
Analogous expressions for $Q_2$, $Q_3$, and $Q_4$ follow by simple permutations of indices.

Finally, we define the supersymmetric ensemble as a grand canonical ensemble
where the operators that are separated in the charge space along the direction
of the preserved supercharge are weighed equally.
The preserved supercharge $Q$ carries quantum numbers
$(E, J, Q_1, Q_2, Q_3, Q_4) =
\left( \frac12, -\frac12, \frac12, \frac12, \frac12, \frac12 \right)$,
so the supersymmetric ensemble corresponds to imposing the relation
\bea\label{4dsusyens}
\frac{xyzw}{p} &=& 1~,
\eea
between the fugacities.

Because of the relation \eqref{4dsusyens} between the $5$ potentials, in the
supersymmetric ensemble the $5$ charges $Q_{1, 2, 3, 4}$ and $J$ are not independent.
The expressions (\ref{4dchgpotrel}) give the relation:
\bea\label{4dccmicro}
(\mathbb{Q}_3)_{\frac92}^2
- (\mathbb{Q}_1)_{\frac92} (\mathbb{Q}_2)_{\frac92} (\mathbb{Q}_3)_{\frac92}
+ (\mathbb{Q}_1)_{\frac92}^2 (\mathbb{Q}_4)_{\frac92} &=& 0~,
\eea
where
\bea
(\mathbb{Q}_1)_{\frac92}
&\equiv& Q_1+Q_2+Q_3+Q_4~, \nn\\
(\mathbb{Q}_2)_{\frac92} 
&\equiv& Q_1Q_2+Q_1Q_3+Q_1Q_4 + Q_2Q_3 + Q_2Q_4 + Q_3Q_4 + 1 ~, \nn\\
(\mathbb{Q}_3)_{\frac92}
&\equiv& Q_1Q_2Q_3 + Q_1Q_2Q_4 + Q_1Q_3Q_4 + Q_2Q_3Q_4 - 2J~, \nn\\
(\mathbb{Q}_4)_{\frac92}
&\equiv& Q_1Q_2Q_3Q_4 + J^2~.
\eea
It is precisely the supersymmetric AdS$_4$ black hole charge constraint (\ref{4dcc})
with the numerical values
\bea\label{4dNvalue}
 \frac{\sqrt{2}}{3} N^{\frac32} = 1 \quad \leftrightarrow \quad G_4 = \frac12~. 
\eea
We interpret this relative scale of all charges as the effective number of free multiplets
needed to account for the constraint. 

The formulae simplify significantly when we do not distinguish between all $4$ electric charges. 
First, let $z=x$ and $w=y$ in (\ref{4dchgpotrel}):
\bea\label{4dchgpotrelurp}
Q_1 = Q_3 &=& \frac{x^2}{1-p} ~, \nn\\
Q_2 = Q_4 &=& \frac{y^2}{1-p} ~, \nn\\
J &=& \frac{p (x^2+y^2)}{(1-p)^2}~.
\eea
The definition of the supersymmetric ensemble \eqref{4dsusyens} simplifies to $p=x^2y^2$, 
and then the charges \eqref{4dchgpotrelurp} satisfy
\bea\label{4dccmicrourp}
Q_1Q_2 (Q_1+Q_2)^2 - (Q_1+Q_2) J - J^2 &=& 0~. 
\eea
This is the pairwise unrefined version of the charge constraint (\ref{4dccurp}) with
$ \frac{\sqrt{2}}{3} N^{\frac32} = 1$.

To treat all $4$ electric charges as identical, we further let $x=y$ in (\ref{4dchgpotrelurp}):
\bea\label{4dchgpotrelur}
Q \equiv Q_{1,2,3,4} &=& \frac{x^2}{1-p} ~, \nn\\
J &=& \frac{2 p x^2}{(1-p)^2}~.
\eea
These charges, with the equation $p=x^4$ defining the supersymmetric ensemble, satisfy
\bea\label{4dccmicrour}
4 Q^4 - 2QJ - J^2 &=& 0~. 
\eea
This is the fully unrefined version of the charge constraint (\ref{4dccur}) with 
$ \frac{\sqrt{2}}{3} N^{\frac32} = 1$.

\section{AdS$_7$}\label{sec:7d}

In this section we derive the charge constraint for the supersymmetric,
rotating and charged black holes in AdS$_7$.
from the dual $(2, 0)$ theory in 6d.

\subsection{The Black Hole and the Charge Constraint}\label{sec:7dBH}

Asymptotically AdS$_7$ black holes arise as solutions to a consistent truncation
of the 11d supergravity on $S^4$.
They carry the mass $E$ and
three angular momenta $J_{1,2,3}$ for the isometry $SO(2,6)$ of AdS$_7$,
and two charges $Q_{1,2}$ for the isometry $SO(5) \sim Sp(4)$ of $S^4$.
Particular solutions with equal angular momenta \cite{Chong:2004dy},
those with equal charges \cite{Chow:2007ts}
and those with two vanishing angular momenta and two independent charges
\cite{Wu:2011gp,Chow:2011fh} were constructed some time ago,
but the solution with all angular momenta and charges independent
was found only recently in \cite{Bobev:2023bxl}.

These black holes are supersymmetric when the unitarity bound between the
mass and the charges
\bea\label{7dunitbound}
E \geq J_1 + J_2 + J_3 + Q_1 + Q_2~,
\eea
is saturated.
However, the saturation is possible only when the charges obey the
additional relation \cite{Larsen:2020lhg,Bobev:2023bxl}\footnote{
The convention for charges differ from that of \cite{Bobev:2023bxl}
by $J_i^{here} = J_i^{there}$ and $Q_i^{here} = \frac{Q_i^{there}}{2}$.}
\bea\label{7dcc}
&& \frac12 (Q_1^2 + Q_2^2 + 4Q_1Q_2) + \frac{N^3}{3} (J_1+J_2+J_3)
- \frac{Q_1Q_2(Q_1+Q_2) \!-\! \frac{N^3}{3} (J_1J_2 + J_2J_3 + J_3J_1)}
{Q_1+Q_2- \frac{N^3}{3}}\nn\\
\!&\!=\!&\! \sqrt{
\left( \frac12 (Q_1^2 + Q_2^2 + 4Q_1Q_2) + \frac{N^3}{3} (J_1+J_2+J_3) \right)^2
- \left( Q_1^2Q_2^2 + \frac{2N^3}{3}J_1J_2J_3 \right)}~.
\eea
In the formulae above, we set the AdS$_7$ radius $\ell_7 = 1$.
We traded the 7d Newton's constant for the field theory variable $N$ via
$$N^3 = \frac{3\pi^2}{16G_7}~,$$
for future convenience,
but we stress that the origin of the charge constraint (\ref{7dcc}) is purely gravitational.
(\ref{7dcc}) is the supersymmetric AdS$_7$ black hole charge constraint.
We also present an unrefined ($J_1=J_2=J_3=J$ and $Q_1=Q_2=Q$,
where $Q$ should not be confused with the preserved supercharge)
version of (\ref{7dcc}) to make the formula more approachable:
\bea\label{7dccur}
&&\left( Q^4 + \frac{2N^3}{3}J^3 \right) \left(Q - \frac{N^3}{6} \right)^2 \nn\\
&=& 2 (3 Q^2 + N^3 J) \left(Q^3 - \frac{N^3}{2} J^2 \right) \left(Q - \frac{N^3}{6} \right)
- \left( Q^3 - \frac{N^3}{2} J^2  \right)^2~. \nn\\
\eea

\subsection{The 6d $(2, 0)$ Free Tensor Multiplet}

The charged, rotating AdS$_7$ black holes introduced in the previous subsection
are dual to the 6d $(2, 0)$ theory.
In this subsection we present the free tensor multiplet of the $(2, 0)$ superconformal algebra
needed to construct the single particle partition function.

The 6d $(2, 0)$ superconformal algebra has maximal bosonic subalgebra
$\mathfrak{so}(2,6) \oplus \mathfrak{sp}(4)$, matching the isometry of AdS$_7 \times S^4$.
Local operators in the theory are organized into representations of this subalgebra.
A super-representation of the 6d $(2, 0)$ superconformal algebra is uniquely specified by
the Dynkin labels of its superconformal primary.
For easy comparison with black hole spacetimes, 
we use $SO(6)$ for the Lorentz group and $SO(5)$ for the R-symmetry group, 
instead of $SU(4)$ for Lorentz and $Sp(4)$ for R-symmetry 
used in \cite{Cordova:2016emh}.\footnote{This amounts to
the interchanges $j_1 \leftrightarrow j_2$ and $R_1 \leftrightarrow R_2$.}
So we write representations of the bosonic subalgebra as
$$[j_1, j_2, j_3]^{[R_1, R_2]}_E~,$$
where $E$ is the conformal weight, $[j_1, j_2, j_3]$ are the $SO(6)$ Dynkin labels
so that $[1,0,0]$ is the vector ${\bf 6}$,
and $[R_1,R_2]$ are the $SO(5)$ Dynkin labels so that $[1,0]$ is the vector ${\bf 5}$.

The black hole charges used in section \ref{sec:7dBH} refer to the
orthogonal basis that is related to the Dynkin basis as
\bea
&& J_1 = j_1 + \frac{j_2+j_3}{2}~, \qquad
J_2 = \frac{j_2+j_3}{2}~, \qquad
J_3 = \frac{-j_2+j_3}{2}~, \nn\\
&& Q_1 = R_1 + \frac{R_2}{2}~, \qquad ~~~
Q_2 = \frac{R_2}{2}~.
\eea

A (not necessarily the highest) weight $[j_1, j_2, j_3]^{[R_1, R_2]}_E$,
is annihilated by our choice of a supercharge $Q$ if it saturates the unitarity bound
\bea\label{7dunitboundmicro}
E &\geq& j_1 + \frac12 j_2 + \frac32 j_3 + 2R_1 + 2R_2 \nn\\
&=& J_1 + J_2 + J_3 + 2Q_1 + 2Q_2~,
\eea
that every weight must satisfy. Such weights correspond to local BPS operators.

\begin{table}\catcode`\-=12
\begin{center}
\begin{tabular}{| c | c | c | c c c c c | c c c c c |}
\hline
& Bosonic Rep. & $E$ & $j_1$ & $j_2$ & $j_3$ & $R_1$ & $R_2$ &
$J_1$ & $J_2$ & $J_3$ & $Q_1$ & $Q_2$ \\
\hline
\hline
\multirow{5}{*}{Free fields} & \multirow{2}{*}{$[0,0,0]^{[1,0]}_2$} & 2 & 0 & 0 & 0 & 1 & 0 & 0 & 0 & 0 & 1 & 0 \\
& & 2 & 0 & 0 & 0 & $-1$ & 2 & 0 & 0 & 0 & 0 & 1 \\
\cline{2-13}
& \multirow{3}{*}{$[0,1,0]^{[0,1]}_\frac52$} & $\frac52$ & 0 & 1 & 0 & 0 & 1 &
$\frac12$ & $\frac12$ & $-\frac12$ & $\frac12$ & $\frac12$ \\
& & $\frac52$ & 1 & $-1$ & 0 & 0 & 1 &
$\frac12$ & $-\frac12$ & $\frac12$ & $\frac12$ & $\frac12$ \\
& & $\frac52$ & $-1$ & 0 & 1 & 0 & 1 &
$-\frac12$ & $\frac12$ & $\frac12$ & $\frac12$ & $\frac12$ \\
\hline
\hline
Eq. of motion & $[0,0,1]^{[0,1]}_\frac72$ & $\frac72$ & 0 & 0 & 1 & 0 & 1 &
$\frac12$ & $\frac12$ & $\frac12$ & $\frac12$ & $\frac12$ \\
\hline
\hline
\multirow{3}{*}{Derivatives} & \multirow{3}{*}{$[1,0,0]^{[0,0]}_1$} & 1 & 1 & 0 & 0 & 0 & 0 & 1 & 0 & 0 & 0 & 0 \\
& & 1 & $-1$ & 1 & 1 & 0 & 0 & 0 & 1 & 0 & 0 & 0 \\
& & 1 & 0 & $-1$ & 1 & 0 & 0 & 0 & 0 & 1 & 0 & 0 \\
\hline
\end{tabular}
\caption{\label{7dfmtable}
Components of the BPS operators in the free tensor multiplet $D_1 [0,0,0]^{[1,0]}_2$.
The first $5$ rows are free fields, followed by the equation of motion
and $3$ derivatives.}
\end{center}
\end{table}

In 6d superconformal theories, a field $[j_1, j_2, j_3]^{[R_1, R_2]}_E$
is free if $j_1 = 0$, at least one of $j_2$ and $j_3$ is zero,
and $E = 2+\frac{j_2+j_3}{2}$.
There is only one multiplet of the 6d (2, 0) superconformal algebra that contains a free field: 
the free tensor multiplet $D_1 [0,0,0]^{[1,0]}_2$ \cite{Cordova:2016emh}.
In Table \ref{7dfmtable} we summarize all weights in the free tensor multiplet that
saturate the unitarity bound (\ref{7dunitboundmicro}).

In Table \ref{7dfmtable}, we have listed $5$ free fields: $2$ scalars and $3$ spinors.
The entry below is an equation of motion that implements a relation between two spinors,
so it can be counted as a negative field.
The three last entries are derivatives that may act on any of the fields
and on the equation of motion, to produce their BPS descendants.
The gradient in $6$ dimensions has $6$ components but only $3$ preserve the BPS-ness of the field. 
The $5$ free fields, modulo the equation of motion, and with possible derivatives taken into account,
are the exhaustive list of supersymmetric operators in the free tensor multiplet.

\subsection{The Supersymmetric Ensemble}

We now compute the single particle BPS partition function as
a trace over the free tensor multiplet states given in Table \ref{7dfmtable},
with fugacities $(p,q,r,x,y)$ conjugate to each charge:
\bea\label{7dsppf}
Z_{\rm sp} &\equiv& {\rm Tr}_{\rm BPS} \left[ p^{2J_1} q^{2J_2} r^{2J_3} x^{2Q_1} y^{2Q_2} \right]\cr
&=& \frac{x^2+y^2+xypqr \left( \frac{1}{p^2} + \frac{1}{q^2} + \frac{1}{r^2} - 1\right)
}{(1-p^2)(1-q^2)(1-r^2)}~.
\eea
The $-1$ inside the parenthesis in the numerator is due to the equation of motion. The 
geometric series in $p^2$, $q^2$, and $r^2$ are from the derivatives. 
(\ref{7dsppf}) is the single particle partition function.

The grand canonical partition function over the full Hilbert space is given by the
ordinary exponential of the single particle partition function: $Z \equiv \exp[Z_{\rm sp}]$.
We use it to compute the macroscopic charges as statistical averages:
\bea\label{7dchgpotrel}
Q_1 &=&  \frac{2x^2+xypqr \left( \frac{1}{p^2} + \frac{1}{q^2} + \frac{1}{r^2} - 1\right)
}{2(1-p^2)(1-q^2)(1-r^2)}~, \nn\\
J_1 &=&  \frac{2p^2(x^2+y^2)+xypqr (1+p^2)\left( -\frac{1}{p^2} + \frac{1}{q^2} + \frac{1}{r^2} - 1\right)
+ 4 xypqr}{2(1-p^2)^2(1-q^2)(1-r^2)}~,
\eea
Analogous expressions for $Q_2$, $J_2$ and $J_3$ follow by permutations of indices. 

Finally, we define the supersymmetric ensemble as a grand canonical ensemble
where the operators that are separated in the charge space along the direction
of the preserved supercharge are weighed equally.
The preserved supercharge $Q$ carries quantum numbers
$(E, J_1, J_2, J_3, Q_1, Q_2) =
\left( \frac12, -\frac12, -\frac12, -\frac12, \frac12, \frac12 \right)$,
so the supersymmetric ensemble corresponds to imposing the relation
\bea\label{7dsusyens}
\frac{xy}{pqr} &=& 1~,
\eea
between the fugacities.

In the supersymmetric ensemble defined by (\ref{7dsusyens}),
there is one relation between the $5$ charges (\ref{7dchgpotrel}):
\bea\label{7dccmicro}
&& \frac12 (J_1+J_2+J_3) + \frac12 (Q_1^2 + Q_2^2 + 4Q_1Q_2)
- \frac{Q_1Q_2(Q_1+Q_2) - \frac12 (J_1J_2 + J_2J_3 + J_3J_1)}
{Q_1+Q_2- \frac12} \nn\\
&=& \sqrt{
\left( \frac12 (J_1+J_2+J_3) + \frac12 (Q_1^2 + Q_2^2 + 4Q_1Q_2) \right)^2
- \left( J_1J_2J_3 + Q_1^2Q_2^2 \right)}~.
\eea
It is precisely the supersymmetric AdS$_7$ black hole charge constraint (\ref{7dcc})
with the numerical values
\bea\label{7dNvalue}
\frac{2}{3}N^3 = 1 \quad \leftrightarrow \quad   \frac{\pi^2}{8G_7} = 1~.
\eea
We interpret this relative scale of all charges as the effective number of free multiplets needed to account for the constraint.
It is satisfying that the numerical factor $\frac{2}{3}<1$ since the interpolation from weak to strong coupling is expected to decrease the effective number of degrees of freedom. 

The formulae simplify significantly when we do not distinguish between
the $3$ angular momenta and between the $2$ electric charges. 
Let $x=y$ and $p=q=r$ in (\ref{7dchgpotrel})):
\bea\label{7dchgpotrelur}
Q &=&  \frac{2x^2+ 3x^2p - x^2p^3}{2(1-p^2)^3}~, \nn\\
J &=&  \frac{x^2p + 4x^2p^2  + 4 x^2p^3 -x^2p^5}{2(1-p^2)^4}~.
\eea
The definition of the supersymmetric ensemble \eqref{7dsusyens} simplifies to $x^2=p^3$, 
and then the charges \eqref{7dchgpotrelur} satisfy
the unrefined charge constraint (\ref{7dccur}) with $\frac{2}{3}N^3 = 1$.

\section{Discussions}\label{sec:ccdiscuss}

In this chapter, we derived the supersymmetric charge constraint
for the AdS$_{4,5,7}$ black holes using the simple prescription given in section \ref{sec:gendim}.
We think the computations are illuminating, especially because they are so simple.
However, we acknowledge that, in its current form, the argument is heuristic
and subject to significant concerns. 
These challenges are the subject of this final section.
We divide them into four issues, even though their possible resolutions are interrelated:
\begin{enumerate}
\item[1)] {\it Coupling dependence.} Unlike the index, the partition function depends on the coupling $g_\mathrm{YM}$. We study an ensemble of states generated by free fields and, even so, we compare the result to black holes that correspond to strong coupling. 

\item[2)] {\it Gauge dynamics.} In each case, we consider a single free field,
rather than the dynamics due to gauge degrees of freedom. 

The dependence on Newton's constant is determined by dimensional analysis in gravity,
while the dependence on the rank in the dual CFT is reproduced by assuming that
it is in its deconfined phase. However, a numerical constant of ${\cal O}(1)$ is put in by hand. 

\item[3)] {\it Classical statistics.} We consider a classical gas of BPS particles.
Technically, we take the multiparticle partition function to be the simple exponential
of the single particle partition function, rather than the plethystic exponential.
We did not justify why this approximation is sufficient. 

\item[4)] {\it The supersymmetric ensemble}: is defined so states that differ by
the charges of the preserved supercharge $Q$ are given equal weight.
This is motivated by the real part of the supersymmetry constraint on complex fugacities,
which are well established in supersymmetric black hole spacetimes
\cite{Hosseini:2017mds,Cabo-Bizet:2018ehj,Choi:2018hmj,Larsen:2019oll,Larsen:2021wnu}.
We did not provide a self-contained justification of this ensemble in the CFT,
except for the CFT$_2$ argument presented in section \ref{sec:3d}. 
\end{enumerate}

The numerical factor mentioned in 2) presents a concrete goal
that involves several of these issues.
If all fields were genuinely free, the number of independent multiplets would be $N^2$
in AdS$_5$/CFT$_4$, from the dimension of the $SU(N)$ gauge group of $\cN=4$ SYM,
and similarly in other dimensions.
This type of a na\"{\i}ve count of multiplets would not even take the projection
onto gauge singlets for the physical Hilbert space into account.
This can in principle be addressed by upgrading to a matrix model and, in particular,
confronting 3) \cite{Aharony:2003sx}.
However, this still leaves 1), the dependence on the coupling constant:
some of the BPS states in the free theory may gain anomalous dimensions
and be lifted from being BPS.
The projection onto singlets and the dependence on the coupling both suggest that
the na\"{\i}ve scaling in $N$ overcounts the microscopic states rather than undercounts.
It is therefore encouraging that all the needed ${\cal O}(1)$ adjustments are smaller than $1$.
Table \ref{summary} records the rescaling factors $\frac{\sqrt{2}}{3} N^{3/2}$, $\frac12 N^2$
and $\frac23 N^3$ for the AdS$_4$, AdS$_5$, and AdS$_7$ charge constraints, respectively.
The situation is reminiscent of the famous $3/4$-renormalization of high temperature
D3-brane entropy as the coupling is taken from weak to strong \cite{Gubser:1996de}.

Our discussion of supersymmetric black holes in AdS$_3$
is on more solid footing than in the higher dimensions. 
That is because the superalgebra is much stronger,
it gives a complete basis of characters for both short and long supermultiplets
of the $\mathcal{N} = 4$ super-Virasoro algebra, and so no free field assumption is needed.
In this context the supersymmetric ensemble is justified by a symmetry,
and the constraint we find agrees precisely with the black hole side,
with no numerical factor put in by hand.
These results offer a template for higher dimensions that we have pursued,
especially when addressing 4), but it is possible that other lessons remain hidden in plain sight. 

Our approach is fundamentally limited by us studying the partition function,
rather than the supersymmetric index.
Therefore, our computation is unavoidably subject to dependence on the
coupling constant that is beyond our control.
On the other hand, although the index is an invaluable tool for circumventing
the coupling dependence, it has its own structural limitations.
Because it is insensitive to many quantum states, it can at best provide a lower bound
on the black hole entropy, and so any agreement is only genuinely successful
if it is understood why cancellations are subleading.
The limitations of the index are especially pertinent in our context,
the constraint on charges that is satisfied by all supersymmetric black holes in AdS spacetimes.
That is because the index is independent of the relevant physical variable,
to the best of our understanding. 

For the future, the vision ultimately is that all the various contributions
to the partition function, in gravity and in CFT, whether boundary conditions correspond
to an index or not, can be disentangled.
Significant strides have been taken towards this goal in the most favorable circumstances,
such as asymptotically flat spacetimes with at least $\frac{1}{8}$ of the supersymmetries
\cite{Sen:2009vz,Sen:2009gy,Sen:2014aja,Iliesiu:2022kny,Iliesiu:2022onk}.
For asymptotically AdS spacetimes with maximal supersymmetry, the setting we have studied,
the current research frontier is at a lower level of understanding,
but recent years have witnessed much progress, using a variety of techniques
\cite{Iliesiu:2020qvm,Heydeman:2020hhw,Boruch:2022tno}.
The work presented in this chapter, including the challenges discussed in this section,
is a contribution to these developments.

\clearpage


\part{Towards Quantum Black Hole Microstates}
\chapter{The Black Hole Cohomology Problem}\label{sec:cohoproblem}

In the second part starting from this chapter,
we work on constructing the explicit expressions for supersymmetric black hole microstates
in the language of the dual weakly coupled superconformal field theory.

As mentioned in the Introduction,
the AdS$_5$ black holes introduced in section \ref{sec:AdS5BH} are solutions
of the supergravity theory in 5 dimensions,
and therefore the dual black hole microstates live in the strongly coupled
4-dimensional CFT with a large gauge group $SU(N)$ with $N\to\infty$.
However, since the black hole microstates are correctly counted by
a coupling independent quantity, namely the index,
there should be as many analogous states in the weakly coupled field theory.
See \cite{Budzik:2023vtr} for the connection between states in
the weakly coupled and the strongly coupled theories.
In a different point of view, one may argue that microstates in the finite-$N$,
weakly coupled regime of the field theory are dual to black hole microstates
in the full quantum gravity theory, rather than its supergravity approximation.
With these arguments in mind, we study the supersymmetric quantum states,
or local BPS operators, in the weakly coupled, finite-$N$ field theory.

We focus on $\frac{1}{16}$-BPS states of
the 4d $\mathcal{N}=4$ Yang-Mills theory with $SU(N)$ gauge group,
dual to type IIB string theory in $AdS_5\times S^5$.
The BPS states can be reformulated as classical cohomologies with respect
to a nilpotent supercharge $Q$.
Our goal in this part is to construct such cohomologies for finite values of $N=2,3,4$
that are not of the graviton type, and therefore potentially
represent the black hole microstates.

This part is based on \cite{Choi:2023znd,Choi:2023vdm}
in collaboration with Jaehyeok Choi, Sunjin Choi,
Seok Kim, Eunwoo Lee, Jehyun Lee and Jaemo Park.

\section{Formulation of the Problem}\label{sec:cohoformulation}

In this section, we reformulate the problem of listing local BPS operators
in the weakly coupled 4d $\mathcal{N}=4$ Yang-Mills theory on $\mathbb{R}^4$
into that of finding classical cohomologies with respect to a supercharge $Q$.
We also review how to systematically construct such cohomologies,
using the BPS letters as building blocks, partly repeating \ref{sec:earlymm}.

The $\mathcal{N}=4$ Yang-Mills theory with $SU(N)$ gauge group
carries a continuous real marginal coupling constant $g_{\rm YM}$,
and enjoys $\mathcal{N}=4$ superconformal symmetry $PSU(2,2|4)$
at any value of $g_{\rm YM}$. 
The theta angle will not be relevant in our discussions.

The theory includes six real scalars, eight fermions and the gauge field,
all in the $SU(N)$ adjoint representation.
To repeat \eqref{N4SYMfields}, we denote them as
\begin{eqnarray}\label{fields-firstappear}
  \textrm{vector}&:&A_\mu\sim A_{\alpha\dot\beta}\ , \hspace{4.3cm}
  (\mu=1,2,3,4\ , \ \alpha=\pm\ ,\ \dot\beta=\dot{\pm}) \nn\\
  \textrm{scalar}&:&\Phi_{ij}(=-\Phi_{ji})\ ,\ \overline{\Phi}^{ij}\sim
  {\textstyle \frac{1}{2}}\epsilon^{ijkl}\Phi_{kl}
  \ ,\qquad (i,j,k,l=1,2,3,4)\nn\\
  \textrm{fermion}&:&\Psi_{i\alpha}\ ,\ \overline{\Psi}^i_{\dot\alpha}\ .
\end{eqnarray}
$\alpha,\dot\alpha$ are the doublet indices of the Lorentz group
$SU(2)_L\times SU(2)_R\sim SO(4)$ which rotate the $S^3$,
and $\mu$ is the vector index.
Superscripts $i,j$ are for the fundamental representation of the $SU(4)$ R-symmetry,
while the subscripts are for the anti-fundamental representation.
For later convenience, we arrange these fields into 
$\mathcal{N}=1$ supermultiplets as follows, with manifest 
covariance only for the $SU(3)\subset SU(4)$ part of R-symmetry,  
\begin{eqnarray}
  \textrm{vector multiplet}&:&A_{\alpha\dot\beta}\ ,\ 
  \lambda_\alpha=\Psi_{4\alpha}\ ,\ \bar\lambda_{\dot\alpha}=\overline{\Psi}^4_{\dot\alpha}~,\\
  3\textrm{ chiral multiplets}&:&\phi_m=\Phi_{4m}\ ,\ \bar\phi^m=\overline{\Phi}^{4m}\ ,\ 
  \psi_{m\alpha}=-i\Psi_{m\alpha}\ ,\ \bar\psi^m_{\dot\alpha}=i\overline{\Psi}^m_{\dot\alpha}
  \nonumber\ ,
\end{eqnarray}
where $m=1,2,3$ is the index for the $SU(3)$ subset of the R-symmetry and labels the chiral multiplets.

We consider the Euclidean CFT on $\mathbb{R}^4$,
related to the Lorentzian CFT on $S^3\times\mathbb{R}$ by radial quantization, 
which regards the radius of $\mathbb{R}^4$ as the exponential of the Euclidean time $\tau$
and makes a Wick rotation $\tau=it$. Here we note the operator-state map, 
in which the local operators at the origin of $\mathbb{R}^4$ map to the states propagating 
in $S^3\times\mathbb{R}$.
We will omit the spacetime arguments of the local operators.

The CFT is invariant under $32$ supersymmetries, represented by the 
$16$ Poincar\'e supercharges $Q^i_\alpha$, $\overline{Q}_{i\dot\alpha}$ and the 
$16$ conformal supercharges $S_{i\alpha}$, $\overline{S}^i_{\dot\alpha}$. 
In the radially quantized theory, $S$'s are Hermitian conjugates of $Q$'s:
$S_i^\alpha=(Q^i_\alpha)^\dag$, $\overline{S}^{i\dot\alpha}=(\overline{Q}_{i\dot\alpha})^\dag$.
Together with other symmetry generators, these supercharges 
form the $PSU(2,2|4)$ superconformal algebra.
The most important part of the algebra for our discussion is \cite{Kinney:2005ej}
\begin{equation}\label{QS-algebra}
  \{Q^i_\alpha,S_j^\beta\}={\textstyle \frac{1}{2}}E\delta^i_j\delta_\alpha^\beta+
  {R^i}_j\delta_\alpha^\beta+{J_\alpha}^\beta\delta^i_j~,
\end{equation}
where $E$ is the dilatation operator
(or the Hamiltonian on $S^3\times\mathbb{R}$ multiplied by the radius of $S^3$),
${R^i}_j$ is the $SU(4)$ R-charges,
and ${J_\alpha}^\beta$ is the left $SU(2) \subset SO(4)$ angular momenta.
We choose two of the supercharges to be preserved:
$Q\equiv Q^4_-$ and $S = Q^\dag\equiv S_4^-$.
These two supercharges satisfy $Q^2=0$, $(Q^\dag)^2=0$,
and from (\ref{QS-algebra}) one obtains
\begin{equation}\label{QS-algebra-chosen}
  2\{Q,Q^\dag\}=E-(Q_1+Q_2+Q_3+J_1+J_2)\ .
\end{equation}
This is the identical choice of supercharges as in \eqref{QSalgebra}.
On the right hand side, we expressed $2{R^4}_4=-Q_1-Q_2-Q_3$ and 
$2{J_-}^-=-J_1-J_2$ in terms of the five charges which rotate the mutually orthogonal 
2-planes on $\mathbb{R}^6\supset S^5$ and $\mathbb{R}^4\supset S^3$, respectively, 
all normalized to have $\pm\frac{1}{2}$ values for spinors.

The $\frac{1}{16}$-BPS states/operators of our interest preserve these
$2$ Hermitian supercharges.
Thus, we are interested in gauge-invariant local operators $O$
that are annihilated by $Q$:
\begin{equation}
  [Q,O\}=0\ ,\ [Q^\dag,O\}=0\ .
\end{equation}
It follows from (\ref{QS-algebra-chosen}) that the BPS operators of our interest
can be arranged to be the eigenstates of $H$, $R_I$ and $J_i$,
with respective eigenvalues $E$, $R_I$ and $J_i$ that satisfy
\begin{equation}\label{BPS-relation}
  E = Q_1+Q_2+Q_3+J_1+J_2~.
\end{equation}
The charges $Q_I$, $J_i$ on the right hand side are part of the
non-Abelian charges and cannot depend on the coupling $g_{\rm YM}$.
However, $E$ is in general a function of $g_{\rm YM}$, so that 
a BPS state may become anomalous as $g_{\rm YM}$ changes.

Let us first consider local BPS operators of the free ($g_{\rm YM} = 0$) theory.
In the free theory, the operators satisfying the BPS relation \eqref{BPS-relation}
can be easily constructed using the BPS elementary fields.
The BPS elementary fields are the members of the free vector multiplet
$B_1\overline{B}_1 [0;0]^{[0,1,0]}_1$ that satisfy the BPS relation \eqref{BPS-relation},
and they have been summarized in Table \ref{5dfmtable}.
We give the following names to the nine free fields and two derivatives:
\begin{equation}\label{BPS-fields}
  \phi^m\ ,\ \ \psi_{m}\ ,\ \ f\ ,\ \ 
  \lambda_{\dot\alpha}\ ,\ \ \partial_{\dot\alpha}\ .
\end{equation}
%
Note that these are a BPS subset of \eqref{fields-firstappear},
but with bars and some indices that are common to BPS fields stripped off
because non-BPS fields will not appear any more.
Also note that $m=1,2,3$ and $\alpha = \pm$.
With these, we construct independent `letters' for the gauge invariant operators. 
Basically, acting any numbers of two derivatives $\partial_{\dot\alpha}$
on a BPS field forms a letter.
In the free theory, the derivatives $\partial_{\dot\alpha}$ acting on the same field commute,
so all $SU(2)_R$ indices appearing in a letter should be symmetrized.
However, the equation of motion operator is null and should not be included.
The only equation of motion constructed using (\ref{BPS-fields}) is
\begin{equation}
  \partial_{\dot\alpha}\lambda^{\dot\alpha}=0 ~~ \Leftrightarrow ~~
  \partial_{[\dot\alpha}\lambda_{\dot{\beta}]}=0~.
\end{equation}
The equation of motion can be imposed by requiring the $SU(2)_R$ indices
carried by the derivatives and the gaugino $\lambda_{\dot\alpha}$
to be symmetrized within a letter.
So for example,
\begin{eqnarray}\label{BPS-letters}
&& \partial_{(\dot\alpha_1}\cdots\partial_{\dot\alpha_n)}\phi^m~, \qquad
\partial_{(\dot\alpha_1}\cdots\partial_{\dot\alpha_n)}\psi_{m}~, \qquad
\partial_{(\dot\alpha_1}\cdots\partial_{\dot\alpha_n)}f~, \nn\\
&& \partial_{(\dot\alpha_1}\cdots\partial_{\dot\alpha_{n-1}}\lambda_{\dot\alpha_{n})}
\end{eqnarray}
are the BPS letters.
Multiplying these letters and contracting all $SU(N)$ indices that are
omitted in (\ref{BPS-letters}),
one can construct general gauge-invariant BPS operators in the free theory.

Moving away from the free theory, 
we want to study how many of these operators remain BPS at the 1-loop level,
i.e. at the order $\mathcal{O}(g_{\rm YM}^2)$.
The dilatation operator $H(g_{\rm YM})$ can be expanded in $g_{\rm YM}^2$, $H(g_{\rm YM})=\sum_{L=0}^\infty
g_{\rm YM}^{2L} H_{(L)}$. At least in perturbation theory, this operator can be 
diagonalized within the subspace of free BPS operators.\footnote{More precisely, 
for the gauge invariance in the interacting theory, the subsector 
is defined at $g_{\rm YM}\neq 0$ by promoting the derivatives $\partial_{\dot\alpha}$ 
appearing in the operators to the covariant derivatives 
$D_{\dot\alpha}\equiv\partial_{\dot\alpha}-i[A_{+\dot\alpha},\ ]$.} 
Within this subspace, $H_{(0)}$ is equal to $\sum_IR_I+\sum_i J_i$.
We want to find the subset of free BPS operators that satisfy (\ref{BPS-relation})
in the next order, so they must be annihilated by $H_{(1)}$.
Within the free BPS sector, one finds that
\begin{equation}\label{QS-algebra-expand}
  \{Q(g_{\rm YM}),Q^\dag(g_{\rm YM})\}=H(g_{\rm YM})-\sum_I R_I-\sum_i J_i=
  \sum_{L=1}^\infty g_{\rm YM}^{2L}H_{(L)}\ .
\end{equation}
$Q$ and $Q^\dag$ also depend on $g_{\rm YM}$. 
Since the free BPS fields are annihilated by $Q$ and $Q^\dag$ at the leading
$\mathcal{O}(g_{\rm YM}^0)$ order,
their coupling expansions start from the $\mathcal{O}(g_{\rm YM}^1)$ `half-loop' order.
Therefore, the leading 1-loop Hamiltonian $H_{(1)}$ in (\ref{QS-algebra-expand})
is given by the anticommutator of $Q$ and $Q^\dag$ at the half-loop order.
In particular, $Q_{(\frac{1}{2})}$ at $\mathcal{O}(g_{\rm YM}^1)$ is
precisely the supercharge of the classical interacting field theory. 
So the 1-loop BPS operators should be annihilated by both
$Q$ and $Q^\dag$ at the classical half-loop order.

The local BPS operators annihilated by $Q$ and $Q^\dag$
are in 1-1 map with cohomology classes of $Q$.
The cohomology class is defined by the set of operators $O$
built from the BPS letters (\ref{BPS-letters}) that are closed under the action of $Q$,
i.e. $[Q,O\}=0$, with the equivalence relation $O\sim O+[Q,\Lambda\}$,
where $\Lambda$ is also an operator constructed from the BPS letters (\ref{BPS-letters}).
We can call this a cohomology because of the nilpotency $Q^2=0$.
These cohomology classes are in 1-to-1 map to the BPS operators $O_{\rm BPS}$
that satisfy $[Q,O_{\rm BPS}\}=0$ and $[Q^\dag,O_{\rm BPS}\}=0$,
because the latter can be understood as harmonic forms \cite{Grant:2008sk}.
Therefore, we shall construct and study the representatives of the
cohomologies of the classical half-loop supercharge $Q$,
which map to the 1-loop BPS operators. 
The actions of classical (half-loop) $Q$ on the free BPS fields are given by 
\bea\label{Q-classical}
  && Q\phi^m=0 ~,\quad Q\lambda_{\dot\alpha}=0~, \quad
  Q\psi_{m}=-{\textstyle \frac{i}{2}}\epsilon_{mnp}[\phi^n,\phi^p]~, \nn\\ 
  && Qf=-i[\phi^m,\psi_{m}]~, \quad
  [Q,D_{\dot\alpha}]=-i[\lambda_{\dot\alpha} \}~,
\eea
where we absorbed the $g_{\rm YM}$ factors on the right hand sides
into the normalization of fields.

It is well known that there are fewer BPS states at the 1-loop level than in the free theory. 
It has been conjectured (for instance, explicitly in \cite{MinwallaTalk2006}) 
that the 1-loop BPS states remain BPS at general non-zero coupling. 
Some perturbative evidence of this conjecture was discussed in \cite{Chang:2022mjp}. 
We will assume this conjecture.

Let us summarize this section.
Equivalently to listing 1-loop BPS local operators of the $\mathcal{N}=4$ SYM,
we shall find classical cohomology classes with respect to the supercharge $Q$.
The cohomology classes are defined as gauge invariant operators
constructed using the BPS letters (\ref{BPS-letters}) by multiplying them
and contracting the gauge indices,
that are annihilated by $Q$ under the rule (\ref{Q-classical}),
up to identification of operators that differ by $Q$-exact operators.

\section{The BMN Sector}\label{sec:bmn}

Combinatorial possibilities of gauge invariant operators constructed
using the BPS letters (\ref{BPS-letters}) grow rapidly with the number of
letters allowed and with the gauge rank $N$,
and the cohomology problem quickly becomes computationally complicated.
In order to reach meaningful results with limited computing power,
we will restrict to a subset of the operators, that we call the BMN sector.
This restriction has been motivated by the observation that the smallest
black hole cohomology found for $N=2$ can be expressed without
any gauginos and derivatives, as we shall show in section \ref{sec:cohosu2}.
In this section we introduce the BMN sector, or truncation.

The radially quantized QFT lives on $S^3\times\mathbb{R}$. 
The fields are expanded in spherical harmonics of the Lorentz group $SO(4)$.
It was shown in \cite{Kim:2003rza} that the \textit{classical} $\mathcal{N}=4$ 
Yang-Mills theory has a consistent truncation which keeps finite degrees of freedom, 
described by the BMN matrix model \cite{Berenstein:2002jq}. 
The modes kept after the truncation are given by: 
(1) s-wave modes $\phi_m(t)$, $\phi^m(t)$ of the scalars, 
(2) lowest spinor harmonics modes $\psi_{m\alpha}(t)$, 
$\lambda_\alpha(t)$ (the spinor indices 
are defined using the labels of Killing spinor fields \cite{Kim:2003rza}), (3) vector 
potential 1-form restricted to $A=A_0(t)dt+A_i(t)\sigma_i$ where $\sigma_i$ with 
$i=1,2,3$ are the right-invariant 1-forms on $S^3$ in our 
convention. This is a 
consistent truncation of the nonlinear equations of motion, and not a quantum 
reduction in any sense. So the full quantum BMN theory is 
a priori unrelated to the 4d Yang-Mills theory. However, since our 1-loop cohomology problem 
uses classical supercharge $Q$ only, it can be truncated to the BMN model.
If the conjecture of \cite{MinwallaTalk2006} is true,
the whole BPS cohomology problem would have a quantum truncation to this model.

In general, the BMN theory and the full Yang-Mills theory behave differently in many ways. 
The difference starts from the number of ground states. The Yang-Mills 
theory on $S^3\times\mathbb{R}$ has a unique vacuum, while the BMN model has 
many ground states labeled semiclassically by the discrete values of $A_i$. 
In the quantum BMN theory, viewed as an M-theory in the plane wave background, these ground 
states describe various M2/M5-brane configurations with zero lightcone energies 
\cite{Maldacena:2002rb}. In the Yang-Mills theory, however, there are large gauge
transformations on $S^3$ which can gauge away these ground states to $A_i=0$. 
So if one wishes to study the Yang-Mills theory using this matrix model, it suffices to 
consider the physics around $A_i=0$.

Recall that our cohomology problem is completely classical, using the classical
supercharge $Q$ at the half-loop order.
Therefore, this problem should have a truncation to the BMN matrix model.
This turns out to be the cohomology problem defined using
\begin{equation}\label{letter-bmn}
  \phi^m\ ,\ \ \psi_{m}\ ,\ \ f\ ,
\end{equation}
without using any gauginos $\lambda_{\dot\alpha}$ or derivatives $D_{\dot\alpha}$.
These operators close under the action of $Q$: 
$[Q,\phi^m]=0$, $\{Q,\psi_{m}\}=-i\epsilon_{mnp}[\phi^m,\phi^n]$, 
$[Q,f]=-i[\psi_{m},\phi^m]$.
So it is possible to restrict the cohomology problem by using operators
constructed only using the BMN letters (\ref{letter-bmn}).
Note that the truncation is also applied to the operator $\Lambda$ when one 
identifies two operators $O_1$ and $O_2$ related as
$O_2-O_1=[Q,\Lambda\}$. This is why the gauginos $\lambda_{\dot\alpha}$ cannot be 
included in this truncation. Although it is $Q$-closed by itself, $\lambda_{\dot\alpha}$ 
can be obtained by acting $Q$ on the covariant derivative, 
$[Q,D_{\dot\alpha}]=-i[\lambda_{\dot\alpha},\ \}$.
So if one had tried to include $\lambda_{\dot\alpha}$ into the truncation and construct 
operators like $O_1$, $O_2$, $\Lambda$, one may incorrectly conclude that certain 
$O_1$ and $O_2$ are different by not including derivatives in $\Lambda$. This truncation of 
the cohomology problem was known in \cite{Chang:2022mjp,Choi:2022caq}, 
although the relation to the BMN truncation was not explicitly addressed.\footnote{We thank 
Nakwoo Kim for first pointing this out to us.} Notice also that this truncation is not 
kinematic, i.e. cannot be inferred without knowing the dynamical information of 
the classical theory. 

The BMN truncation is the $SU(2)_R$ invariant truncation. 
In our cohomology problem, this means that no ingredients include the $\dot\alpha$ 
indices for $SU(2)_R$. This is why $\lambda_{\dot\alpha}$ and $D_{\dot\alpha}$ 
are excluded. Similarly, in the representation theory, only a small subset of 
$PSU(1,2|3)$ generators can be used to generate a multiplet. Among the $PSU(1,2|3)$ 
generators $Q^m_+$, $\overline{Q}_{m\dot\alpha}$ and $P_{+\dot\alpha}$, only the 
three supercharges $Q^m_+$ which belong to $SU(1|3)$ act 
within BMN cohomologies.
It serves as a great combinatorial advantage that the derivatives are disallowed,
because it only allows a finite number of BPS letters (\ref{BPS-letters})
to be used for construction of the BPS operators.

\section{The Index over Cohomologies}\label{sec:cohoindex}

Now that we have defined cohomologies in the full and the BMN sector
of the 4d $\mathcal{N}=4$ Yang-Mills theory,
it is useful to introduce a tool to count them.
As always, it is the index.

As we have argued in section \ref{sec:cohoformulation},
the cohomologies with respect to the supercharge $Q$ are in 1-1 map
with the local BPS operators, or the $\frac{1}{16}$-BPS states
in the 1-loop level of the field theory.
That being said, an index over the cohomologies is identical to that over
the $\frac{1}{16}$-BPS states that have been defined in section \ref{sec:earlyindex}.
We repeat the definition \eqref{4dN4index2} while getting rid of tildes.
Also, in this section we use the letter $Z$ to indicate the index.
\bea\label{cohoindexdef}
Z (\Delta_I,\, \omega_i) &=&
\mathrm{Tr} \left[(-1)^F 
e^{\Delta_I Q_I+ \omega_i J_i} \right]~,
\hspace{0.6cm} \mathrm{where} ~~~
e^{\frac{\Delta_1+\Delta_2+\Delta_3-\omega_1-\omega_2}{2}} = 1~.~~
\eea
Note from \eqref{cohoindexdef} and from discussions in section \ref{sec:earlyindex}
that due to the relations between five chemical potentials, the index is
only able to distinguish cohomologies in 4 directions in the 5-dimensional charge space.
Specifically, the index does not distinguish two cohomologies whose charges
$Q_I$ differ by $\frac{n}{2}$ and $J_i$ by $-\frac{n}{2}$, where $n$ is an integer.

It is often useful to unrefine the chemical potentials to make the index
a function of only one variable.
We have done this unrefinement in section \ref{sec:ccp}, see \eqref{4dN4indexur}:
\bea\label{cohourindexdef}
&& e^{\Delta_1} = e^{\Delta_2} = e^{\Delta_3} = t^2~, \qquad
e^{\omega_1} = e^{\omega_2} = t^3~, \nn\\
\Rightarrow && Z (t) =
\mathrm{Tr} \left[(-1)^F t^{2(Q_1+Q_2+Q_3) + 3(J_1+J_2)} \right]
\equiv \mathrm{Tr} \left[(-1)^F x^{\cJ} \right] ~,
\eea
where we defined a combination of the charges
\be\label{defcJ}
\cJ \equiv 2Q_1+2Q_2+2Q_3+3J_1+3J_2~.
\ee
This $\cJ$ can be thought of as an `overall' charge,
and we shall expand in $t$, in powers of this overall charge
to truncate results throughout this part of the dissertation.

The index as defined \eqref{cohoindexdef} can be taken over all cohomologies,
but it is possible at will to take it only over cohomologies in the BMN sector.
However, in the BMN sector all cohomologies have $J = J_1=J_2$,
so it is allowed yet redundant to keep both chemical potentials $\omega_{1,2}$.
Therefore, we take $\omega = \omega_1 = \omega_2$ for the BMN index.
Then, we substitute $\omega = \frac{\Delta_1+\Delta_2+\Delta_3}{2}$
to satisfy the condition in \eqref{cohoindexdef}.
As a result, the BMN index is a function of 3 chemical potentials: 
\bea\label{cohoBMNindexdef}
Z_\mathrm{BMN} (\Delta_I) &=&
\mathrm{Tr}_\mathrm{BMN} \left[(-1)^F e^{\Delta_I (Q_I+ J)} \right]~,
\eea
that only distinguishes 3 combinations of 4 charges $Q_I+J$ where $I=1,2,3$.
We often denote these three charges as
\be
q_I \equiv Q_I + J~.
\ee

Similarly to the full index, we can unrefine the BMN index via
$e^{\Delta_1} = e^{\Delta_2} = e^{\Delta_3} = t^2$,
resulting in the unrefined BMN index as a function of $t$ only:
\bea\label{cohourBMNindexdef}
Z_\mathrm{BMN} (t) &=&
\mathrm{Tr}_\mathrm{BMN} \left[(-1)^F t^{2(Q_1+Q_2+Q_3)+6J} \right]~.
\eea

Since the BMN sector is a restriction of the BPS cohomologies,
the entropy of BMN cohomologies will be smaller than the entropy of all cohomologies. 
Despite, the large $N$ BMN entropy will still exhibit the black hole like growth.
Taking $j$ (schematically) to be the charges, 
the black hole like entropy growth is 
\begin{equation}\label{entropy-bh-scaling}
  S(j,N)=N^2f({\textstyle \frac{j}{N^2}})~,
\end{equation}
where $f(x)$ is a generic function that does not explicitly depend on $N$, 
$N\gg 1$, $j\gg 1$ and the ratio $\epsilon\equiv\frac{j}{N^2}$ does not scale in $N$. 
Roughly, the scaled charge parameter 
$\epsilon$ measures the size of the black hole in the AdS unit.
In the rest of this section, we show that the BMN entropy scales as
(\ref{entropy-bh-scaling}) when $\epsilon$ is parametrically small (but not scaling in $N$),
i.e. for small black holes.
We expect without a proof the same to be true at general $\epsilon$,
see \cite{Choi:2023vdm} for some comments on the BMN entropy of large black holes,
and \cite{Chang:2024lkw} for a more general recent work on the BMN matrix model.

Recall the matrix integral expression \eqref{4dN4index3} for the index \eqref{cohoindexdef}.
From this the BMN index can be obtained via truncation:
\begin{eqnarray}\label{BMN-index-integral}
  Z_\mathrm{BMN} (\Delta_I) &=&\frac{1}{N!}\int_0^{2\pi}
  \prod_{a=1}^N \frac{d\alpha_a}{2\pi}
  \frac{\prod_{a\neq b}(1-e^{i\alpha_{ab}})\prod_{a,b=1}^N\prod_{I<J}
  (1-e^{\Delta_I+\Delta_J}e^{i\alpha_{ab}})}
  {\prod_{a,b=1}^N\left[(1-e^{\Delta_1+\Delta_2+\Delta_3}e^{i\alpha_{ab}})
  \prod_{I=1}^3(1-e^{\Delta_I}e^{i\alpha_{ab}})\right]}\nonumber\\
  &&\hspace{2.7cm}
  \times\ \frac{(1-e^{\Delta_1+\Delta_2+\Delta_3})\prod_{I=1}^3(1-e^{\Delta_I})}
  {\prod_{I<J}(1-e^{\Delta_I+\Delta_J})}~,
\end{eqnarray}
where the second line (inverse of the $U(1)$ index) is multiplied 
to make it an $SU(N)$ index rather than $U(N)$.
This integral can be computed either exactly using the residue sum 
or in a series expansion in $t$ defined by 
$(e^{\Delta_1},e^{\Delta_2},e^{\Delta_3})=t^2(x,y^{-1},x^{-1}y)$.

Using (\ref{BMN-index-integral}),
let us compute the large $N$ entropy in the small black hole regime: 
$j\gg 1$, $N\gg 1$ and $\epsilon\equiv\frac{j}{N^2}$ fixed and much smaller 
than $1$ (but not scaling in $N$).\footnote{The term `small black hole' 
has at least three different meanings in the literature. It sometimes denotes 
string scale black holes, for which 2-derivative gravity description breaks down 
near the horizon. 
In our example, since $\epsilon$ does not scale in $N$, the 2-derivative 
gravity is reliable everywhere. Also, small black holes
sometimes mean AdS black holes with negative specific heat or susceptibility. 
What we call `small black holes' belong to this class, but are more specific. 
Our notion is precisely the same as \cite{Bhattacharyya:2010yg,Choi:2021lbk}.} 
This regime is reached by taking all $\Delta_I$'s to be small. 
The approximate large $N$ calculation of the entropy 
can be done by following all the calculations in section 5.3 of \cite{Choi:2021lbk} 
with minor changes in the setup. In particular, the calculations from  (5.88) 
to (5.91) there can be repeated by simply replacing all 
$2-(-e^\gamma)^n-(-e^{\gamma})^{-n}$ by $1$ (which are the denominators 
of the letter indices in the two setups) and remembering 
that $\beta_I$ there are $-\frac{\Delta_I}{2}$ here. The resulting 
eigenvalue distribution is along the interval $\alpha\in (-\pi,\pi)$ on the real axis 
(the gap closes in the small black hole limit), with the distribution function
\begin{equation}
  \rho(\alpha)=\frac{3}{4\pi^3}(\pi^2-\alpha^2)\ .
\end{equation}
The free energy $\log Z$ 
of this saddle point is given by
\begin{equation}
  \log Z = \frac{3N^2}{\pi^2}\Delta_1\Delta_2\Delta_3\ .
\end{equation}
(For small black holes with negative susceptibility, the grand canonical index is not 
well defined. Whenever we address $\log Z$, a Laplace transformation to the micro-canonical ensemble is assumed.)
The entropy at given charges $q_I\equiv R_I+J$ is given by extremizing 
\begin{equation}
  S_{\rm BMN}(q_I;\Delta_I)=\log Z-\sum_I q_I\Delta_I~,
\end{equation}
in $\Delta_I$'s, which is given by
\begin{equation}
  S_{\rm BMN}(q_I)=2\pi\sqrt{\frac{q_1q_2q_3}{3N^2}}\ .
\end{equation}
This expression is valid when $q_I=N^2\epsilon_I$ with $\epsilon_I\ll 1$. 
The entropy $\sim N^2\sqrt{\epsilon_1\epsilon_2\epsilon_3}\propto N^2$ 
exhibits a black hole like scaling (\ref{entropy-bh-scaling}).
$S_{\rm BMN}$ is smaller than the full entropy 
$S(q_I)=2\pi\sqrt{\frac{2q_1q_2q_3}{N^2}}$ 
in the small black hole regime \cite{Choi:2021lbk} by a factor of $\frac{1}{\sqrt{6}}$.
This is natural since the BMN truncation loses cohomologies. 
However, the fact that $S_{\rm BMN}(q_I)$ scales like $S(q_I)$ implies 
that the truncation provides a good simplified model for black holes, 
at least for the small black holes $\epsilon_I \ll 1$.

\section{Graviton Cohomologies}\label{sec:gravitons}

There is a well known family of cohomology classes in $\mathcal{N}=4$ SYM,
known as the multi-graviton cohomologies.
These graviton-type cohomologies are well-defined and completely classified \cite{Kinney:2005ej},
but it was shown that there are not enough of them to account for the black hole entropy.
We want to exclude them in our discussions, and instead find cohomologies
that are not of the graviton type.
We use the terms non-graviton cohomology and black hole cohomology interchangably,
simply because the non-graviton cohomologies at least include
the black hole cohomologies for sure (from counting) and
we are not aware of any further classification among the
non-graviton cohomologies that is applicable.
We also refer to a recent work \cite{Chang:2024zqi} in this direction.
In this section, we review the notion of graviton cohomologies,
especially at finite $N$, and also explain how to list and count them.

We first distinguish the multi-graviton and single-graviton cohomologies.
The multi-graviton cohomologies are the same as non-graviton cohomologies
and they include the single-graviton cohomologies.

The multi-graviton cohomologies are defined to be the polynomials of single-graviton cohomologies. 
(This definition naturally yields the familiar large $N$ cohomologies for the supergravitons.) 
Single-graviton cohomologies are completely understood
\cite{Kinney:2005ej,Janik:2007pm,Chang:2013fba}, as we shall review in a moment:
they are nontrivial cohomologies at arbitrary $N$ by definition,
in the sense that no trace relations exist between single-graviton operators.
Polynomials of these single-trace cohomologies define the multi-trace cohomologies.
Some polynomials may be trivial, i.e. $Q$-exact at finite $N$.
However, they are $Q$-closed at arbitrary $N$ without using any trace relations. 
This will be in contrast to the black hole cohomologies, which should become 
$Q$-closed only after applying trace relations at particular $N$.

When $N$ is larger than the energy, the multi-graviton operators 
defined above are all nontrivial cohomologies since no trace relations can 
be applied to make them $Q$-exact. So in this setup, the `graviton cohomologies' 
defined abstractly in the previous paragraph actually map to the familiar 
$\frac{1}{16}$-BPS graviton states in $AdS_5\times S^5$. Trace number of the operator 
is regarded as the particle number.

At finite $N$, all the multi-trace operators mentioned in the previous paragraphs 
are still $Q$-closed. However, some of their linear combinations may be zero or 
$Q$-exact only when $N$ takes a particular value smaller than their energies,
due to the trace relations. So the independent graviton cohomologies 
reduce at finite $N$. Such reductions of states are a well known finite $N$ effect in 
the gravity dual. It is called the stringy exclusion principle \cite{Maldacena:1998bw}, 
which happens because gravitons polarize into D-brane giant gravitons 
\cite{McGreevy:2000cw,Grisaru:2000zn,Hashimoto:2000zp}. 
The reduction/exclusion mechanism is the same for any $N$ 
in QFT, making it natural to call them `finite $N$ gravitons' at general finite $N$.

Now we concretely explain the list of the graviton cohomologies.
One starts by listing the single-trace graviton cohomologies. These are completely found 
and collected into supermultiplets. The relevant algebra for these multiplets is the
$PSU(1,2|3)$ subset of the superconformal symmetry $PSU(2,2|4)$ that commutes 
with $Q,Q^\dag$. The multiplets for single-trace graviton cohomologies  
are called $S_n$ with $n=2,3,\cdots$ \cite{Kinney:2005ej}. $S_n$ is obtained by 
acting the Poincar\'e supercharges $Q^m_+$, $\overline{Q}_{m\dot\alpha}$ and the 
translations $P_{+\dot\alpha}$ in $PSU(1,2|3)$ on 
the following primary operators
\begin{equation}\label{Sn-primary}
  u^{i_1i_2\cdots i_n}={\rm tr}(\phi^{(i_1}\phi^{i_2}\cdots\phi^{i_n)})\ .
\end{equation}
See \cite{Kinney:2005ej} for more details.
In the notation of \cite{Cordova:2016emh}, it is also a subrepresentation
of the short representation $B_1\bar{B}_1[0;0]^{[0,n,0]}$.
At large $N$, multiplying the operators in $S_n$'s yields independent multi-trace cohomologies.
At finite $N$, trace relations reduce the independent single-trace 
and multi-trace operators. Following \cite{Choi:2023znd}, we first identify 
the dependent single-trace operators as follows. Using the 
Cayley-Hamilton identity, one can show that all single-trace operators in $S_{n\geq N+1}$ 
can be expressed as polynomials of operators in $S_{n\leq N}$ \cite{Choi:2023znd}. So
it suffices to use only the operators in $S_{n\leq N}$ to 
generate graviton cohomologies. The remaining single-trace generators in $S_{n\leq N}$ 
are not independent when we multiply 
them. In other words, there are further trace relations for gravitons within $S_{n\leq N}$.
The last trace relations are not systematically understood, to the best of our 
knowledge.

To simplify the discussions, let us temporarily consider the BMN sector.
The subset of $PSU(2,2|4)$ that acts within the BMN sector is $SU(2|4)$. 
The subset $SU(1|3)\subset SU(2|4)$ commutes with $Q,Q^\dag$ and generates 
the supermultiplets of BMN cohomologies.
In each $S_n$, there is a finite number of single-trace generators in the BMN sector. 
They are given by 
\begin{eqnarray}\label{BMN-mesons}
  (u_n)^{i_1\cdots i_n}&=&{\rm tr}(\phi^{(i_1}\cdots \phi^{i_n)})
  \nn \\
  {(v_n)^{i_1\cdots i_{n\!-\!1}}}_j&=&{\rm tr}(\phi^{(i_1}\cdots\phi^{i_{n\!-\!1})}\psi_j)
  -\textrm{`trace'}
  \nn \\
  (w_n)^{i_1\cdots i_{n\!-\!1}}&=&{\rm tr}(\phi^{(i_1\cdots i_{n\!-\!1})}f+
  {\textstyle \frac{1}{2}}\epsilon^{jk(i_p}{\textstyle \sum_{p=1}^{n\!-\!1}}
  \phi^{i_1}\cdots \phi^{i_{p\!-\!1}}\psi_j\phi^{i_{p\!+\!1}}\cdots\phi^{i_{n\!-\!1})}\psi_k)\ .
\end{eqnarray}
Here, `trace' denotes the terms to be subtracted to ensure that 
the contractions of the upper/lower $SU(3)$ indices are zero.
They are completely determined by this condition from the first term,
but the general expression is cumbersome to write.
The BMN multi-graviton cohomologies are 
polynomials of $u_n$, $v_n$, $w_n$. These polynomials are subject to trace relations. 
These trace relations hold up to $Q$-exact 
terms.\footnote{In principle there might be relations which hold  
without any $Q$-exact terms. In practice, with extensive studies of the $SU(2)$ and 
$SU(3)$ graviton operators in the BMN sector, all trace relations of this 
sort that we found have nontrivial $Q$-exact terms.} 
For instance, the lowest trace relations for $N=2$ are 
\begin{equation}\label{relation-example}
  R_{ij}\equiv\epsilon_{ikm}\epsilon_{jln}(u_2)^{kl}(u_2)^{mn}
  =Q\left[-i\epsilon_{a_1a_2(i}{\rm tr}(\psi_{j)}\phi^{a_1}\phi^{a_2})\right]\ .
\end{equation}
More concretely, some components of these relations are
\begin{equation}
  {\rm tr}(X^2){\rm tr}(Y^2)-[{\rm tr}(XY)]^2\sim 0\ ,\ 
  {\rm tr}(XY){\rm tr}(XZ)-{\rm tr}(X^2){\rm tr}(YZ)\sim 0~,
\end{equation}
where $\sim$ hold up to $Q$-exact terms.
Such $Q$-exact combinations are zeros in 
cohomology. Of course multiplying gravitons to such relations yields further 
relations. Trace relations cannot be seen if one does not know that the `meson' or
`glueball' operators $u_n,v_n,w_n$ are made of `gluons' $\phi,\psi,f$.
To enumerate graviton cohomologies without overcounting, 
we first consider the Fock space made by the operators $\{u_n,v_n,w_n\}$ with
$n=2,\cdots, N$ and then take care of the trace relations to eliminate the dependent 
states.

It is important to find all fundamental trace relations of the polynomials of $u_n,v_n,w_n$, 
which cannot be decomposed into linear combinations of smaller relations. 
Let us denote by $R_a(\{u_n,v_n,w_n\})$ the fundamental trace relations, 
with $a$ being the label. 
Non-fundamental trace relations are obtained by linear combinations of $R_a$'s,
\begin{equation}\label{non-fundamental-relation}
  \sum_a f_a(\{u_n,v_n,w_n\})R_a(\{u_n,v_n,w_n\})\ .
\end{equation}
In general, (\ref{non-fundamental-relation}) is nonzero and $Q$-exact. 
However, for some choices of $f_a$'s, the combination (\ref{non-fundamental-relation}) 
may be exactly zero. If (\ref{non-fundamental-relation}) exactly vanishes, 
this yields a `relation of relations.' 
In terms of the mesonic variables $u_n,v_n,w_n$, they are trivial expressions, meaning 
that various terms just cancel to zero. They just represent the ways 
in which fundamental relations $R_a$ can be redundant at higher orders. 
For example, consider the relations $R_{ij}$ of (\ref{relation-example}) in the 
$SU(2)$ gauge theory. 
Some relations of these relations are given by
\begin{equation}
  u^{ik}R_{jk}(u_2)-{\textstyle \frac{1}{3}}\delta^i_ju^{kl}R_{kl}(u_2)=0~,
\end{equation}
in the $[1,1]$ representation. For instance, one can immediately see for $i=1$, $j=2$ 
that 
\begin{equation}\label{rel-of-rel-example1}
  u^{1i}R_{2i}=u^{11}[u^{23}u^{13}-u^{12}u^{33}]
  +u^{12}[u^{33}u^{11}-(u^{13})^2]+u^{13}[u^{12}u^{13}-u^{11}u^{23}]=0\ .
\end{equation}
This is a trivial identity if expanded in mesons.  
$u^{11}R_{21}$ and $-u^{12}R_{22}-u^{13}R_{23}$ represent same constraint 
$u^{11}(u^{23}u^{13}-u^{12}u^{33})=Q[\cdots]$, 
implying that $R_{ij}$'s are not independent.

The graviton cohomologies in the full theory, instead of restricting to the BMN sector,
can be understood similarly.
In (\ref{BMN-mesons}), we have listed the finite number of single-trace generators,
which are the BPS operators in the multiplet $S_n$, in the BMN sector.
The full list of the single-trace generators in the multiplet $S_n$ is shown in Table \ref{tab:Sn}.
They consist of the superconformal primary $|n\rangle = {\rm tr}(\phi^{(i_1}\cdots \phi^{i_n)})$
in the first row, actions of supercharges that commute with $Q$ on the primary
as displayed in the other rows, and any number of derivatives $\partial_{\dot\alpha}$ of them.
Polynomials of these single-trace generators constitute the multi-graviton cohomologies
in the full $SU(N)$ theory.

\begin{table}[t]
\begin{center}
\begin{tabular}{c|c|c|c|c}
	\hline
    $(-1)^F E^\prime$ & $J^\prime$ & $R_1^\prime$ & $R_2^\prime$ & construction\\
	\hline $n$ & $0$ & $n$ & $0$ & $|n\rangle$ \\
    $-(n+\frac{1}{2})$ & $\frac{1}{2}$ & $n-1$ & $0$ &
    $\overline{Q}_{m\dot\alpha}|n\rangle$ \\
    $n+1$ & $0$ & $n-2$ & $0$
    & $\overline{Q}_{m\dot{+}}\overline{Q}_{n\dot{-}}|n\rangle$ \\
	$-(n+1)$ & $0$ & $n-1$ & $1$ &
    $Q^m_+|n\rangle$\\
    $n+\frac{3}{2}$ & $\frac{1}{2}$ & $n-2$ & $1$ &
    $Q^m_+\overline{Q}_{n\dot\alpha}|n\rangle$ \\
    $-(n+2)$ & $0$ & $n-3$ & $1$ &
    $Q^m_+\overline{Q}_{n\dot{+}}\overline{Q}_{p\dot{-}}|n\rangle$ \\
    $n+2$ & $0$ & $n-1$ & $0$ & $Q^m_+Q^n_+|n\rangle $ \\
    $-(n+\frac{5}{2})$ & $\frac{1}{2}$ & $n-2$ & $0$ &
    $Q^m_+Q^n_+\overline{Q}_{p\dot\alpha}|n\rangle$ \\
    $n+3$ & $0$ & $n-3$ & $0$ &
    $Q^m_+Q^n_+\overline{Q}_{p\dot{+}}\overline{Q}_{q\dot{-}}|n\rangle$ \\
    \hline
\end{tabular}
\caption{The state contents of the $PSU(1,2|3)$ supergraviton multiplet $S_n$. 
For low $n$'s, the rows with negative $R_1^\prime$ are absent. $|n\rangle$ schematically 
denotes the superconformal primaries.}\label{tab:Sn}
\end{center}
\end{table}

Interestingly, trace relations described so far will be used in section \ref{sec:constructcoho}
to construct the ansatz for the non-graviton cohomologies.
In the meantime, we shall exploit a more practical way of enumerating the
graviton cohomologies, as we explain in section \ref{sec:ngindex}.

\section{Strategies for Finding Black Hole Cohomologies}\label{sec:cohostrategy}

At this point, the problem has been well-defined.
Our goal is to list cohomologies with respect to $Q$ in the $\mathcal{N}=4$ SYM
with gauge group $SU(N)$ where $N$ is finite,
that do not belong to the graviton-type cohomologies as defined
in section \ref{sec:gravitons}.
For computational advantage, we will often restrict to the BMN sector of
the cohomologies, as introduced in section \ref{sec:bmn}.
In this last section of the chapter, let us outline our strategy for solving this problem.

We summarize our strategy for solving this problem as follows:
\begin{enumerate}
\item Compute the non-graviton index to locate non-graviton cohomologies. 

\item List non-graviton $Q$-closed operators in the target charge sector.

\item Find a subset of operators found in step 1 which are not $Q$-exact.
\end{enumerate}

First, we compute the index over the non-graviton cohomologies at finite $N$,
with or without restriction to the BMN sector.
This is done by computing the index over the graviton cohomologies,
and subtracting from the index over all cohomologies.
With chemical potentials refining the charges,
one can identify the charge sectors that contain non-graviton states.
Because of the property of the index that bosons and fermions are counted
with opposite signs, one could miss pairs of non-graviton cohomologies
which cancel in the index.
Due to this limitation, we give up finding all cohomologies and search only
for those captured by the index.

Exactly computing the full index is relatively easy at not too large $N$,
especially in the BMN sector, using the matrix integral.
More difficult is to count the finite $N$ gravitons to be subtracted,  
taking into account the trace relations between finite $N$ matrices
which produce extremely nontrivial linear dependence between
the multi-graviton operators.
To make the computation of the graviton index more feasible,
we use the property of the graviton cohomologies that they are faithfully
counted by substituting each elementary field with corresponding diagonal matrix,
rather than a more general traceless matrix.
Using this property, we compute the graviton index in the
BMN sector of $SU(2)$ theory analytically by hand,
and that of the full $SU(2)$ theory manually on a computer.

For the BMN sectors of $SU(3)$ and $SU(4)$ theories,
we boost the computation of the graviton index using Gr\"obner basis.
As we will explain in section \ref{sec:ngiSU3}, counting finite $N$ gravitons
reduces to counting certain class of polynomials,
whose generators are subject to certain relations.
In principle, these relations can be systematically studied using the  Gr\"obner basis. 
However in practice, finding Gr\"obner basis can be computationally very difficult.
So we use a hybrid method of the Gr\"obner basis
(in a subsector in which this basis can be found easily)
and a more brutal counting of independent polynomials by computer,
order by order in the charges.
The results for the non-graviton indices are presented in chapter \ref{sec:ngindex}.

With charge sectors where non-graviton cohomologies exist are located using the index,
we would now like to construct the non-graviton cohomologies.
For the $SU(2)$ theory, it was possible to find the expression for the
smallest non-graviton cohomology \cite{Chang:2022mjp,Choi:2022caq},
and for all non-graviton cohomologies within the BMN sector \cite{Choi:2023znd},
by trials and errors and extensive searches.
This result will be presented in section \ref{sec:cohosu2}.

For the $SU(3)$ theory, even with restriction to the BMN sector,
the naive approach has proved to be infeasible due to combinatorial complexity.
Therefore we take a more streamlined approach of steps 2 and 3 described above.

We present a class of ansatz for the $Q$-closed operators. 
In order for the final cohomology not to be of graviton type,
$Q$ acting on the operator should vanish by trace relations.  
We find a method of constructing a class of operators which become $Q$-closed
only after imposing trace relations. 
Our ansatz uses the trace relations of the graviton cohomologies that we
detected while computing the index.
Trace relations of gravitons mean that certain polynomials of
single-graviton cohomologies are $Q$-exact. 
These relations satisfy `relations of relations',
i.e. certain linear combinations of trace relations (with the coefficients
being graviton cohomologies) are identically zero. 
In other words, relations of relations are linear combinations of $Q$-exact terms which vanish.
So they provide operators which become $Q$-closed thanks to the trace relations,
which validate our ansatz for the non-graviton cohomologies. 

Some $Q$-closed operators mentioned in the previous paragraph 
are not $Q$-exact, providing new cohomologies, while others are $Q$-exact. 
Determining whether a $Q$-closed operator is $Q$-exact or not is very hard. 
We developed a numerics-assisted approach to make this step affordable on the computer,
by ordering the Grassmann variables and then inserting many random integers to the matrix elements. 
As a result, we construct the smallest non-graviton cohomology
in the BMN sector of the $SU(3)$ theory in \ref{sec:cohosu3}.
Furthermore, by extending the numerics-assisted approach,
we prove that the smallest non-graviton cohomology that we construct
is the only one in its charge sector,
denying the possibility that the index may have missed a boson-fermion pair
of non-graviton cohomologies in the charge sector.

\chapter{The Non-Graviton Index}\label{sec:ngindex}

As the first step of solving the cohomology problem,
we compute the index over graviton cohomologies
and subtract from the full index to obtain the
index over non-graviton cohomologies.
It identifies the charges of some of the non-graviton cohomologies.

\section{Methods for the Graviton Index}\label{sec:gindex}

Recall from section \ref{sec:gravitons} that graviton cohomologies are polynomials of 
single-trace graviton cohomologies, which are elements of supermultiplets $S_{n\leq N}$.
They are listed in Table \ref{tab:Sn}, and in particular those in the BMN sector
are the mesons listed in \eqref{BMN-mesons}.
We wish to enumerate linearly independent operators among these, 
i.e. we wish to mod out by linear relations between them.
There are two main strategies that we exploit to ease this computation:
the eigenvalue counting and the Gr\"obner basis.
We only employ the first strategy for computations in $SU(2)$,
while for $SU(3)$ and $SU(4)$ we employ both strategies.
In this section we explain these two strategies.

Let us explain the first idea, the eigenvalue counting.
We first review how the multi-gravitons 
made only of the chiral primaries $u_n$ of (\ref{Sn-primary}) are enumerated. 
Based on rather physical arguments, \cite{Kinney:2005ej} proposed to count 
them by taking all three scalars $\phi^m$ to be diagonal matrices.\footnote{The argument 
is often dubbed `quantizing the moduli space' of the QFT. For exact quantum states, 
it relies on the protection of the moduli space against  
quantum corrections. At the level of classical cohomologies, its proof 
should be elementary, although we do not pursue it here.}
With this restriction, the problem of enumerating independent
gauge-invariant operators, which are multi-trace operators of the matrices $u_n$,
reduces to enumerating independent Weyl-invariant polynomials of the eigenvalues.

Our interest is in counting the finite $N$ graviton cohomologies 
involving all the descendants in $S_n$, not only the chiral primaries $u_n$.
The descendants are obtained from $u_n$ by acting 
the supercharges in $PSU(1,2|3)$.
Since the single-graviton states belong to absolutely protected multiplets $S_n$,
and since their multiplications trivially remain in cohomology both for free
and 1-loop calculations,
we can generate the descendants by acting the supercharges of the
strictly free theory \cite{Choi:2023znd}. 
The actions of free supercharges are linear so that 
diagonal $\phi^m$'s transform to diagonal $\psi_m$, $f$, and other descendants.
The covariant derivatives on the fields also reduce to ordinary 
derivatives since $g_{\rm YM}=0$. 
Therefore, the descendant BPS letters can be taken to be diagonal matrices as well,
for the purpose of enumerating graviton operators.
This is an especially big advantage for the $SU(2)$ theory,
where each elementary field is represented by only one eigenvalue.

So the counting of graviton operators is reduced to the counting 
of certain polynomials of the eigenvalues.
We have $N-1$ eigenvalues for each field $\phi^m$, $\psi_m$,
$\lambda_{\dot\alpha}$, $f$ and their derivatives.
As we truncate by the overall order $\cJ$ of operators,
only a finite number of derivatives are allowed and thus the number of
variables that are needed to describe graviton operators is also finite.
In the BMN sector where there are no gauginos nor derivatives,
$7(N-1)$ variables are needed to describe graviton operators.
Let us denote these eigenvalues collectively as $\lambda_I$,
not to be confused with the gauginos.
Let us also denote the single-trace graviton operators collectively as $g_i$'s.
They are the members of Table \ref{tab:Sn} and their descendants
with $n=2,\cdots,N$ for the full sector,
and the `mesonic generators' $\{u_n,v_n,w_n\}$ (\ref{BMN-mesons}) for the BMN sector.
These are now regarded as polynomials $g_i(\lambda_I)$ of the eigenvalues $\lambda_I$.
Then, we want to count the polynomials $p(g_i)$ of the mesons $g_i$,
which can be written as polynomials $p(g_i(\lambda_I))$ of eigenvalues $\lambda_I$.

These polynomials are not all independent because certain polynomials $p(g_i)$ of $g_i$'s
may be zero when written as polynomials $p(g_i(\lambda_I))$ of $\lambda_I$.
Such polynomials can be thought of as constraints on the space of polynomials.
These are remnants of the trace relations 
of the $N\times N$ matrices. Had we been keeping all the $N\times N$ matrix elements, 
trace relation would have been zero up to a $Q$-exact term. Since the action of $Q$ 
yields a commutator, the $Q$-exact term vanishes when the fields are diagonal.
So general trace relations up to $Q$-exact terms reduce to exact polynomial constraints.

For the $SU(2)$ theory, it was possible to count the number of linearly independent
polynomials given a set of polynomials $p(g_i(\lambda_I))$ of $\lambda_I$.
However, as the number of variables grows, a more systematic treatment became inevitable.
Therefore, we further develop the strategy for enumerating independent graviton
cohomologies in the BMN sectors of the $SU(3)$ and $SU(4)$ theories,
as we now explain.

Counting constrained polynomials is a classic mathematical problem,
with known solution.
This brings us to the second strategy that we exploit: the Gr\"obner basis. 
See e.g. \cite{Cox_2015}.
Let us briefly explain a flavor of its properties and 
how it is used to solve the enumeration problem.

Recall that the multi-graviton operators are
given by the set of all polynomials $p(g_i)$ of $g_i$'s.
However, this set is overcomplete and therefore not suitable for the counting purpose, because of the constraints.
That is, some of the polynomials are zero and consequently some of the polynomials are equivalent to each other. 

We want to better understand the constraints, i.e. polynomials of $g_i$ that are zero.
The constraints appear because each meson $g_i$ 
is not an independent variable but instead made of the gluons $\lambda_I$,
i.e. $g_i=g_i(\lambda_I)$ where the right hand side is a polynomial of $\lambda_I$ that corresponds to the meson $g_i$.
All constraints are derived from the fact that
\begin{equation}\label{meson-def}
 G_i(g_i, \lambda_I) \equiv g_i-g_i(\lambda_I) = 0\ ,
\end{equation}
for each meson labeled by $i$.
Therefore, the set of all polynomials of the mesons $g_i$ and the eigenvalues $\lambda_I$ 
that are zero (also known as the ideal) is \emph{generated} by (\ref{meson-def}), 
in the sense that any element of this set can be written as 
\begin{equation}\label{zero-basis} 
  \sum_i q_i(g_i,\lambda_I) G_i (g_i,\lambda_I)~,
\end{equation}
where $q_i(g_i,\lambda)$ are polynomials of $g_i$ and $\lambda_I$.
If we restrict to elements of this `set of zeroes’ that only involve $g_i$ but not $\lambda_I$, those will be precisely the constraints that mod out the set of all polynomials $p(g_i)$.

Although (\ref{meson-def}) is the most intuitive basis that generates the set of zeroes 
like (\ref{zero-basis}), it is often not the most convenient basis. 
The same set of zeroes can be generated by many different choices of the basis,
possibly with different numbers of generators.
Gr\"obner basis is one of these choices with the following special property.
Let $\{G_a(g_i,\lambda_I)\}$ be a basis of the set of zero polynomials of $(g_i,\lambda_I)$.
Then, for any polynomial $p (g_i,\lambda_I)$,
suppose one tries to `divide’ this polynomial by the basis $\{G_a (g_i,\lambda_I)\}$.
This is a process of writing the polynomial as
\begin{equation}\label{poly-div}
  p (g_i,\lambda_I) =\sum_a q_a (g_i,\lambda_I) G_a (g_i,\lambda_I)  +r (g_i,\lambda_I)\ ,
\end{equation}
where $r (g_i,\lambda_I)$ can no longer be `divided by’ $\{G_a (g_i,\lambda_I)\}$, which can be well-defined by setting an ordering scheme between variables and their monomials.
Naturally, $r (g_i,\lambda_I)$ can be thought of as the remainder of the division.
In general, there can be multiple ways --- with different $q_a$ and $r$ --- to write $p (g_i,\lambda_I)$ as (\ref{poly-div}).
The special property of the Gr\"obner basis is that if $\{G_a (g_i,\lambda_I)\}$
were the Gr\"obner basis of the set of zeroes, then the remainder
$r (g_i,\lambda_I)$ is unique for each given $p (g_i,\lambda_I)$.
Note that since $\{G_a (g_i,\lambda_I)\}$ generates the set of zeroes,
(\ref{poly-div}) implies that the polynomial $p (g_i,\lambda_I)$ is equivalent to its remainder $r (g_i,\lambda_I)$.
It follows that the set of all polynomials $p (g_i,\lambda_I)$ is identical to the set of 
all possible remainders $r (g_i,\lambda_I)$ under division by the Gr\"obner basis.
However, unlike in the set of all polynomials $p (g_i,\lambda_I)$,
there are no polynomials in the set of all remainders that are equivalent due to the constraints, because otherwise one of them should have been divided once more to yield the other as the remainder.
Therefore, the set of remainders can be used to count the number of independent polynomials of $(g_i,\lambda_I)$ under constraints.

There is a canonical procedure to find the Gr\"obner basis of the set of zeroes
given one choice of basis \eqref{meson-def},
known as \emph{Buchberger's algorithm}.
Many computer algebra softwares implement this algorithm or its improved versions.
The Gr\"obner basis depends wildly on the ordering scheme between
variables and monomials, so it is important to choose
a nice ordering scheme which eases the calculations. 
This ordering is difficult to know in advance,
so some amount of trials and errors is involved in finding the Gr\"obner basis.

By setting an appropriate ordering scheme,
it is possible to consistently truncate the Gr\"obner basis for zero polynomials
of $(g_i,\lambda_I)$, into that for zero polynomials of $g_i$ only.
Then, the set of all possible remainders $r(g_i)$ under division
by the truncated Gr\"obner basis form a faithful --- complete but not overcomplete ---
set of all independent polynomials of $g_i$, and therefore the set of
all independent graviton operators.
Moreover, one can easily construct a monomial basis for this set of remainders,
from which it is straightforward to compute both the partition function and the index over graviton operators.

Although the graviton index for the BMN sector of the $SU(2)$ theory
can be computed analytically by hand using the first strategy of eigenvalue counting,
as we will show in section \ref{sec:ngiSU2bmn},
we easily reproduce this result by employing both strategies --- the eigenvalue
counting and the Gr\"obner basis --- explained so far.
This is done by finding a Gr\"obner basis of relations between $SU(2)$
BMN gravitons that consists of 66 generators (after truncation),
and counting the set of all possible remainders under division by those.

Unfortunately, the computation of the Gr\"obner basis quickly becomes very cumbersome if 
the generators of the constraints $\{ g_i - g_i(\lambda_I)\}$ are numerous and complicated. 
For relations between a subset of $SU(3)$ BMN gravitons that do not involve $f$,
i.e. $u_n$ and $v_n$ in \eqref{BMN-mesons},
we found the Gr\"obner basis with $1170$ generators (after truncation) after several hours
of computation on a computer.
For the complete set of $SU(3)$ BMN gravitons including $w_n$,
we were unable to find the Gr\"obner basis due to lack of computing resources:
it takes months at least and it is tricky to parallelize.
Therefore, we have devised a hybrid method to take maximal advantage of the
Gr\"obner basis obtained for the non-$f$ subsector as we now describe.

We first list the complete and independent monomial basis of graviton operators,
i.e. set of monomials of the mesons $g_i$,
that consist of $u_n,v_n$ but not of $w_n$ ($n=2,3$),
up to the charge order $\cJ=54$.
This can be done for any order $\cJ$ because the Gr\"obner basis for the non-$f$ subsector
has been obtained.
Then, one can construct an overcomplete set of all graviton operators by multiplying
each basis from the previous step by arbitrary numbers of $w_2$ and $w_3$,
again up to $\cJ=54$.
Note that $w_2$ and $w_3$ include 3 and 6 different species of single-graviton operators,
respectively, so the size of the overcomplete set grows quickly.

It is helpful to fragment the problem by classifying the operators according to their charges.
Namely, each charge sector is specified by 4 non-negative integers
$2J$ and $q_I = Q_I + J$ (where $I=1,2,3$).
The overall order $\cJ=2(q_1+q_2+q_3)$, defined in \eqref{defcJ}, 
is always even in the BMN sector.
This classification is useful because all single-graviton operators $u_n,v_n,w_n$
and therefore all multi-graviton operators have definite charges,
and operators with different sets of charges can never have a linear relation between them.
Moreover, different charge sectors with merely permuted charges $(q_1, q_2, q_3)$
should contain the same number of independent graviton operators.
Therefore, we separately consider the overcomplete basis of gravitons in each charge sector
with $q_1 \leq q_2 \leq q_3$.

In order to count linearly independent operators among the overcomplete set in any charge sector,
we rewrite each operator as a polynomial of the eigenvalues.
This is done by substituting the mesons with corresponding eigenvalue polynomials
$u_n(\lambda_I)$, $v_n(\lambda_I)$ and $w_n(\lambda_I)$,
which are obtained by writing the gluons in terms of their eigenvalues.
For the eigenvalues of the $SU(3)$ traceless elementary fields, we use the convention
\begin{eqnarray}\label{SU3diag}
f = \begin{pmatrix} f_1 & 0 & 0 \\ 0 & f_2 & 0 \\ 0 & 0 & -f_1-f_2 \end{pmatrix}~,
\end{eqnarray}
and likes.

The number of independent polynomials within each charge sector
is determined as the rank of their coefficient matrix.
We have used the software \texttt{Singular} \cite{DGPS} for finding the Gr\"obner basis,
writing each operator as an eigenvalue polynomial, and extracting the coefficient matrix within each charge sector,
and \texttt{numpy} for computing the rank of the matrix.

The computation of indices for the $SU(3)$ theory have been performed up to $\cJ=54$
on personal computers.
For example, the computation for the charge sector $(2J, q_1, q_2, q_3) = (7,9,9,9)$,
which turns out to be the largest, 
the coefficient matrix was $31026 \times 20940$ with rank 3242.

For the counting of $SU(4)$ BMN gravitons, we take a similar hybrid approach.
Separation into charge sectors works identically to the $SU(3)$ theory.
However, computation of the Gr\"obner basis is even more heavy, both time-wise and memory-wise,
so we were only able to obtain the Gr\"obner basis for a subsector of $SU(4)$
BMN gravitons involving $u_n$ ($n=2,3,4$), i.e. the chiral primaries.
We first list the complete and independent monomial basis of the chiral primaries
$u_n$ using the Gr\"obner basis, up to the order $\cJ=30$.
Then we construct an overcomplete set of all multi-graviton operators
within each charge sector by multiplying each independent basis
by appropriate numbers of $v_2$, $v_3$, $v_4$, $w_2$, $w_3$ and $w_4$, again up to $\cJ=30$.

We write each operator in the overcomplete basis as a polynomial of the eigenvalues.
For the traceless elementary fields in the $SU(4)$ theory,
we used the following convention for the diagonal entries:
\begin{eqnarray}\label{SU4diag}
f = \begin{pmatrix} f_1 & 0 & 0 & 0 \\ 0 & f_2 & 0 & 0 \\ 0 & 0 & f_3-f_1 & 0 \\
0 & 0 & 0 & - f_2 - f_3 \end{pmatrix}~,
\end{eqnarray}
which slightly simplifies the polynomials compared to the more canonical convention
$f = {\rm diag}(f_1,~ f_2,~ f_3,~ -f_1-f_2-f_3)$.

The computation of indices for the $SU(4)$ theory have been performed up to
$\cJ=30$ on personal computers.
For example, the computation for the charge sector $(2J, q_1, q_2, q_3) = (3,5,5,5)$,
which turns out to be the largest, 
the coefficient matrix was $12079 \times 116042$ with rank 3788.

\section{$SU(2)$, BMN Sector}\label{sec:ngiSU2bmn}

In this section, we compute the graviton index, and thus the non-graviton index,
for the BMN sector of the $SU(2)$ theory.
This is done by employing the first of two strategies explained above,
namely the eigenvalue counting.
We represent each of the seven elementary fields by a single eigenvalue,
so all graviton operators can be written as polynomials of 7 variables,
3 of which are Grassmannian.

In terms of eigenvalues, BPS graviton polynomials are
arbitrary products of the following single-gravitons:
\begin{eqnarray}\label{BMNsggrav}
  {\bf 6}&:& x^2\ ,\ y^2\ ,\ z^2\ ,\ xy\ ,\ yz\ ,\ zx~, \nn\\
  {\bf 8}&:&\psi_1\cdot(y,z)\ ,\ \psi_2\cdot(z,x)\ ,\ \psi_3\cdot(x,y)\ ,\ 
  \psi_1 x-\psi_2 y\ ,\ \psi_2 y-\psi_3 z~, \nn\\
  {\bf 3}&:&xf-\textstyle{\frac{1}{2}}\psi_2\psi_3\ ,\ yf-\textstyle{\frac{1}{2}}\psi_3\psi_1\ ,\ zf-\textstyle{\frac{1}{2}}\psi_1\psi_2\ . 
\end{eqnarray}
and the goal of this section is to count independent polynomials among them.
$\psi_{1,2,3}$ are Grassmann variables while $x$, $y$, $z$, $f$ are bosonic. 

In the third line of (\ref{BMNsggrav}), $xf,yf,zf$ are accompanied by two-fermion terms,
but for the purpose of counting independent graviton polynomials,
these terms can be omitted, as we prove now.

Let $\mathfrak{V}$ be the infinite set of all possible products of $6+8=14$ polynomials
in the first two lines of (\ref{BMNsggrav}).
Define two series of vector spaces $V_k$ and $\tilde{V}_k$ as
\begin{eqnarray}
    V_k &=& {\rm span} \left\{ \mathfrak{v} \times (xf)^a(yf)^b(zf)^c
    ~|~ \mathfrak{v} \in \mathfrak{V}~,~ a+b+c \leq k \right\}~, \\
    \tilde{V}_k &=& {\rm span} \left\{ \mathfrak{v} \times
    (xf-\textstyle{\frac{1}{2}}\psi_2\psi_3)^a(yf-\textstyle{\frac{1}{2}}\psi_3\psi_1)^b
    (zf-\textstyle{\frac{1}{2}}\psi_1\psi_2)^c
    ~|~ \mathfrak{v} \in \mathfrak{V}~,~ a+b+c \leq k \right\}~. \nonumber
\end{eqnarray}
We want to show that rank of $V_\infty$ and rank of $\tilde{V}_\infty$ are equal.
We do this by induction.
Clearly ${\rm rank}(V_0) = {\rm rank}(\tilde{V}_0)$.
Now, suppose that ${\rm rank}(V_{k-1}) = {\rm rank}(\tilde{V}_{k-1})$
and let us show that ${\rm rank}(V_k) = {\rm rank}(\tilde{V}_k)$.
The equivalent statement is the following:
\begin{itemize}
\item
Consider a pair of polynomials
\begin{eqnarray}
    v &=& \sum_{i=1}^n r_i (xf)^{a_i} (yf)^{b_i} (zf)^{c_i}~, \nonumber \\
    \tilde{v} &=& \sum_{i=1}^n r_i (xf-\textstyle{\frac{1}{2}}\psi_2\psi_3)^{a_i}
(yf-\textstyle{\frac{1}{2}}\psi_3\psi_1)^{b_i}(zf-\textstyle{\frac{1}{2}}\psi_1\psi_2)^{c_i}~, \nonumber
\end{eqnarray}
where $r_i \in V_0 = \tilde{V}_0$ and $a_i + b_i + c_i = k$ for all $i$
so that $v \in V_k$ and $\tilde{v} \in \tilde{V}_k$.

Then $v \in V_{k-1}$ if and only if $\tilde{v} \in \tilde{V}_{k-1}$.
\end{itemize}

The $\leftarrow$ part is easy. If $\tilde{v} \in \tilde{V}_{k-1}$,
then $\tilde{v}$ equals a linear combination of polynomials that are at most of degree $k-1$
in $xf-\textstyle{\frac{1}{2}}\psi_2\psi_3$ and the likes.
Collecting terms with degree $k$ in $f$, the equality becomes $v=0 \in V_{k-1}$.

To show the $\rightarrow$ part, first note that $v \in V_{k-1}$ implies $v=0$,
since $v$ is homogeneous in $f$ with degree $k$. Now,
\begin{eqnarray}
    \tilde{v} ~=~ - \frac12 \sum_i^n r_i
    \Big[ & a_i \psi_2 \psi_3 (xf-\textstyle{\frac{1}{2}}\psi_2\psi_3)^{a_i-1}
    (yf-\textstyle{\frac{1}{2}}\psi_3\psi_1)^{b_i} (zf-\textstyle{\frac{1}{2}}\psi_1\psi_2)^{c_i}& \\
    & +b_i \psi_3 \psi_1 (xf-\textstyle{\frac{1}{2}}\psi_2\psi_3)^{a_i}
    (yf-\textstyle{\frac{1}{2}}\psi_3\psi_1)^{b_i-1} (zf-\textstyle{\frac{1}{2}}\psi_1\psi_2)^{c_i}& \nonumber \\
    & +c_i \psi_1 \psi_2 (xf-\textstyle{\frac{1}{2}}\psi_2\psi_3)^{a_i}
    (yf-\textstyle{\frac{1}{2}}\psi_3\psi_1)^{b_i} (zf-\textstyle{\frac{1}{2}}\psi_1\psi_2)^{c_i-1}& \Big]~. \nonumber
\end{eqnarray}
If $r_i \psi_j \psi_{j+1}$ for all $i$ and $j=1,2,3$ all belong to $V_0 = \tilde{V}_0$,
it will establish $\tilde{v} = \tilde{V}_{k-1}$.
Indeed, if $r_i$, which is a product of 14 polynomials in the first two lines of (\ref{BMNsggrav}),
contains any of the {\bf 6} in the first line, 
this factor can combine with two $\psi$'s and $r_i \psi_j \psi_{j+1} \in V_0$.
For example, $y^2 \psi_2\psi_3 = (\psi_2 y - \psi_3 z) (\psi_3 y)$.
On the other hand, if $r_i$ contains two or more factors of the {\bf 8} in the second line,
after multiplication by two $\psi$'s it will vanish due to Grassmannian nature of $\psi$,
so automatically $r_i \psi_j \psi_{j+1}= 0 \in V_0$.

Therefore the only possibility that remains in concern is when $r_i$ is precisely one of the {\bf 8}.
This leaves only a finite number of exceptions that one can explicitly work out.
That is, if $v = 0$ with the eight $r_i$ (they cannot mix with other $r_i$ due to homogeneity)
with appropriate numerical coefficients $\alpha_i$:
\begin{eqnarray}
r_1 = \alpha_1 \psi_1 y~,~ r_2 = \alpha_2 \psi_1 z~,~ \cdots
~,~ r_8 = \alpha_8 (\psi_2 y-\psi_3 z)~, \nonumber
\end{eqnarray}
it follows that $\tilde{v} = 0$ as well.
This completes the proof that ${\rm rank}(V_k) = {\rm rank}(\tilde{V}_k)$
given ${\rm rank}(V_{k-1}) = {\rm rank}(\tilde{V}_{k-1})$,
and by induction the number of independent products of (\ref{BMNsggrav})
is not affected by the $\psi \psi$ terms in the third line.

With this rule established, we now count the number of independent graviton polynomials in the BMN sector.
This task is greatly simplified by the fact that all $6+8+3 = 17$ but only two single-graviton generators
are monomials, because linear independence between monomials is rather transparent.
Our strategy will be to order the counting problem carefully 
so that we can work with the monomial basis as far as possible,
and treat the contribution from the two polynomial generators later.

Since there are 3 Grassmann variables, it is convenient to classify 
the graviton operators into $2^3 = 8$ sectors according to their Grassmannian contents.
\bigskip

\hspace*{-.65cm}\underline{\bf 0-fermion sector}

\hspace*{-.65cm}We first focus on the 0-fermion sector: graviton operators that do not contain 
any $\psi$'s.
It is clear that such operators are created by multiplying bosonic single-gravitons on 
the first and third lines of (\ref{BMNsggrav}).
Since all of them are monomials, we may simply write down a list of distinct monomials that can be
obtained by multiplying bosonic single-gravitons, then their linear independence is guaranteed.
The first six single-gravitons can be used to create any monomial $x^ay^bz^c$,
where $a,b,c$ are non-negative integers and $a+b+c$ is even.
Including $xf$, $yf$, $zf$, an eligible monomial may contain any number of $f$
as long as it is supported by at least as many $x$, $y$, or $z$.
Therefore, multi-gravitons in the 0-fermion sector are precisely described as
\begin{eqnarray}\label{BMNgrav0f}
G_0 &=& \{ x^ay^bz^cf^d ~|~ a,b,c,d \in \mathbb{Z}^{\geq 0}~,
a+b+c \geq d~,~ a+b+c+d = 0~(\text{mod } 2) \}~.~~~
\end{eqnarray}

Because we can attribute to each of $x$, $y$, $z$ and $f$ a unit of their own quantum numbers,
the partition function for $G_0$ can be simply defined by the sum over monomials,
\begin{eqnarray}\label{BMNdefgf}
Z_0(x,y,z,f) &=& \sum_{g \in G_0} g~.
\end{eqnarray}
It can be computed as follows.
If there were no restrictions to $a,b,c,d$ except being non-negative integers,
the generating function would be $\frac{1}{(1-x)(1-y)(1-z)(1-f)}$.
From this, we subtract the sum of monomials for which $d > a+b+c$, which is
\begin{eqnarray}\label{BMNgrav0f1}
\frac{1}{(1-xf)(1-yf)(1-zf)} \cdot \frac{f}{1-f}~.
\end{eqnarray}
Then we project to the even part under $(x,y,z,f) \to (-x,-y,-z,-f)$, obtaining 
\begin{eqnarray}\label{BMNgrav0f2}
Z_0 &=& \left[\frac{1}{(1-x)(1-y)(1-z)(1-f)}- \frac{1}{(1-xf)(1-yf)(1-zf)} \cdot \frac{f}{1-f} 
\right]_{\rm even} \nonumber \\
&=& \left[ \frac{1-f(xy+yz+zx-xyz)+f^2xyz}{(1-x)(1-y)(1-z)(1-xf)(1-yf)(1-zf)} 
\right]_{\rm even} \nonumber \\
&=& \frac{1+\chi_2+f (\chi_3 - \chi_1 \chi_2) + f^2 (\chi_3^2+\chi_1 \chi_3)}{(1-x^2)(1-y^2)(1-z^2)(1-xf)(1-yf)(1-zf)}~.
\end{eqnarray}
Abbreviations for cyclic polynomials
\begin{eqnarray}\label{defchi}
\chi_1 &=& x+y+z~, \nonumber \\
\chi_2 &=& xy+yz+zx~, \nonumber \\
\chi_3 &=& xyz~,
\end{eqnarray}
will be used from now on.
\bigskip

\hspace*{-.65cm}\underline{\bf 1-fermion sector}

\hspace*{-.65cm}Now we list (independent) operators with one fermion, 
either $\psi_1$, $\psi_2$ or $\psi_3$. These are obtained by multiplying any operator in 
0-fermion sector $G_0$ by a generator on the second line of (\ref{BMNsggrav}). As mentioned
earlier, the last two of these may create non-monomial operators, so let us first
proceed without them. 

Operators with one $\psi_1$ can only be obtained by multiplying operators in $G_0$
by either $y\psi_1$ or $z\psi_1$. As a result, the list of such operators is simply the following monomials:
\begin{equation}\label{BMNgrav1f1}
\{ x^ay^bz^cf^d \psi_1 ~|~ a,b,c,d \in \mathbb{Z}^{\geq 0}~,~b+c \geq 1~,~a+b+c-1 \geq d~,~ a+b+c+d = 1~(\text{mod } 2) \}~.~~~
\end{equation}
Operators containing one $\psi_2$ or one $\psi_3$ can be listed by 
cyclic permutations of letters.

Next, we ask what new operators arise when multiplying an operator in the 0-fermion sector 
$G_0$ by $x\psi_1-y\psi_2$.
If $x\psi_1-y\psi_2$ multiplies $x^ay^bz^cf^d \in G_0$ such that (i) $c \geq 1$ 
or (ii) $a \geq 1$ and $b \geq 1$,
both monomials $x^{a+1}y^bz^cf^d \psi_1$ and $x^ay^{b+1}z^cf^d \psi_2$ that appear in the product
are already counted in (\ref{BMNgrav1f1}) and corresponding $\psi_2$ sector respectively. 
So no new independent operators arise. Therefore, new operators that are obtained using
$x\psi_1-y\psi_2$ are classified as follows:
\begin{enumerate}
\item $(x^{a \geq 1}y^0z^0f^d)\cdot (x\psi_1-y\psi_2)$: In this case, the second monomial $x^ay^1z^0f^d \psi_2$
is already counted in $\psi_2$ sector corresponding to (\ref{BMNgrav1f1}),
while the first monomial is not counted in the $\psi_1$ sector.
Therefore, these can be regarded new monomials $x^{a+1}y^0z^0f^d \psi_1$ in $\psi_1$ sector.

\item $(x^0y^{b \geq 1}z^0f^d)\cdot (x\psi_1-y\psi_2)$: In this case, the first monomial $x^1y^bz^0f^d \psi_1$
is already counted in $\psi_1$ sector (\ref{BMNgrav1f1}),
while the second monomial is not counted in the $\psi_2$ sector.
Therefore, these can be regarded new monomials $x^0y^{b+1}z^0f^d \psi_2$ in $\psi_2$ sector.

\item $(1) \cdot (x\psi_1-y\psi_2)$: In this case, both monomials $x\psi_1$ and $y\psi_2$ have not been counted
in respective sectors. Therefore, this cannot be regarded as a new monomial in one of $\psi_1$ or $\psi_2$ sector.
Instead, this should be understood as an exceptional non-monomial operator. 
\end{enumerate}
Similar arguments can be made for multiplication by $y\psi_2 - z\psi_3$.

As a result, the list of monomials in $\psi_1$ sector is now extended to
\begin{equation}\label{BMNgrav1f2}
G_{\psi_1} = \{ x^ay^bz^cf^d \psi_1 ~|~ a,b,c,d \in \mathbb{Z}^{\geq 0}~,
a+b+c-1 \geq d~, a+b+c+d = 1 ~ (\text{mod } 2) \}
\backslash \{x \psi_1 \}~.
\end{equation}
List of monomials $G_{\psi_2}$ in $\psi_2$ sector and $G_{\psi_3}$ in $\psi_3$ sector 
are defined by cyclicity. In addition, there are two exceptional operators 
$x\psi_1-y\psi_2$ and $y\psi_2 - z\psi_3$ that are not monomials 
and do not belong to any of $G_{\psi_m}$. So the whole set $G_1$ of 1-fermion 
BPS gravitons is given by
\begin{eqnarray}\label{BMNgrav1f3}
G_1 &=& G_{\psi_1} \cup G_{\psi_2} \cup G_{\psi_3} \cup \{ x\psi_1-y\psi_2~,~y\psi_2 - z\psi_3\}~.
\end{eqnarray}
Alternatively, one can take $G_{\psi_1}$ to \emph{not} exclude $x \psi_1$,
similarly $G_{\psi_2}$ and $G_{\psi_3}$ to \emph{not} exclude $y \psi_2$ and $z \psi_3$ respectively,
but instead exclude just  $x \psi_1 + y \psi_2 + z\psi_3$ at the end.

The existence of such non-monomial operators forbids us from attributing individual quantum numbers to $\psi$'s.
Instead, they carry a negative unit of respective scalar quantum numbers,
and a positive unit of overall $\psi$-number:
\begin{eqnarray}\label{BMNdefqn}
x \to [x]~, y \to [y]~, z \to [z]~, f \to [f]~, \psi_1 \to \frac{[\psi]}{[x]}~, \psi_2 \to \frac{[\psi]}{[y]}~, \psi_3 \to \frac{[\psi]}{[z]}~.
\end{eqnarray}
The partition function of the 1-fermion sector is given by a function of $x$, $y$, $z$, $f$ 
and $\psi$.
The partition function in $\psi_1$ sector (and of the rest of the 1-fermion sector) can be computed
analogously to the 0-fermion sector.
Starting from $\frac{1}{(1-x)(1-y)(1-z)(1-f)} \cdot \frac{\psi}{x}$, we implement the restriction $a+b+c-1 \geq d$
by subtracting its complement, extract the odd part under $(x,y,z,f) \to (-x,-y,-z,-f)$, and further subtract $x \psi_1 \to \psi$.
\begin{eqnarray}\label{BMNgrav1f4}
Z_{\psi_1} &=& \left[\frac{1}{(1-x)(1-y)(1-z)(1-f)}- \frac{1}{(1-xf)(1-yf)(1-zf)} \cdot \frac{1}{1-f} \right]_{\rm odd}
\cdot \frac{\psi}{x} - \psi \nonumber \\
&=& \left[ \frac{x+y+z - (xy+yz+zx)(1+f) + xyz(1+f+f^2)}{(1-x)(1-y)(1-z)(1-xf)(1-yf)(1-zf)} 
\right]_{\rm odd}
\cdot \frac{\psi}{x} - \psi \nonumber \\
&=& \frac{\chi_1 + \chi_3 -f (\chi_2 +\chi_2^2 - \chi_1 \chi_3 - \chi_3^2) + f^2 \chi_3 (1+ \chi_2)}
{(1-x^2)(1-y^2)(1-z^2)(1-xf)(1-yf)(1-zf)} \cdot \frac{\psi}{x} - \psi~.
\end{eqnarray}
Note that $Z_{\psi_2}$ and $Z_{\psi_3}$ can be computed similarly.
Further including $x\psi_1-y\psi_2~,~y\psi_2 - z\psi_3$, one obtains the following 
partition function for $G_1$:
\begin{eqnarray}\label{BMNgrav1f5}
Z_{1}  &=& \frac{\chi_1 + \chi_3 -f (\chi_2 +\chi_2^2 - \chi_1 \chi_3 - \chi_3^2) + f^2 \chi_3 (1+ \chi_2)}
{(1-x^2)(1-y^2)(1-z^2)(1-xf)(1-yf)(1-zf)} \cdot \frac{\chi_2}{\chi_3} \cdot \psi - \psi~.
\end{eqnarray}
\bigskip

\hspace*{-.65cm}\underline{\bf 2-fermion sector}

\hspace*{-.65cm}We consider operators that contain two of three $\psi$'s.
These are obtained by multiplying a generator on the second line of (\ref{BMNsggrav})
to an operator in $G_1$.
Focusing on the $\psi_1 \psi_2$ sector, we first note there are three ways to obtain an operator in this sector.
\begin{enumerate}
\item Multiply either $x\psi_2$ or $z\psi_2$ to an operator in $G_{\psi_1}$ (\ref{BMNgrav1f2}).
Such a set of operators are
\bea\label{BMNgrav2f1}
\{ x^ay^bz^cf^d \psi_1\psi_2 ~|~ a,b,c,d \in \mathbb{Z}^{\geq 0}~,~a+c \geq 1~,~
a+b+c-2 \geq d~, &&\nn\\
a+b+c+d = 0~(\text{mod } 2) \}
&\backslash& \{x^2 \psi_1 \psi_2 \}~. \hspace{1cm}
\eea

\item Multiply either $y\psi_1$ or $z\psi_1$ to an operator in $G_{\psi_2}$, analogous to (\ref{BMNgrav1f2}):
\bea\label{BMNgrav2f2}
\{ x^ay^bz^cf^d \psi_1\psi_2 ~|~ a,b,c,d \in \mathbb{Z}^{\geq 0}~,~b+c \geq 1~,~
a+b+c-2 \geq d~, && \nn\\
a+b+c+d = 0~(\text{mod } 2) \}
&\backslash& \{y^2 \psi_1 \psi_2 \}~. \hspace{1cm}
\eea

\item Multiply $x\psi_2$, $z\psi_2$, $y\psi_1$ or $z\psi_1$ to $x\psi_1 - y\psi_2$.
These supplement $x^2 \psi_1 \psi_2$ and $y^2 \psi_1 \psi_2$ excluded in (\ref{BMNgrav2f1}) and (\ref{BMNgrav2f2}).
\end{enumerate}
Taking the union of the three sets above, we arrive at
\begin{equation}\label{BMNgrav2f3}
G_{\psi_1\psi_2} = \{ x^ay^bz^cf^d \psi_1 \psi_2 ~|~ a,b,c,d \in \mathbb{Z}^{\geq 0}~,~a+b+c-2 \geq d~,~
a+b+c+d = 0~(\text{mod } 2) \}~,
\end{equation}
and similarly for $\psi_2\psi_3$ and $\psi_3\psi_1$ sectors.

Note that we have not explicitly considered multiplying, for example, 
$x\psi_3$ or $y\psi_3$ to $x\psi_1-y\psi_2$. Both monomials obtained this way are already
included in $G_{\psi_3\psi_1}$ and $G_{\psi_2 \psi_3}$,
so they do not add any new independent operators.
Furthermore, there is a possibility of multiplying $x\psi_1-y\psi_2$ or $y\psi_2-z\psi_3$ to the operators in
the 1-fermion sector.
These may give rise to
\begin{eqnarray}
(x\psi_1-y\psi_2)(x\psi_1-y\psi_2) &\sim& xy \psi_1 \psi_2~,\nonumber \\
(y\psi_2-z\psi_3)(y\psi_2-z\psi_3) &\sim& yz \psi_2 \psi_3~, \nonumber \\
(x\psi_1-y\psi_2)(y\psi_2-z\psi_3) &\sim& xy \psi_1 \psi_2 + yz \psi_2 \psi_3 + zx \psi_3 \psi_1~,
\end{eqnarray}
but again, all of the monomials are already counted in respective 2-fermion sectors.
Therefore, we conclude that the 2-fermion sectors can be written completely in monomial basis,
by (\ref{BMNgrav2f3}) and its cyclic versions:
\begin{eqnarray}\label{BMNgrav2f4}
G_2 &=& G_{\psi_1\psi_2} \cup G_{\psi_2\psi_3} \cup G_{\psi_3\psi_1}~.
\end{eqnarray}

The partition function of 2-fermion sector can be computed as before.
The result is:
\begin{eqnarray}\label{BMNgrav2f5}
Z_{\psi_1\psi_2} &=& \frac{\chi_1^2-\chi_2 -\chi_2^2 +2 \chi_1 \chi_3+\chi_3^2 +f(\chi_3 -\chi_1\chi_2)
+f^2 \chi_3 (\chi_1+\chi_3)}{(1-x^2)(1-y^2)(1-z^2)(1-xf)(1-yf)(1-zf)} \cdot \frac{\psi^2}{xy}~,
\end{eqnarray}
for the individual sector, and
\begin{eqnarray}\label{BMNgrav2f6}
Z_{2} &=& Z_{\psi_1\psi_2} +Z_{\psi_2\psi_3} +Z_{\psi_3\psi_1} \nonumber \\
&=& \frac{\chi_1^2-\chi_2 -\chi_2^2 +2 \chi_1 \chi_3+\chi_3^2 +f(\chi_3 -\chi_1\chi_2)
+f^2 \chi_3 (\chi_1+\chi_3)}{(1-x^2)(1-y^2)(1-z^2)(1-xf)(1-yf)(1-zf)} \cdot \frac{\chi_1}{\chi_3} \cdot \psi^2~,
\end{eqnarray}
for the entire 2-fermion sector.
\bigskip

\hspace*{-.65cm}\underline{\bf 3-fermion sector}

\hspace*{-.65cm}We finally investigate the 3-fermion sector, i.e. operators 
that contain all $\psi_1$, $\psi_2$ and $\psi_3$.
One way to obtain 3-fermion operators is to multiply $x\psi_3$ or $y\psi_3$ to the $\psi_1\psi_2$-sector
(\ref{BMNgrav2f3}). Set of such operators is
\begin{equation}\label{BMNgrav3f1}
\{ x^ay^bz^cf^d \psi_1 \psi_2 \psi_3 ~|~ a,b,c,d \in \mathbb{Z}^{\geq 0}~,~a+b \geq 1~,~a+b+c-3 \geq d~,~
a+b+c+d = 1~(\text{mod } 2) \}~.
\end{equation}
By cyclicity, there are two more sets of 3-fermion operators that are obtained by $x \to y \to z \to x$ from
(\ref{BMNgrav3f1}). Their union is,
\begin{equation}\label{BMNgrav3f2}
G_{\psi_1 \psi_2 \psi_3} = \{ x^ay^bz^cf^d \psi_1 \psi_2 \psi_3 ~|~ a,b,c,d \in \mathbb{Z}^{\geq 0}~,~a+b+c-3 \geq d~,~
a+b+c+d = 1~(\text{mod } 2) \}~.
\end{equation}
One can easily check that multiplying non-monomial blocks $x\psi_1-y\psi_2$ or $y\psi_2-z\psi_3$
to 2-fermion sector does not produce any new operator.

Partition function of the 3-fermion sector (\ref{BMNgrav3f1}) is
\begin{equation}\hspace{-0.5cm}\label{BMNgrav3f3}
Z_{3} = \left[ \frac{ \splitfrac{-1+\chi_1^2-2\chi_2-\chi_2^2+2\chi_1\chi_3+\chi_3^2
+f ( \chi_1+\chi_3)}{-f^2 ( \chi_2 +\chi_2^2 -\chi_1\chi_3-\chi_3^2)
+f^3 \chi_3 ( 1+\chi_2)}}{(1-x^2)(1-y^2)(1-z^2)(1-xf)(1-yf)(1-zf)} +1\right]
\cdot \frac{\psi^3}{f \chi_3}~.
\end{equation}
\bigskip

\hspace*{-.65cm}\underline{\bf The index}

\hspace*{-.65cm}The complete list of BPS multi-graviton operators in BMN sector of 
the $SU(2)$ theory is given by (\ref{BMNgrav0f}), (\ref{BMNgrav1f3}), (\ref{BMNgrav2f4}) 
and (\ref{BMNgrav3f2}). Corresponding partition function is $Z_0 + Z_1 + Z_2 + Z_3$,
each of which is presented in (\ref{BMNgrav0f2}), (\ref{BMNgrav1f5}), (\ref{BMNgrav2f6}) and (\ref{BMNgrav3f3}). Attributing minus sign to the fermion number $\psi$ in the partition function and further setting $\psi,f\rightarrow xyz$ will yield the index, 
where $(x,y,z)=(e^{\Delta_1},e^{\Delta_2},e^{\Delta_3})$.

To facilitate comparison with the other parts of this paper, 
we compute the unrefined index of the graviton partition function.
This is obtained simply by substituting
\begin{eqnarray}\label{unrindex}
x,y,z \to t^2~, \qquad f \to t^6 ~, \qquad \psi \to -t^6~.
\end{eqnarray}
in to the partition function. The result is
\begin{equation}\label{BMNgravindex}
\hspace*{-.3cm}Z_{\rm grav} = \frac{\splitfrac{1+3t^4-8t^6-6t^{10}+10t^{12}+
9t^{14}-9t^{16}+16t^{18}}{-18t^{20}-3t^{22}+t^{24}-3t^{26}+9t^{28}-2t^{30}+3t^{32}-3t^{34}}}
{(1-t^4)^3 (1-t^8)^3}~.
\end{equation}

Meanwhile, the full index over all cohomologies in the BMN sector of the $SU(2)$ theory
can be computed via residue sum of the matrix integral (\ref{BMN-index-integral}).
We only present the unrefined ($e^{\Delta_1}=e^{\Delta_2}=e^{\Delta_3}\equiv t^2$)
version, as our main focus will be on the non-graviton index.
\begin{eqnarray}
  Z&=&\left[\frac{}{}\right.
  1+3t^2+12t^4 + 20t^6 + 42t^8 + 48t^{10} + 75t^{12} + 66t^{14} 
  + 81t^{16} \nn\\
  &&\hspace*{.3cm}+ 55t^{18} + 54t^{20} + 27t^{22} + 19t^{24} + 6t^{26} + 
  3t^{28}\left.\frac{}{}\right]\frac{(1-t^2)^3}{ (1-t^{12})(1 - t^8 )^3 }\ .
\end{eqnarray}

The difference $Z - Z_{\rm grav}$ will be the index that counts non-graviton operators.
We find a simple analytic formula for the difference:
\begin{equation}\label{bh-index-bmn}
  Z-Z_{\rm grav}=\left[-\frac{e^{4(\Delta_1+\Delta_2+\Delta_3)}}{1-e^{2(\Delta_1+\Delta_2+\Delta_3)}}\right]\cdot 
  \left[\prod_{I=1}^3(1-e^{\Delta_I})\right]\cdot \left[\prod_{I=1}^3
  \frac{1}{1-e^{\Delta_I}e^{\Delta_1+\Delta_2+\Delta_3}}\right]~.
\end{equation}
Its unrefined version ($e^{\Delta_1}=e^{\Delta_2}=e^{\Delta_3}\equiv t^2$)
is also informative:
\begin{eqnarray}\label{BMNindexdiff}
Z - Z_{\rm grav} &=& - \frac{t^{24}}{1-t^{12}} \cdot \frac{(1-t^2)^3}{(1-t^8)^3}~.
\end{eqnarray}

From this formula, one finds the first black hole cohomology at 
$j=24$. This `threshold' black hole cohomology was already identified 
in \cite{Chang:2022mjp,Choi:2022caq},
as we shall review and rewrite in a more compact form in the next chapter.
It may look like there are many black hole states beyond this threshold, 
but most of them are rather trivial.
To make this point clear, we would like to first interpret various factors of
(\ref{bh-index-bmn}), which will be extensively justified later.

(\ref{bh-index-bmn}) is a multiplication of three factors.
We interpret the first factor as the `core' black hole primary operators.
Constructing this part of the cohomologies will be the goal of section \ref{sec:cohosu2}.
The second factor comes from the $SU(1|3)$ descendants obtained from
the first factor by acting $Q^m_+$. 
The supercharge $Q^m_+$ carries charges $Q_I=\delta_{I,m}-\frac{1}{2}$
and $J=\frac{1}{2}$, so is weighted by  $e^{\Delta_I}$.
So the second factor comes from the Fock space obtained by acting three $Q^m_+$'s.
Finally, the third factor comes from multiplying
certain multi-gravitons to the core black hole cohomologies.
Among the $17$ graviton states listed in  
(\ref{BMNsggrav}), only $3$ types on the third line can contribute. 
The remaining $14$ gravitons multiplying the core
black hole operators do not appear in the index.
This aspect too will be discussed further in section \ref{sec:cohosu2}.

\section{$SU(2)$}

For the $SU(2)$ theory but without restriction to the BMN sector,
it is not possible to compute the non-graviton index analytically,
partly because the graviton operators are polynomials of
an infinite number of variables.
Note that derivatives of the eigenvalues should be considered
as different variables in the algebraic point of view.
Therefore, truncation by the order of the operator is inevitable.

Counting the graviton cohomologies with a computer using the eigenvalue 
setup explained in section \ref{sec:gindex} ,
we have obtained the graviton index $Z_{\rm grav}$ for the $SU(2)$
theory until $t^{40}$ order.
Substracting from the full index, the non-graviton index for the $SU(2)$ theory
is given by 
\begin{eqnarray}\label{su2-general-bh}
  Z-Z_{\rm grav}&=&\left[-t^{24}-\chi_{(1,3)}t^{32}
  -(\chi_{(1,\bar{3})}+\chi_{(3,6)})t^{34}-\chi_{(2,3)}t^{35}
  +(\chi_{(3,1)}+\chi_{(3,8)})t^{36}\right.\nonumber\\
  &&\left.-(\chi_{(2,\bar{3})}+\chi_{(4,6)})t^{37}+\chi_{(5,3)}t^{38}
  +(\chi_{(2,1)}+2\chi_{(4,1)}+\chi_{(4,8)})t^{39}\right.\nonumber\\
  &&\left.-(2\chi_{(1,6)}+\chi_{(3,\bar{3})}+\chi_{(5,\bar{3})}
  +\chi_{(5,6)})t^{40}\right]\chi_{\rm D}+\mathcal{O}(t^{41})\ .
\end{eqnarray}
We have organized the result into $SU(2)_R\times SU(3) \subset PSU(1,2|3)$ characters.
We have also factored out by $\chi_{D}$ which is given by
\begin{eqnarray}\label{characters-t40}
  \chi_{(2J^\prime+1,{\bf{\rm R}})}\!&\!\equiv\!&\!
  \chi^{SU(2)_R}_{J^\prime}(p)\chi^{SU(3)}_{\bf\rm R}(x,y)~, \\
  \chi_{D}\!&\!\equiv\!&\! \frac{(1\!-\!t^2z_1)(1\!-\!\frac{t^2}{z_2})(1\!-\!\frac{t^2z_2}{z_1})
  (1\!-\!\frac{tp}{z_1})(1\!-\!\frac{t}{pz_1})(1\!-\! tz_2p)(1\!-\!\frac{tz_2}{p})
  (1\!-\!\frac{tz_1p}{z_2})(1\!-\!\frac{tz_1}{z_2p})}{(1-t^3p)(1-\frac{t^3}{p})}~, \nn
\end{eqnarray}
where $t^6=e^{\Delta_1+\Delta_2+\Delta_3}=e^{\omega_1+\omega_2}$, 
$z_1=e^{\frac{2\Delta_1-\Delta_2-\Delta_3}{3}}$, $z_2^{-1}=
e^{\frac{-\Delta_1+2\Delta_2-\Delta_3}{3}}$, 
$p=e^{\frac{\omega_1-\omega_2}{2}}$.
This is a factor for superconformal descendants.
Since the non-gravitons should appear in representations of $PSU(1,2|3)$,
the subset of the $\mathcal{N}=4$ superconformal group $PSU(2,2|4)$
that commutes with the supercharge $Q$,
it is economical to write only the superconformal primaries
from which the descendants automatically follow.
It is very likely that all non-graviton operators belong to the $A_1\bar{L}$-type
supermultiplets in the notation of \cite{Cordova:2016emh},
and $\chi_D$ is the factor that yields the character of the supermultiplet
when multiplied to the character of the superconformal primary.
So each term in the square bracket of (\ref{su2-general-bh})
should represent a $PSU(1,2|3)$ supermultiplet whose superconformal
primary transforms under the denoted representations under the bosonic subalgebra.

\section{$SU(3)$, BMN Sector}\label{sec:ngiSU3}

Following the computational procedures explained earlier in this chapter,
including the eigenvalue counting and the Gr\"obner basis,
we have computed the $SU(3)$ graviton index $Z_{\rm grav}$ until $t^{54}$ order.
We write the difference $Z-Z_{\rm grav}$ with the full index $Z$,
which is the index over non-graviton cohomologies or the `black hole' cohomologies,
in the form of
\begin{equation}\label{difference-factored}
  Z-Z_{\rm grav}=Z_{\rm core}(\Delta_I)\cdot 
  \prod_{I=1}^3\frac{1}{1-e^{\Delta_I}e^{\Delta_1+\Delta_2+\Delta_3}}
  \cdot \prod_{I<J}(1-e^{\Delta_I+\Delta_J})\ .
\end{equation}
The factors that dress the index over \emph{core} non-graviton cohomologies
will be explained shortly.
$Z_{\rm core}(\Delta_I)\equiv f(t,x,y)$ with 
$e^{\Delta_1}=t^2x$, $e^{\Delta_2}=t^2y^{-1}$, $e^{\Delta_3}=t^2x^{-1}y$ 
can be expanded as 
\begin{equation}
  f(t,x,y)=\sum_{\cJ=0}^{54}\sum_{{\bf R}_j}
  (-1)^{F({\bf R}_\cJ)}\chi_{{\bf R}_\cJ}(x,y)t^{\cJ}+\mathcal{O}(t^{56})\ ,
\end{equation}
where ${\bf R}_\cJ$ runs over the $SU(3)$ irreducible representations which 
appear at $t^\cJ$ order ($\cJ$ is even in the BMN sector), $\chi_{{\bf R}_\cJ}(x,y)$ is 
its character, and 
$F({\bf R}_\cJ)$ is its fermion number. 
The representations ${\bf R}_\cJ$ appearing in the expansion of $f$, together 
with their bosonic/fermionic natures, are shown in Table \ref{tower}.
We have classified the representations into several groups, i.e. what we suspect 
to be the fermionic towers $F_0,...,F_4$, the bosonic towers $B_1,...,B_3$, 
and the remainders $F_{\rm exc}$, $B_{\rm exc}$ for which we 
do not see particular patterns (thus named `exceptional').
Entries that appear in cyan may be related to the towers $F_1$ and $B_1$
of core primaries by dressing of $w_3$ gravitons.
We will comment on the dressings later.
Entries in gray are not observed in the non-graviton index,
but we included them because if we assume that they appear in boson/fermion pairs,
then the tower structure is reinforced.

\begin{table}[t]
\begin{center}
\begin{tabular}{c||c|c|c|c|c|c||c|c|c|c}
	\hline
    $j$ & $F_0$ & $F_1$  & $F_2$ & $F_3$ & $F_4$ & $F_{\rm exc}$ & $B_1$ & $B_2$ 
    & $B_3$ & $B_{\rm exc}$ \\
	\hline 
    $24$ & $[0,0]$ &&&&&&&&& \\
    $26$ &&&&&&&&&& \\
    $28$ &&&&&&&&&& \\
    $30$ & $[0,0]$ & $[3,0]$ &&&&&&&& \\
    $32$ && $[4,0]$ &&&&&&&& \\
    $34$ && $[5,0]$ &&&&& $[3,1]$ &&& \\
    $36$ & $[0,0]$ & $[6,0]$ &&&&& $[4,1]$ &&& $[3,0]$ \\
    $38$ && $[7,0]$ &&&& $[1,0]$ & $[5,1]$ &&& \\
    $40$ && $[8,0]$ & $\color{cyan} [5,0]$ && $\color{cyan}[3,1]$ && $[6,1]$ &&& \\
    $42$ & $[0,0]$ & $[9,0]$ & $\color{cyan}[6,0]$ && $\color{cyan}[4,1]$ && $[7,1]$ &&& $[1,1]$ \\
    $44$ && $[10,0]$ & $\color{cyan}[7,0]$ && $\color{lightgray} [5,1]$ && $[8,1]$ & $\color{lightgray} [5,1]$ && \\
    $46$ && $[11,0]$ & $\color{cyan}[8,0]$ && $\color{lightgray} [6,1]$ & $[2,0]$ & $[9,1]$ & $\color{lightgray} [6,1]$ && $[5,0]$ \\
    $48$ & $[0,0]$ & $[12,0]$ & $\color{cyan}[9,0]$ && $\color{lightgray} [7,1]$ & $[3,0]$ & $[10,1]$ & $\color{lightgray} [7,1]$ && $[4,1]$ \\
    $50$ && $[13,0]$ & $\color{cyan}[10,0]$ & $\color{cyan}[7,0]$ & $\color{lightgray} [8,1]$ && $[11,1]$ & $\color{lightgray} [8,1]$ && $[4,0]$ \\
    $52$ & & [14,0] & \color{cyan} [11,0] & \color{cyan} [8,0] & \color{lightgray} [9,1] & 
    [2,0] & $[12,1]$ & \color{lightgray} [9,1] && [3,1]\\ 
    $54$ & & [15,0] & \color{cyan} [12,0] & \color{cyan} [9,0] & \color{lightgray} [10,1] & 
    [4,1] & $[13,1]$ & \color{lightgray} [10,1] & \color{cyan} [7,1] & \\
    \hline
\end{tabular}
\end{center}
\caption{$SU(3)$ Dynkin labels of fermionic/bosonic black hole cohomologies 
after factoring out the descendants and the conjectured graviton hairs of $w_2$,
organized into towers by empirical reasons.
}\label{tower}
\end{table}

We comment on the factors which we have taken out in (\ref{difference-factored}).
The factor $\prod_{I<J}(1-e^{\Delta_1+\Delta_2})$ accounts for $SU(1|3)$ descendants.
For each non-graviton cohomology in ${\bf R}_j$ that contributes to $Z_{\rm core}$,
the entire $SU(1|3)$ multiplet obtained by acting the three fermionic generators $Q_+^m$
must also be non-graviton cohomologies.
Every such multiplet is a long multiplet of the $SU(1|3)$,
so the corresponding character is simply the contribution from the primary
times the factor $\prod_{I<J}(1-e^{\Delta_1+\Delta_2})$.
This fact can be argued using the embedding supergroup
$PSU(2,2|4)$ of the 4d $\mathcal{N}=4$ theory.
For any of the three generators $Q_+^m$ to annihilate the $SU(1|3)$ primary,
the primary of a bigger representation of $PSU(2,2|4)$ that includes the $SU(1|3)$ multiplet
must be annihilated by $Q_+^4$ \emph{and} by the $SU(4)_R$ lowering operator
that is not part of the $SU(3) \subset SU(4)_R$.
The only $PSU(2,2|4)$ representations that satisfy this property are
$B_1 \bar{B}_1 [0;0]^{[0,n,0]}$, namely the graviton operators, or the identity.
For details on the relevant representation theory, we refer to
\cite{Cordova:2016emh}, particularly its section 2.2.4,
or to appendix B of \cite{Choi:2023znd}.

The second factor of (\ref{difference-factored}) was taken out for an empirical 
reason, with an expectation that they come from the graviton hairs of $w_2$'s 
in (\ref{BMN-mesons}). 
Namely, we conjecture that $w_2$ gravitons multiplying 
the core black hole cohomologies represented by $Z_{\rm core}$ provide 
nontrivial product cohomologies.
Although we have little logical justification of the last claim (except that similar 
hairs are allowed in the $SU(2)$ theory), we think that the phenomenological evidence 
of this claim is compelling since various simple patterns in 
Table \ref{tower} are clear only after factoring it out.

We refer to section 3.1 of \cite{Choi:2023vdm} for discussions on the tower structures.
Various scenarios and suggestions are presented there,
as to how and which graviton cohomologies may multiply to some of the
black hole cohomologies displayed in Table \ref{tower}
to yield other cohomologies also displayed in Table \ref{tower}.
The discussion on partial no-hair behavior that will be explained in
section \ref{sec:cohosu2} extends with various complications to Table \ref{tower}.

\section{$SU(4)$, BMN Sector}\label{sec:ngiSU4}

In the $SU(4)$ case, using similar strategies as for the $SU(3)$ case,
we computed $Z_{\rm grav}$ until $j=30$ level. 
The index $Z-Z_{\rm grav}$ over non-graviton cohomologies is given by
\begin{equation}
  Z-Z_{\rm grav}=\left[-\chi_{[2,0]}(x,y)t^{28}
  -\chi_{[3,0]}(x,y)t^{30}+\mathcal{O}(t^{32})
  \right]\cdot \prod_{I<J}(1-e^{\Delta_I+\Delta_J})\ .
\end{equation}
The second factor generates the Fock space of each $SU(1|3)$ multiplet, 
while the first factor in the square parenthesis represents the primary non-gravitons.
One finds that the BMN index predicts an apparent threshold of non-graviton 
cohomologies at $\cJ=2(Q_1+Q_2+Q_3) + 6J = 28$.
Again, conservatively, this is an upper bound for the threshold for
two different reasons: first because the index may miss a 
pair of canceling threshold cohomologies at lower charges, and also because the true 
threshold might lie outside the BMN sector (carrying nonzero $SU(2)_R$ spin $J_1-J_2$).
Anyway, the above apparent threshold is higher than the $SU(3)$ threshold. 
So it is natural to expect that it was an exception that 
the $SU(2)$ and $SU(3)$ thresholds were the same: the (apparent) 
thresholds for $\cJ$ are $24,24,28,\cdots$ for $N=2,3,4\cdots$. 
To obtain the threshold level in terms of energy $E=\sum_I Q_I+ \sum_i J_i$, 
one should construct the actual cohomologies which account for the 
$t^{28}$ term. This will not be done in this thesis.

\chapter{Constructing the Cohomologies}\label{sec:constructcoho}

The non-graviton indices computed in the previous chapter
guide us to focus on certain charge sectors to construct the
simplest non-graviton cohomologies.
For example, the indices for BMN sectors of the $SU(2)$ and $SU(3)$ theories
suggest that we attempt to construct a fermionic black hole cohomology
only using the BMN letters, that is a singlet under the $SU(3)$ subgroup
of the R-symmetry group at the order $\cJ=24$,
equivalently $q_1 = q_2 = q_3 = 4$.

For the $SU(2)$ theory, such a cohomology will indeed be the
threshold black hole cohomology, i.e. one with the lowest order.
It has been shown in \cite{Chang:2022mjp} through an extensive search in the
space of all cohomologies that in the $SU(2)$ theory,
the fermionic singlet cohomology at $\cJ=24$ is the first and the only one
until the order $\cJ=25$.
For the $SU(3)$ theory, the extensive study of \cite{Chang:2022mjp}
has been performed only until $\cJ=19$,
so the possibility that a black hole cohomology exists between $\cJ=20$
and $\cJ=24$ but outside of the BMN sector,
or the possibility that a boson-fermion pair of black hole cohomologies
exists between the same order, are not ruled out.
However, it is unlikely that the threshold cohomology for the $SU(3)$ theory
appears at a lower order than for the $SU(2)$ theory,
so we are somewhat confident that the $\cJ=24$ black hole cohomology
that we shall present is indeed the threshold black hole cohomology
of the $SU(3)$ theory.

\section{$SU(2)$, BMN Sector}\label{sec:cohosu2bmn}

The threshold black hole cohomology for the $SU(2)$ theory
was shown to exist at the order $\cJ=24$ through an extensive search in the
space of all cohomologies \cite{Chang:2022mjp},
and its explicit form was written down shortly after in \cite{Choi:2022caq}.
Meanwhile, our result on the BMN non-graviton index of the $SU(2)$ theory,
in particular the first factor of (\ref{bh-index-bmn}),
\begin{equation}\label{index-core-primary}
  -\frac{t^{24}}{1-t^{12}}=-t^{24}-t^{36}-t^{48}-t^{60}-\cdots\ ,
\end{equation}
suggests that there is one fermionic black hole cohomology at every
12 value of $\cJ$ starting from the threshold at $\cJ=24$.
In this section, we present the explicit form \eqref{On-summary}
of these core black hole cohomologies.
These, together with the factors in \eqref{bh-index-bmn} who interpretation
was given below the equation,
account for all black hole cohomologies detected by the index
in the BMN sector of the $SU(2)$ theory.

The index, in particular the core factor (\ref{index-core-primary}),
predicts unique fermionic cohomology at each order
$\cJ=24+12n$ ($n=0,1,2,\cdots$), all singlets of $SU(3)\subset SU(4)$.
For the $SU(2)$ gauge group, we use the 3-dimensional vector notation for the adjoint fields.
In the remaining part of this section,
$\phi^m=(X,Y,Z)$, $\psi_{m}$ ,$f$ will denote 3 dimensional vectors,
and inner/outer products will replace the trace/commutators. 
The $Q$-transformations of these $3$-vectors are given by
\begin{equation}\label{Q-bmn-SU(2)}
  Q\phi^m=0\ ,\ \ Q\psi_m={\textstyle \frac{1}{2}}\epsilon_{mnp}\phi^n\times \phi^p
  \ ,\ Qf=\phi^m\times\psi_m\ .
\end{equation}

\hspace*{-0.65cm}\underline{\bf $O_0$ operator at $t^{24}$}

\hspace*{-0.65cm}This operator has charges 
$E=\frac{19}{2}$, $Q_1=Q_2=Q_3=\frac{3}{2}$, $J_1=J_2=\frac{5}{2}$. 
A representative of this cohomology \cite{Choi:2022caq} is given by
\begin{eqnarray}\label{n=0-old}
  O_0^\prime&=&(X\cdot \psi_1-Y\cdot \psi_2)(X\cdot \psi_3)(\psi_2\cdot \psi_1\times\psi_1)
  +(Y\cdot \psi_2-Z\cdot \psi_3)(Y\cdot \psi_1)(\psi_3\cdot \psi_2\times\psi_2)\nonumber\\
  &&+(Z\cdot \psi_3-X\cdot \psi_1)(Z\cdot \psi_2)(\psi_1\cdot \psi_3\times\psi_3)\ .
\end{eqnarray} 
Note that the second and third terms are obtained by making cyclic permutations 
of $(X,\psi_1)$, $(Y,\psi_2)$, $(Z,\psi_3)$ on the first term. The cyclic permutations 
are part of the $SU(3)$ symmetry, thus symmetries of the cohomology problem,
On the other hand, odd permutations accompanied by the sign flips of all 
$\psi_m$'s and $\phi^m$'s are part of $SU(4)\times SU(2)_L$ symmetry 
which leave $Q$ invariant, thus being symmetries of the cohomology problem. 
To construct a better representative of this cohomology, consider the following operator 
obtained by permuting $(X,\psi_1)\leftrightarrow(Y,\psi_2)$ and flipping signs of all 
$\phi^m,\psi_m$ on (\ref{n=0-old}):
\begin{eqnarray} 
  O_0^{\prime\prime}&=&(X\cdot \psi_1-Y\cdot \psi_2)(Y\cdot \psi_3)(\psi_1\cdot \psi_2\times\psi_2)
  +(Y\cdot \psi_2-Z\cdot \psi_3)(Z\cdot \psi_1)(\psi_2\cdot \psi_3\times\psi_3)\nonumber\\
  &&+(Z\cdot \psi_3-X\cdot \psi_1)(X\cdot \psi_2)(\psi_3\cdot \psi_1\times\psi_1)\ .
\end{eqnarray}
One can show 
\begin{eqnarray}\label{n=0}
  O_0^\prime-O_0^{\prime\prime}&=&-2Q[(\psi_1\cdot\psi_2)(\psi_2\cdot\psi_3)(\psi_3\cdot\psi_1)]
  \ ,\\
  O_0&\equiv&-5(O_0^\prime+O_0^{\prime\prime})=
  \epsilon^{p_1p_2p_3}v^{m}_{\ \ p_1}v^{n}_{\ \ p_2}(\psi_m\cdot\psi_n\times \psi_{p_3})\ ,
  \nonumber
\end{eqnarray}
where 
\begin{equation}
  v^m_{\ \ n}\equiv (\phi^m\cdot\psi_n)-{\textstyle \frac{1}{3}}\delta^m_n(\phi^p\cdot\psi_p)
\end{equation}
are graviton cohomologies in the $S_2$ multiplet. $O_0$ is manifestly an $SU(3)$ singlet. 
Note that the second term of $v$ proportional to $\delta^m_n$ drops out when $v$ is
inserted into (\ref{n=0}), because of the symmetry of $\psi_m\cdot\psi_n\times\psi_{p_3}$ 
and the antisymmetry of $\epsilon^{p_1p_2p_3}$. So we can write
\begin{equation}\label{n=0-alternative}
  O_0=\epsilon^{p_1p_2p_3}(\phi^m\cdot \psi_{p_1})(\phi^n\cdot \psi_{p_2})
  (\psi_m\cdot\psi_n\times \psi_{p_3})\ .
\end{equation}

To show that $O_0$ is a black hole cohomology, one should check 
that it is $Q$-closed, not $Q$-exact, and not of graviton type. 
The first and third are trivial. $O_0$ is not graviton-like 
because it consists of seven (odd) letters: since $SU(2)$ gravitons are 
made of operators in $S_2$, they always have an even number of letters. 
To check $Q$-closedness, first note that $Q$ acts only 
on $\psi_m\cdot\psi_n\times \psi_{p_3}$ because 
$v^m_{\ \ n}$ are $Q$-closed. One finds
\begin{equation}\label{Q-psi-3}
  Q(\psi_m\cdot \psi_n\times \psi_p)={\textstyle \frac{3}{2}}
  \epsilon_{(m|qr}(\phi^q\times \phi^r)\cdot(\psi_{|n}\times\psi_{p)})
  =3\epsilon_{(m|qr}(\phi^q\cdot\psi_{|n})(\phi^r\cdot \psi_{p)})=
  3\epsilon_{(m|qr}v^{q}_{\ \ |n}v^{r}_{\ \ p)}\ .
\end{equation}
At the last step, the second term of $\phi^q\cdot\psi_n=v^q_{\ \ n}+\delta^q_n(\cdots)$ 
etc. does not survive after the index contractions. Inserting it to $QO_0$ and
replacing the product of two $\epsilon$'s by three $\delta$'s, $QO_0$ is given by 
various row/column contractions of 
four $3\times 3$ traceless matrices $v^m_{\ \ n}$. Possible terms are
${\rm tr}(v^4)$ and ${\rm tr}(v^2){\rm tr}(v^2)$, but the fermionic nature of 
$v$ and the cyclicity of trace ensure that they are all zero. 
So $QO_0$ is zero because there are no nonzero terms that can contribute.

The non-$Q$-exactness was originally shown after a calculation using computer
\cite{Chang:2022mjp,Choi:2022caq}. Here we provide an analytic argument. 
We assume $Q$-exactness, narrow down the possible 
$Q$-exact terms and then show 
that no combination of them works.
$O_0$ is at the $\mathcal{O}(\phi^2\psi^5)$ order. If this is $Q$-exact, the schematic 
structure should be as follows:
\begin{equation}\label{n=0-exact}
  \phi^2\psi^5= Q_{f}[f\phi\psi^4]+Q[\psi^6]\ .
\end{equation}
$Q_f$ means the part of $Q$ acting on $f$. $Q$ may also act on $\psi$ 
in this term to produce a term at $\mathcal{O}(f\phi^3\psi^3)$ order, 
and if $O_0$ is completely $Q$-exact, $Q_\psi(f\phi\psi^4)$ 
should cancel $Q_f[f^2\phi^2\psi^2]$. We shall only consider the 
$Q$-exactness of $O_0$ within the $\phi^2\psi^5$ order and find a contradiction.
The terms on the right hand side should respect all the $SU(3)\times SU(2)$ tensor structures 
of the left hand side.
There might be terms violating some of these structures separately on the first and second 
terms, but they should cancel by themselves and we do not care about this part. 
We consider the terms which respect them and the equation should hold 
within this sector separately if (\ref{n=0-exact}) is generally true.   
(\ref{n=0-alternative}) is given by multiplying 
the following scalar and fermion factors,
\begin{eqnarray}\label{n=0-bootstrap}
  \phi^{(m}_{(i}\phi^{n)}_{j)}&:&({\bf 6},{\bf 5}+{\bf 1})\in SU(3)\times SU(2)\\
  \psi_{p_1(i|}\psi_{p_2|j)}(\psi_m\cdot\psi_n\times\psi_{p_3})&:&
  ({\bf 3},{\bf 5}+{\bf 1})\otimes(\overline{\bf 10},{\bf 1})\ ,\nonumber
\end{eqnarray}
where $i,j=1,2,3$ are $SU(2)\sim SO(3)$ indices. 
The operators on the right hand side of (\ref{n=0-exact}) should  
respect these structures.

We shall first write down all possible terms on the right hand 
side satisfying several consequences of (\ref{n=0-bootstrap}) after contracting 
all the indices, 
obtaining only a small number of terms. Some useful requirements 
are: (1) $SU(3)$ singlet condition of $O_0$, 
(2) exchange symmetry of the two $SU(3)$ indices carried by the scalars. 
We first consider the term $Q[\psi^6]$. $Q$ acting on any $\psi$ produces a 
term of the form $\phi^m\times\phi^n$, violating  
the condition (2). So there are no terms of the form $Q[\psi^6]$ that we can write down.
Now we try to write down all the gauge-invariant operators at $f\phi\psi^4$ order
which can appear inside $Q_f$ in (\ref{n=0-exact}). 
Since it consists of six letters, we take three pairwise inner products. (Contractions 
by two $\epsilon$ tensors can also be written as three inner products.) The possible terms are
\begin{equation}
  (f\cdot \phi^m)(\psi_{[n_1}\cdot\psi_{n_2]})(\psi_{[p_1}\cdot \psi_{p_2]})\ \ ,\ \ 
  (\phi^m\cdot\psi_{n_1})(f\cdot \psi_{n_2})(\psi_{[p_1}\cdot \psi_{p_2]})\ .
\end{equation}
$Q_f$ transformation of the first term violates the 
condition (2) since
$Q[f\cdot\phi^m]=(\phi^n\times \psi_n)\cdot\phi^m=(\phi^m\times\phi^n)\cdot\psi_n$.
Now imposing the condition (1) on the second term, one should contract the $SU(3)$ 
indices to form singlets. One finds
\begin{equation}
  ({\bf 8}\oplus{\bf 1})\otimes{\bf 3}\otimes\overline{\bf 3}\rightarrow
  {\bf 27}\oplus{\bf 10}\oplus\overline{\bf 10}\oplus{\bf 8}\oplus{\bf 8}
  \oplus{\bf 8}\oplus{\bf 8}\oplus{\bf 1}\oplus{\bf 1}
  \nonumber
\end{equation}
so there are two possible singlets. They are
\begin{equation}\label{n=0-two-terms}
  (\phi^m\cdot\psi_m)\epsilon^{npq}(f\cdot \psi_n)(\psi_p\cdot\psi_q)\ \ ,\ \ 
  (\phi^m\cdot \psi_n)(f\cdot\psi_m)\epsilon^{npq}(\psi_p\cdot \psi_q)\ .
\end{equation}
Acting $Q_f$ on them and separating the $\phi^2$ and $\psi^5$ parts as we did 
in (\ref{n=0-bootstrap}), we obtain a part consistent with (\ref{n=0-bootstrap}) 
and the rest. Focusing on the former part, they are given by
$\phi^{(m}_{(i}\phi^{r)}_{j)}$ times
\begin{equation}
  \psi_{(m}^{(i}(\psi_{r)}\times\psi_{n})^{j)}\epsilon^{npq}(\psi_p\cdot\psi_q)\ \ ,\ \ 
  \psi_{n}^{(i}(\psi_{r}\times\psi_{m})^{j)}\epsilon^{npq}(\psi_p\cdot\psi_q)
\end{equation}
respectively. If $O_0$ is $Q$-exact, a suitable linear combination of these two terms 
should yield $O_0$. The agreement should happen for every coefficient of 
$\phi^{(m}_{(i}\phi^{r)}_{j)}$ separately, demanding
\begin{equation}
  \epsilon^{pqr}\psi_p^{(i}\psi_q^{j)}(\psi_m\cdot\psi_n\times \psi_r)=
  A\psi_{(m}^{(i}(\psi_{n)}\times\psi_{r})^{j)}\epsilon^{pqr}(\psi_p\cdot\psi_q)
  +B\psi_{r}^{(i}(\psi_{n}\times\psi_{m})^{j)}\epsilon^{pqr}(\psi_p\cdot\psi_q)
\end{equation}
for suitable $A,B$. Inserting two different sets of $m, r, i, j$, we found that there are no solutions for $A$ and $B$.
This proves that $O_0$ is not $Q$-exact.

One can also easily show the non-$Q$-exactness 
by studying the $SU(1|3)$ descendants obtained by acting $Q_+^a Q_+^b$. 
For instance, one obtains
\begin{eqnarray}
  &&Q_+^2Q_+^1 O_0^\prime=\\
  &&-(Y\cdot f+\psi_3\cdot\psi_1)^2\psi_3\cdot(\psi_2\times\psi_2)
 -(X\cdot f+\psi_2\cdot\psi_3)(Z\cdot f+\psi_1 \cdot \psi_2)\psi_1\cdot(\psi_3\times\psi_3)\nonumber\\
 &&-(X\cdot f+\psi_2\cdot\psi_3)(X\cdot \psi_3)f \cdot(\psi_1\times\psi_1)
 +2(Y\cdot\psi_2-Z\cdot\psi_3)(Y\cdot f+\psi_3\cdot\psi_1)\psi_3\cdot(\psi_2\times f)
 \nonumber\\
  &&-2(Y\cdot f+\psi_3\cdot \psi_1)(X\cdot \psi_3)\psi_2\cdot(\psi_1\times f)
-(Z\cdot\psi_3-X\cdot\psi_1)(Z\cdot f+\psi_1\cdot \psi_2)f\cdot(\psi_3\times\psi_3)
\nonumber
\end{eqnarray}
which contains uncanceled $\phi^0\psi^7$ terms on the second line.
Since acting $Q$ always creates one or more $\phi$ factors, 
these terms cannot be $Q$-exact. Since a descendant of $O_0^\prime$
is not $Q$-exact, $O_0$ cannot be $Q$-exact either, providing a simpler proof. 
Or alternatively, one can prove non-$Q$-exactness by acting three $Q_+$'s to $O_0^\prime$ 
and check that it contains nonzero term at $f\psi^6$ order,
\begin{eqnarray}\label{QQQ-on-O0}
  &&Q^1_+ Q^2_+ Q^3_+ O_{0}^\prime=Q^1_+Q^2_+Q^3_+O_0^{\prime\prime}=\\
  &&(X\cdot f+\psi_2\cdot\psi_3)^2 f \cdot(\psi_1\times\psi_1)
  +2(X\cdot f+\psi_2\cdot\psi_3)(Y\cdot f+\psi_3\cdot\psi_1) f\cdot(\psi_1\times\psi_2) 
  \nonumber\\
  &&+(1,2,3\rightarrow 2,3,1)+(1,2,3\rightarrow 3,1,2)=
  G^mG^n f\cdot(\psi_m\times\psi_n)\nonumber
\end{eqnarray}
where $G^m\equiv\phi^m\cdot f+\frac{1}{2}\epsilon^{mnp}\psi_n\cdot\psi_p$.
Proof of this sort will sometimes be useful later. For instance, one can show that 
$(Z\cdot f+\psi_1\cdot\psi_2)O_0^\prime$ is not $Q$-exact, since 
its descendant
\begin{equation}\label{QQ-on-t36}
  Q^2_+Q^1_+\left[(Z\cdot f+\psi_1\cdot\psi_2)O_0^\prime\right]
  =(Z\cdot f+\psi_1\cdot\psi_2)Q^2_+Q^1_+O_0^\prime
\end{equation}
contains a term at $\phi^0\psi^9$ order.

\hspace*{-0.65cm}\underline{\bf $O_1$ operator at $t^{36}$}

\hspace*{-0.65cm}Now we construct the cohomology
which accounts for the $-t^{36}$ term of (\ref{index-core-primary}).
It should be fermionic, has charge $\cJ=2(Q_1+Q_2+Q_3)+6J=36$,
and should be an $SU(2)_R\times SU(3)$ singlet because we expect unique cohomology
(unless there is a cancellation at this order which obscures the true degeneracy). 
We call this operator $O_1$.
From the last condition, we set three $Q_I$ equal and two $J_i$ equal.
Still, we do not know the individual $Q$ and $J$ so we should make a guess. 
Our first guess was to add extra $\Delta J=2$ to the charges $Q=\frac{3}{2}$, $J=\frac{5}{2}$ 
of $O_0$. We listed all operators in this sector and found the cohomology 
by computer.
Then we made several trials until we found the following $SU(3)$-invariant 
representative:
\begin{eqnarray}
  O_1&=&(f\cdot f) \epsilon^{c_1c_2c_3}(\phi^a \cdot\psi_{c_1})(\phi^b \cdot \psi_{c_2})(\psi_a\cdot \psi_b\times\psi_{c_3})\\
  &&+\epsilon^{b_1b_2b_3}\epsilon^{c_1c_2c_3}(f\cdot \psi_{b_1})(\phi^a\cdot \psi_{c_1})(\psi_{b_2}\cdot \psi_{c_2})
  (\psi_a\cdot \psi_{b_3}\times \psi_{c_3})\nonumber\\
  &&-{\textstyle \frac{1}{72}}\epsilon^{a_1a_2a_3}\epsilon^{b_1b_2b_3}\epsilon^{c_1c_2c_3}
  (\psi_{a_1}\cdot \psi_{b_1}\times \psi_{c_1})(\psi_{a_2}\cdot \psi_{b_2}\times \psi_{c_2})
  (\psi_{a_3}\cdot \psi_{b_3}\times \psi_{c_3})\ .\nonumber
\end{eqnarray}
It is not graviton type since it is made of 
nine (odd) letters. One can also easily check that it is not $Q$-exact. 
This is because the last term contains no scalars. 
Since $Q$ transformations (\ref{Q-bmn-SU(2)}) always yield
scalars, the last term cannot be made $Q$-exact. So $O_1$ is not $Q$-exact. 

Now we discuss the $Q$-closedness. $O_1$ takes the form of
\begin{equation}
  O_1=(f\cdot f)O_0+f\cdot \xi+\chi\ ,
\end{equation}
where the $SU(2)$ triplet $\vec{\xi}$ and the singlet $\chi$ are given by
\begin{eqnarray}
  \vec{\xi}&=&\epsilon^{b_1b_2b_3}\epsilon^{c_1c_2c_3}\vec{\psi}_{b_1}
  (\phi^a\cdot \psi_{c_1})(\psi_{b_2}\cdot \psi_{c_2})
  (\psi_a\cdot \psi_{b_3}\times\psi_{c_3})\nonumber\\
  \chi&=&-{\textstyle \frac{1}{72}}\epsilon^{a_1a_2a_3}\epsilon^{b_1b_2b_3}\epsilon^{c_1c_2c_3}
  (\psi_{a_1}\cdot \psi_{b_1}\times \psi_{c_1})(\psi_{a_2}\cdot \psi_{b_2}\times \psi_{c_2})
  (\psi_{a_3}\cdot \psi_{b_3}\times \psi_{c_3})\nonumber\\
  &=&-120\psi_1^1\psi_1^2\psi_1^3\psi_2^1\psi_2^2\psi_2^3\psi_3^1\psi_3^2\psi_3^3\ .
\end{eqnarray}
$Q$-closedness is equivalent to the following equations:
\begin{equation}\label{n=1-closed-summary}
  2(\vec{\phi}^m\times\vec{\psi}_m)O_0+Q_\psi\vec{\xi}=0\ \ ,\ \ \vec{\phi}^m\cdot(\vec{\psi_m}\times\vec{\xi})
  +Q_\psi \chi=0\ .
\end{equation}
 Note that $\vec\xi$ is related to 
$O_0$ by
\begin{equation}
  \vec{\xi}=-{\textstyle \frac{1}{2}}\epsilon^{mnp}\vec{\psi}_m
  \psi_n\cdot{\textstyle \frac{\partial}{\partial\phi^p}}O_0\ .
\end{equation}
So the first equation can be written as the following 
equations of $O_0$:
\bea\label{O1-O0}
  && 4(\vec\phi^m\times\vec\psi_m)O_0 \\
  &=& \left[(\vec\phi^a\times\vec\phi^b)
  \psi_a\cdot{\textstyle \frac{\partial}{\partial\phi^b}}+\vec\psi_a
 (\phi^a\times\phi^b)\cdot{\textstyle \frac{\partial}{\partial\phi^b}}
 -\vec\psi_{a}(\psi_{b}\times\phi^a)\cdot
 {\textstyle \frac{\partial}{\partial\psi_b}}+
 \vec\psi_{b}(\psi_{a}\times\phi^a)\cdot
 {\textstyle \frac{\partial}{\partial\psi_b}}\right]O_0\ . \nn
\eea
This is a property of $O_0$. 
The second/third terms cancel due to 
$\left(\phi^b\times\frac{\partial}{\partial\phi^b}+
\psi_b\times\frac{\partial}{\partial\psi_b}\right)O_0=0$, which holds 
because it is the $SU(2)$ gauge transformation on a gauge invariant operator $O_0$. 
One can further simplify (\ref{O1-O0}) using various properties of $O_0$. Obvious ones are
\begin{eqnarray}
  &&{\textstyle \phi^m\cdot\frac{\partial}{\partial\phi^m}O_0=n_{\rm B}O_0\ \ ,\ \ 
  \psi_m\cdot\frac{\partial}{\partial\psi_m}O_0=n_{\rm F}O_0}\ \ \ 
  (n_{\rm B},n_{\rm F})=(2,5)\nonumber\\
  &&\vec{\varepsilon}_a^{\ b}\left[{\textstyle \phi^a\cdot\frac{\partial}{\partial\phi^b}
  -\psi_b\cdot\frac{\partial}{\partial\psi_a}}\right]O_0=0\ \ \ (\vec\varepsilon_a^{\ a}=0)\ .
\end{eqnarray}
The first two equations count the numbers of bosonic/fermionic fields in $O_0$. 
The last equation is the $SU(3)$ invariance of $O_0$, which holds for any 
$\varepsilon$. Equivalently, one obtains
\begin{equation}
  {\textstyle \left[\phi^a\cdot\frac{\partial}{\partial\phi^b}
  -\psi_b\cdot\frac{\partial}{\partial\psi_a}\right]O_0=\frac{1}{3}
  (n_{\rm B}-n_{\rm F})}\delta^a_bO_0\ .
\end{equation}
Finally, note that $\delta_{ij}$ contracts the $SU(2)$ gauge triplet indices 
only between boson-fermion pairs in $O_0$, while fermion indices are 
contracted only with $\epsilon_{ijk}$. This effectively promotes $SU(2)\sim SO(3)$ to 
$SL(3)$ within $O_0$, where bosons/fermions transform in the fundamental and 
anti-fundamental representations, respectively. This leads to the following property:
\begin{equation}
  {\textstyle \left[\phi^a_i\cdot\frac{\partial}{\partial\phi^a_j}
  -\psi_a^j\cdot\frac{\partial}{\partial\psi_a^i}\right]O_0=\frac{1}{3}
  (n_{\rm B}-n_{\rm F})}\delta^j_iO_0\ .
\end{equation}
Using these properties, (\ref{O1-O0}) can be written as
\begin{equation}\label{O1-O0-simpler}
  (\vec{\psi}_a\times\vec{\phi}^b)(\phi^a\cdot{\textstyle \frac{\partial}{\partial\phi^b}})
  O_0=({\textstyle 4-\frac{n_{\rm F}+2n_{\rm B}}{3}})(\vec{\phi}^m\times\vec{\psi}_m)O_0=
  (\vec{\phi}^m\times\vec{\psi}_m)O_0\ .
\end{equation}

Both (\ref{O1-O0-simpler}) and the second equation of (\ref{n=1-closed-summary}) can be 
easily checked on a computer. We have no extra analytic insights on why (\ref{O1-O0-simpler}) 
this holds, except that using complicated representation analysis of $SU(2)\times SU(3)$ should 
provide the analytic proof. (We tried to simplify the equation for $O_0$ as much as possible
since they might provide insights on the generalization 
to higher $N$'s in the future.) On the other hand, one can easily prove the second equation of 
(\ref{n=1-closed-summary}). First note that 
$\psi_m\times\xi$ is an $SU(2)$ vector involving $8$ $\psi$'s. 
There are nine independent operators involving eight $\psi$'s, depending on 
which of the $9$ components is lacking. So it is proportional to
$\frac{\partial}{\partial\psi_m^i}\chi$. Since it has to form a gauge-invariant 
by contracting with two scalars $\phi^m_i$, $\phi^a_j$, one should be able to write
$\frac{\delta}{\delta\psi_m^i}\chi$ as
an object with two $SU(3)$ antifundamental and two $SU(2)$ triplet indices by multiplying 
invariant tensors. The only possible term is $\epsilon_{man}\epsilon_{ijk}\frac{\partial}{\partial\psi_n^k}\chi$. One can compute the proportionality constant by computing a 
term, e.g. at $m=1,a=2,i=1,j=2$, finding $-\frac{1}{2}$. So one obtains
\begin{equation}
  \phi^m\cdot(\psi_m\times\xi)=-{\textstyle \frac{1}{2}\phi^{m}_i\phi^a_j
  \epsilon_{man}\epsilon_{ijk}\frac{\partial\chi}{\partial\psi_n^k}}=
  -{\textstyle \frac{1}{2}\epsilon_{man}(\phi^m\times\phi^a)\cdot 
  \frac{\partial\chi}{\partial\psi_n}}
  =-Q_\psi\chi\ ,
\end{equation}
proving the second equation of (\ref{n=1-closed-summary}).

One may wonder if $O_1$ is a descendant of $O_0$, or a lower black hole operator times 
graviton operators appearing in (\ref{bh-index-bmn}). Since $O_{1}$ is at $t^{36}$ order, 
the only possible way of getting operators at this order from $O_0$ is 
$(Q^m_+Q^n_+O_0)(\phi^p \cdot f+\frac{1}{2}\epsilon^{pqr}\psi_q\cdot\psi_r)$. 
However, during our numerical construction of the cohomologies at this order, we separately 
constructed the last operator which is not cohomologous to $O_1$. See also the end of 
this subsection for an analytic proof (applicable to all $O_n$'s with $n\geq 1$).

\hspace*{-0.65cm}\underline{\bf $O_n$ operator at $t^{24+12n}$ ($n\geq 2$)}

\hspace*{-0.65cm}
We can use the structures of the operators $O_0$ and $O_1$ to analytically construct an infinite tower 
of cohomologies $O_n$ accounting for (\ref{index-core-primary}). Consider 
\begin{equation}\label{On}
  O_n\equiv (f\cdot f)^nO_0+n(f\cdot f)^{n-1}f\cdot \xi
  +{\textstyle \frac{2n^2+n}{3}}(f\cdot f)^{n-1}\chi
\end{equation}
for $n\geq 2$. At $n=1$, this is just $O_1$ that we discussed above. 
We will now show that these are new black hole like cohomologies at 
$t^{24+12n}$ order. It is again easy to show that these are not graviton type because
they are made of odd letters. It is not $Q$-exact because the last term does not 
contain scalars.

Now we derive the $Q$-closedness. Its $Q$-action is given by 
\begin{eqnarray}
  QO_n\!&\!=\!&\!(f\cdot f)^{n-1}\left[\vec{f}\cdot\left(2n(\vec{\phi}^m\times\vec{\psi}_m)O_0
  +nQ_\psi\vec{\xi}\right)+n(\vec{\phi}^m\times\vec{\psi}_m)\cdot\vec{\xi}
  +{\textstyle \frac{2n^2+n}{3}}Q_\psi \chi\right]\\
  &&\!+2(n^2-n)(f\cdot f)^{n-2}\vec{f}\cdot(\vec{\phi}^m\times\vec{\psi}_m)(f\cdot \xi)
  +{\textstyle \frac{2n(n-1)(2n+1)}{3}}(f\cdot f)^{n-2}\vec{f}\cdot(\vec{\phi}^m\times\vec{\psi}_m)\chi\ .
  \nonumber
\end{eqnarray}
The first two terms on the first line cancel due to the first equation of 
(\ref{n=1-closed-summary}). The last term on the second line is zero because it 
includes $10$ fermions.
Inserting the second equation of (\ref{n=1-closed-summary}) to the last term on 
the first line, one obtains
\begin{equation}\label{Q-On}
  QO_n={\textstyle \frac{2(n^2-n)}{3}}(f\cdot f)^{n-2}
  \left[-(f\cdot f)(\phi^m\times\psi_m)\cdot\xi
  +3(f\times\phi^m)\cdot\psi_m(f\cdot \xi)\right]\ .
\end{equation}
The second term contains $8$ fermions, where the fermions carry $ma$ indices for $SU(3)$ 
and three $SU(2)$ triplet indices to be contracted with 
$(f\times\phi^m)_k$, $f_i$, $\phi^a_j$. From the contraction structures 
of $\xi$, one finds that $b_1,c_1$ are antisymmetric so the 
corresponding $i,j$ indices should be symmetric. The only possible $8$-fermion 
terms satisfying these conditions are
\begin{equation}\label{Q-On-2nd}
  \epsilon_{man}\delta_{ij}\frac{\partial}{\partial \psi_n^k}\chi\ \ ,\ \ 
  \epsilon_{man}\delta_{k(i}\frac{\partial}{\partial \psi_n^{j)}}\chi\ .
\end{equation}
Explicitly computing two components in the second term of (\ref{Q-On}), 
one finds that the linear combination is
\begin{equation}
  \epsilon_{man}\left[\delta_{ij}{\textstyle \frac{\partial}{\partial \psi_n^k}}
  -{\textstyle \delta_{k(i} \frac{\partial}{\partial \psi_n^{j)}} }\right]\chi\ .
\end{equation}
Contracting this with $f_i$, $\phi^a_j$, $(f\times\phi^m)_k$, one obtains 
\begin{eqnarray} 
  &&\epsilon_{man}\left[(f\cdot\phi^a)(f\times\phi^m)\cdot
  {\textstyle \frac{\partial}{\partial \psi_n}}
  -{\textstyle \frac{1}{2}}[(f\times\phi^m)\cdot\phi^a]
  f\cdot{\textstyle \frac{\partial}{\partial \psi_n}}\right]\chi\\
  &&={\textstyle \frac{1}{2}}\epsilon_{man}\left[
  \left[f\times(f\times(\phi^m\times\phi^a))\right]\cdot
  {\textstyle \frac{\partial}{\partial \psi_n}}
  -[(f\times\phi^m)\cdot\phi^a]
  f\cdot{\textstyle \frac{\partial}{\partial \psi_n}}\right]\chi\nonumber\\
  &&=-{\textstyle \frac{1}{2}\epsilon_{man}(f\cdot f)(\phi^m\times\phi^a)\cdot 
  {\textstyle \frac{\partial}{\partial \psi_n}}}\chi=-(f\cdot f)Q_\psi\chi
  =(f\cdot f)(\phi^m\times\psi_m)\cdot\xi\ .\nonumber
\end{eqnarray}
So the second term of (\ref{Q-On}) cancels the first term, ensuring that $O_n$ is 
$Q$-closed.  So we have shown that the operator
\begin{eqnarray}\label{On-summary}
  O_n&=&(f\cdot f)^n \epsilon^{c_1c_2c_3}(\phi^a \cdot\psi_{c_1})(\phi^b \cdot \psi_{c_2})(\psi_a\cdot \psi_b\times\psi_{c_3})\\
  &&+n(f\cdot f)^{n-1}\epsilon^{b_1b_2b_3}\epsilon^{c_1c_2c_3}(f\cdot \psi_{b_1})(\phi^a\cdot \psi_{c_1})(\psi_{b_2}\cdot \psi_{c_2})
  (\psi_a\cdot \psi_{b_3}\times \psi_{c_3})\nonumber\\
  &&-{(\textstyle\frac{n}{72}+\frac{n^2-n}{108})}
  (f\cdot f)^{n-1}\epsilon^{a_1a_2a_3}\epsilon^{b_1b_2b_3}\epsilon^{c_1c_2c_3} \nn\\
  && \hspace{4cm} \cdot
  (\psi_{a_1}\cdot \psi_{b_1}\times \psi_{c_1})(\psi_{a_2}\cdot \psi_{b_2}\times \psi_{c_2})
  (\psi_{a_3}\cdot \psi_{b_3}\times \psi_{c_3})\nonumber
\end{eqnarray}
at $t^{24+12n}$ order is a black hole cohomology.

One may wonder if these are primaries captured in the first factor of (\ref{bh-index-bmn}), 
or if they are related to other $O_{n^\prime}$ with $n^\prime <n$ by acting some 
$Q^m_+$'s and/or gravitons on the third factor. One can show that the latter possibilities 
are all impossible. Suppose $O_n$ is obtained by acting acting $p$ $Q$'s on $O_{n^\prime}$ 
and multiplying $q$ gravitons. Then $p,q$ should satisfy
\begin{equation}
  2p+8q=12(n-n^\prime)\ ,\ \ p=0,1,2,3\ ,\ q\geq 0\ .
\end{equation}
Possible solutions are 
\begin{equation}
  (p,q,n-n^\prime)=(2,1,1)\ ,\ (0,3,2)\ ,\ (2,4,3)\ ,\ (0,6,4)\ ,\ 
  (2,7,5)\ ,\ (0,9,6)\ ,\ \cdots\ .
\end{equation}
The cases with even $n-n^\prime$ and $p=0$ yield operators 
at $t^{24+12n}$ order obtained by multiplying $O_{n^\prime}$ and 
$\frac{3}{2}(n-n^\prime)$ graviton operators of the form 
$\phi^m\cdot f+\frac{1}{2}\epsilon^{mnp}\psi_n\psi_p$. However, these cannot 
be cohomologous to $O_n$ because they do not have a term at 
$\mathcal{O}(f^{2n-2}\phi^0\psi^9)$ order that $O_n$ has, 
which cannot be changed by adding $Q$-exact terms. Now we consider the cases with 
odd $n-n^\prime$ and $p=2$, $q=\frac{3}{2}(n-n^\prime)-\frac{1}{2}$, and again 
consider whether the operator 
$(Q^a_+Q^b_+O_{n^\prime})(\phi\cdot f+\psi\cdot\psi)^q$ has a term 
at $f^{2n-2}\phi^0\psi^9$ order. Let us 
first study how the actions of $Q^a_+$ and $Q^b_+$ on $O_{n^\prime}$ can produce 
a term with no scalars. $Q^a_+$ either act as $\phi\rightarrow\psi$ or 
$\psi\rightarrow f$, so there are following possibilities:
\begin{equation}
  f^{2n^\prime}\phi^2\psi^5\rightarrow f^{2n^\prime}\psi^7\ ,\ 
  f^{2n^\prime-1}\phi\psi^7\rightarrow f^{2n^\prime}\psi^7\ ,\ \ 
  f^{2n^\prime-2}\psi^9\rightarrow f^{2n^\prime}\psi^7\ .
\end{equation}
In all three cases, we multiply gravitons of the form $(\phi \cdot f+\psi\cdot \psi)^q$ 
and see whether there can be a term at $f^{2n-2}\phi^0 \psi^9$ order.
This is possible only if $n=n^\prime+1$, $p=2$, $q=1$. That is, 
the only possible relations between different $O_n$'s are 
\begin{equation}\label{On-hairy?}
  O_n\stackrel{?}{\sim} \epsilon_{abc}(Q^a_+Q^b_+ O_{n-1})(\phi^c\cdot f+{\textstyle \frac{1}{2}}
  \epsilon^{cde}\psi_d\cdot\psi_e)\ ,
\end{equation}
where $\sim$ means up to a multiplicative factor and addition of $Q$-exact terms. 
We act three $Q^a_+$'s on (\ref{On-hairy?}) 
and show that this equation cannot hold. Acting $Q^1_+Q^2_+Q^3_+$ on the right hand side 
yields zero, so if this equation is true,  
$Q^1_+Q^2_+Q^3_+O_n$ should be $Q$-exact. However, this cannot be the case 
since it contains a term at $f^{2n+1}\psi^6$ order, which does not contain 
scalars so cannot be $Q$-exact. More concretely, one starts from
\begin{eqnarray}
O_n&=&(f\cdot f)^n O_{0}+\frac{20n}{3}(f\cdot f)^{n-1}\sum_{\textrm{cyclic}} 
(f\cdot \psi_3)(\psi_3\cdot \psi_2)(X\cdot\psi_2)(\psi_1\cdot \psi_1\times \psi_1)\nonumber\\
&&-\frac{10}{3}(\frac{n}{6}+\frac{n^2-n}{9})(f\cdot f)^{n-1}(\psi_1\cdot\psi_1\times\psi_1)(\psi_2\cdot\psi_2\times\psi_2)(\psi_3\cdot\psi_3\times\psi_3) 
\end{eqnarray}
where $\sum_{\textrm{cyclic}}$ means summation over the cyclic permutations 
of $(X,\psi_1)$, $(Y,\psi_2)$, $(Z,\psi_3)$. Acting $Q^1_+Q^2_+Q^3_+$, 
one obtains the following terms without scalars,
\begin{eqnarray}\label{QQQOn}
  && \hspace{-1.4cm} Q^+_1 Q^+_2 Q^+_3 O_{0} \\
  &=& -10(X\cdot f+\psi_2\cdot\psi_3)^2 f \cdot(\psi_1\times\psi_1) \nn\\
  && -20(X\cdot f+\psi_2\cdot\psi_3)(Y\cdot f+\psi_3\cdot\psi_1) f\cdot(\psi_1\times\psi_2) +\textrm{cyclic,}\nn\\
  &\rightarrow&  -20(\psi_2\cdot \psi_3)^2 (f\cdot \psi_1\times \psi_1) + 
  \textrm{cyclic}~, \nn\\
  && \hspace{-1.4cm} Q^+_1 Q^+_2 Q^+_3 (f\cdot \psi_3)(\psi_3\cdot \psi_2)(X\cdot\psi_2)(\psi_1\cdot \psi_1\times \psi_1)+ \textrm{cyclic}\nonumber\\
  &\rightarrow& -3(f\cdot f)(\psi_2\cdot \psi_3)^2 (f\cdot \psi_1\times \psi_1)+6(f\cdot \psi_2)(f\cdot \psi_3)(\psi_2\cdot \psi_3)(f\cdot \psi_1\times \psi_1) + 
  \textrm{cyclic,}\nonumber\\
  && \hspace{-1.4cm} Q^+_1 Q^+_2 Q^+_3 (\psi_1\cdot\psi_1\times\psi_1)(\psi_2\cdot\psi_2\times\psi_2)(\psi_3\cdot\psi_3\times\psi_3) \nn\\
  &=& -27 (f \cdot\psi_1\times\psi_1)(f \cdot\psi_2\times\psi_2)(f\cdot\psi_3\times\psi_3)\nonumber\\
  &=&18((f\cdot f)(\psi_2\cdot\psi_3)^2(f\cdot\psi_1\times \psi_1))-2(f\cdot \psi_2)(f\cdot \psi_3)(\psi_2\cdot \psi_3)(f\cdot \psi_1\times \psi_1) + \textrm{cyclic})\ . \nn
\end{eqnarray}
These terms at $f^{2n+1}\phi^0\psi^6$ order do not cancel, implying that 
$Q^1_+Q^2_+Q^3_+O_0$ cannot be $Q$-exact. 
So at least among the possibilities visible in the 
index (\ref{bh-index-bmn}), we have checked that different $O_n$'s are not related 
in trivial manners. 

Note also that the product of two $O_n$'s vanishes, $O_m O_n=0$. 
This is because each operator includes $5$ or more $\psi$'s, 
so the product involves $10$ or more $\psi$'s which vanishes by Fermi 
statistics.  

\section{$SU(2)$ and Partial No-Hair Behavior}\label{sec:cohosu2}

The non-graviton index (\ref{su2-general-bh}) for the full $SU(2)$ theory
shows that there are many more black hole cohomologies
outside of the BMN sector.
We are interested in the BPS cohomologies contained 
in the square bracket of (\ref{su2-general-bh}),
because we believe that the $\chi_D$ factor addresses the $PSU(1,2|3)$
descendants of those in the square bracket which are not really new.
In this section, we shall see that many of the cohomologies in the square bracket
are actually the threshold cohomology at $\cJ=24$
multiplied by some graviton operators.
Furthermore, we show that multiplication by many other graviton operators lead to
$Q$-exact operators and therefore do not create a new cohomology.
We refer to this phenomenon as the partial no-hair behavior.

In principle, constructing all the cohomologies order by order as done in
\cite{Chang:2022mjp} will confirm that the $\chi_D$ factor indeed stands
for the $PSU(1,2|3)$ descendants,
but we shall not comprehensively do this job in this thesis.
Rather, we shall proceed by considering possible superconformal representation
structures of the $\frac{1}{16}$-BPS states compatible with this index,
finding many illuminating structures. 
As emphasized, we may miss some BPS states in case their
multiplets completely cancel in the index. 

The index can be written as a sum over the short $\mathcal{N}=4$
representations. Equivalently, it can be written as a sum over 
$\frac{1}{16}$-BPS multiplets of $PSU(1,2|3)\subset PSU(2,2|4)$. 
The last multiplets are embedded in the 
short representations of $PSU(2,2|4)$ in canonical manners:
see appendix B of \cite{Choi:2023znd}.
Knowing this representation sum is equivalent to knowing the primary contents.
We will study this expansion order by order in $t$.  
As already mentioned in section \ref{sec:cohosu2bmn}, there are two classes 
of black hole cohomologies: those which can be written as products of other black hole 
cohomologies and gravitons which we call `hairy' and 
the rest which we call `core.' 

We start by studying the black hole cohomologies in the BMN sector that we identified in 
section \ref{sec:cohosu2bmn}.
Among these, two of them $O_0$, $O_1$ appear within the $t^{40}$ order.
We can show that all $O_n$'s are core black hole primaries of 
the $\frac{1}{16}$-BPS multiplets.
The coreness of $O_n$ is already shown in section \ref{sec:cohosu2bmn},
at least within the states visible in the index (\ref{bh-index-bmn}),
since it suffices to show this within the BMN sector.
We only need to show that they are 
$\frac{1}{16}$-BPS primaries in their full $PSU(1,2|3)$ representations.
$O_0$ is clearly a $\frac{1}{16}$-BPS primary 
since it is the lowest black hole cohomology.
Since $j \equiv J_1+J_2=5$ is too large,
$O_0$ can only belong to the $\mathcal{N}=4$ multiplet
$A_1\overline{L}[4;0]^{[2,0,0]}_{9}$.
The primary $O_0$ of the $\frac{1}{16}$-BPS multiplet is obtained by acting
$Q^\prime\equiv Q^4_+$ on a primary of this $\mathcal{N}=4$ multiplet.
The index over this multiplet is
\begin{equation}
  \chi_{24}\equiv-t^{24}\chi_D(t,x,y,p)\ ,
\end{equation}
where $\chi_D$ is defined in (\ref{characters-t40}).
So the first term $-t^{24}$ in the square bracket of (\ref{su2-general-bh}) 
corresponds to the contribution of this multiplet. 

Next we consider other $O_n$'s.
We can prove that they are also primaries by showing that acting any of
the nine $Q$'s in $PSU(1,2|3)$ yields nontrivial and independent cohomologies.
(This is because $O_n$ does not contain derivatives and cannot be 
a conformal descendant.)
We have shown in section \ref{sec:cohosu2bmn} that the action of any 
$Q^m_+$ on $O_n$ is nontrivial and independent because acting all
three of them yields a nontrivial cohomology.
One can also show that $\overline{Q}_{m\dot\alpha}O_n$ are all nontrivial and independent. 
It suffices to show that the six $\overline{Q}_{m\dot\alpha}$'s 
acting on $Q^1_+Q^2_+Q^3_+O_n$ are independent. 
This is easily shown by studying the terms obtained 
by acting $\overline{Q}_{m\dot\alpha}$ on the $\mathcal{O}(f^{2n+1}\phi^0\psi^6)$ 
order terms of $Q^1_+Q^2_+Q^3_+O_n$ in (\ref{QQQOn}). 
In particular, one obtains terms at $f^{2n}\phi^0\psi^6D\psi$ by acting 
$\overline{Q}_{m\dot\alpha}$ on $f$.
These terms cannot be $Q$-exact since it involves neither
$\phi^m$ or $\lambda_{\dot\alpha}$. 
This proves that all $6$ operators
$\overline{Q}_{m\dot\alpha}Q^1_+Q^2_+Q^3_+O_n$ are nontrivial.
They are also independent since their $SU(2)_R\times SU(3)$
quantum numbers are different.
This shows that $O_{n\geq 1}$ are $\frac{1}{16}$-BPS primaries. 
$O_n$ belongs to the $\mathcal{N}=4$ multiplet 
$A_1\overline{L}[4+4n;0]^{[2,0,0]}_{9+4n}$, 
which contributes to to the index as $-t^{24+12n}\chi_D(t,x,y,p)$.

Now with the nature of $O_n$ understood,
we come back to study the series (\ref{su2-general-bh}) until $t^{40}$ order,
trying to better characterize other cohomologies order by order in $t$.
Once the lowest operator $O_0$ is identified, 
all the states in its $\frac{1}{16}$-BPS multiplet are not really new operators. 
So we subtract $\chi_{24}$ from $Z-Z_{\rm grav}$ and see what are left: 
\begin{eqnarray}\label{grav-thres-deficit}
  Z-Z_{\rm grav}-\chi_{24}&=&\left[-\chi_{(1,3)}t^{32}
  -(\chi_{(1,\bar{3})}+\chi_{(3,6)})t^{34}-\chi_{(2,3)}t^{35}
  +(\chi_{(3,1)}+\chi_{(3,8)})t^{36}\right.\nonumber\\
  &&-(\chi_{(2,\bar{3})}+\chi_{(4,6)})t^{37}+\chi_{(5,3)}t^{38}
  +(\chi_{(2,1)}+2\chi_{(4,1)}+\chi_{(4,8)})t^{39}\nonumber\\
  &&\left.-(2\chi_{(1,6)}+\chi_{(3,\bar{3})}+\chi_{(5,\bar{3})}
  +\chi_{(5,6)})t^{40}\right]\chi_D+\mathcal{O}(t^{41})\ .
\end{eqnarray}
Somewhat surprisingly, after subtracting the multiplet of $O_0$, one finds that 
the remaining index starts from $t^{32}$ order.
Namely, in the range $t^{25}\sim t^{31}$, the index does not capture any
new black hole cohomologies except the trivial descendants of $O_0$.
At first sight this may look like a boring result, but the triviality of the 
index in this range has a nontrivial implication.

Recall that cohomologies multiply to yield new cohomologies.
This is because of the Leibniz rule of the classical $Q$ acting on product operators.
So apparently, one can multiply light graviton cohomologies 
to $O_0$ or its descendants to obtain many new cohomologies
in the range $t^{25}\sim t^{31}$.
The possible product cohomologies of $O_0$ and gravitons below
$t^{32}$ order are
\begin{eqnarray}\label{no-hair-till-31}
  &&O_0 \cdot (\phi^{(m}\cdot\phi^{n)})\ ,\ \ 
  O_0 \cdot (\phi^m\cdot\lambda_{\dot\alpha})\ ,\ \ 
  O_0 \cdot (\lambda_{\dot{+}}\cdot\lambda_{\dot{-}})\ ,\nonumber\\
  &&O_0 \cdot (\phi^m\cdot\psi_{n}-{\textstyle \frac{1}{3}}\delta^m_n
  \phi^p\cdot\psi_{p})\ ,\ \ 
  O_0 \cdot (\lambda_{\dot\alpha}\cdot\psi_{m}
  -{\textstyle \frac{1}{2}}\epsilon_{mnp}\phi^n\cdot D_{\dot\alpha}\phi^p)\ ,\nonumber\\
  &&O_0 \cdot \partial_{\dot{\alpha}}(\phi^{(m}\cdot\phi^{n)})\ .
\end{eqnarray}
Other possible products below $t^{32}$ 
involving the descendants of $O_0$ are  
\begin{eqnarray}\label{no-hair-2}
  \overline{Q}O_0&\times&
  (\phi^m\cdot \phi^n\ ,\ 
  \phi^m\cdot\lambda_{\dot\alpha}\ ,\ 
  \lambda_{\dot{+}}\cdot\lambda_{\dot{-}}\ ,\ 
  \phi^m\cdot\psi_{n}-{\textstyle \frac{1}{3}}\delta^m_n
  \phi^p\cdot\psi_{p})\ ,
  \nonumber\\
  (Q,\overline{Q}\overline{Q})O_0&\times&
  (\phi^m\cdot \phi^n\ ,\ 
  \phi^m\cdot\lambda_{\dot\alpha})\ ,
  \nonumber\\
  (Q\overline{Q},\overline{Q}\overline{Q}\overline{Q},\partial)O_0&\times&
  (\phi^m\cdot\phi^n)\ .
\end{eqnarray}
The triviality of the index (\ref{grav-thres-deficit}) in 
this range implies two possibilities for these product cohomologies. 
The first possibility is that these product cohomologies are $Q$-exact, 
i.e. absent in the BPS spectrum. 
Another possibility is that these product cohomologies are nontrivial but there 
are cancellations in the index, either among themselves or with new 
core black hole cohomologies.\footnote{We have checked that cancellations 
cannot happen within the product cohomologies listed above.  
It is logically possible (although a bit unnatural) that some new core black
hole primaries appear in this range, precisely canceling with some of the product operators
above if they are not $Q$-exact. Although in different contexts, certain black holes 
are known not to appear in the index. For instance, asymptotically flat 
multi-center BPS black holes or BPS black rings are not captured by the index 
\cite{Dabholkar:2009dq}.}
Among (\ref{no-hair-till-31}) and (\ref{no-hair-2}), we explicitly show that
\begin{equation}\label{Qextshown}
  O_0 \cdot (\phi^m\cdot \phi^n)\ ,\ O_0 \cdot (\phi^m\cdot\lambda_{\dot\alpha})\ ,\ 
 O_0\cdot (\phi^m\cdot \psi_{n}-{\textstyle \frac{1}{3}}\delta^m_n\phi^p\cdot\psi_{p})
\end{equation}
are all $Q$-exact.

Six operators $O_0(\phi^{(m}\cdot \phi^{n)})$ at $t^{28}$ order are 
all $Q$-exact. An $SU(3)$ covariant expression is 
\begin{equation}
    \begin{aligned}
  O_0 \cdot (\phi^{(m}\cdot \phi^{n)}) = -\frac{1}{14} Q [&20 \epsilon^{rs(m} (\phi^{n)} \cdot \psi_{p}) (\phi^p \cdot \psi_{r}) (\phi^q \cdot \psi_{q})(f\cdot \psi_{s})   \\
  -& 20\epsilon^{prs} (\phi^{(m} \cdot \psi_{p}) (\phi^{n)} \cdot \psi_{r}) (\phi^q \cdot \psi_{q}) (f\cdot \psi_{s}) \\
  +&30 \epsilon^{prs} (\phi^{(m} \cdot \psi_{p}) (\phi^{n)} \cdot \psi_{r})(\phi^{q} \cdot \psi_{s}) (f\cdot \psi_{q})  \\
  -& 7 \epsilon^{a_1 a_2p} \epsilon^{b_1 b_2(m}  (\phi^{n)} \cdot \psi_{p}) (\phi^q \cdot \psi_{q}) (\psi_{a_1} \cdot \psi_{a_2}) (\psi_{b_1} \cdot \psi_{b_2}) \\
  +& 18 \epsilon^{a_1 a_2p} \epsilon^{b_1 b_2(m} (\phi^{n)} \cdot \psi_{q}) (\phi^q \cdot \psi_{p}) (\psi_{a_1} \cdot \psi_{a_2}) (\psi_{b_1} \cdot \psi_{b_2}) ]\ .
  \end{aligned}
\end{equation}
Six operators $O_0 \cdot(\phi^m\cdot\lambda_{\dot\alpha})$ at $t^{29}$ order
are also all $Q$-exact. 
An $SU(2)_R \times SU(3)$ covariant expression is  
\begin{equation}
    \begin{aligned}
        O_0 \cdot ( \phi^m\cdot\lambda_{\dot\alpha}) 
        = \frac{1}{8}Q [& 40 \epsilon^{m np}  (f \cdot \psi_{q}) (\lambda_{\dot\alpha} \cdot \psi_{r})(\phi^{q} \cdot \psi_{n}) (\phi^{r} \cdot \psi_{p})\\
        -&4 \epsilon^{ma_1a_2} \epsilon^{nb_1b_2} (\lambda_{\dot\alpha} \cdot \psi_{n}) (\phi^p \cdot \psi_{p}) (\psi_{a_1} \cdot \psi_{a_2}) (\psi_{b_1} \cdot \psi_{b_2})\\
        +&6 \epsilon^{ma_1a_2} \epsilon^{nb_1b_2} (\lambda_{\dot\alpha} \cdot \psi_{p}) (\phi^p \cdot \psi_{n}) (\psi_{a_1} \cdot \psi_{a_2}) (\psi_{b_1} \cdot \psi_{b_2})\\
        +&\epsilon^{na_1a_2} \epsilon^{pb_1b_2} (\lambda_{\dot\alpha} \cdot \psi_{n}) (\phi^m \cdot \psi_{p}) (\psi_{a_1} \cdot \psi_{a_2}) (\psi_{b_1} \cdot \psi_{b_2}) ]\ .
    \end{aligned}
\end{equation}
Eight operators $O_0 \cdot \left(\phi^m\cdot\psi_{n}-{\textstyle \frac{1}{3}}\delta^m_n\phi^p\cdot\psi_{p}\right)$ at $t^{30}$ order are all $Q$-exact. An $SU(3)$ covariant expression is 
\begin{eqnarray}
  &&O_0 \cdot \left(\phi^m\cdot\psi_{n}-{\textstyle \frac{1}{3}}\delta^m_n\phi^p\cdot\psi_{p}\right)\\ &&=\frac{1}{4} Q\left[\epsilon_{npq}\epsilon^{ra_1a_2} \epsilon^{qb_1b_2} \epsilon^{mc_1c_2}(\phi^{p} \cdot \psi_{r})  (\psi_{a_1} \cdot \psi_{a_2})(\psi_{b_1} \cdot \psi_{b_2}) (\psi_{c_1} \cdot \psi_{c_2}) \right]\ .\nonumber
\end{eqnarray}

We did not manage to prove the $Q$-exactness of operators other than \eqref{Qextshown}.
Since these operators do not appear at all in the index, all of them may be 
$Q$-exact until $t^{31}$ order. More robustly/modestly, we can say that our index 
exhibits a no-hair behavior for $O_0$ until $t^{31}$ order.
It will be interesting to clarify this issue in the future.

The $Q$-exactness of these product operators implies that $O_0$ abhors
the dressings by certain gravitons, reminiscent of the black hole no-hair theorem. 
Especially, $(\phi^m\cdot\phi^n)\sim {\rm tr}(\phi^m\phi^n)$ multiplied to 
$O_0$ are $Q$-exact. This is interesting because these operators correspond to 
bulk scalar fields which have been discussed in the context of hairy AdS$_5$ black holes
\cite{Bhattacharyya:2010yg,Markeviciute:2018yal,Markeviciute:2018cqs}.
More precisely, it is the `s-wave' modes of these scalars that have been used 
to construct hairy black holes, precisely dual to the conformal primary operator 
${\rm tr}(\phi^m\phi^n)$.
Here, note that the BPS limits of the hairy black holes 
constructed this way all exhibit substantial back reactions to the core black holes, 
at least near the horizon, no matter how small the hair parameter is
\cite{Markeviciute:2018yal,Markeviciute:2018cqs}. 

Now we consider the lowest term $-\chi_{(1,3)}t^{32}$  of (\ref{grav-thres-deficit}). 
In fact, this term comes from the following product of $O_0$ and gravitons:
\begin{equation}\label{t32-non-Q-exact}
  O_0 \cdot (\phi^m\cdot f +{\textstyle \frac{1}{2}}\epsilon^{mnp}\psi_{n}\cdot\psi_{p})\ .
\end{equation}
It is easy to show that this is not $Q$-exact, e.g. by acting two 
$Q^m_+$ as shown in (\ref{QQ-on-t36}). These operators contain 
terms at $f^0\phi^0\psi^9$ order, which cannot be $Q$-exact. 
So the operators (\ref{t32-non-Q-exact}) themselves are not $Q$-exact either. 
Therefore, the no-hair interpretation that we made so far holds 
only for certain low-lying gravitons, at best.
Among the conformal primaries of $S_2$ (see Table \ref{tab:Sn}),
these three gravitons are the only ones which explicitly appear in the 
index when multiplied to $O_0$.
At this stage, it may seem that two more gravitons 
$f \cdot \lambda_{\dot\alpha}+{\textstyle\frac{2}{3}}
  \psi_{m}\cdot D_{\dot\alpha}\phi^m
  -{\textstyle \frac{1}{3}}\phi^m \cdot D_{\dot\alpha}\psi_{m}$
at $\mathcal{O}(t^9)$ might multiply $O_0$ to show up at $t^{33}$ order,
but we will see below that the index does not capture them. 
Therefore, out of the $32$ particle species of conformal primary particles 
in the $S_2$ multiplet, $29$ gravitons except $\phi^m\cdot f +
{\textstyle \frac{1}{2}}\epsilon^{mnp}\psi_{n}\cdot\psi_{p}$
do not appear in the index when they multiply $O_0$. 
In the BMN sector, our studies in section 3.1 imply a similar theorem for 
all $O_n$, at least as seen by the index.
Among the $17$ particle species of gravitons in the BMN sector,
all $14$ particles except $\phi^m\cdot f +
{\textstyle \frac{1}{2}}\epsilon^{mnp}\psi_{n}\cdot\psi_{p}$
do not appear in the index when they multiply any $O_n$.

The $3$ product coholomologies at $t^{32}$ order violating the no-hair theorem 
should be the primaries of $PSU(1,2|3)$. This is again contained in a short multiplet 
of $A_1\overline{L}$ type, whose contribution to the index is given by 
$\chi_{32}=t^{32}\chi_{(1,3)}\chi_{ D}$. 
We subtract this from $Z-Z_{\rm grav}-\chi_{24}$, and study the remaining cohomologies. 
We can then try to interpret the lowest order term of the remainder and judge whether it
comes from new core black hole primaries or products of already known core primaries and
gravitons. If one can clarify the nature of the cohomologies at this lowest order, one 
can again subtract the characters of their supermultiplets and keep exploring even higher
orders. Since it becomes more and more difficult to judge the $Q$-exactness of the 
possible product operators, we shall only make much simpler and structural studies 
until the $t^{40}$ order. Namely, we shall try to see if the surviving index can be 
explained as the products of known gravitons and core primaries $O_n$, without the 
need of any new core black hole primaries. Studies we made so far showed that 
this is possible until $t^{32}$ order. Namely, the index until this order is compatible 
with having no more new core primaries and only three more product cohomologies 
(\ref{t32-non-Q-exact}). We shall show that the graviton spectrum is such that 
new core black hole primaries should appear at $t^{39}$ order at the latest. This 
not only proves from the index the existence of new core black hole primaries, 
but will also show scenarios of possible hairy black holes.

After eliminating the contribution of the multiplet $\chi_{32}$ to the index, the 
remaining index vanishes at $t^{33}$ order.
In principle, there are two possible product operators at this order
that completely cancel each other in the index even if they are not $Q$-exact.
They are
\begin{equation}
  O_0 \cdot \partial_{\dot{\alpha}}(\lambda_{\dot\beta}\cdot\lambda^{\dot\beta})\ \ ,\ \ 
  O_0 \cdot \left(f \cdot\lambda_{\dot\alpha}+{\textstyle\frac{2}{3}}
  \psi_{m}\cdot D_{\dot\alpha}\phi^m
  -{\textstyle \frac{1}{3}}\phi^m \cdot D_{\dot\alpha}\psi_{m}\right)\ .
\end{equation}
So these product operators, even if they exist, do not appear in the index. 

The lowest nonzero term of $Z-Z_{\rm grav}-\chi_{24}-\chi_{32}$ is 
$-(\chi_{(1,\bar{3})}+\chi_{(3,6)})t^{34}$.
The only possible product operators at this order which may account for these two terms,
unless they are $Q$-exact, are 
\begin{equation}
  O_0 \cdot {\partial}^{\dot\alpha}\left(\lambda_{\dot\alpha}\cdot\psi_{m}
  -\textstyle{\frac{1}{2}}\epsilon_{mnp}\phi^n\cdot D_{\dot\alpha}\phi^p\right)
  \ \ ,\ \ 
  O_0 \cdot \partial_{\dot\alpha}\partial_{\dot{\beta}}(\phi^{m}\cdot\phi^{n})\ .
\end{equation}
If they are nontrivial, they are in the $\mathcal{N}=4$ representations 
$A_1\overline{L}[6;2]^{[2,2,0]}_{13}$ and $A_1\overline{L}[6;0]^{[3,0,1]}_{13}$,
respectively. Assuming that they are both non-$Q$-exact,
the order $t^{34}$ is accounted for by these hairy black hole operators.
Their multiplets will contribute $-(\chi_{(1,\bar{3})}+\chi_{(3,6)})t^{34}\chi_D$ to the index,
because both are of type $A_1\overline{L}$.

Subtracting them, now the leading term is $-\chi_{(2,3)}t^{35}$.
The only possible product cohomologies which can account for this term are
\begin{equation}
  O_0 \cdot \partial_{\dot\alpha}(f\cdot\phi^m+{\textstyle \frac{1}{2}}
  \epsilon^{mnp}\psi_{n}\cdot\psi_{p})\ ,
\end{equation} 
provided they are not $Q$-exact.
In this case, its multiplet is again $A_1\overline{L}$ type and contributes
$-\chi_{(2,3)}t^{35}\chi_D$ to the index.

Subtracting this, now the leading term is $+(\chi_{(3,1)}+\chi_{(3,8)})t^{36}$.
Since there is one fermionic black hole primary $O_1$ that we now from
section \ref{sec:cohosu2bmn}, we study whether the product cohomologies
may account for $+(1+\chi_{(3,1)}+\chi_{(3,8)})t^{36}$. 
The only possible set is  
\bea\label{su2-t36-contribution}
  && O_0\cdot \partial_{\dot\alpha}(f\cdot\lambda_{\dot\beta}+{\textstyle\frac{2}{3}}
  \psi_{m}\cdot D_{\dot\beta}\phi^m-{\textstyle \frac{1}{3}}\phi^m \cdot 
  D_{\dot\beta}\psi_{m})~, \nn\\
  && O_0 \cdot \partial_{\dot\alpha}\partial_{\dot\beta}(\phi^m\cdot\psi_{n}-
  {\textstyle \frac{1}{3}}\delta^m_n\phi^p\cdot\psi_{p})\ .
\eea
Provided they are not $Q$-exact, they are again the primaries of
$A_1\overline{L}$ type multiplets,
so they contribute to the index by \eqref{su2-t36-contribution} times $\chi_D$.

Subtracting the contributions of these multiplets,
the leading term is $-(\chi_{(2,\bar{3})}+\chi_{(4,6)})t^{37}$.
The only possible product cohomologies that can account for this term are 
\begin{equation}
  O_0\cdot \partial_{\dot\alpha} \partial^{\ \ \dot\beta}
  (\lambda_{\dot\beta}\cdot\psi_{m}-\textstyle{\frac{1}{2}}
  \epsilon_{mnp}\phi^n\cdot D_{\dot\beta}\phi^p)\ \ ,\ \ 
  O_0\cdot \partial_{\dot\alpha}\partial_{\dot\beta}\partial_{\dot\gamma}
  (\phi^m\cdot\phi^n)\ .
\end{equation}

Further processing to subtract the contributions of their multiplets,
again $A_1\overline{L}$ type, the lowest term is $+\chi_{(5,3)}t^{38}$.
one possible set of product cohomologies which can account for this is 
\begin{equation}
  O_0\cdot \partial_{(\dot\alpha}\partial_{\dot\beta}\partial_{\dot\gamma}
  (\lambda_{\dot\delta)}\cdot\phi^m)\ .
\end{equation}
Apart from these, the following two sets of product cohomologies  
\begin{equation}
  O_0\cdot \partial_{\dot\alpha}\partial_{\dot\beta}
  {\partial}^{\dot\gamma}(\lambda_{\dot\gamma}\cdot\phi^m)\ \ ,\ \ 
  O_0\cdot \partial_{\dot\alpha}\partial_{\dot\beta}
  (f\cdot\phi^m+{\textstyle \frac{1}{2}}\epsilon^{mnp}\psi_{n}\cdot\psi_{p})
\end{equation}
exactly cancel in the index, so there are two possible ways in which product hairy 
cohomologies can account for this order. In either case, they are all in the 
$A_1\overline{L}$ type multiplets,
so their contribution to the index is again just $\chi_{(5,3)}t^{38} \cdot \chi_D$.

Subtracting the last multiplets, the lowest term is 
$+(\chi_{(2,1)}+2\chi_{(4,1)}+\chi_{(4,8)})t^{39}$.
All possible product cohomologies at this order are 
\begin{eqnarray}
  ({\bf4},{\bf1})^\textrm{F}&:&O_0\partial_{\dot\alpha}\partial_{\dot\beta}\partial_{\dot\gamma}
  (\lambda_{\dot\delta}\cdot\lambda^{\dot\delta})\ ,\\
  ({\bf2},{\bf1})^\textrm{B}&:&O_0\partial_{\dot\alpha}\partial^{\ \ \dot\beta}
  (f\cdot\lambda_{\dot\beta}+{\textstyle\frac{2}{3}}
  \psi_{m}\cdot D_{\dot\beta}\phi^m
  -{\textstyle \frac{1}{3}}\phi^m \cdot D_{\dot\beta}\psi_{m})\ ,\nonumber\\
  ({\bf4},{\bf1})^\textrm{B}&:&O_0\partial_{(\dot\alpha}\partial_{\dot\beta}
  (f\cdot\lambda_{\dot\gamma)}+{\textstyle\frac{2}{3}}
  \psi_{m}\cdot D_{\dot\gamma)}\phi^m-{\textstyle \frac{1}{3}}\phi^m \cdot D_{\dot\gamma)}\psi_{m})\ ,\nonumber\\
  ({\bf4},{\bf8})^\textrm{B}&:&O_0\partial_{\dot\alpha}\partial_{\dot\beta}\partial_{\dot\gamma}
  (\phi^m\cdot\psi_{n}-{\textstyle \frac{1}{3}}
  \delta^m_n\phi^p\cdot\psi_{p})\ .\nonumber
\end{eqnarray}
We used the superscripts B/F to mark their bosonic/fermionic statistics, respectively. 
With these candidates, we find that the closest one can get to the index
at this order is the case in which all three classes of bosonic operators
are nontrivial while the fermionic operators are $Q$-exact.
In this case, their contribution at this order 
is maximal and becomes $+(\chi_{(2,1)}+\chi_{(4,1)}+\chi_{(4,8)})t^{39}$.
There is still one factor of $\chi_{(4,1)} \cdot t^{39}$ remaining to be addressed.
Therefore, there should be at least $4$ core black hole primaries
in the $SU(2)_R$ representation (4,1),
to account for the remaining $+\chi_{(4,1)}t^{39}$.
Of course this is only the latest order in which new core black hole primaries
should appear, because it may as well appear at lower orders
due to some non-$Q$-exactness assumptions we made for
product cohomologies being invalid.

So we have shown that, from the index data until $t^{40}$ order,
there should exist more core primary operators other than $O_n$ in the BMN sector.
This conclusion is obtained by supposing otherwise,
and trying to explain the index as product cohomologies of $O_n$ 
and gravitons but finding a contradiction at $t^{39}$.
We should also emphasize that the structure of the index admits natural explanations
in terms of hairy product operators in a wide range $t^{33}\sim t^{38}$.
Note also that most of the gravitons appearing in this 
range are conformal descendants in the $S_2$ multiplet.

\section{$SU(3)$, BMN Sector}\label{sec:cohosu3}

We turn to constructing the threshold black hole cohomology of the $SU(3)$ theory.
From the index we computed in section \ref{sec:ngiSU3}, we know that
it is a singlet under the $SU(3)$ subgroup of the R-symmetry group $SU(4)$,
has the order $\cJ=24$, and is fermionic.
Whereas the analogous threshold black hole cohomology presented in
section \ref{sec:cohosu2bmn} could be found by some clever trials and errors
in \cite{Choi:2022caq}, such an approach is not viable for the $SU(3)$ theory
where the elementary fields are represented by larger matrices.
Therefore we take a more strategic approach,
that we organize into four subsections to explain.

In subsection \ref{sec:cohoansatz}, we will introduce an ansatz for
constructing $Q$-closed non-graviton operators,
that takes advantage of various trace relations that were obtained
as byproducts of the graviton index computation of section \ref{sec:ngindex}.
Based on the ansatz, we will present in subsection \ref{sec:Qclosed}
many $Q$-closed non-graviton operators in the target charge sector,
i.e. $q_1 = q_2 = q_3 = 4$ leading to $\cJ=24$.
Among these, we comment that all but \eqref{Q-coho} are $Q$-exact.
Then in subsection \ref{sec:cohoexactness},
we explain the numerics-assisted method that we have used
to determine the $Q$-exactness of the $Q$-closed operators.
Utilizing our check of $Q$-exactness, it is possible to prove that
\eqref{Q-coho} is in fact the only cohomology in the target sector,
denying the possibility that the index may have missed a boson-fermion pair
of non-graviton cohomologies in the target sector.
This is explained in subsection \ref{sec:coho24}

\subsection{An Ansatz for Closed Non-Graviton Operators}\label{sec:cohoansatz}

The cohomologies we would like to construct should be, by definition,
$Q$-closed and not $Q$-exact. 
Unlike gravitons, the $Q$-closedness of the black hole
cohomologies should be ensured by the trace relations.
(Otherwise, that is if it is a cohomology at given energy and at arbitrary values of $N$,  
it is a graviton cohomology.)
So it is important to know what kind of nontrivial trace relations are available
for $N\times N$ matrices when the number of fields is larger than $N$.

It seems to be widely believed that all $SU(N)$ trace relations are
derived from the Cayley-Hamilton identity.
For instance, see \cite{Ebertshauser:2001nj} (p.7, below eqn.(19))
and \cite{Dempsey:2022uie}.
But in practice it is inefficient to search for the trace relations
that we need just from this identity.
Fortunately, we already implicitly know many trace relations
from the calculations reported in section \ref{sec:ngindex}.
Namely, when enumerating finite $N$ gravitons, 
we have counted multi-graviton operators subject to various trace relations
between the generators $g_i$. 
So one can take advantage of these trace relations to construct black hole cohomologies. 
This leads to our `ansatz' for black hole cohomologies, which we explain now.

We can motivate the ideas with a simple example in the $SU(2)$ theory \cite{Chang:2022mjp,Choi:2022caq,Choi:2023znd}. 
A representative of the threshold non-graviton cohomology in $SU(2)$
can be written using the BMN mesons \eqref{BMN-mesons} by
\begin{equation}\label{SU2-threshold}
 O_0\equiv\epsilon^{abc}{(v_2)^m}_a{(v_2)^n}_b{\rm tr}(\psi_{(c}\psi_{m}\psi_{n)})
\end{equation}
where $v_2$ is the graviton operator in the $S_2$ multiplet. 
Let us see how this operator becomes $Q$-closed.
Acting $Q$ on $O_0$, $Q$ acts only on ${\rm tr}(\psi_{(c}\psi_m\psi_{n)})$
since $v_2$ is $Q$-closed. One obtains
\begin{equation}\label{trace-relation-example}
  Q{\rm tr}(\psi_{(c}\psi_m\psi_{n)})\propto \epsilon_{ab(c}{(v_2)^a}_m{(v_2)^b}_{n)}
  \equiv R(v_2)_{cmn}
\end{equation}
after using $SU(2)$ trace relations. 
Plugging this into $QO_0$, one obtains
\begin{equation}\label{syzygy-example}
  QO_0\propto \epsilon^{abc}{(v_2)^m}_a{(v_2)^n}_bR(v_2)_{cmn}=0\ .
\end{equation}
At the last step, one can show that the quartic mesonic polynomial 
$\epsilon^{abc}{(v_2)^m}_a{(v_2)^n}_bR(v_2)_{cmn}$ 
is identically zero \cite{Choi:2023znd}.
From the viewpoint of section \ref{sec:ngindex},
(\ref{trace-relation-example}) are graviton trace relations
and the last step of (\ref{syzygy-example}) is a relation of relations.
So the operator $O_0$ is shown to be $Q$-closed by using
the trace relations and a relation of relations of the finite $N$ graviton operators.

This idea can be extended to construct operators which become $Q$-closed 
only after using trace relations. Namely, for each relation of relations 
such as (\ref{syzygy-example}),
we can construct a $Q$-closed operator such as (\ref{SU2-threshold}).
We still need to check that they are not $Q$-exact for them to represent 
nontrivial $Q$-cohomologies, which we will do in section \ref{sec:cohoexactness}.
Also, there are non-graviton cohomologies which are not constructed
in this way \cite{Choi:2023znd}.
For these reasons, the $Q$-closed operators constructed in this way are
mere ans\"atze for the non-graviton cohomologies.

In appendix \ref{sec:trrel}, we have collected all $SU(3)$
fundamental trace relations that involve $u_n,v_n$ only, 
and manifestly wrote them in $Q$-exact forms.
We have found trace relations involving $u_n,v_n,w_n$ until $\cJ=20$ order.
We have also found all relations between the fundamental graviton 
trace relations at $\cJ=24$ and some more at $\cJ=30$ orders in the
$SU(3)\subset SO(6)_R$ singlet sector,
where the index predicts non-graviton cohomologies (see Table \ref{tower}). 
In other charge sectors, one can immediately write down $Q$-closed operators 
if one finds new relations of the fundamental trace relations.

When we write a fundamental trace relation $R_a$ in a $Q$-exact form
as $R_a \sim Qr_a$, there is an ambiguity in $r_a$ by addition of 
arbitrary $Q$-closed operators.
We partly fix it so that $r_a$ vanishes when all the letters are restricted
to diagonal matrices.
Since the $Q$-closed operators constructed from relations of relations
are linear combinations of $r_a$'s, they vanish with diagonal letters.
This makes it impossible for our ansatz to be gravitons. 
So our ansatz is guaranteed to yield a non-graviton cohomology
unless it is $Q$-exact.

\subsection{$Q$-Closed Non-Graviton Operators}\label{sec:Qclosed}

Based on the ansatz, we now list the non-graviton $Q$-closed operators
at the threshold level $\cJ=24$, which are singlets under the
$SU(3) \subset SU(4)_R$ global symmetry,
in the BMN sector of the $SU(3)$ gauge theory.

At $\cJ \equiv 2(Q_1+Q_2+Q_3)+6J=24$,
operators are further distinguished by the overall R-charge
$R \equiv \frac{Q_1+Q_2+Q_3}{3}$.
The BMN operators which are $SU(3) \subset SU(4)_R$ singlets 
satisfy $Q_1=Q_2=Q_3$ and $J_1=J_2$.
Then the possible charges of the operators are
$(R,J) = (\frac{n}{2},\frac{8-n}{2})$ where $n=0,\cdots , 8$.
In each charge sector, the number of letters is fixed to $n+4$.
However, our ansatz further restricts the charges since acting $Q$
on our ansatz should become a polynomial of $u_{2,3}, v_{2,3}, w_{2,3}$.
As a result, there exist in total 7 possible charge sectors within our ansatz:
$(R,J) = (\frac{n}{2},\frac{8-n}{2})$ where $n=1,\cdots , 7$.

When $(R,J) = (\frac{1}{2}, \frac{7}{2})$ or $(1,3)$, there are no $Q$-closed 
operators within our ansatz using the trace relations in the appendix.
One can understand it heuristically as follows. At these charges, 
$R$ is so small that only a small number of scalars is admitted.  
As the graviton generators contain at least one scalar field, only few types of graviton polynomials 
exist in these sectors, which are not enough to host relations 
of relations. Therefore, these charge sectors are incompatible with our ansatz. 
The other $5$ charge sectors host $Q$-closed operators in our ansatz, 
whose explicit forms will be presented below.

We now present the $Q$-closed non-graviton 
operators in each of the five charge sectors,
$(R,J) = (\frac{n}{2},\frac{8-n}{2})$ where $n=3,\cdots ,7$. For convenience, we 
rewrite here the definition of the single-trace generators of the $SU(3)$ 
BMN gravitons $u_{2,3}, v_{2,3}, w_{2,3}$:
\begin{equation}
    \begin{aligned}
        u^{ij} \equiv & \; \textrm{tr}  \left(\phi^{(i} \phi^{j)}\right)\ ,\ 
        u^{ijk} \equiv \; \textrm{tr} \left(\phi^{(i} \phi^{j} \phi^{k)}\right)\ , \\
        {v^{i}}_j \equiv & \; \textrm{tr} \left( \phi^i \psi_{j}\right) - {\textstyle \frac{1}{3} }
        \delta^i_j\textrm{tr}\left(\phi^a \psi_{a}\right)\ , \ 
        {v^{ij}}_k \equiv \; \textrm{tr} \left(\phi^{(i} \phi^{j)} \psi_{k} \right) 
        - \! {\textstyle \frac{1}{4}} \delta^{i}_{k}  
        \textrm{tr} \left( \phi^{(j} \phi^{a)} \psi_{a} \right) 
        - \! {\textstyle \frac{1}{4}} \delta^{j}_{k}  \textrm{tr} \left( \phi^{(i} \phi^{a)} \psi_{a} \right)\ , \\
        w^i \equiv & \; \textrm{tr} \left(f \phi^i + {\textstyle \frac{1}{2}} \epsilon^{ia_1a_2} \psi_{a_1} \psi_{a_2}\right)\ ,\
        w^{ij} \equiv \; \textrm{tr}\left(f \phi^{(i} \phi^{j)} +\epsilon^{a_1a_2(i}\phi^{j)} \psi_{a_1} \psi_{a_2}\right)\ .
    \end{aligned}
\end{equation}

\paragraph{i) $(R,J) = (\frac{3}{2}, \frac{5}{2})$.}
The operators in this sector are made of 7 letters.
The possible numbers $(n_\phi,n_\psi,n_f)$ of scalars, fermions and $f$ in each term are
$(n_\phi, n_\psi, n_f) = (4,1,2)$, $(3,3,1)$ and $(2,5,0)$.
We find one $Q$-closed operator in this sector from the trace relations and a relation of 
relations in appendix A. This $Q$-closed operator is given by
\bea\label{O1} 
        O^{(2,1)} &\equiv&
        65u^{ij}  (r_{20}^{(2,1)})_{ij} -39w^{ij}   (r_{14}^{(1,1)})_{ij} +5w^i   (r_{16}^{(1,1)})_i  \nn\\
        && +312{v^{jk}}_i   (r_{16}^{(1,2)})^i_{jk} +26{v^j}_i  (r_{18}^{(1,2)})^i_j
        +6w^i   (r_{16}^{(0,3)})_i~.
\eea
The superscripts denote $(n_f, n_\psi)$ of the terms with maximal $n_f$ in the operator. 
$r_j^{(n_f,n_\psi)}$'s are given in \eqref{tr-r}, \eqref{tr-r-f} where $R_j^{(n_f,n_\psi-1)} \equiv i \, Q\, r_j^{(n_f,n_\psi)}$'s are 
the fundamental trace relations.
The $Q$-closed operator \eqref{O1} turns out to be $Q$-exact.
In fact, \eqref{O1} is even under the parity transformation of \cite{Beisert:2004ry}.
It is already known that all such even operators in this charge sector
are $Q$-exact for all $N\geq 3$ \cite{Budzik:2023vtr}, which we confirm.

\paragraph{ii) $(R,J) = (2, 2)$.}
The operators in this sector are made of 8 letters.
Allowed $(n_\phi, n_\psi, n_f)$ are $(6,0,2), (5,2,1),(4,4,0)$.
We find 4 $Q$-closed operators in this sector given by
\begin{equation}\label{O2}
    \begin{aligned}
        O_1^{(1,2)} \equiv & -3{v^{(j}}_{i} w^{k)}   (r_{10}^{(0,1)})^i_{jk} -3u^{(ij}w^{k)}   (r_{12}^{(0,2)})_{ijk} +\epsilon_{a_1a_2i} u^{a_1j} w^{a_2}  (r_{12}^{(0,2)})^i_j\ , \\
        O_2^{(1,2)} \equiv & -9u^{a(i} {v^{j)}}_{a}   (r_{14}^{(1,1)})_{ij} +10\epsilon_{a_1a_2(i}u^{a_1k} {v^{a_2}}_{j)}   (r_{14}^{(1,1)})^{ij}_k \\
        & + 30 {v^{(j}}_{i} w^{k)}   (r_{10}^{(0,1)})^i_{jk}
        +60u^{(jk} {v^{l)}}_i   (r_{14}^{(0,3)})^i_{jkl}\ , \\
        O_3^{(1,2)} \equiv & -3u^{a(i} {v^{j)}}_{a}   (r_{14}^{(1,1)})_{ij} 
        +6\epsilon_{a_1a_2(i}u^{a_1k} {v^{a_2}}_{j)}   (r_{14}^{(1,1)})^{ij}_k
        +4u^{ijk}   (r_{18}^{(1,2)})_{ijk} 
        +14{v^{(j}}_{i} w^{k)}   (r_{10}^{(0,1)})^i_{jk} \\
        & -6w^{ij}   (r_{14}^{(0,2)})_{ij} 
        -12\epsilon^{a_1a_2(i} {v^{j}}_{a_1} {v^{k)}}_{a_2}   (r_{12}^{(0,2)})_{ijk} 
        -4{v^j}_a {v^a}_i   (r_{12}^{(0,2)})^i_j\ , \\
        O_4^{(1,2)} \equiv & -3u^{a(i} {v^{j)}}_{a}   (r_{14}^{(1,1)})_{ij} 
        +14\epsilon_{a_1a_2(i}u^{a_1k} {v^{a_2}}_{j)}   (r_{14}^{(1,1)})^{ij}_k
        +8{v^{jk}}_{i}   (r_{16}^{(1,1)})^i_{jk} 
        +42{v^{(j}}_{i} w^{k)}   (r_{10}^{(0,1)})^i_{jk} \\
        &+12 u^{(ij}w^{k)}   (r_{12}^{(0,2)})_{ijk} 
        -24 w^{ij}   (r_{14}^{(0,2)})_{ij} 
        -36\epsilon^{a_1a_2(i} {v^{j}}_{a_1} {v^{k)}}_{a_2}   (r_{12}^{(0,2)})_{ijk} 
        - 8 {v^{jk}}_i   (r_{16}^{(0,3)})^i_{jk}\ .
    \end{aligned}
\end{equation}
All operators in \eqref{O2} are $Q$-exact.

\paragraph{iii) $(R,J) = (\frac{5}{2}, \frac{3}{2})$} The operators in this sector are made of 9 letters. Allowed $(n_\phi, n_\psi, n_f)$ are $(7,1,1),(6,3,0)$. We find 13 $Q$-closed operators in this sector given by
\begin{equation}\label{O3}
    \begin{aligned}
    &O_1^{(1,1)} \equiv \epsilon_{a_1a_2 i } u^{a_1 (j} w^{k) a_2}   (r_{10}^{(0,1)})^i_{jk} \ , \\
    &O_2^{(1,1)} \equiv \epsilon_{a_1a_2 i } u^{a_1 jk} w^{a_2}   (r_{10}^{(0,1)})^i_{jk} \ , \\
    &O_3^{(1,1)} \equiv \epsilon_{a_1a_2i}\epsilon_{b_1b_2j} u^{a_1b_1}u^{a_2b_2k}   (r_{14}^{(1,1)})^{ij}_k 
    +5{v^a}_i {v^{jk}}_a (r_{10}^{(0,1)})^i_{jk} -2{v^{(j}}_a {v^{k)a}}_i  (r_{10}^{(0,1)})^i_{jk} \ , \\
    &O_1^{(0,3)}  = -\epsilon_{i a_1 a_2} \left( 4 u^{a_1 b} {v^{j a_2}}_{b} + 3u^{j a_1b} {v^{a_2}}_{b}\right)   (r_{12}^{(0,2)})^i_j = \frac{1}{2} i\, Q ((r_{12}^{(0,2)})^i_j (r_{12}^{(0,2)})^j_i) \ , \\
    &O_2^{(0,3)}  = -\epsilon_{a_1 a_2 (i}\left(u^{a_1(k}{v^{l)a_2}}_{j)}+
        u^{kl a_1}{v^{a_2}}_{j)}\right)   (r_{12}^{(0,2)})^{ij}_{kl} = \frac{1}{2} i\, Q ((r_{12}^{(0,2)})^{kl}_{ij} (r_{12}^{(0,2)})^{ij}_{kl})\ , \\
    &O_3^{(0,3)} \equiv -u^{a(i}{v^{jk)}}_a   (r_{12}^{(0,2)})_{ijk} \ , \\
    &O_4^{(0,3)} \equiv -\epsilon_{a_1 a_2 i} u^{a_1 b} {v^{a_2}}_b   (r_{14}^{(0,2)})^i\ , \\
    &O_5^{(0,3)} \equiv 6{v^a}_i {v^{jk}}_a  (r_{10}^{(0,1)})^i_{jk}
    +6u^{a(ij}{v^{k)}}_a   (r_{12}^{(0,2)})_{ijk} 
    +\epsilon_{a_1a_2 i} u^{a_1 bj}{v^{a_2}}_{b}   (r_{12}^{(0,2)})^i_j\ ,\\
    &O_6^{(0,3)} \equiv 24 {v^{(j}}_a {v^{k)a}}_i  (r_{10}^{(0,1)})^i_{jk} 
    +6 u^{a(i} {v^{j)}}_{a}   (r_{14}^{(0,2)})_{ij} 
    - \epsilon_{a_1a_2(i} u^{a_1 k} {v^{a_2}}_{j)}   (r_{14}^{(0,2)})^{ij}_k \ ,\\
    &O_7^{(0,3)} \equiv {v^a}_i v^{jk}_{~~a}  (r_{10}^{(0,1)})^i_{jk} 
    \!-\!10 v^{(j}_{~~a} v^{k)a}_{~~~i}  (r_{10}^{(0,1)})^i_{jk} 
    \!+\!6u^{a(ij} v^{k)}_{~a}  (r_{12}^{(0,2)})_{ijk}  
    \!+\!10 \epsilon_{a_1a_2(i} u^{a_1 kl}v^{a_2}_{~j)}   (r_{12}^{(0,2)})^{ij}_{kl}  \ ,\\
    &O_8^{(0,3)} \equiv  5{v^a}_i v^{jk}_{~~a}  (r_{10}^{(0,1)})^i_{jk} 
    \!-\!2 v^{(j}_{~~a} v^{k)a}_{~~~i} (r_{10}^{(0,1)})^i_{jk} 
    \!+\!9u^{a(ij} v^{k)}_{~a}   (r_{12}^{(0,2)})_{ijk}  
    \!+\!6\epsilon_{a_1 a_2 i} u^{a_1 (j}u^{kl) a_2}   (r_{14}^{(0,3)})^{i}_{jkl}\ ,\\
    &O_9^{(0,3)} \equiv 6{v^a}_i v^{jk}_{~~a}  (r_{10}^{(0,1)})^i_{jk} 
    \!+\! 12v^{(j}_{~~a} v^{k)a}_{~~~i}  (r_{10}^{(0,1)})^i_{jk} 
    \!+\!18u^{a(ij} v^{k)}_{~a}  (r_{12}^{(0,2)})_{ijk}  
    \!-\!\epsilon_{a_1a_2(i} u^{a_1 k} {v^{a_2}}_{j)}   (r_{14}^{(0,2)})^{ij}_k \ ,\\
    &O_{10}^{(0,3)} \equiv 38{v^a}_i {v^{jk}}_a  (r_{10}^{(0,1)})^i_{jk} 
    +4{v^{(j}}_a {v^{k)a}}_i  (r_{10}^{(0,1)})^i_{jk} 
    +24u^{a(ij}{v^{k)}}_a  (r_{12}^{(0,2)})_{ijk}  
    +5u^{(jk}{v^{l)}}_i   (r_{14}^{(0,2)})^i_{jkl}  \ .
    \end{aligned}
\end{equation}
All except for $O^{(0,3)}_6$ in \eqref{O3}  are $Q$-exact. 
Therefore, a representative of the cohomology in this sector can be written as
\begin{eqnarray}\label{Q-coho}
    O &\equiv &-6O_6^{(0,3)} \\
    &=& \; 288 {v^{j}}_a {v^{ka}}_i \epsilon_{c_1c_2(j} \tr \left( \phi^{c_1}\phi^{c_2}\phi^{i}\psi_{k)} \right) -72 {v^{a}}_b {v^{bk}}_a
    \epsilon_{c_1c_2(k} \tr \left( \phi^{c_1}\phi^{c_2}\phi^{d}\psi_{d)} \right)\nonumber\\
        &&+36\epsilon_{a_1a_2i} u^{a_1 k} {v^{a_2}}_{j} 
        \left[ 2\tr\left(\phi^{(i} \phi^{c} \phi^{j)} \psi_{(c} \psi_{k)}\right)+2\tr\left(\phi^{(i|} \phi^{c} \phi^{|j)} \psi_{(c} \psi_{k)}\right) \right.\nonumber\\
        &&\qquad \qquad \qquad \qquad \quad \left. 
        +9 \tr\left( \phi^{(i} \phi^{j} \psi_{(c} \phi^{c)} \psi_{k)} \right)-6 \tr\left( \phi^{(i} \phi^{j)} \psi_{(c} \phi^c \psi_{k)} \right)\right] \nonumber\\
        &&-9\epsilon_{a_1a_2j} u^{a_1 b} {v^{a_2}}_{b} \left[ 2\tr\left(\phi^{(j} \phi^{c} \phi^{d)} \psi_{(c} \psi_{d)}\right)+2\tr\left(\phi^{(j|} \phi^{c} \phi^{|d)} \psi_{(c} \psi_{d)}\right) \right.\nonumber\\
        &&\qquad \qquad \qquad \quad \;\;\, \left. \, +9 \tr\left( \phi^{(j} \phi^{d} \psi_{(c} \phi^{c)} \psi_{d)} \right)-6 \tr\left( \phi^{(j} \phi^{d)} \psi_{(c} \phi^c \psi_{d)} \right)\right] \nonumber\\
        &&-20 u^{ai} {v^{j}}_{a} \epsilon_{b_1b_2b_3}\left[2 \tr\left(\psi_{(i}\psi_{j)} \phi^{b_1} \phi^{b_2}\phi^{b_3}\right) + \tr\left(\psi_{(i} \phi^{b_1}\psi_{j)}\phi^{b_2}\phi^{b_3}\right)\right]  \nonumber\\
        &&-36 u^{ai} {v^{j}}_{a} \epsilon_{b_1b_2(i}\left[\tr\left(\psi_{j)} \psi_{c}\phi^{b_1}\phi^{b_2}\phi^{c}\right)+\tr\left(\psi_{j)} \psi_{c}\phi^{b_1}\phi^{c}\phi^{b_2}\right)+\tr\left(\psi_{j)} \psi_{c}\phi^{c}\phi^{b_1}\phi^{b_2}\right)\right]\nonumber \\
        &&-36 u^{ai} {v^{j}}_{a} \epsilon_{b_1b_2(i}\left[\tr\left(\psi_{j)} \phi^{b_1}\psi_{c}\phi^{b_2}\phi^{c}\right)+\tr\left(\psi_{j)} \phi^{b_1}\psi_{c}\phi^{c}\phi^{b_2}\right)+\tr\left(\psi_{j)} \phi^{c}\psi_{c}\phi^{b_1}\phi^{b_2}\right)\right] \nonumber\\
        &&-36 u^{ai} {v^{j}}_{a}  \epsilon_{b_1b_2(i}\left[\tr\left(\psi_{j)} \phi^{b_1}\phi^{b_2}\psi_{c}\phi^{c}\right)+\tr\left(\psi_{j)} \phi^{b_1}\phi^{c}\psi_{c}\phi^{b_2}\right)+\tr\left(\psi_{j)} \phi^{c}\phi^{b_1}\psi_{c}\phi^{b_2}\right)\right] \nonumber\\
        &&-36 u^{ai} {v^{j}}_{a}  \epsilon_{b_1b_2(i}\left[\tr\left(\psi_{j)} \phi^{b_1}\phi^{b_2}\phi^{c}\psi_{c}\right)+\tr\left(\psi_{j)} \phi^{b_1}\phi^{c}\phi^{b_2}\psi_{c}\right)+\tr\left(\psi_{j)} \phi^{c}\phi^{b_1}\phi^{b_2}\psi_{c}\right)\right] \nonumber\\
        &&+12 u^{ai} {v^{j}}_{a}  \epsilon_{b_1b_2(i}
        \big[ 5\tr\left(\psi_{j)}\phi^{b_1}\phi^{b_2}\right)\tr\left(\psi_{c}\phi^{c}\right)
        +2\tr\left(\psi_{j)}\phi^{(b_1}\phi^{c)}\right)\tr\left(\psi_{c}\phi^{b_2}\right) \nn\\
        && \hspace{3.2cm} -2 \tr\left(\psi_{j)}\phi^{b_2}\right)\tr\left(\psi_{c}\phi^{(b_1}\phi^{c)}\right)\big]\ .
        \nonumber
\end{eqnarray}
The scaling dimension of this cohomology $O$ is $E=3R+2J=\frac{21}{2}$. Note that the 
representative found above does not contain the letter $f$.

\paragraph{iv) $(R,J) = (3, 1)$.}
The operators in this sector are made of 10 letters.
Allowed $(n_\phi, n_\psi, n_f)$ are $(9,0,1)$ and $(8,2,0)$.
We find 6 $Q$-closed operators in this sector given by
\begin{equation}\label{O4}
    \begin{aligned}
        &O^{(0,2)}_1 \equiv - \epsilon_{a_1a_2i} u^{a_1 b} u^{jk} {v^{a_2}}_{b}   (r_{10}^{(0,1)})^i_{jk}
        +2\epsilon_{a_1a_2 i}u^{a_1 b} u^{a_2 (j} {v^{k)}}_{b}   (r_{10}^{(0,1)})^i_{jk} \ , \\
        &O^{(0,2)}_2 \equiv -6\epsilon_{a_1a_2i} u^{a_1b(j} {v^{k)a_2}}_{b}   (r_{10}^{(0,1)})^i_{jk} 
        - \epsilon_{a_1a_2(i} u^{a_1(k} {v^{l)a_2}}_{j)}   (r_{12}^{(0,1)})^{ij}_{kl} \ , \\
        &O^{(0,2)}_3 \equiv -\epsilon_{a_1a_2i} u^{a_1 b} u^{jk} {v^{a_2}}_{b}   (r_{10}^{(0,1)})^i_{jk} 
        - \epsilon_{a_1a_2(i} u^{a_1kl} {v^{a_2}}_{j)}   (r_{12}^{(0,1)})^{ij}_{kl} \ , \\
        &O^{(0,2)}_4 \equiv -\epsilon_{a_1a_2i} u^{a_1 b} u^{jk} {v^{a_2}}_{b}   (r_{10}^{(0,1)})^i_{jk} 
        + \epsilon_{a_1a_2(i} \epsilon_{j)b_1b_2} u^{a_1b_1} u^{a_2b_2} u^{kl}   (r_{12}^{(0,2)})^{ij}_{kl} \ , \\
        &O^{(0,2)}_5 \equiv -4\epsilon_{a_1a_2i} u^{a_1 b} u^{jk} {v^{a_2}}_{b}   (r_{10}^{(0,1)})^i_{jk} 
        - 24\epsilon_{a_1a_2i} u^{a_1b(j} {v^{k)a_2}}_{b}   (r_{10}^{(0,1)})^i_{jk} \\
        & \hspace{1.44cm} -\epsilon_{a_1a_2(i} \epsilon_{j)b_1b_2} u^{a_1b_1} u^{a_2b_2k}   (r_{14}^{(0,2)})^{ij}_k  \ , \\
        &O^{(0,2)}_6 \equiv -\epsilon_{a_1a_2i} u^{a_1 b} u^{jk} {v^{a_2}}_{b}   (r_{10}^{(0,1)})^i_{jk} 
        \!+\! 12\epsilon_{a_1a_2i} u^{a_1b(j} {v^{k)a_2}}_{b}   (r_{10}^{(0,1)})^i_{jk} 
        \!+\! 3\epsilon_{a_1a_2i} u^{a_1(j} u^{kl)a_2}   (r_{14}^{(0,2)})^i_{jkl} \ .
    \end{aligned}
\end{equation}
All the operators in \eqref{O4} are $Q$-exact.

\paragraph{v) $(R,J) = (\frac{7}{2}, \frac{1}{2})$.}
The operators in this sector are made of 11 letters.
The only allowed $(n_\phi, n_\psi, n_f)$ is $(9,1,0)$.
We find 1 $Q$-closed operator in this sector given by
\begin{eqnarray}\label{O5}
    O^{(0,1)} & \equiv& 36 \epsilon_{a_1a_2a_3} \epsilon_{b_1b_2 i} u^{a_1b_1} u^{a_2b_2} u^{a_3jk}   (r_{10}^{(0,1)})^i_{jk} 
    +5 \epsilon_{a_1 a_2 a_3} \epsilon_{b_1 b_2 b_3} u^{a_1 b_1}u^{a_2 b_2}u^{a_3 b_3}   r_{12}^{(0,1)} 
    \nonumber\\
    &&-6 \epsilon_{a_1a_2(i} \epsilon_{j)b_1b_2} u^{a_1b_1} u^{a_2b_2} u^{kl}   (r_{12}^{(0,1)})^{ij}_{kl}\ . 
\end{eqnarray}
The operator \eqref{O5} is $Q$-exact.

In summary, we have found 1 fermionic black hole cohomology using our ansatz,
which is a singlet under $SU(3) \subset SU(4)_R$ at order $\cJ=24$.
It is represented by (\ref{Q-coho}).
Its charges and scaling dimension are given by
$(R,J,E) = \left(\frac{5}{2},\frac{3}{2},\frac{21}{2}\right)$.

\subsection{Filtering Exact Operators}\label{sec:cohoexactness}

In this subsection, we sketch how to determine $Q$-exactness of various
$Q$-closed operators listed in the previous subsection.

To check whether a given operator is $Q$-exact or not,
especially to check non-$Q$-exactness,
one has to rule out all possible ways of writing the operator as $Q$ of `something'.
That being said, one needs to construct all possible operators that can
participate in `something' (the meaning of which will be made clear shortly)
and show that the \emph{target} operator is linearly independent of $Q$-actions of them.
More specifically, we divide the check of $Q$-exactness into 4 steps,
that we summarize as follows.
\begin{enumerate}
    \item Construct all gauge-invariant operators whose $Q$-action may participate
    in reproducing the \emph{target}.
    \item Count the number of linearly independent operators from step 1,
    and extract the maximal subset of linearly independent operators.
    This is called the \emph{basis}.
    \item Act $Q$ on the basis operators, then again count and extract
    the maximal subset of linearly independent ones between them.
    \item Check if the target is linearly independent of the result of step 3.
\end{enumerate}

Now we explain what operators `may participate in reproducing the target' in step 1.
This consists of two criteria: the charges and the parity under permutation.

First, the charges of the target operator constrain the charges,
thus the letter contents of the basis operators.
Note that the action of $Q$ increases $Q_{I=1,2,3}$ by $\frac12$
and decreases $J = J_1 = J_2$ by $\frac12$.
Therefore, the basis operators must have the set of charges
that differ by the corresponding amount from the target,
otherwise their $Q$-actions are disjoint from the target.
Note that all of our targets are $SU(3)$ singlets, so we always have $R=Q_1=Q_2=Q_3$.

Second, all of our targets being singlets under the $SU(3)$ subgroup
of the $SU(4)$ R-symmetry group,
imposes a stronger constraint than just restricting to the
charge sectors with $Q_1=Q_2=Q_3$.
Each basis operator must be invariant under cyclic permutation
$\phi^{i} \to \phi^{i+1}$ and simultaneously $\psi_{i} \to \psi_{i+1}$,
where $i=1,2,3$ mod 3.
Moreover, if there are even/odd number of $\phi$'s and $\psi$'s combined,
which carries one $SU(3)$ index each,
it requires even/odd number of Levi-Civita symbols to write the operator
covariantly while contracting all indices.
Therefore, we may restrict to
i) operators with even number of $\phi$'s and $\psi$'s combined,
that are even under all $3!$ permutations of $SU(3)$ indices, and
ii) operators with odd number of $\phi$'s and $\psi$'s combined,
that are even under even (cyclic) permutations of $SU(3)$ indices
and odd under odd (swap) permutations of $SU(3)$ indices.
Also note that this permutation property commutes with the action of $Q$,
so that $Q$ of a non-trivial operator satisfies this property
if and only if the original operator does.
This permutation property is necessary but not sufficient for an operator
to be an $SU(3)$ singlet.
However, we impose this property on the basis instead of requiring $SU(3)$ singlets,
because the latter requires many sums over dummy indices
and thus the former is computationally more efficient.
Our conclusions on the singlet sector will be valid despite.

For example, suppose that the target operator is \eqref{Q-coho},
which has charges $(R,J) = (\frac52,\frac32)$.
Operators whose $Q$-action may reproduce this target operator
must then have $(R,J) = (2,2)$.
Possible choices of letter contents are
$(n_\phi, n_\psi, n_f) = (6,0,2), (5,2,1)$, or $(4,4,0)$,
and numbers of $\phi^i$ minus numbers of $\psi_i$ must be equal between $i=1,2,3$.
Further taking into account the permutation property,
the basis operators whose $Q$-action `may participate in reproducing the target'
\eqref{Q-coho} can be classified into the following 7 subsectors.
($(-1)^\epsilon$ in subsectors 5 and 6 indicates minus sign for odd permutations,
because there are odd number of $\phi$'s and $\psi$'s in those subsectors.)
\begin{itemize}
    \item Subsector 1: $(\phi^1)^4(\psi_1)^4 + $ (permutations)
    \item Subsector 2: $(\phi^1)^3(\phi^2)^1(\psi_1)^3(\psi_2)^1 +$ (permutations) 
    \item Subsector 3: $(\phi^1)^2(\phi^2)^2(\psi_1)^2(\psi_2)^2 +$ (permutations)
    \item Subsector 4: $(\phi^1)^2(\phi^2)^1(\phi^3)^1(\psi_1)^2(\psi_2)^1(\psi_3)^1 +$ (permutations) 
    \item Subsector 5: $(\phi^1)^3(\phi^2)^1(\phi^3)^1(\psi_1)^2f^1 + (-1)^\epsilon$ (permutations)
    \item Subsector 6: $(\phi^1)^2(\phi^2)^2(\phi^3)^1(\psi_1)^1(\psi_2)^1f^1+ (-1)^\epsilon$ (permutations)
    \item Subsector 7: $(\phi^1)^2(\phi^2)^2(\phi^3)^2f^2 +$ (permutations) 
\end{itemize}
Appropriate sums over permutations of single- and multi-trace operators
in each of these subsectors are the result of step 1,
some of which we write down below to help visualize:
\begin{eqnarray}\label{QEx-basis-eg}
    &&{\rm tr}(\phi^1\phi^1\psi_1\phi^1\psi_1\psi_1\phi^1\psi_1)
    + {\rm tr}(\phi^2\phi^2\psi_2\phi^2\psi_2\psi_2\phi^2\psi_2)
    + {\rm tr}(\phi^3\phi^3\psi_3\phi^3\psi_3\psi_3\phi^3\psi_3)~, \nn \\
    &&
    {\rm tr}(\phi^1\phi^1\phi^2\psi_2){\rm tr}(\psi_1\psi_2){\rm tr}(\phi^2\psi_1) +
    {\rm tr}(\phi^2\phi^2\phi^3\psi_3){\rm tr}(\psi_2\psi_3){\rm tr}(\phi^3\psi_2) \nn\\
    && \hspace{1cm} + {\rm tr}(\phi^3\phi^3\phi^1\psi_1){\rm tr}(\psi_3\psi_1){\rm tr}(\phi^1\psi_3)
    + {\rm tr}(\phi^3\phi^3\phi^2\psi_2){\rm tr}(\psi_3\psi_2){\rm tr}(\phi^2\psi_3) \nn\\
    && \hspace{1cm} + {\rm tr}(\phi^1\phi^1\phi^3\psi_3){\rm tr}(\psi_1\psi_3){\rm tr}(\phi^3\psi_1) +
    {\rm tr}(\phi^2\phi^2\phi^1\psi_1){\rm tr}(\psi_2\psi_1){\rm tr}(\phi^1\psi_2)~, \nn \\
    && {\rm tr}(\phi^2\phi^2\psi_1\psi_2\phi^3){\rm tr}(f\phi^1\phi^1) +
    {\rm tr}(\phi^3\phi^3\psi_2\psi_3\phi^1){\rm tr}(f\phi^2\phi^2) \nn\\
    && \hspace{1cm} + {\rm tr}(\phi^1\phi^1\psi_3\psi_1\phi^2){\rm tr}(f\phi^3\phi^3)
    - {\rm tr}(\phi^3\phi^3\psi_1\psi_3\phi^2){\rm tr}(f\phi^1\phi^1) \nn\\
    && \hspace{1cm} - {\rm tr}(\phi^1\phi^1\psi_2\psi_1\phi^3){\rm tr}(f\phi^2\phi^2) -
    {\rm tr}(\phi^2\phi^2\psi_3\psi_2\phi^1){\rm tr}(f\phi^3\phi^3)~.
\end{eqnarray}

Given the operators from step 1, the rest is relatively straightforward, at least conceptually.
There are non-trivial trace relations between operators from step 1,
so in step 2 we extract linearly independent basis operators.
Then in step 3, we consider $Q$-actions of the basis operators,
and again count the number of linearly independent ones among them.
These should form a complete basis of all $Q$-exact operators in the
target charge sector and with the aforementioned permutation property.
Therefore, the target operator is $Q$-exact if and only if it is a linear combination
of the $Q$-actions of the basis operators.
More generally, if there are multiple target operators,
the number of cohomologies among them would be equal to
the number of linearly independent ones among the basis \emph{and} all target operators,
minus the number of linearly independent ones among the basis only.

Each of step 2-4 involves counting and/or finding linearly independent operators
among a given set of gauge-invariant operators.
Each operator is a sum over single- and multi-trace operators
written in terms of seven species of fields $\phi^m$, $\psi_m$ and $f$.
To completely account for trace relations between them,
we first convert the operators written in terms of adjoint fields
into polynomials of their matrix elements, by substituting
\begin{eqnarray}
f &=& \begin{pmatrix} f_1 & f_2 & f_3 \\ f_4 & f_5 & f_6 \\ f_7 & f_8 & -f_1-f_5 \end{pmatrix}~,
\end{eqnarray}
and likes for 6 other fields. In this way, every operator is now written as
a polynomial of $8 \times 7 = 56$ variables, 24 of which are Grassmannian.
So the problem boils down to finding linear dependence between a set of polynomials.
Although this is the same problem that was encountered while computing the
graviton index in section 3, the same method of extracting the
coefficient matrix is extremely unpractical here.
It is because there are four times as many variables (recall that for
counting gravitons, we substituted each field with a diagonal matrix),
and therefore exponentially larger number of monomials appear in polynomials.
As a result, the coefficient matrix will have a huge number of columns
that is not viable for computers.

For this reason, we have devised a numerics-assisted approach to find
linear dependence between the polynomials with large number of variables.
The approach stems from the basic fact that if some linear combination
of certain polynomials vanishes,
it must also be zero if we attribute any specific number to each variable.
So let us represent each polynomial by an array of numbers, i.e. a row vector,
by substituting each variable with a set of randomly chosen integers.
Then we examine the linear dependence between vectors, instead of polynomials.

The substitution can be repeated for arbitrarily many sets of integers,
so the row vector can be made arbitrarily long.
Obviously, the length of the row vectors, i.e. number of columns,
must be at least as many as there are independent polynomials.
Otherwise, it will be always possible to find a relation between the row vectors
even if the polynomials they represent are independent.
On the other hand, the length of the row vectors need not be much more than
the number of independent polynomials, as we will explain shortly.

This makes it clear why this method is efficient.
It naturally realizes the basic principle that in order to distinguish
$n$ different entities, one needs at least $n$ data for each entity,
whereas extracting the coefficient matrix for the polynomials with so many variables
will equivalently convert each polynomial into an unnecessarily long vector.

There are two issues with this approach that we need to address.
The first is that 24 of 56 variables are Grassmannian,
which cannot be properly substituted with c-numbers.
The second is that randomness is involved in this approach,
and it may lead to errors albeit unlikely.

The issue with Grassmann variables can be easily addressed by ordering them in a
definite manner within each monomial.
That is, we fully expand each polynomial (which includes eliminating squares of Grassmann variables),
and let variables be multiplied only in a certain order within each monomial.
During this process the coefficients may flip signs, but the result of this process
is unique for each polynomial.
Once we have done this, none of the Grassmann properties will be used
when finding linear relations between the polynomials,
because each monomial is now compared verbatim with monomials in other polynomials.
Therefore, it is now safe to substitute Grassmann variables with c-numbers.
This principle was also implied while extracting the coefficient matrix of
graviton operators in section 3.

As for the randomness, first note that substituting (sufficiently many sets of)
random integers never miss the true dependence between polynomials.
If there is a true linear dependence between polynomials,
i.e. a linear combination that vanishes,
the same linear combination must be zero for whatever numbers are put in,
so the row vectors corresponding to the polynomials must be linearly dependent.
Note that all polynomials have rational coefficients and we put in random integers,
so there is no issue with machine precision.

However, the converse is possible: this method may find false linear dependence between polynomials.
This is simply because a non-vanishing polynomial may evaluate to zero when
certain values are put into variables.
That is, the randomly chosen values could miraculously be the roots of the polynomial.
This type of error can be made arbitrarily more unlikely by increasing
the number of columns, i.e. number of sets of random integers that are put in.
Let us roughly estimate the unlikelihood.

\begin{figure}
    \centering
    \includegraphics[width=0.45\textwidth]{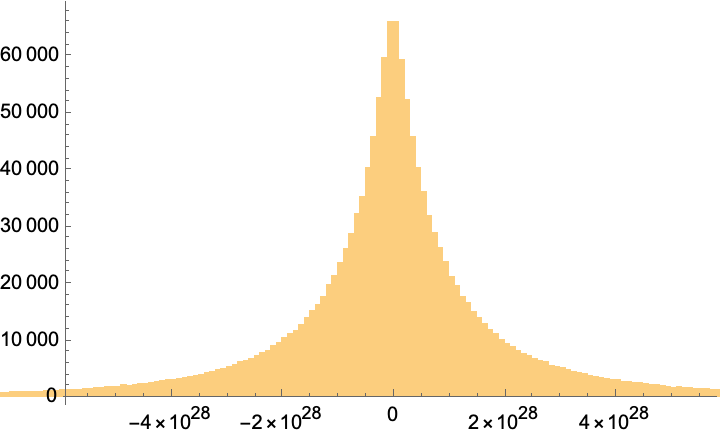}
    \caption{An example distribution of 1.5$\times 10^6$ evaluated values
    of $Q$-actions of basis polynomials for \eqref{O3}.
    Width of each bin is $10^{27}$.}
    \label{fig:egdistribution}
\end{figure}

Suppose that the number of columns is $m+n$ where $m$ is the true number of independent polynomials.
For this method to find a false dependence, both of the followings must happen:
i) there exists a non-trivial linear combination of the polynomials
that vanishes for the first $m$ sets of random integers, and
ii) this polynomial further vanishes for the additional $n$ sets of random integers.
The probability of i) is relatively difficult to estimate, since it involves
intricate tuning of $m-1$ coefficients in a linear combination of the polynomials.
Therefore we only estimate the probability of ii) as follows.
A typical basis polynomial such as $Q$-action of those in (\ref{QEx-basis-eg})
\begin{footnote}{These are used in step 3 and 4 of determining $Q$-exactness
of $Q$-closed operators in the charge sector $(R,J)=(\frac52,\frac32)$,
of which one is the non-graviton cohomology \eqref{Q-coho}}\end{footnote}
evaluates to $\sim 10^{28}$ when a random integer between $1$ and $1000$
are substituted into each variable.
(See Fig. \ref{fig:egdistribution}. for an example.)
This is a natural scale considering that the typical polynomial is a sum over
$\sim 10^{6}$ monomials (with both signs) that each consists of 9 letters,
so for example $10^{6} \times (10^{2.5})^{9} \sim 10^{28}$.
This value is far smaller than the number of all possible random choices
--- which is $(10^3)^{56}$ if all 7 gluons, thus $7\times (3^2-1)$ variables,
are involved ---
so each integer value within magnitude $\sim 10^{28}$ will be sufficiently populated.
Furthermore, since a typical polynomial consists of many$\sim 10^{6}$ monomials,
we assume that the evaluation of the polynomial is like a random walk with
sufficient iterations, and thus the factorization property of integers is blurred.
For these reasons, let us assume that the distribution of the evaluated values
is continuous.
Then, the probability that this value falls within $O(1)$ is estimated to be
$\sim 10^{-28}$, even accounting for the shape of the distribution.
For ii), this must happen for $n$ independent sets of random variables,
so the probability of ii) is estimated to be $10^{-28n}$.
In step 3, $n$ was taken to be 175, so the estimated probability of ii)
is $10^{-4900}$.

This method of detecting linear dependence was used between numerous sets
of polynomials while determining $Q$-exactness of various operators
in different charge sectors.
Numbers that appeared in the previous paragraph slightly differ between occasions.
Typical values of the polynomials differ because they consist of different numbers
of letters, and $n$ is inevitably different because the number of columns are set
before we know the number of true independent polynomials.
However, in any case, we use at least $n \geq 30$ and the estimated probability
of ii) has order of magnitude of a few negative hundreds at the worst.
Furthermore, when a $Q$-closed operator is determined to be $Q$-exact,
we checked analytically the relation between the target and basis polynomials
to further eradicate the margin for error.

Employing the method explained so far, we have constructed the basis operators
in each and all charge sectors with $Q_1 = Q_2 = Q_3$ at the order $\cJ=24$,
with the aforementioned permutation property.
We have also evaluated the $Q$-actions of the bases,
that should form the basis of $Q$-exact operators.
Then we have determined $Q$-exactness of all $Q$-closed
non-graviton operators obtained from our ansatz in the previous subsection.
The result is that all operators in section \ref{sec:Qclosed}
except for the fermionic \eqref{Q-coho} are $Q$-exact.

\subsection{Ansatz-Independent Studies}\label{sec:coho24}

\begin{table}
\begin{center}
\begin{tabular}{|c|c|c|c|c|c|c|c|c|}
     \hline
     $R$ & $J$ & \#letters & \#basis & \#closed & \#exact & \#coh. & \#gravitons & \#BH coh. \\
     \hline
     0 & 4 & 4 & 1 & 0 & 0 & 0 & 0 & 0 \\
     $\frac12$ & $\frac72$ & 5 & 9 & 1 & 1 & 0 & 0 & 0 \\
     1 & 3 & 6 & 91 & 8 & 8 & 0 & 0 & 0 \\
     $\frac32$ & $\frac52$ & 7 & 511 & 85 & 83 & 2 & 2 & 0 \\
     2 & 2 & 8 & 1369 & 445 & 426 & 19 & 19 & 0 \\
     $\frac52$ & $\frac32$ & 9 & 1898 & 953 & 924 & 29 & 28 & 1 \\
     3 & 1 & 10 & 1456 & 961 & 945 & 16 & 16 & 0 \\
     $\frac72$ & $\frac12$ & 11 & 633 & 505 & 495 & 10 & 10 & 0 \\
     4 & 0 & 12 & 136 & 136 & 128 & 8 & 8 & 0 \\
     \hline
\end{tabular}
\caption{ 
For each charge sector $R=Q_1=Q_2=Q_3$ and $J$ at level $\cJ=24$,
we present the numbers of operators discussed in the text.
The last column shows that \eqref{Q-coho} is the only black hole cohomology
in the target charge sector.}
\label{tab:countall}
\end{center}
\end{table}

From the fact that we have constructed and counted all
operators and their $Q$-actions in the $Q_1 = Q_2 = Q_3$ charge sectors
at $\cJ=24$ order with the permutation property,
we can also prove the non-existence of any other $SU(3)$ singlet non-graviton
cohomology at $\cJ=24$ order.
Recall that the result of step 2 in the previous subsection is
a complete basis of all operators, in a given charge sector $(R,J)$
and with permutation property that is designed to include all $SU(3)$ singlets.
There are further linear relations between $Q$-actions of these basis operators, 
reducing the number of independent $Q$-exact operators at charge sector 
$(R+\frac{1}{2},J-\frac{1}{2})$ in step 3.
The reduced operators correspond to the $Q$-closed operators at charge sector $(R,J)$:
$$
\text{(\#closed)}_{(R,J)} = \text{(\#basis)}_{(R,J)} - \text{(\#exact)}_{(R+\frac12,J-\frac12)}~.
$$
Then the number of $Q$-cohomologies is given by
$$
\text{(\#coh.)}_{(R,J)} = \text{(\#closed)}_{(R,J)}-\text{(\#exact)}_{(R,J)}~.
$$
Meanwhile, we can also count the number of independent graviton cohomologies
in these charge sectors and with the same permutation property,
like we counted the full set of gravitons in subsection 3.1.
The number of non-graviton cohomologies is given by
$$
\text{(\#BH coh.)}_{(R,J)} = \text{(\#coh.)}_{(R,J)}-\text{(\#gravitons)}_{(R,J)}~.
$$
We present all the numbers mentioned in this paragraph in Table \ref{tab:countall}.
We find only one non-graviton cohomology in the $(R,J) = (\frac52,\frac32)$ sector,
which is the fermionic cohomology presented in \eqref{Q-coho}.
Since the operators with the permutation property in the $Q_1=Q_2=Q_3$
charge sectors include all $SU(3)$ singlets,
we conclude that \eqref{Q-coho} is the only non-graviton cohomology
that is an $SU(3)$ singlet at order $\cJ=24$.

The computation presented in this subsection, of constructing the basis operators
and counting independent ones between them and their $Q$-actions,
is essentially the sort of computation that was performed in \cite{Chang:2022mjp},
although we find our numerics-assisted approach to be more efficient.
Moreover, we have only performed this computation in the $Q_1=Q_2=Q_3$
charge sectors at $\cJ=24$ order in the BMN sector,
and further restricted to operators with certain permutation property.
This is because we focused on the $SU(3)$ singlet sector at order $\cJ=24$,
where the non-graviton index indicated the existence of a non-graviton cohomology.


\clearpage

\chapter{Conclusion}\label{sec:conc}

Throughout this dissertation, we have discussed various properties
of the supersymmetric states, or local BPS operators,
of superconformal field theories dual to AdS black holes in different dimensions
including AdS$_{3,4,5,7}$.
In particular we focussed on the AdS$_3$ black holes and
the dual $(4, 4)$ SCFT$_2$ with $\frac14$ of the supersymmetries,
and on the AdS$_5$ black holes and the dual 4d $\cN=4$
Super-Yang-Mills theory with $\frac{1}{16}$ of the supersymmetries.
The supersymmetric states subject to our study have been enumerated
using the index to account for the Bekenstein-Hawking entropy of the
dual black holes, which is a thermodynamic quantity,
in a statistical sense \`a la Boltzmann.
This dissertation went beyond the enumeration to study the macroscopic charges
of the supersymmetric ensemble, and to identify the black hole states.

In part I, we have done the followings.
\begin{itemize}
\item We argued that the AdS$_3$ black holes may be unstable under
decay into some particles.

\item We demonstrated clearly in a simple example with the $U(1)$ gauge group
for the 4d $\cN=4$ Super-Yang-Mills theory,
how complexification of chemical potentials may overcome the boson/fermion
cancellations in the index.

\item We addressed that the derivation of the black hole entropy by treating
the index as a BPS limit of the partition function can be applied to
AdS$_3$ black holes under intricate limiting procedures.

\item We gave a heuristic derivation of the supersymmetric charge constraints
on the AdS$_{3,4,5,7}$ black holes as the relations between macroscopic charges
of the supersymmetric ensembles of the free field theories.
\end{itemize}

In part II, we have done the followings.
Recall that the black hole cohomologies
i) are related to the BPS states in the strongly coupled field theory,
which are the dual black hole microstates, or
ii) are dual states to the smallest, the most quantum black holes in the quantum gravity theory.
\begin{itemize}
\item We wrote explicit expressions for possibly all black hole cohomologies
in the BMN sector of the 4d $\cN=4$ Super-Yang-Mills theory with the gauge group $SU(2)$.

\item We showed that there must be another set of four black hole cohomologies
in the $SU(2)$ theory at the charge $\cJ = 39$.

\item We observed the black hole partial no-hair behavior, that the black holes
abhor dressing by gravitons, especially by chiral primary black holes with little rotation.

\item We constructed a fermionic singlet black hole cohomology in the $SU(3)$ theory
at $\cJ = 24$, which is likely the smallest one in the theory.

\item We found that a fermionic triplet black hole cohomology in the $SU(4)$ theory
must appear at $\cJ = 28$.
\end{itemize}

Related either directly or indirectly to this dissertation,
we suggest several future research directions.
\begin{itemize}
\item Our heuristic derivation of the supersymmetric charge constraints
on the AdS$_{4,5,7}$ black holes may be made more rigorous.
We have already discussed several directions in section \ref{sec:ccdiscuss}.
For examples, we expect that inclusion of the gauge degrees of freedom
in a principled manner may explain the scaling of the charges by
appropriate powers of $N$.
Furthermore, it is possible that the $O(1)$ numerical factors appear while
connecting the free field theory results to the strongly coupled theory.

\item It was observed recently
\cite{Imamura:2021ytr,Gaiotto:2021xce,Murthy:2022ien,Imamura:2022aua}
that the index for the $\frac{1}{16}$-BPS states of the 4d $\cN=4$ Yang-Mills theory
admits the structure of an expansion of expansions.
There has been progress \cite{Aharony:2021zkr,Lee:2022vig,Beccaria:2023hip,
Eleftheriou:2023jxr,Beccaria:2024vfx,Kim:2024ucf,Deddo:2024liu}
in interpreting this expansion in terms of giant gravitons, or D3-branes in string theory
\cite{McGreevy:2000cw,Grisaru:2000zn,Hashimoto:2000zp}.
Given the coupling independence of the index,
this opens a new window to understand the structure of the black holes in the
quantum gravity theory, not limited to its supergravity approximation.

\item To complete our argument on the instability of the AdS$_3$ black hole,
we may need to address existence and abundance of particles
with the assumed charges.
Giant gravitons in the AdS$_3$ black hole background
\cite{Mandal:2007ug,Raju:2007uj}
can potentially play the role of these particles.
This is a work in progress with Finn Larsen.

\item It is interesting to slightly lift the focus from strictly supersymmetric black holes.
There has been considerable progress
\cite{Iliesiu:2020qvm,Heydeman:2020hhw,Boruch:2022tno}
regarding the spectrum of nearly extremal and nearly BPS black holes.
These works are based on the effective theory of the near-extremal black holes,
a gravity theory in the near-horizon geometry AdS$_2$
that is being slightly broken by a dilaton \cite{Larsen:2018iou,Ghosh:2019rcj,
Choi:2021nnq,Choi:2023syx,Turiaci:2023jfa,Cabo-Bizet:2024gny}.
This theory is approximated by the Jackiw-Teitelboim (JT) gravity 
\cite{Jackiw:1984je,Teitelboim:1983ux}
and its supersymmetric generalizations
\cite{Fu:2016vas,Forste:2017kwy,Forste:2017apw}
leading to the (super-)Schwarzian actions.
The fact that the (super-)JT gravity is solved at quantum level
\cite{Stanford:2017thb,Mertens:2017mtv} is leveraged into quantum corrections
to the low energy spectrum of the near-extremal and near-BPS black holes.
See \cite{Mertens:2022irh} for a review.
It will be valuable to work out the details of the effective theory,
including scrutinizing the validity of approximations used in the process
and examining the higher order effects,
and sketching out the pattern of symmetry breaking.
This is a work in progress with Sangmin Choi and Finn Larsen.

\item One may consider extending the results of Part II into
gauge groups with higher ranks and/or into higher levels of the charge,
to detect and construct more examples of black hole cohomologies.
This is certainly a desired progress, but limitations on the computing power
strongly suggests that one take a completely different approach.

\item Since the black hole cohomologies represent BPS states in the
weakly coupled gauge theory, it is important to make connections between
operators in the weakly and strongly coupled theories.
It is one of the major goals of the field of integrability.
For an example, see \cite{Budzik:2023vtr} that followed our work.
\end{itemize}

We hope that through the enormous effort from generations of physicists
that this dissertation joins, 
black holes shed bright light to quantum gravity.

\clearpage

%

\appendix 

\addcontentsline{toc}{chapter}{Appendix}

\let\oldaddtoc\addtocontents
\renewcommand{\addtocontents}[2]{%
    \expandafter\ifstrequal\expandafter{#1}{toc}%
    {
    \oldaddtoc{loa}{#2}}%
    {
    \oldaddtoc{#1}{#2}}
}
\chapter{Graviton Trace Relations}\label{sec:trrel}

In this appendix that is related to section \ref{sec:cohosu3} of the main part,
we first list the trace relations between the graviton cohomologies
in the BMN sector of the $SU(3)$ theory.
Then we construct the relations of relations at $\cJ=24$ which are singlets under 
the $SU(3) \subset SU(4)_R$ global symmetry.
These are the two sectors in which the index predicted
fermionic cohomologies in the $SU(3)$ singlet.
The results at $\cJ=24$ are used in section \ref{sec:Qclosed}
to construct the threshold cohomology.

The trace relations are the linear dependence between the multi-trace operators,
up to $Q$-exact operators, due to the finite size of the matrices.
In this appendix, we shall only consider the trace relations between gravitons.
Let us first arrange the trace relations by their level $\cJ$ and distinguish
them into two types; fundamental ones and the others.
The fundamental trace relations at level $\cJ$ cannot be written as
linear combinations of the trace relations at lower levels $\cJ' (< \cJ)$,
multiplied by the gravitons at level $\cJ-\cJ'$. 
All trace relations of gravitons can be expressed as linear combinations
of the fundamental trace relations with the coefficients being graviton cohomologies. 
We explicitly constructed the fundamental trace relations until certain levels, 
which will be presented below.

The single-trace generators of the $SU(3)$ BMN gravitons are given by
\begin{equation}\label{BMN-gen}
    \begin{aligned}
        u^{ij} \equiv & \; \textrm{tr}  \left(\phi^{(i} \phi^{j)}\right)\ ,\ 
        u^{ijk} \equiv \; \textrm{tr} \left(\phi^{(i} \phi^{j} \phi^{k)}\right)\ , \\
        {v^{i}}_j \equiv & \; \textrm{tr} \left( \phi^i \psi_{j}\right) - {\textstyle \frac{1}{3} }
        \delta^i_j\textrm{tr}\left(\phi^a \psi_{a}\right)\ , \
        {v^{ij}}_k \equiv \; \textrm{tr} \left(\phi^{(i} \phi^{j)} \psi_{k} \right) 
        -\! {\textstyle \frac{1}{4}} \delta^{i}_{k}  
        \textrm{tr} \left( \phi^{(j} \phi^{a)} \psi_{a} \right) 
        -\! {\textstyle \frac{1}{4}} \delta^{j}_{k}  \textrm{tr} \left( \phi^{(i} \phi^{a)} \psi_{a} \right)\ , \\
        w^i \equiv & \; \textrm{tr} \left(f \phi^i + {\textstyle \frac{1}{2}} \epsilon^{ia_1a_2} \psi_{a_1} \psi_{a_2}\right)\ ,\
        w^{ij} \equiv \; \textrm{tr}\left(f \phi^{(i} \phi^{j)} +\epsilon^{a_1a_2(i}\phi^{j)} \psi_{a_1} \psi_{a_2}\right)\ ,
    \end{aligned}
\end{equation}
where we suppressed the subscript of $u_n, v_n, w_n$
since it can be easily read off from the number of the indices.
Note that the $Q$-actions on $\phi, \psi, f$ are given by 
\begin{equation}
  Q \phi^m=0\ ,\quad Q\psi_{m}=-\frac{i}{2}\epsilon_{mnp}[\phi^n,\phi^p]\ ,\quad Qf=-i[\phi^m,\psi_{m}]\ .
\end{equation}
We would like to find the fundamental trace relations of \eqref{BMN-gen}.

It is helpful to start from the Gr\"obner basis for the trace relations. 
The Gr\"obner basis contains all fundamental trace relations. In general, the Gr\"obner basis 
also contains some non-fundamental trace relations. 
We shall obtain the fundamental trace relations from the Gr\"obner basis by induction.

At the lowest level of the trace relations, all of them are fundamental.
Namely, every generator of the Gr\"obner basis at such level are the fundamental relations.
For the $SU(3)$ theory, the lowest level is $\cJ=10$. 
In order to organize them into covariant forms in the $SU(3)$ global symmetry, 
we use the following computational strategy (which also proves useful at higher orders).
We list the polynomials of \eqref{BMN-gen} which have the same representations 
as the lowest fundamental trace relations at $\cJ=10$.
Among them, we should find particular linear combinations which vanish
when all off-diagonal elements of $\phi^m,\psi_m, f$ are turned off,
since the graviton trace relations vanish with diagonal fields. 
Once such combinations are identified, keeping $\phi,\psi,f$ general in this combination 
will yield the $Q$-exact operators for the lowest fundamental trace relations.
This way, we can find the fundamental trace relations at the lowest level.\footnote{There 
can be linear combinations which vanish even when the off-diagonal elements are turned on. 
In principle, they can also be the trace relations but most of them are just the identities that 
hold at arbitrary $N$. In practice, we only find them as mesonic identities between 
\eqref{BMN-gen} rather than the trace relations.}

Now, suppose that we found all fundamental trace relations until the level $\cJ$.
We can construct the fundamental trace relations at $\cJ+2$ as follows.
We first construct all non-fundamental trace relations at level $\cJ+2$ by
multiplying suitable graviton cohomologies to the fundamental ones below the level $\cJ$.
Not all of them are linearly independent so we should extract
a linearly independent set among them.
This lets us to compute the $SU(3)$ character of the non-fundamental trace relations
at level $\cJ+2$.
Next, we consider a union of the non-fundamental trace relations and
the Gr\"obner bases at level $\cJ+2$.
Note that the Gr\"obner basis will contain all fundamental trace relations
and some non-fundamental ones.
We extract a linearly independent set among such union,
which contains all fundamental and non-fundamental relations.
We also compute the $SU(3)$ character over them.
Finally, we subtract the former character from the latter,
which yields the $SU(3)$ character of the fundamental trace relations at level $\cJ+2$.
Then we list the multi-trace operators using \eqref{BMN-gen}
which can account for it as before. 
Among them, we find particular linear combinations which vanish
when all off-diagonal elements of $\phi,\psi, f$ are turned off,
and which are linearly independent from the non-fundamental trace relations
we constructed above.
The final results are the fundamental trace relations at level $\cJ+2$.
In this way, one can construct the fundamental trace relations inductively.

In principle, one can obtain all fundamental trace relations of gravitons
from the above induction. For the $SU(2)$ theory, it can be easily done.
We found a 66-dimensional Gr\"obner basis,
and there exist 48 fundamental trace relations among them.
However, for the $SU(3)$ theory, we could not do a similar calculation
since the construction of the Gr\"obner basis is time-consuming.
We constructed it only in two subsectors: 
(1) all trace relations between $u_2, u_3, v_2, v_3$, and 
(2) trace relations between $u_2, u_3, v_2, v_3, w_2, w_3$ until $\cJ \leq 20$. 
From the subsector (1), which has 1170 generators,
we obtained all fundamental trace relations between $u_2, u_3, v_2, v_3$,
i.e. the relations which do not involve $f$'s.
There are in total 287 relations whose lowest level is $\cJ=10$
and the highest level is $\cJ=30$.
On the other hand, from the subsector (2),
we could generate the fundamental trace relations involving $f$'s until $\cJ \leq 20$. 
There are in total 130 relations involving $f$'s between $14 \leq \cJ \leq 20$.
These are enough to construct relations of relations at $\cJ=24$.

Before presenting their explicit forms, we first explain our notation.
When we write down certain operator in the irreducible representation
$\mathbf{R}$ under $SU(3) \subset SU(4)_R$ as $O^{i_1i_2i_3...}_{j_1j_2j_3...}$,
the actual form of such an operator should be understood as
$O^{i_1i_2i_3...}_{j_1j_2j_3...}$ subtracted by its trace part to make it traceless, like 
\begin{equation}
    \begin{aligned}
        &[n,0] : O^{i_1i_2i_3\cdots i_n} \to O^{i_1i_2i_3\cdots i_n}\ , \qquad [0,n] : O_{i_1i_2i_3\cdots i_n} \to O_{i_1i_2i_3\cdots i_n}\ , \\
        &[1,1]: O^i_j \to O^i_j - {\textstyle \frac{1}{3}} \delta^i_j O^a_a\ , \\
        &[2,1]: O^{ij}_k \to O^{ij}_k -{\textstyle \frac{1}{2}} \delta^{(i}_kO^{j)a}_a 
        \ , \qquad [1,2]: O^{i}_{jk} \to 
        O^{i}_{jk} -{\textstyle \frac{1}{2}} \delta^{i}_{(j}O^{a}_{k)a} \ , \\
        &[3,1]: O^{ijk}_l \to O^{ijk}_l -{\textstyle \frac{3}{5}} \delta^{(i}_lO^{jk)a}_a \ , \qquad [1,3]: O^{i}_{jkl} \to O^{i}_{jkl} -{\textstyle \frac{3}{5}} \delta^{i}_{(j}O^{a}_{kl)a}\ , \\
        &[2,2]: O^{ij}_{kl} \to O^{ij}_{kl} - {\textstyle \frac{4}{5}} \delta^{(i}_{(k} O^{j)a}_{l)a} 
        + {\textstyle \frac{1}{10}} \delta^{(i}_{(k} \delta^{j)}_{l)} O^{a_1a_2}_{a_1a_2}\ , 
    \end{aligned}
\end{equation}
and so on. Here, $[\cdot ,\cdot ]$ are the Dynkin labels for $SU(3)$.

Below, we list the explicit forms of the fundamental trace relations according to
their level $\cJ$ and representation under $SU(3) \subset SU(4)_R$ as $t^\cJ [R_1',R_2']$.
The relations which do not involve $f$'s are given as follows:
{\allowdisplaybreaks
    \begin{align}\label{tr-rel}
        &t^{10} [1,2](u_2u_3): (R_{10}^{(0,0)})^i_{jk} = \epsilon_{a_1 a_2 (j} \epsilon_{k) b_1 b_2} u^{a_1 b_1} u^{i a_2 b_2} \nonumber\\
        &t^{12} [0,0](u_2u_2u_2) : R_{12}^{(0,0)} =  \epsilon_{a_1 a_2 a_3} \epsilon_{b_1 b_2 b_3} u^{a_1 b_1}u^{a_2 b_2}u^{a_3 b_3} \nonumber\\
        &t^{12} [2,2](u_2u_2u_2,u_3u_3) : (R_{12}^{(0,0)})^{ij}_{kl} = \epsilon_{a_1 a_2 (k} \epsilon_{l) b_1 b_2} \left( u^{a_1 b_1} u^{a_2 b_2} u^{ij} + 6 u^{a_1 b_1 (i} u^{j) a_2 b_2} \right)  \nonumber\\
        &t^{12} [0,3](u_2v_3): (R_{12}^{(0,1)})_{ijk} = \epsilon_{(i|a_1a_2}\epsilon_{|j|b_1b_2} u^{a_1b_1} {v^{a_2 b_2}}_{|k)} \nonumber\\
        &t^{12} [1,1](u_2v_3, u_3v_2) : (R_{12}^{(0,1)})^i_j =  \epsilon_{j a_1 a_2} \left( 4 u^{a_1 b} {v^{i a_2}}_{b} + 3u^{i a_1b} {v^{a_2}}_{b}\right) \nonumber\\
        &t^{12} [2,2](u_2v_3, u_3v_2) : (R_{12}^{(0,1)})^{ij}_{kl} =\epsilon_{a_1 a_2 (k}\left(u^{a_1(i}{v^{j)a_2}}_{l)}+
        u^{ij a_1}{v^{a_2}}_{l)}\right) \nonumber\\
        &t^{14} [1,0](u_2u_2v_2) : (R_{14}^{(0,1)})^i = \epsilon_{a_1a_2a_3} u^{i a_1} u^{b a_2} {v^{a_3}}_{b}  \nonumber\\
        &t^{14} [0,2](u_2u_2v_2, u_3v_3) : (R_{14}^{(0,1)})_{ij} = \epsilon_{a_1a_2 (i|}\left(\epsilon_{b_1b_2b_3} u^{a_1b_1} u^{a_2b_2}{v^{b_3}}_{|j)} -2 \epsilon_{|j) b_1b_2 } u^{a_1 b_1 c} {v^{a_2b_2}}_c \right) \nonumber\\
        &t^{14} [2,1](u_2u_2v_2, u_3v_3) : (R_{14}^{(0,1)})^{ij}_k =  \epsilon_{k a_1a_2} \left(3 u^{(a_1 b} u^{ij)} {v^{a_2}}_{b} + 4u^{a_1b} u^{a_2 (i} {v^{j)}}_b +24 u^{a_1 b (i} {v^{j) a_2}}_b \right)  \nonumber\\
        &t^{14} [1,3](u_2u_2v_2, u_3v_3) : (R_{14}^{(0,1)})^i_{jkl} =  \epsilon_{(j|a_1 a_2} \epsilon_{|k| b_1 b_2} \left(u^{a_1 b_1} u^{a_2 b_2} {v^i}_{|l)} + 6 u^{ia_1b_1} {v^{a_2b_2}}_{|l)}\right) \nonumber\\
        &t^{14} [3,2] (u_2u_2v_2, u_3v_3) : (R_{14}^{(0,1)})^{ijk}_{lm} = \epsilon_{a_1a_2(l} \left( u^{(a_1 i} u^{jk)} {v^{a_2}}_{m)} + 6 u^{a_1 (ij} {v^{k)a_2}}_{m)}  \right)  \nonumber\\
        &t^{14} [1,3](v_2v_3) : (R_{14}^{(0,2)})^{i}_{jkl} = \epsilon_{a_1 a_2 (j} {v^{a_1}}_{k} {v^{i a_2}}_{l)} \nonumber\\
        &t^{16}[0,1](u_2v_2v_2, v_3v_3):(R_{16}^{(0,2)})_i = \epsilon_{i a_1 a_2} \left(12u^{bc}{v^{a_1}}_b{v^{a_2}}_c +13u^{a_1 b}{v^{a_2}}_c{v^c}_b + 12{{v}^{a_1b}}_c{v^{a_2c}}_b \right) \nonumber\\
        &t^{16}[1,2] (u_2v_2v_2, v_3v_3): (R_{16}^{(0,2)})^i_{jk} = 
        \epsilon_{a_1a_2(j} \big(3u^{ib}{v^{a_1}}_{k)}{v^{a_2}}_b
        -7u^{ia_1}{v^b}_{k)}{v^{a_2}}_b \nn\\
        & \hspace{0.6\textwidth} + 6u^{a_1b}{v^i}_{k)}{v^{a_2}}_b 
        + 24{v^{a_1b}}_{k)}{v^{ia_2}}_b \big) \nonumber\\
        &t^{16}[2,3] (u_2v_2v_2,v_3v_3): (R_{16}^{(0,2)})^{ij}_{klm} = \epsilon_{a_1a_2(k}\left({u^{a_1(i}{v^{j)}}_{l}v^{a_2}}_{m)}+ 3{v^{a_1(i}}_{l}{v^{j)a_2}}_{m)}\right) \nonumber\\
        &t^{18}[0,0](u_3v_2v_2): R_{18}^{(0,2)} = \epsilon_{a_1a_2a_3} u^{a_1bc} {v^{a_2}}_b {v^{a_3}}_c \nonumber\\
        &t^{20}[1,0](v_2v_2v_3) : (R_{20}^{(0,3)})^i = 2{v^a}_c{v^b}_a{v^{ic}}_b - 3{v^i}_a{v^c}_b{v^{ab}}_c \nonumber\\
        &t^{22}[2,0](u_2v_2v_2v_2) : (R_{22}^{(0,3)})^{ij} = u^{ij}{v^a}_b{v^b}_c{v^c}_a -3 u^{a(i}{v^{j)}}_b{v^b}_c{v^c}_a +3 u^{ab}{v^{(i}}_a{v^{j)}}_c{v^c}_b \nonumber\\
        &t^{24}[0,0] (u_2v_2v_2v_3) : R_{24}^{(0,3)} =  \epsilon_{a_1a_2a_3}u^{a_1b}{v^{a_2}}_b{v^{a_3c}}_{d}{v^{d}}_c \nonumber\\
        &t^{26}[1,0] (v_2v_2v_2v_3): (R_{26}^{(0,4)})^i= {v^{i}}_a {v^a}_b{v^d}_c{v^{bc}}_d 
        \nonumber\\
        &t^{30}[0,0] (v_2v_2v_2v_2v_2) : R_{30}^{(0,5)} =  {v^{a}}_b{v^{b}}_c{v^{c}}_d{v^{d}}_e{v^{e}}_a \nonumber\\
        &t^{30}[3,0] (v_2v_2v_2v_2v_2) : (R_{30}^{(0,5)})^{ijk}= \epsilon^{a_1a_2(i}{v^{j}}_{a_1}{v^{k)}}_{a_2} {v^{b}}_c{v^{c}}_d{v^{d}}_b\ .
\end{align}
}
Here, the superscripts of $R$ denote $(n_f, n_\psi)$ of the terms with
maximal $n_f$ in the trace relations and the subscripts denote their $\cJ$.
Their $SU(3)$ representations can be read off from the number of upper and lower indices.
The listed trace relations vanish up to $Q$-exact operators whose explicit form
will be discussed below.
As explained before, this is the exhaustive set of the fundamental trace relations
of gravitons which do not involve $f$'s.
The fundamental trace relations involving $f$'s until $\cJ = 20$ are given by
{\allowdisplaybreaks
    \begin{align}\label{tr-rel-f} 
        &t^{14}[0,2](v_2v_3,u_2w_3) : (R_{14}^{(1,0)})_{ij}  = \epsilon_{a_1a_2(i} \left(8{v^{a_1 b}}_{j)} {v^{a_2}}_b + 5\epsilon_{j)b_1b_2}u^{a_1b_1}w^{a_2b_2}\right) \nonumber\\
        &t^{14}[2,1](v_2v_3,u_2w_3,u_3w_2) : \nn\\
        & \hspace{.5cm} (R_{14}^{(1,0)})^{ij}_k =  2 {v^{(i}}_a{v^{j)a}}_k -5{v^{a}}_k{v^{ij}}_a +3\epsilon_{ka_1a_2}u^{a_1(i}w^{j)a_2} + 3\epsilon_{ka_1a_2}u^{ija_1}w^{a_2} \nonumber\\
        &t^{16}[0,1] (v_3v_3, u_2v_2v_2, u_2u_2w_2) : \nonumber\\
        &\hspace{.5cm} (R_{16}^{(1,0)})_i = 
        \epsilon_{ia_1a_2}\left(48 {v^{a_1b_1}}_{b_2} {v^{a_2b_2}}_{b_1} 
        + 9 u^{b_1b_2} {v^{a_1}}_{b_1} {v^{a_2}}_{b_2} 
        - 13 \epsilon_{b_1b_2b_3} u^{a_1b_1} u^{a_2b_2} w^{b_3}\right) \nonumber\\
        &t^{16}[1,2] (v_3v_3, u_2v_2v_2, u_3w_3, u_2u_2w_2) :\nonumber\\
        &\hspace{.5cm}
        (R_{16}^{(1,0)})^i_{jk}= 
        \epsilon_{a_1a_2(j|} \left(24{v^{i a_1}}_b {v^{ba_2}}_{|k)} +
        2u^{i a_1} {v^{a_2}}_b {v^{b}}_{|k)} \!-\! 6u^{a_1 b} {v^{a_2}}_b {v^{i}}_{|k)}
        \right.\nonumber\\ 
        &\hspace{2.5cm}
        \left.+6\epsilon_{|k)b_1b_2}u^{ia_1b_1} w^{a_2b_2} +\epsilon_{|k)b_1b_2}u^{a_1b_1} u^{a_2b_2} w^i \right)  \nonumber\\
        &t^{16}[3,1] (v_3v_3, u_2v_2v_2, u_3w_3, u_2u_2w_2) :\nonumber\\
        &\hspace{.5cm}
        (R_{16}^{(1,0)})^{ijk}_l = 24 {v^{(ij}}_a {v^{k)a}}_l + 7 u^{(ij}{v^{k)}}_a {v^{a}}_l -6u^{a(i}{v^{j}}_a {v^{k)}}_l \nn\\
        &\hspace{2.5cm} + 18 \epsilon_{la_1a_2} u^{a_1(ij} w^{k)a_2} +3\epsilon_{la_1a_2} u^{(ij} u^{k)a_1} w^{a_2} \nonumber\\
        &t^{16}[1,2](v_2w_3, v_3w_2) : (R_{16}^{(1,1)})^i_{jk} = \epsilon_{a_1a_2(j}\left({v^{a_1}}_{k)}w^{a_2i} + {v^{a_1i}}_{k)}w^{a_2} \right) \nonumber\\
        &t^{18}[0,0](v_2v_2v_2,u_2v_2w_2) : R_{18}^{(1,1)} = {v^{a_1}}_{a_2}{v^{a_2}}_{a_3}{v^{a_3}}_{a_1} -3\epsilon_{a_1a_2a_3} u^{a_1 b}{v^{a_2}}_{b} w^{a_3} \nonumber\\
        &t^{18}[1,1](v_2v_2v_2,v_3w_3,u_2v_2w_2) : \nonumber\\
        &\hspace{.5cm}
        (R_{18}^{(1,1)})^i_j = 9{v^{i}}_{a_1} {v^{a_1}}_{a_2} {v^{a_2}}_{j}-24\epsilon_{ja_1a_2} {v^{ia_1}}_{b} w^{b a_2} \nonumber\\
        &\hspace{2.5cm} -13 \epsilon_{ja_1a_2} u^{ia_1} {v^{a_2}}_{b} w^b -16\epsilon_{ja_1a_2} u^{ib} {v^{a_1}}_{b} w^{a_2} +5\epsilon_{ja_1a_2} u^{a_1b} {v^{i}}_{b} w^{a_2}  \nonumber\\
        &t^{18}[0,3](v_2v_2v_2,v_3w_3,u_2v_2w_2) : \nonumber\\
        &\hspace{.5cm} 
        (R_{18}^{(1,1)})_{ijk} =  
        \epsilon_{a_1a_2(i|}\! \left(3{v^{a_1}}_{|j|}{v^{a_2}}_{b} {v^{b}}_{|k)} 
        \!-\! 3\epsilon_{b_1b_2|j}{v^{a_1b_1}}_{k)} w^{a_2b_2} 
        \!-\! \epsilon_{b_1b_2|j} u^{a_1b_1} {v^{a_2}}_{k)} w^{b_2}\right) \nonumber\\
        &t^{18}[2,2](v_2v_2v_2,v_3w_3,u_2v_2w_2) : \nn\\
        &\hspace{.5cm} (R_{18}^{(1,1)})_{ij}^{kl} = 2{v^{(i}}_{a}{v^{j)}}_{(k} {v^{a}}_{l)} -6 \epsilon_{a_1a_2(k} {v^{a_1 (i}}_{l)}w^{j)a_2}-\epsilon_{a_1a_2(k} u^{ij}{v^{a_1}}_{l)}w^{a_2}  \nonumber\\
        &t^{20}[0,2](v_2v_2w_2,u_2w_2w_2,w_3w_3): \nn\\
        &\hspace{.5cm} (R_{20}^{(2,0)})_{ij} = 2\epsilon_{a_1a_2(i|}{v^{a_1}}_{b}{v^{b}}_{|j)} w^{a_2} - 3 \epsilon_{a_1a_2a_3} {v^{a_1}}_{i}{v^{a_2}}_{j}w^{a_3} 
        \nonumber\\
        &\hspace{2.5cm} + \epsilon_{ia_1a_2}\epsilon_{jb_1b_2}u^{a_1b_1}w^{a_2} w^{b_2} + 3 \epsilon_{ia_1a_2}\epsilon_{jb_1b_2}w^{a_1b_1}w^{a_2b_2} \ .
    \end{align}
}
The relations involving one $f$ appear from $\cJ=14$ and those involving
two $f$'s appear from $\cJ=20$.
We do not find any relations involving three $f$'s until $\cJ = 20$.

As explained before, the trace relations \eqref{tr-rel}, \eqref{tr-rel-f} vanish
up to $Q$-exact operators, which we now construct explicitly.
In principle, one should first construct the complete basis of the $Q$-exact operators,
which have the same level $\cJ$ and $SU(3)$ representation with the target trace relation.
(The $Q$-action does not change $\cJ$ and $SU(3)$ representation.)
However, in practice, we can make some ans\"atze for the $Q$-exact form to
reduce the dimension of $Q$-exact basis.
One of our working assumptions is that the maximal number of $f$'s appearing
before the $Q$-action is the same as that of the trace relation.
There is a priori no reason to assume that but it turns out to be true for our examples.
After imposing this assumption (and a couple of extra practical assumptions),
we find a particular linear combination of the $Q$-exact operators in our basis
for the target trace relation.
In general, when we write $R_I\sim Qr_I$ for a trace relation $R_I$,
there exist ambiguities of $r_I$ since we can add arbitrary $Q$-closed operators to $r_I$.
We partly fix them by requiring $r_I$ to vanish when $\phi,\psi,f$ are restricted 
to diagonal matrices.
We do not know whether such a requirement can be satisfied in general,
but it does work for our examples.
The purpose of this requirement will be explained later.
The other ambiguities are fixed by hand to get compact expressions.

Below, we list the operators $r_j^{(n_f,n_\psi)}$ related to the fundamental trace relations
$R_j^{(n_f,n_\psi-1)}$ by $i \, Q r_j^{(n_f,n_\psi)} = R_j^{(n_f,n_\psi-1)}$.
We will not list all $r_j^{(n_f,n_\psi)}$'s, but only those which are
used in section \ref{sec:cohosu3}.
For the relations without $f$'s in \eqref{tr-rel}, we obtain
{\allowdisplaybreaks
\begin{align}\label{tr-r}
        &(r_{10}^{(0,1)})^i_{jk} =  -2\, \epsilon_{a_1a_2(j} \tr \left( \phi^{a_1}\phi^{a_2}\phi^{i}\psi_{k)} \right)\ ,  \nonumber\\
        &r_{12}^{(0,1)} = \epsilon_{a_1a_2a_3} \left[ 6 \tr\left( \psi_b \phi^{a_1} \right) \tr \left( \phi^b \phi^{a_2} \phi^{a_3}\right) - \tr\left( \psi_b \phi^{a_1} \phi^{a_2}\right) \tr \left( \phi^b  \phi^{a_3}\right)  \right] \nonumber\\
        &\qquad \quad  -3\epsilon_{a_1a_2a_3} \big[
        \tr\left(\psi_b \phi^b \phi^{a_1} \phi^{a_2} \phi^{a_3}\right)
        +\tr\left(\psi_b \phi^{a_1} \phi^b \phi^{a_2} \phi^{a_3}\right) \nn\\
        &\hspace{5cm}+\tr\left(\psi_b \phi^{a_1} \phi^{a_2}\phi^b  \phi^{a_3}\right)
        +\tr\left(\psi_b\phi^{a_1} \phi^{a_2} \phi^{a_3} \phi^b \right)\big] \ , 
        \nonumber\\
        &(r_{12}^{(0,1)})^{ij}_{kl} =  -2\epsilon_{a_1a_2(k}\left[ \tr\left(\psi_{l)}\phi^{(i}\phi^{j)}\phi^{a_1}\phi^{a_2}\right) +7\tr\left(\psi_{l)}\phi^{(i|}\phi^{a_1}\phi^{|j)}\phi^{a_2}\right) \right]\ , 
        \nonumber\\
        &(r_{12}^{(0,2)})_{ijk} = \frac{1}{2} \epsilon_{a_1 a_2 (i} \tr \left( \phi^{a_1} \psi_{j}\phi^{a_2} \psi_{k)} \right) \ , \nonumber\\
        &(r_{12}^{(0,2)})^{i}_{j} =  6 \tr \left( \phi^{(i} \phi^{a)} \psi_{(a} \psi_{j)}\right) - 5 \tr\left(\phi^{[i} \psi_{a}\phi^{a]} \psi_{j}\right)\ ,
        \nonumber\\
        &(r_{12}^{(0,2)})^{ij}_{kl} =   \tr \left(\phi^{(i}\phi^{j)} \psi_{(k}\psi_{l)}\right)\ , \nonumber\\
        &(r_{14}^{(0,2)})^i  =3\, \tr \left( \phi^i \psi_{a_1} \phi^{a_1} \phi^{a_2} \psi_{a_2} \right) + 2\, \tr \left( \phi^i \phi^{a_1} \right) \tr\left( \phi^{a_2} \psi_{(a_1} \psi_{a_2)} \right) \nonumber\\ 
        & \hspace{4cm} -6\, \tr\left( \phi^i \psi_{a_1} \right) \tr \left( \phi^{[a_1} \phi^{a_2]} \psi_{a_2} \right) -\, \tr \left(\phi^i \psi_{a_1}\psi_{a_2} \right) \tr\left(\phi^{a_1} \phi^{a_2}\right)\ , \nonumber\\
        &(r_{14}^{(0,2)})_{ij}  =   \frac{5}{9}\epsilon_{a_1a_2a_3}\left[2 \tr\left(\psi_{(i}\psi_{j)} \phi^{a_1} \phi^{a_2}\phi^{a_3}\right) + \tr\left(\psi_{(i} \phi^{a_1}\psi_{j)}\phi^{a_2}\phi^{a_3}\right)\right] \nonumber\\
        &\qquad \quad +\epsilon_{a_1a_2(i}\left[\tr\left(\psi_{j)} \psi_{a_3}\phi^{a_1}\phi^{a_2}\phi^{a_3}\right)+\tr\left(\psi_{j)} \psi_{a_3}\phi^{a_1}\phi^{a_3}\phi^{a_2}\right)+\tr\left(\psi_{j)} \psi_{a_3}\phi^{a_3}\phi^{a_1}\phi^{a_2}\right)\right] \nonumber\\
        &\qquad \quad +\epsilon_{a_1a_2(i}\left[\tr\left(\psi_{j)} \phi^{a_1}\psi_{a_3}\phi^{a_2}\phi^{a_3}\right)+\tr\left(\psi_{j)} \phi^{a_1}\psi_{a_3}\phi^{a_3}\phi^{a_2}\right)+\tr\left(\psi_{j)} \phi^{a_3}\psi_{a_3}\phi^{a_1}\phi^{a_2}\right)\right] \nonumber\\
        &\qquad \quad +\epsilon_{a_1a_2(i}\left[\tr\left(\psi_{j)} \phi^{a_1}\phi^{a_2}\psi_{a_3}\phi^{a_3}\right)+\tr\left(\psi_{j)} \phi^{a_1}\phi^{a_3}\psi_{a_3}\phi^{a_2}\right)+\tr\left(\psi_{j)} \phi^{a_3}\phi^{a_1}\psi_{a_3}\phi^{a_2}\right)\right] \nonumber\\
        &\qquad \quad +\epsilon_{a_1a_2(i}\left[\tr\left(\psi_{j)} \phi^{a_1}\phi^{a_2}\phi^{a_3}\psi_{a_3}\right)+\tr\left(\psi_{j)} \phi^{a_1}\phi^{a_3}\phi^{a_2}\psi_{a_3}\right)+\tr\left(\psi_{j)} \phi^{a_3}\phi^{a_1}\phi^{a_2}\psi_{a_3}\right)\right] \nonumber\\
        &\qquad \quad -\frac{1}{3}\epsilon_{a_1a_2(i}\big[
        5\tr\left(\psi_{j)}\phi^{a_1}\phi^{a_2}\right)\tr\left(\psi_{a_3}\phi^{a_3}\right)
        +2\tr\left(\psi_{j)}\phi^{(a_1}\phi^{a_3)}\right)\tr\left(\psi_{a_3}\phi^{a_2}\right) \nn\\
        & \hspace{9cm} -2 \tr\left(\psi_{j)}\phi^{a_2}\right)\tr\left(\psi_{a_3}\phi^{(a_1}\phi^{a_3)}\right) \big]\ , \nonumber\\
        &(r_{14}^{(0,2)})^{ij}_{k}  =
        12\tr\left(\phi^{(i} \phi^{a} \phi^{j)} \psi_{(a} \psi_{k)}\right)
        +12\tr\left(\phi^{(i|} \phi^{a} \phi^{|j)} \psi_{(a} \psi_{k)}\right) \nn\\
        & \hspace{4cm} +54 \tr\left( \phi^{(i} \phi^{j} \psi_{(a} \phi^{a)} \psi_{k)} \right)
        -36 \tr\left( \phi^{(i} \phi^{j)} \psi_{(a} \phi^a \psi_{k)} \right) \ ,\nonumber\\
        &(r_{14}^{(0,2)})^i_{jkl}  = 2\epsilon_{a_1a_2(j}\left[\tr\left(\phi^i \phi^{a_1} \phi^{a_2}\psi_k \psi_{l)} \right) +3\tr\left(\phi^i \phi^{a_1} \psi_k \phi^{a_2} \psi_{l)} \right)-2\tr\left(\phi^i  \psi_k \phi^{a_1} \phi^{a_2} \psi_{l)}\right)\right]\ , 
        \nonumber\\
        &(r_{14}^{(0,3)})^i_{jkl}  =  -\frac{1}{2} \tr \left(\phi^i \psi_{(j}\psi_{k}\psi_{l)} \right)\ , \nonumber\\
        &(r_{16}^{(0,3)})_i = 
        \frac{39}{4} \tr \left( \psi_i \{\psi_{b_1}\psi_{b_2}, \phi^{b_1}\phi^{b_2}\} \right)
        +2\tr \left( \psi_i \psi_{b_1} \phi^{b_1}\psi_{b_2}\phi^{b_2} \right)
        - \frac{61}{4}\tr \left( \psi_i \psi_{b_1} \phi^{b_2}\psi_{b_2}\phi^{b_1} \right) \nn\\
        & \hspace{1.8cm}
        + \frac{97}{4}\tr \left( \psi_i  \phi^{b_1}\psi_{b_1}\psi_{b_2}\phi^{b_2} \right)  
        -\frac{41}{4} \tr \left( \psi_i  \phi^{b_2}\psi_{b_1}\psi_{b_2}\phi^{b_1} \right)
        -5 \tr \left( \psi_i  \psi_{b_1}\phi^{b_1}\phi^{b_2}\psi_{b_2} \right) \nn\\
        & \hspace{1.8cm}
        -\frac{25}{2} \tr \left( \psi_i  \psi_{b_1}\phi^{b_2}\phi^{b_1}\psi_{b_2} \right)
        +2\tr \left( \psi_i  \phi^{b_1}\psi_{b_1}\phi^{b_2}\psi_{b_2} \right)
        - \frac{61}{4}\tr \left( \psi_i  \phi^{b_2}\psi_{b_1}\phi^{b_1}\psi_{b_2} \right) \nn\\
        & \hspace{1.8cm}
        - \frac{11}{4} \tr \left(  \phi^{b_1}\phi^{b_2}\right) \tr \left(\psi_i \psi_{b_1}\psi_{b_2} \right)
        - \frac{27}{2} \tr \left( \psi_{b_1}\psi_{b_2} \right) \tr \left(\psi_i \phi^{b_1}\phi^{b_2} \right) \nn\\
        & \hspace{1.8cm}
        + \frac{29}{4} \tr \left(\phi^{b_2} \psi_{b_2} \right) \tr \left(\psi_i [\psi_{b_1},\phi^{b_1}] \right)\ , \nonumber\\
        &(r_{16}^{(0,3)})^i_{jk} =
        2 \tr \left( \psi_{(j} \psi_{k)} \psi_b \phi^b \phi^i \right)
        -4\tr \left( \psi_{(j} \psi_{k)} \psi_b \phi^i \phi^b \right)
        - \tr \left( \psi_{(j|} \psi_b \psi_{|k)} \{\phi^b ,\phi^i\} \right) \nn\\
        & \hspace{1.8cm}
        -4 \tr \left( \psi_{(j} \psi_{k)} \phi^{(b}\psi_b  \phi^{i)} \right)
        +7\tr \left( \psi_{(j|} \{\psi_b, \phi^{b}\}\psi_{|k)}  \phi^{i} \right)
        -11\tr \left( \psi_{(j|} \{\psi_b, \phi^{i}\}\psi_{|k)}  \phi^{b} \right) \nn\\
        & \hspace{1.8cm}
        -4\tr \left( \psi_{(j} \psi_{k)}  \phi^b \phi^i \psi_b\right)
        +2\tr \left( \psi_{(j} \psi_{k)}  \phi^i \phi^b \psi_b\right) \nn\\
        & \hspace{1.8cm}
        +3 \tr \left( \psi_{(j|} \psi_b \right) \tr \left( \psi_{|k)} [\phi^b, \phi^i] \right)
        +6 \tr \left( \psi_{(j} \phi^{[b} \right) \tr \left( \{\psi_{k)},\psi_b\} \phi^{i]} \right) \ .
\end{align}
}
For the relations involving $f$'s in \eqref{tr-rel-f}, we find
{\allowdisplaybreaks
    \begin{align}\label{tr-r-f}
        (r_{14}^{(1,1)})_{ij} =&
        5 \epsilon_{a_1a_2(i} \tr \left(f \phi^{a_1}\psi_{j)}\phi^{a_2}\right) + \tr \left( \phi^a \left\{\psi_a , \psi_{(i} \psi_{j)}\right\} \right) -4 \, \tr \left(\phi^a \psi_{(i|} \psi_{a} \psi_{|j)}\right) \ , \nn \\
        (r_{14}^{(1,1)})^{ij}_k =&
        3 \, \tr \left(f\phi^{(i} \phi^{j)}\psi_k\right)
        - 3 \, \tr \left(f\psi_k\phi^{(i} \phi^{j)}\right)
        \nn\\ &
        + \epsilon^{a_1 a_2 (i} \tr \left( \phi^{j)} \psi_k \psi_{a_1} \psi_{a_2}\right)
        - \epsilon^{a_1 a_2 (i} \tr \left( \phi^{j)} \psi_{a_1} \psi_{a_2} \psi_k \right) \ , \nn \\
        (r_{16}^{(1,1)})_i =&
        13 \epsilon_{a_1a_2a_3} \tr \left(f \psi_i\right) \tr \left(\phi^{a_1}\phi^{a_2}\phi^{a_3}\right)
        + \frac{10}{3} \epsilon_{a_1a_2a_3} \tr \left(f \phi^{a_1}\right) \tr \left(\psi_i\phi^{a_2}\phi^{a_3}\right)
        \nn\\ &
        + \frac{10}{3} \epsilon_{a_1a_2a_3} \tr \left(f \phi^{a_1}\phi^{a_2}\right) \tr \left(\psi_i\phi^{a_3}\right)
        +46 \epsilon_{ia_1a_2} \tr \left(f \phi^{b}\right) \tr \left(\psi_b\phi^{a_1}\phi^{a_2}\right)
        \nn\\ &
        - 7\epsilon_{ia_1a_2} \tr \left(f \phi^{a_1}\right) \tr \left(\psi_b\phi^{a_2}\phi^{b}\right)
        -7\epsilon_{ia_1a_2} \tr \left(f \phi^{b}\phi^{a_1}\right) \tr \left(\psi_b\phi^{a_2}\right)
        \nn\\ &
        +6\epsilon_{ia_1a_2} \tr \left(f \phi^{a_1}\phi^{a_2}\right) \tr \left(\psi_b\phi^{b}\right)
        - \frac{115}{3} \epsilon_{a_1a_2a_3} \tr\left(f\psi_i \phi^{a_1} \phi^{a_2} \phi^{a_3}\right)
        \nn\\ &
        - \frac{95}{3}\epsilon_{a_1a_2a_3} \tr\left(f\phi^{a_1}\psi_i  \phi^{a_2} \phi^{a_3}\right)
        +5 \epsilon_{a_1a_2a_3} \tr\left(f\phi^{a_1} \phi^{a_2} \psi_i \phi^{a_3}\right)
        \nn\\ &
        +36 \epsilon_{ia_1a_2} \tr\left(f\psi_{b} \phi^{a_1} \phi^{a_2} \phi^{b}\right)
        -43\epsilon_{ia_1a_2} \tr\left(f\psi_{b} \phi^{a_1} \phi^{b} \phi^{a_2}\right)
        \nn\\ &
        +39\epsilon_{ia_1a_2} \tr\left(f \phi^{a_1}\psi_{b} \phi^{a_2} \phi^{b}\right)
        -68\epsilon_{ia_1a_2} \tr\left(f \phi^{a_1}\phi^{a_2}\psi_{b}  \phi^{b}\right)
        \nn\\ &
        + 39\epsilon_{ia_1a_2}\tr\left(f \phi^{a_1}\phi^{b}\psi_{b}  \phi^{a_2}\right) 
        + 13\tr \left( \psi_i\{\psi_{b_1}\psi_{b_2},\phi^{b_1}\phi^{b_2}\} \right)
        \nn\\ &
        -31\tr \left( \psi_i\{\psi_{b_1}\psi_{b_2},\phi^{b_2}\phi^{b_1}\} \right)
        + 14\tr \left( \psi_i\psi_{b_1}\phi^{b_1}\psi_{b_2}\phi^{b_2} \right)
        \nn\\ &
        -22\tr \left( \psi_i\psi_{b_1}\phi^{b_2}\phi^{b_1}\psi_{b_2} \right)
        +14\tr \left( \psi_i\phi^{b_1}\psi_{b_1}\phi^{b_2}\psi_{b_2} \right) \ ,
        \nn \\
        (r_{16}^{(1,1)})^i_{jk} =&
        \epsilon_{a_1a_2(j}\left[-4 \tr \left( f \phi^i\right) \tr \left( \psi_{k)} \phi^{a_1} \phi^{a_2}\right)
        - \tr \left( \phi^i \phi^{a_2} \right) \tr \left( f [\psi_{k)}, \phi^{a_1}]\right) \right]
        \nn\\ &
        +\epsilon_{a_1a_2(j} \big[
        3 \tr \left( f\phi^{a_1}\{ \psi_{k)} ,\phi^i\} \phi^{a_2} \right)
        +5 \tr \left(f \{\psi_{k)}, \phi^{a_1} \phi^i \phi^{a_2}\}\right)
        \nn\\ & \hspace{2cm}
        -4 \tr\left( f \psi_{k)} \phi^i \phi^{a_1} \phi^{a_2}\right)
        -4 \tr\left( f \phi^{a_1} \phi^{a_2} \phi^i \psi_{k)}\right)\big]
        \nn\\ &
        +2\tr\left( \psi_{(j} \psi_{k)} \psi_b [\phi^b, \phi^i] \right)
        -3 \tr \left( \psi_{(j|} \psi_b \psi_{|k)} \{\phi^b, \phi^i\} \right)
        +6 \tr \left( \psi_{(j|}\{\psi_b, \phi^b\} \psi_{|k)} \phi^i\right)
        \nn\\ &
        -9 \tr \left(\psi_{(j|}\{\psi_b, \phi^i\} \psi_{|k)} \phi^b  \right)
        -2\tr \left( \psi_{(j} \psi_{k)} \left[\phi^b, \phi^i\right] \psi_b \right)
        \nn\\ &
        + \tr \left( \psi_{(j|} \psi_b\right) \tr \left( \psi_{|k)} [\phi^b, \phi^i] \right)
        + \tr \left(\psi_{(j|}\phi^b\right)\tr\left(\{\psi_{|k)},\psi_b\}\phi^i\right)\ , \nn \\
        (r_{16}^{(1,2)})^i_{jk} =& -\frac{1}{2} \tr\left(f\phi^i\psi_{(j} \psi_{k)} \right)
        -\frac{1}{2} \tr\left(f\psi_{(j}\phi^i \psi_{k)} \right)
        \nn\\ &
        -\frac{1}{2} \tr\left(f\psi_{(j} \psi_{k)}\phi^i \right)
        -\frac{1}{4} \epsilon^{ia_1a_2}\tr \left(\psi_{a_1}\psi_{a_2}\psi_{(j}\psi_{k)}\right)  \ , \nn \\
        (r_{18}^{(1,2)})^i_j =&
        -4\tr\left(f \phi^i \phi^a\right)\tr\left(\psi_j\psi_a\right)
        -5 \tr\left(f \phi^a \phi^i\right)\tr\left(\psi_j\psi_a\right) 
        -\frac{53}{2}\tr\left(f \phi^i \psi_j\right)\tr\left(\phi^a\psi_a\right)
        \nn\\ &
        + 7\tr\left(f \phi^i \psi_a\right)\tr\left(\phi^a\psi_j\right)
        + \frac{15}{2}\tr\left(f \phi^a \psi_j\right)\tr\left(\phi^i\psi_a\right)
        +12 \tr\left(f \phi^a \psi_a\right)\tr\left(\phi^i\psi_j\right)
        \nn\\ &
        +2  \tr\left(f \psi_j\phi^i \right)\tr\left(\phi^a\psi_a\right)
        -13\tr\left(f \psi_a\phi^i \right)\tr\left(\phi^a\psi_j\right) 
        +4\tr\left(f \psi_j\phi^a \right)\tr\left(\phi^i\psi_a\right)
        \nn\\ &
        +6 \tr\left(f  \psi_j\psi_a\right)\tr\left(\phi^i\phi^a\right)
        + \frac{13}{2}\tr\left(f  \psi_a\psi_j\right)\tr\left(\phi^i\phi^a\right)
        -4 \tr\left(f \phi^i \right)\tr\left(\phi^a\psi_j\psi_a\right)
        \nn\\ &
        + 14\tr\left(f \phi^i \right)\tr\left(\phi^a\psi_a\psi_j\right)
        -8\tr\left(f \phi^a \right)\tr\left(\phi^i\psi_j\psi_a\right)
        -8 \tr\left(f \phi^a \right)\tr\left(\phi^i\psi_a\psi_j\right)
        \nn \\ &
        -4 \tr\left(f \psi_j \right)\tr\left(\psi_a\phi^i\phi^a\right)
        -9\tr\left(f \psi_a \right)\tr\left(\psi_j\phi^i\phi^a\right)
        +6 \tr\left(f \psi_a \right)\tr\left(\psi_j\phi^a\phi^i\right)
        \nn \\ &
        +3 \tr\left(f \phi^i \phi^a\psi_j\psi_a\right)
        - \frac{31}{2} \tr\left(f \phi^i \phi^a\psi_a\psi_j\right)
        +3\tr\left(f \phi^a \phi^i\psi_j\psi_a\right) 
        \nn \\ &
        +\frac{5}{2}\tr\left(f \phi^a \phi^i\psi_a\psi_j\right)
        +12\tr\left(f \phi^i \psi_j\phi^a\psi_a\right)
        -\frac{13}{2}\tr\left(f \phi^i \psi_a\phi^a\psi_j\right)
        \nn \\ &
        -6\tr\left(f \phi^a \psi_j\phi^i\psi_a\right)
        -\frac{13}{2}\tr\left(f \phi^a \psi_a\phi^i\psi_j\right)
        +18 \tr\left(f \phi^i \psi_j\psi_a\phi^a\right)
        \nn \\ &
        -12 \tr\left(f \psi_j\phi^i \phi^a \psi_a\right)
        +\frac{17}{2}\tr\left(f \psi_a\phi^i \phi^a \psi_j\right)
        -\frac{43}{2} \tr\left(f \psi_a\phi^a \phi^i \psi_j\right)
        \nn \\ &
        +\frac13 \epsilon^{a_1a_2a_3}\tr\left( \phi^{i} \psi_{j} \right) \tr\left( \psi_{a_1}\psi_{a_2}\psi_{a_3} \right)
        -2\epsilon^{a_1a_2i}\tr\left( \phi^{b} \psi_{a_1} \right) \tr\left( \psi_{b}\psi_{j}\psi_{a_2} \right)
        \nn \\ &
        - 10\epsilon^{a_1a_2a_3} \tr\left( \phi^i \psi_j \psi_{a_1}\psi_{a_2}\psi_{a_3}\right)
        + 8\epsilon^{a_1a_2a_3} \tr\left( \phi^i  \psi_{a_1} \psi_j \psi_{a_2}\psi_{a_3}\right)
        \nn\\ &
        -2\epsilon^{a_1a_2a_3} \tr\left( \phi^i  \psi_{a_1}  \psi_{a_2}\psi_j\psi_{a_3}\right) \ , \nn \\
        (r_{18}^{(1,2)})_{ijk} =&
        -\epsilon_{a_1a_2(i} \bigg[
        \tr \left(f \phi^{a_1}\right) \tr\left( \phi^{a_2} \psi_j \psi_{k)}\right)
        -\frac{3}{2} \tr \left(f \psi_j \right) \tr \left(\psi_{k)} \phi^{a_1} \phi^{a_2}\right)
        \nn\\ & \hspace{3cm}
        +3  \tr \left(f \phi^{a_1} \psi_j \phi^{a_2} \psi_{k)} \right)
        -3 \tr \left(f \psi_j \phi^{a_1}  \psi_{k)} \phi^{a_2} \right) \bigg] \nn \\
        &-\frac{1}{2} \tr \left(\phi^a \psi_a\right) \tr \left(\psi_{(i}\psi_j\psi_{k)}\right)
        +\frac{3}{2} \tr \left(\phi^a \psi_{(i|}\right) \tr \left(\psi_{a}\psi_{|j}\psi_{k)}\right)
        \nn \\ &
        + \frac{1}{2} \tr \left(\phi^a \psi_{(i} \psi_{j|}\right) \tr \left(\psi_a \psi_{|k)}\right)
        +\frac{3}{2} \tr \left(\phi^a \psi_{(i|}\psi_a\psi_{|j}\psi_{k)}\right)
        - \frac{3}{2} \tr \left(\phi^a \psi_{(i}\psi_{j|}\psi_a\psi_{|k)}\right)\ ,
        \nn\\
        (r_{20}^{(2,1)})_{ij} =& -\epsilon_{a_1a_2(i} \bigg[
        \tr\left( ff \right) \tr \left( \phi^{a_1} \phi^{a_2} \psi_{j)} \right)
        +\frac{1}{2}\tr\left( f \psi_{j)}\right) \tr \left( f\phi^{a_1} \phi^{a_2}  \right)
        \nn\\ & \hspace{3cm}
        +2\tr\left( f \phi^{a_1} \right) \tr \left( f[\phi^{a_2}, \psi_{j)} ]\right)\bigg]
        \nn\\ &
        +\epsilon_{a_1a_2(i} \left[ 4 \tr\left( ff  \phi^{a_1} \phi^{a_2} \psi_{j)} \right)
        -\tr\left( f  \phi^{a_1} \phi^{a_2}f \psi_{j)} \right) \right]
        +2\tr\left( f \phi^a \psi_{(i} \right) \tr \left( \psi_{j)} \psi_a  \right)
        \nn\\ &
        -4\tr\left( f  \psi_{(i} \phi^a \right) \tr \left( \psi_{j)} \psi_a  \right)
        -\frac{1}{2} \tr\left(f \psi_{(i}\right) \left( \phi^a \psi_{j)} \psi_a \right)
        - \frac{5}{2}  \tr\left(f \psi_{(i|}\right) \left( \phi^a \psi_{a} \psi_{|j)} \right) 
        \nn\\ &
        + 2\tr\left(f \psi_a\right) \left( \phi^a \psi_{(i} \psi_{j)} \right)
        -4\tr\left(f \psi_{(i|}\psi_a\right) \left( \phi^a  \psi_{|j)} \right)
        +2\tr\left( f \phi^a \psi_{(i} [\psi_{j)}, \psi_a]  \right)
        \nn\\ &
        +4\tr\left( f \phi^a \psi_a  \psi_{(i} \psi_{j)} \right)
        +4 \tr\left( f \psi_{(i} \phi^a  \psi_{j)} \psi_a  \right)
        -3\tr\left( f \psi_{(i|} \phi^a  \psi_a \psi_{|j)}   \right)
        \nn\\ &
        -2 \tr\left( f \psi_{a} \phi^a  \psi_{(i} \psi_{j)}  \right) 
        - \tr\left( f \psi_{(i|}   \psi_a \phi^a \psi_{|j)}  \right)
        +4 \tr\left( f    \psi_a \psi_{(i} \phi^a \psi_{j)}  \right)
        \nn \\&
        +\frac{2}{5} \epsilon^{a_1a_2a_3}\left[2 \tr\left(\psi_{a_1} \psi_{a_2}\right)\tr\left(\psi_{a_3}\psi_{(i}\psi_{j)}\right) -3\tr\left(\psi_{(i|}\psi_{a_1} \psi_{|j)}\psi_{a_2}\psi_{a_3}\right)\right]\ .
    \end{align}
}

Finally, we construct relations of these trace relations.
Consider a linear combination of the trace relations with
coefficients being the graviton cohomologies.
If it vanishes identically, we call it a relation of relations.
While the trace relations are identities that can be seen at the
level of `gluons' $\phi,\psi,f$,
the relations of relations are the identities of mesons
$u_2, u_3, v_2, v_3, w_2, w_3$.
We do not need to know how $u_2, u_3, v_2, v_3, w_2, w_3$ are
made of $\phi,\psi,f$ to obtain the relations of relations. 
After constructing relations of relations,
one can write them as the $Q$-action on certain operators
using \eqref{tr-r}, \eqref{tr-r-f}.
They are the $Q$-closed operators since their $Q$-actions vanish
due to the relations of relations.
This is the way we obtain the $Q$-closed operators in section \ref{sec:Qclosed}.
They can be either $Q$-exact or not and there is no trivial way to judge it easily. 
If they are not $Q$-exact, they are the non-graviton cohomologies
since they are made of the linear combinations of $r_I$'s,
which vanish with diagonal $\phi,\psi,f$.
For the check of the (non-)$Q$-exactness, refer to section \ref{sec:cohoexactness}.

Now we will construct relations of relations at the threshold level $\cJ=24$
which are singlets under $SU(3) \subset SU(4)_R$,
from the trace relations \eqref{tr-rel}, \eqref{tr-rel-f}. 
There are 5 choices of $(R,J)$ in this sector in which relations of relations exist. 

\paragraph{i) $(R,J) = (2,2)$.}
Let us first enumerate all $SU(3) \subset SU(4)_R$ singlets in this sector
made by the product of the trace relations in \eqref{tr-rel}, \eqref{tr-rel-f}
and the graviton cohomologies. There are following 6 singlets:
\begin{equation}
    \begin{aligned}
         &s_1^{(2,0)} = u^{ij}  (R_{20}^{(2,0)})_{ij}\ , \ \ 
         s_2^{(2,0)} = w^{ij}  (R_{14}^{(1,0)})_{ij}\ , \ \ 
         s_3^{(2,0)}= w^i  (R_{16}^{(1,0)})_i \ , \\
        &s_1^{(1,2)} = {v^{jk}}_i  (R_{16}^{(1,1)})^i_{jk}\ , \ \ 
        s_2^{(1,2)}= {v^j}_i   (R_{18}^{(1,1)})^i_j\ , \ \
        s_3^{(1,2)}= w^i  (R_{16}^{(0,2)})_i\ .
    \end{aligned}
\end{equation}
The superscripts denote ($n_f, n_\psi$) of the terms with maximal $n_f$
in the operator, as before.
There is one relation of these relations given by
\begin{equation}
    i \, QO^{(2,1)} \equiv 65s_1^{(2,0)} -39s_2^{(2,0)} +5s_3^{(2,0)}
    -312s_1^{(1,2)} -26s_2^{(1,2)} +6s_3^{(1,2)} = 0\ .
\end{equation}
This is the $Q$-action on the $Q$-closed operator \eqref{O1}.

\paragraph{ii) $(R,J) = (\frac{5}{2},\frac{3}{2})$.}
There exist 12 $SU(3)$ singlets in this sector given by
\begin{equation}
    \begin{aligned}
        &s_1^{(1,1)} = u^{a(i} {v^{j)}}_{a}  (R_{14}^{(1,0)})_{ij} \ , \ \
        s_2^{(1,1)}= \epsilon_{a_1a_2(i}u^{a_1k} {v^{a_2}}_{j)}  (R_{14}^{(1,0)})^{ij}_k\ , \ \
        s_3^{(1,1)} = {v^{jk}}_{i}  (R_{16}^{(1,0)})^i_{jk}\ , \\
        &s_4^{(1,1)} = u^{ijk}  (R_{18}^{(1,1)})_{ijk}\ , \ \
        s_5^{(1,1)} = {v^{(j}}_{i} w^{k)}  (R_{10}^{(0,0)})^i_{jk}\ , \ \
        s_6^{(1,1)} = u^{(ij}w^{k)}  (R_{12}^{(0,1)})_{ijk}\ , \\
        &s_7^{(1,1)} = \epsilon_{a_1a_2i} u^{a_1j} w^{a_2}  (R_{12}^{(0,1)})^i_j\ , \ \
        s_8^{(1,1)} = w^{ij}  (R_{14}^{(0,1)})_{ij}\ , \\
        &s_{1}^{(0,3)} = \epsilon^{a_1a_2(i} {v^{j}}_{a_1} {v^{k)}}_{a_2} (R_{12}^{(0,1)})_{ijk}\ , 
        \ \
        s_{2}^{(0,3)} = {v^j}_a {v^a}_i  (R_{12}^{(0,1)})^i_j\ , \\
        &s_{3}^{(0,3)} = u^{(jk} {v^{k)}}_i  (R_{14}^{(0,2)})^i_{jkl}\ , \ \
        s_{4}^{(0,3)} = {v^{jk}}_i  (R_{16}^{(0,2)})^i_{jk}\ .
\end{aligned}
\end{equation}
There are 4 relations of these relations, given by
\begin{equation}
    \begin{aligned}
        &i \, Q O_1^{(1,2)} \equiv 3s_5^{(1,1)} -3s_6^{(1,1)} +s_7^{(1,1)}= 0\ , \\
        &i \, Q O_2^{(1,2)} \equiv 9s_1^{(1,1)} -10s_2^{(1,1)} - 30 s_5^{(1,1)}-60s_{3}^{(0,3)}= 0\ , \\
        &i \, Q O_3^{(1,2)} \equiv 3s_1^{(1,1)} -6s_2^{(1,1)}+4s_4^{(1,1)} -14s_5^{(1,1)}-6s_8^{(1,1)} -12s_1^{(0,3)} -4s_{2}^{(0,3)}= 0\ , \\
        &i \, Q O_4^{(1,2)} \equiv 3s_1^{(1,1)} -14s_2^{(1,1)}-8s_3^{(1,1)} -42s_5^{(1,1)} +12 s_6^{(1,1)} -24 s_8^{(1,1)} -36s_1^{(0,3)} + 8 s_4^{(0,3)}= 0\ .
    \end{aligned}
\end{equation}
They are the $Q$-actions on \eqref{O2}.

\paragraph{iii) $(R,J) = (3,1)$.}
There exist 16 $SU(3)$ singlets in this sector given by
{\allowdisplaybreaks
    \begin{align}
        &s_1^{(1,0)} = \epsilon_{a_1a_2i}\epsilon_{b_1b_2j} u^{a_1b_1}u^{a_2b_2k}   (R_{14}^{(1,0)})^{ij}_k\ ,\nn\\
        &s_2^{(1,0)} = \epsilon_{a_1a_2 i } u^{a_1 (j} w^{k) a_2} (R_{10}^{(0,0)})^i_{jk}\ , \nn\\
        &s_3^{(1,0)}  = \epsilon_{a_1a_2 i } u^{a_1 jk} w^{a_2}  (R_{10}^{(0,0)})^i_{jk}\ , \nn\\
        &s_1^{(0,2)}  = {v^a}_i {v^{jk}}_a  (R_{10}^{(0,0)})^i_{jk} \ ,\nn\\
        &s_2^{(0,2)}  = {v^{(j}}_a {v^{k)a}}_i  (R_{10}^{(0,0)})^i_{jk}\ ,\nn\\
        &s_3^{(0,2)}  = u^{a(i}{v^{jk)}}_a   (R_{12}^{(0,1)})_{ijk}\ ,\nn\\
        &s_4^{(0,2)}  = u^{a(ij}{v^{k)}}_a   (R_{12}^{(0,1)})_{ijk} \ ,\nn\\
        &s_5^{(0,2)}  = \epsilon_{a_1a_2 i} u^{a_1 b}{v^{a_2 j}}_{b}   (R_{12}^{(0,1)})^i_j \ ,\nn\\
        &s_6^{(0,2)}  = \epsilon_{a_1a_2 i} u^{a_1 bj}{v^{a_2}}_{b}  (R_{12}^{(0,1)})^i_j \ ,\nn\\
        &s_7^{(0,2)}  = \epsilon_{a_1a_2(i} u^{a_1 (k}{v^{l)a_2}}_{j)} (R_{12}^{(0,1)})^{ij}_{kl} 
        \ ,\nn\\
        &s_8^{(0,2)}  = \epsilon_{a_1a_2(i} u^{a_1 kl}{v^{a_2}}_{j)}  (R_{12}^{(0,1)})^{ij}_{kl} 
        \ ,\nn\\
        &s_9^{(0,2)}  = \epsilon_{a_1 a_2 i} u^{a_1 (j}u^{kl) a_2}  (R_{14}^{(0,2)})^{i}_{jkl} \ ,\nn\\
        &s_{10}^{(0,2)}  = \epsilon_{a_1 a_2 i} u^{a_1 b} {v^{a_2}}_b  (R_{14}^{(0,1)})^i\ ,\nn\\
        &s_{11}^{(0,2)}  = u^{a(i} {v^{j)}}_{a}   (R_{14}^{(0,1)})_{ij} \ ,\nn\\
        &s_{12}^{(0,2)}  = \epsilon_{a_1a_2(i} u^{a_1 k} {v^{a_2}}_{j)} (R_{14}^{(0,1)})^{ij}_k \ ,\nn\\
        &s_{13}^{(0,2)}  = u^{(jk}{v^{l)}}_i  (R_{14}^{(0,1)})^i_{jkl} \ .
    \end{align}
}
There are 13 relations of these relations, given by
{\allowdisplaybreaks
    \begin{align}
    &i \, Q O_1^{(1,1)} \equiv s_2^{(1,0)} =0\ , \nonumber\\
    &i \, Q O_2^{(1,1)} \equiv s_3^{(1,0)} =0\ , \nonumber\\
    &i \, Q O_3^{(1,1)} \equiv s_1^{(1,0)} +5s_1^{(0,2)} -2s_2^{(0,2)} =0\ , \nonumber\\
    &i\, QO_1^{(0,3)} \equiv4s_5^{(0,2)}+3s_6^{(0,2)} = (R_{12}^{(0,1)})^i_j  
    (R_{12}^{(0,1)})^j_i = i\, Q\left[\frac{1}{2} i\, Q ((r_{12}^{(0,2)})^i_j (r_{12}^{(0,2)})^j_i) \right] = 0\ , \nonumber\\
    &i\, QO_2^{(0,3)} \equiv  s_7^{(0,2)}+s_8^{(0,2)}= (R_{12}^{(0,1)})^{ij}_{kl} 
    (R_{12}^{(0,1)})^{kl}_{ij}= i\, Q\left[\frac{1}{2} i\, Q ((r_{12}^{(0,2)})^{ij}_{kl} (r_{12}^{(0,2)})^{kl}_{ij}) \right]= 0\ , \nonumber\\
    &i\, QO_3^{(0,3)} \equiv s_3^{(0,2)} = 0\ , \nonumber\\
    &i\, QO_4^{(0,3)} \equiv s_{10}^{(0,2)}= 0\ , \nonumber\\
    &i\, QO_5^{(0,3)} \equiv 6s_1^{(0,2)}-6s_4^{(0,2)}-s_6^{(0,2)}= 0\ ,\nonumber\\
    &i\, QO_6^{(0,3)} \equiv 24 s_2^{(0,2)} -6 s_{11}^{(0,2)} + s_{12}^{(0,2)}= 0\ ,\nonumber\\
    &i\, QO_7^{(0,3)} \equiv s_1^{(0,2)} -10s_2^{(0,2)} -6s_4^{(0,2)} -10 s_8^{(0,2)} = 0\ ,\nonumber\\
    &i\, QO_8^{(0,3)} \equiv  5s_1^{(0,2)} -2s_2^{(0,2)} -9s_4^{(0,2)} +6s_9^{(0,2)}= 0\ ,\nonumber\\
    &i\, QO_9^{(0,3)} \equiv 6s_1^{(0,2)} +12s_2^{(0,2)} -18s_4^{(0,2)} +s_{12}^{(0,2)}= 0\ ,\nonumber\\
    &i\, QO_{10}^{(0,3)} \equiv 38s_1^{(0,2)} +4s_2^{(0,2)} -24s_4^{(0,2)} -5s_{13}^{(0,2)}  = 0\ .
\end{align}
}
They are the $Q$-actions on \eqref{O3}. 
Here, $O^{(0,3)}_1$ and $O^{(0,3)}_2$ are explicitly shown to be $Q$-exact.

\paragraph{iv) $(R,J) = (\frac{7}{2},\frac{1}{2})$.}
There exist 8 $SU(3)$ singlets in this sector given by
\begin{equation}
    \begin{aligned}
        &s^{(0,1)}_1  =  \epsilon_{a_1a_2i} u^{a_1 b} u^{jk} {v^{a_2}}_{b}   
        (R_{10}^{(0,0)})^i_{jk}\ , \ \
        s^{(0,1)}_2  = \epsilon_{a_1a_2 i}u^{a_1 b} u^{a_2 (j} {v^{k)}}_{b} 
        (R_{10}^{(0,0)})^i_{jk}\ , \\
        &s^{(0,1)}_3  = \epsilon_{a_1a_2i} u^{a_1b(j} {v^{k)a_2}}_{b} 
        (R_{10}^{(0,0)})^i_{jk}\ , \ \
        s^{(0,1)}_4  = \epsilon_{a_1a_2(i} u^{a_1(k} {v^{l)a_2}}_{j)}  
        (R_{12}^{(0,0)})^{ij}_{kl} \ , \\
        &s^{(0,1)}_5  = \epsilon_{a_1a_2(i} u^{a_1kl} {v^{a_2}}_{j)} 
        (R_{12}^{(0,0)})^{ij}_{kl}\ , \ \
        s^{(0,1)}_6  = \epsilon_{a_1a_2(i} \epsilon_{j)b_1b_2} u^{a_1b_1} u^{a_2b_2} u^{kl} 
        (R_{12}^{(0,1)})^{ij}_{kl} \ , \\
        &s^{(0,1)}_7  = \epsilon_{a_1a_2(i} \epsilon_{j)b_1b_2} u^{a_1b_1} u^{a_2b_2k}   (R_{14}^{(0,1)})^{ij}_k \ , \ \
        s^{(0,1)}_8  = \epsilon_{a_1a_2i} u^{a_1(j} u^{kl)a_2}  (R_{14}^{(0,1)})^i_{jkl} \ .
    \end{aligned}
\end{equation}
There are 6 relations of these relations, given by
\begin{equation}
    \begin{aligned}
        &i\, QO^{(0,2)}_1 \equiv s_1^{(0,1)}-2s_2^{(0,1)} = 0\ , \\
        &i\, QO^{(0,2)}_2 \equiv 6s_3^{(0,1)} + s_4^{(0,1)} = 0\ , \\
        &i\, QO^{(0,2)}_3 \equiv s_1^{(0,1)} + s_5^{(0,1)} = 0\ , \\
        &i\, QO^{(0,2)}_4 \equiv s_1^{(0,1)} + s_6^{(0,1)} = 0\ , \\
        &i\, QO^{(0,2)}_5 \equiv 4s_1^{(0,1)} + 24s_3^{(0,1)} - s_7^{(0,1)} = 0\ , \\
        &i\, QO^{(0,2)}_6 \equiv s_1^{(0,1)} - 12s_3^{(0,1)} + 3s_8^{(0,1)} = 0\ .
    \end{aligned}
\end{equation}
They are the $Q$-actions on \eqref{O4}.

\paragraph{v) $(R,J) = (4,0)$} There exist 4 $SU(3)$ singlets in this sector given by
\begin{equation}
    \begin{aligned}
        &s^{(0,0)}_1  = \epsilon_{a_1a_2a_3} \epsilon_{b_1b_2 i} u^{a_1b_1} u^{a_2b_2} u^{a_3jk} 
        (R_{10}^{(0,0)})^i_{jk}\ , \\
        &s^{(0,0)}_2  = R_{12}^{(0,0)}   R_{12}^{(0,0)}\ , \\
        &s^{(0,0)}_3  = \epsilon_{a_1a_2(i} \epsilon_{j)b_1b_2} u^{a_1b_1} u^{a_2b_2} u^{kl} 
        (R_{12}^{(0,0)})^{ij}_{kl}\ ,\\
        &s^{(0,0)}_4  = \epsilon_{a_1 a_2 (i} \epsilon_{j) b_1 b_2}  u^{a_1 b_1 (k} u^{l) a_2 b_2} 
        (R_{12}^{(0,0)})^{ij}_{kl}\ .
    \end{aligned}
\end{equation}
There is 1 relation of these relations, given by
\begin{equation}
    i \, QO^{(0,1)} \equiv 36s_1^{(0,0)} +5s_2^{(0,0)} -6s_3^{(0,0)} = 0\ .
\end{equation}
This is the $Q$-action on \eqref{O5}. 
\clearpage
\renewcommand{\addtocontents}[2]{\oldaddtoc{#1}{#2}}


\backmatter

\addcontentsline{toc}{chapter}{Bibliography} 
\begin{singlespace} 
	\bibliographystyle{JHEP}
	\bibliography{dissertation}   
\end{singlespace}

\end{document}